\documentclass[aps,prb,superscriptaddress]{revtex4}
\usepackage{epsfig}
\usepackage{graphicx}
\usepackage{amsfonts}
\usepackage{amsmath}
\usepackage{amssymb}
\usepackage{latexsym}
\usepackage{bm}
\usepackage{pstricks}

\def\8{\infty}

\def\oh{\frac{1}{2}}

\def\undertext#1{\vtop{\hbox{#1}\kern 1pt \hrule}}

\def\VEV#1{\left\langle\,#1\,\right\rangle}

\def\pp#1{\frac{\partial}{\partial#1}}

\def\be{\begin{equation}}
\def\ee{\end{equation}}
\def\bea{\begin{eqnarray} & &}
\def\eea{\end{eqnarray}}

\def\rf#1{(\ref{#1})}

\def\rf#1{(\ref{#1})}

\begin{document}


\title{Plasma Analogy and Non-Abelian Statistics for Ising-type Quantum Hall States}


\author {Parsa Bonderson}
\affiliation{Microsoft Station Q, Elings Hall, University of California
at Santa Barbara, Santa Barbara CA 93106, USA}
\author{Victor Gurarie}
\affiliation{Department of Physics, CB390, University of Colorado,
Boulder CO 80309, USA}
\author {Chetan Nayak}
\affiliation{Microsoft Station Q, Elings Hall, University of California
at Santa Barbara, Santa Barbara CA 93106, USA}
\affiliation{Deparment of Physics, University of California at
Santa Barbara, Santa Barbara CA 93106, USA}


\date{\today}

\begin{abstract}
We study the non-Abelian statistics of quasiparticles
in the Ising-type quantum Hall states which are likely candidates to explain the observed Hall conductivity plateaus in the second Landau level, most notably the one at filling fraction $\nu=5/2$. We complete the program started in Nucl. Phys. B {\bf 506}, 685 (1997)
and show that the degenerate four-quasihole and six-quasihole
wavefunctions of the Moore-Read Pfaffian state
are orthogonal with equal constant norms in the basis given by conformal blocks
in a $c=1+\frac{1}{2}$ conformal field theory.
As a consequence, this proves that the non-Abelian statistics of
the excitations in this state are given by the explicit analytic continuation of these
wavefunctions. Our proof is based on
a plasma analogy derived from the Coulomb gas construction of
Ising model correlation functions involving both order
and (at most two) disorder operators.
We show how this computation also determines the
non-Abelian statistics of collections of more than six
quasiholes and give an explicit expression for the corresponding
conformal block-derived wavefunctions for an arbitrary
number of quasiholes.
Our method also applies to the anti-Pfaffian wavefunction
and to Bonderson-Slingerland hierarchy states constructed over the Moore-Read and anti-Pfaffian states.
\end{abstract}


\maketitle

\section{Introduction}

Non-Abelian braiding statistics~\cite{Leinaas77,Goldin85,Fredenhagen89,Imbo89,Froehlich90,Imbo90,Bais92} is currently the subject of
intense study, partly because the experimental observation
of a non-Abelian anyon would be a remarkable milestone
in fundamental science and partly because of its potential application
to topologically fault-tolerant quantum information processing~\cite{Kitaev2003,Freedman98,Preskill98,Freedman02a,Freedman02b,Freedman03a,Preskill04,Nayak2008,Bonderson08a}. At present, the state
which is the best candidate to support quasiparticles
with non-Abelian braiding statistics is the experimentally-observed
$\nu=5/2$ fractional quantum Hall
state~\cite{Willett87,Pan99b,Eisenstein02,Xia04,Choi08}.
Efforts to observe
non-Abelian anyons in this state~\cite{Fradkin98,DasSarma05,Stern06a,Bonderson06a,Dolev08,Radu08,Willett09a,Bishara09a} and harness them
for quantum computation~\cite{DasSarma05,Bravyi06,Freedman06a,Bonderson08b,Bonderson10a} are predicated
entirely on two assumptions: (i) The observed state
is in the same universality class as either the
Moore-Read (MR) Pfaffian state~\cite{Read1991}
or the anti-Pfaffian state~\cite{Lee07,Levin07}, an assumption
which is supported by numerical studies~\cite{Morf98,Rezayi00,Feiguin08,Peterson08}.
(There is another non-Abelian candidate,
the so-called SU(2)$_2$ NAF state~\cite{Wen91a}, for this plateau,
but it is not supported by numerics.)
(ii) Quasiparticle excitations above these
ground states are non-Abelian anyons.
In order for this assumption to hold, it is necessary
for there to be a degenerate set of $n$-quasiparticle
states {\it and} for quasiparticle braiding to transform these
states into each other in such a way that different braiding
transformations do not commute.

Moore and Read~\cite{Read1991} conjectured that the MR Pfaffian state is non-Abelian
while Greiter, Wen, and Wilczek~\cite{Greiter1992} argued that
it is Abelian. It was subsequently shown by Nayak and Wilczek
~\cite{Nayak1996} and by Read and Rezayi~\cite{Read96}
that there is a $2^{\lfloor\frac{n}{2}\rfloor-1}$-fold degenerate
set of $n$ quasiparticle states.
To show that assumption (ii) is correct, it is further necessary
to show that these degenerate states are transformed into
each other by non-commuting transformations enacted by
quasiparticle braiding. Several different arguments~\cite{Nayak1996,Gurarie1997,Read2000,Ivanov2001,Tserkovnyak03,Stern04,Stone06,Seidel08,Read2008,Baraban09,Prodan09}
strongly support this hypothesis,
but a proof has been missing until now.
By ``proof,'' we mean an argument that relies
on no unproven assumptions beyond the existence of
an excitation gap and the existence of a screening phase
for particular classical two-dimensional ($2$D) plasmas at a particular temperature
and, therefore, is at the same level of rigor as
the Berry's phase calculation for quasiparticles in
the $\nu=1/M$ Laughlin states~\cite{Arovas84}. In this paper,
we supply such a proof by mapping matrix elements
of the MR Pfaffian state to the partition function
of a classical multi-component $2$D plasma, possibly
with magnetic charges. Our derivation extends and
completes a partial result obtained in Ref.~\onlinecite{Gurarie1997}.
Numerical studies provide very strong evidence that the plasmas corresponding to the $\nu=1/M$ Laughlin states with $M \lesssim 70$ are in the screening phase~\cite{Caillol82}. Similar numerical evidence confirming that the plasma (described in our paper) corresponding to the $\nu=1/2$ MR state is in the screening phase has recently also been obtained~\cite{Herland-unpublished}.

One approach to the calculation of the braiding statistics
of quasiparticles in fractional quantum
Hall states is based on an idea due to Moore and Read~\cite{Read1991}. These authors proposed to use the conformal blocks
of conformal field theories~\cite{Belavin1984,DiFrancesco97} (CFTs) as trial wavefunctions for
fractional quantum Hall effect states. The conformal blocks
are the holomorphic parts of correlation functions.
Unlike correlation functions, conformal blocks are not
single-valued. The conformal blocks which are used
as trial wavefunctions for fractional quantum Hall effect states
are single-valued in electron coordinates but are
not single-valued in the coordinates of the quasiparticles,
and it was conjectured that the
properties of the conformal blocks under analytic continuation of the quasiparticle coordinates
define their non-Abelian statistics.

However, as emphasized by Blok and Wen~\cite{Blok92},
the analytic continuation properties of wavefunctions are only part of the story.
An additional contribution to the statistics is given by the Berry's
matrix~\cite{Berry84,Simon83,Wilczek1984,Arovas84,Blok92}.
Wavefunction analytic continuation
only gives the quasiparticle statistics correctly if the conformal
blocks, as electron wavefunctions, have matrix elements
which are independent of the quasiparticle coordinates (when they are well-separated).
This includes, but is not limited to, the diagonal matrix elements,
which are the wavefunctions' norms. When this condition is satisfied,
the Berry's matrix is trivial, apart from a term which accounts
for the Aharonov-Bohm phase due to the (charged) quasiparticles'
motion in the magnetic field. This is because the wavefunctions
are holomorphic in the quasihole coordinates, except for the Gaussian factors
(which give rise to the resulting Aharonov-Bohm terms).

The effective field theory of a fractional quantum Hall state
is expected to be a Chern-Simons theory. Chern-Simons theories
are related to conformal field theories \cite{Witten89}: the Hilbert
space of a Chern-Simons theory with fixed non-dynamical charges
at points ${\eta_1}, \ldots,{\eta_n}$
is equal to the vector space of conformal blocks in
an associated CFT with primary fields at ${\eta_1}, \ldots,{\eta_n}$.
Thus, if the multi-quasiparticle wavefunctions of a
fractional quantum Hall state can be identified with
the conformal blocks of a CFT, it is very natural to conclude that
this fractional quantum Hall state is in the universality class of
the associated Chern-Simons theory. In fact, one can hardly imagine
any other possibility. However, this identification is only correct
if the braiding properties of the multi-quasiparticle wavefunctions
are equal to those of the Chern-Simons theory. This, in turn,
requires the Berry matrix (in the basis given by the conformal blocks)
to be trivial.

Thus, the logic may be summarized as follows~\cite{Nayak1996,Read2008}.
Let us suppose that the quasiparticles of some quantum
Hall state have the special property that when
$n$ quasiparticles are present at arbitrary
fixed positions ${\eta_1},\ldots,\eta_{n}$,
there is a $q$-dimensional space ${\cal V}_{n}$
of degenerate states of the system.
Now let us suppose that $\Psi_\alpha(\eta_\mu;z_i)$ with
$\alpha= 0,1,\ldots,q-1$ are the $q$ conformal blocks
of a correlation function in a CFT, where $z_1,\ldots,z_N$ are coordinates of the $N$ electrons. (We choose the
CFT and the operators in the conformal block so that
they are single-valued in the $z_i$, but possibly
multi-valued in the $\eta_\mu$.) If the
$\Psi_\alpha(\eta_\mu;z_i)$ form a basis for
${\cal V}_{n}$,
then we wish to show that the overlap integrals
\be
\label{eqn:intro-Gs}
G_{\alpha,\beta}({\bar \eta_\mu},{\eta_\mu})\equiv \int \prod_{k=1}^{N} d^2 z_k \, \bar \Psi_\alpha({\bar \eta_\mu}; {\bar z_i})   \Psi_\beta({\eta_\mu};{z_i}).
\ee
are proportional to $\delta_{\alpha \beta}$ diagonal and independent of the quasiparticle positions $\eta_\mu$,
in the limit where the $\eta_\mu$ are far apart from each other. If we can show this,
then the braiding properties of the $n$ quasiparticles
are determined by the analytic continuation properties of the wavefuctions
$\Psi_\alpha(\eta_\mu; z_i)$.

There is significant previous literature which
addresses this problem by analytic or numerical methods
\cite{Nayak1996,Gurarie1997,Tserkovnyak03,Seidel08,Read2008,Baraban09,Prodan09}.
In Section~\ref{sec:Discussion}, we will discuss these previous results and clarify their relation to the result of this paper.

In this paper, we prove, for the MR Pfaffian state, that the overlap integrals of Eq.~\rf{eqn:intro-Gs} are diagonal
and independent of the quasiparticle positions $\eta_\mu$,
in the limit in which the $\eta_\mu$ are far apart. Specifically, we show that
\be
\label{eqn:matrix-elements-constant}
G_{\alpha,\beta}({\bar \eta_\mu},{\eta_\mu})= C \delta_{\alpha \beta} + O \left( e^{- \left| \eta_\mu - \eta_\nu \right|/\ell} \right)
,
\ee
which allows us to define orthonormal states $\left|\Psi_{\alpha} \left( \eta_\mu \right) \right\rangle$ by dividing by the common normalization constant
\be
\left\langle z_i |\Psi_{\alpha} \left( \eta_\mu \right) \right\rangle \equiv G_{\alpha,\alpha}^{-1/2}({\bar \eta_\mu},{\eta_\mu}) \Psi_{\alpha} ({\eta_\mu};{z_i})
.
\ee
We obtain Eq.~\rf{eqn:matrix-elements-constant} by expressing the desired matrix elements
in the form of the partition function of a classical plasma and relying on the screening property of a plasma,
thereby extending Laughlin's plasma analogy~\cite{Laughlin1983} arguments
to these non-Abelian states. Our derivation
completes the program started in Ref.~\onlinecite{Gurarie1997}, where
such a plasma representation was used to prove that the diagonal sum
of norms in Eq.~\rf{eqn:intro-Gs}, $\sum_{\alpha} G_{\alpha,\alpha}$ is a constant independent of the quasiparticle positions
(so long as they are well-separated). The methods used there
did not, however, allow one to prove that their norms
are independently constant and equal,
nor that off-diagonal matrix elements $G_{\alpha,\beta}$
are zero. We accomplish this by extending and elaborating on the methods proposed in Ref.~\onlinecite{Gurarie1997}. One of the important steps in our approach is the explicit construction, via the Coulomb
gas formalism~\cite{Dotsenko1984,Felder1989,Mathur1992}, of
Ising model correlation functions including both order and disorder operators, shown in Eq.~\rf{eq:disorderco}. This equation is one of the significant results of our paper and is interesting in its own right in the context of the Ising CFT.

Although we can directly calculate the Berry's matrix only
for the two-, four-, and six-quasiparticle wavefunctions in this way,
our results determine the braiding
properties of arbitrary numbers of quasiparticles.
We show that the enumeration of multi-quasihole
states~\cite{Nayak1996} [which can be done without
computing the integrals in Eq.~\rf{eqn:intro-Gs}]
allows us to compute the braiding statistics of an
arbitrary number of quasiparticles, given a mild assumption
of locality. This derivation uses special properties
of the MR Pfaffian state and works in a particular basis (the ``qubit basis''), but does not need any
further assumptions beyond the existence of a gap
in the energy spectrum.

We can also utilize similar locality assumptions in the form of the more refined formalism of anyon models, which describes a topological phase with a braided tensor category. For this, the topological structure is specified by the number of topologically-distinct quasiparticle species,
their fusion algebra, the $F$-symbols (which encode associativity of fusion), and the $R$-symbols (which encode braiding).
As we discuss, the $F$- and $R$-symbols can be
determined merely from the two- and four-quasihole wavefunctions.
Thus, the underlying structure of
a topological phase allows us to bootstrap from the four-quasiparticle
case to an arbitrary number of quasiparticles.
In contrast to the previous derivation in the qubit basis, this derivation economizes on the necessary input, i.e. not requiring six-quasiparticle wavefunctions, because it allows (in fact incorporates) changes of basis, in the form of the $F$-symbol transformations.

The results of our paper also apply to the anti-Pfaffian wavefunction,
constructed as the particle-hole conjugate of the MR Pfaffian
state~\cite{Lee07,Levin07}. They similarly apply to
the Bonderson-Slingerland (BS) hierarchical states~\cite{Bonderson08} constructed over these, which provide candidates for all the (other) observed quantum Hall plateaus in the second Landau level. In particular, this includes BS candidate states for $\nu=12/5$, for which there is also some numerical evidence~\cite{Bonderson09a}.

The methods we develop here should also be generalizable to other quantum Hall states, most importantly to the Read-Rezayi (RR) series of parafermion states~\cite{Read1999}. Doing this in practice requires a careful development of the Coulomb gas construction for these states, which has not yet been accomplished, and overcoming additional obstacles that do not exist for the Ising-type states analyzed in this paper. This will remain the subject of future work.

This paper is organized as follows. In Section~\ref{sec:Berry}, we review the derivation of the Berry's matrix for adiabatic processes involving degenerate states. In Section~\ref{sec:two}, we discuss adiabatic transport of quasiparticles in the MR state, and describe the problem to be solved. In Section~\ref{sec:three}, we discuss Laughlin's plasma analogy arguments. In Section~\ref{sec:four}, we review the Coulomb gas construction of the Ising CFT, following Ref.~\onlinecite{Felder1989}. In Section~\ref{sec:five}, we reproduce the result of Ref.~\onlinecite{Gurarie1997} on the sum of the norms of multi-quasiparticle wavefunctions.
In Section~\ref{sec:six}, we extend this Coulomb gas representation
to arbitrary matrix elements of the four-quasihole and six-quasihole
wavefunctions, thus proving that they are orthogonal with equal norms. In Section~\ref{sec:many-quasiparticles}, we show
how these results determine the non-Abelian statistics
for an arbitrary number of quasiparticles.
In Section~\ref{sec:qporthog}, we
use the plasma analogy to show that two wavefunctions (with quasiparticles) are orthogonal
if they do not have matching types of quasiparticles at the same coordinates.
In Section~\ref{sec:anti-Pfaffian}, we use the previous results to determine the statistics of
quasiparticles in the anti-Pfaffian state and BS states. In Section~\ref{sec:Others}, we briefly discuss the application of the methods we have developed to other candidate states based on other CFTs. Finally, in Section~\ref{sec:Discussion}, we discuss previous works that have made progress toward establishing the braiding statistics of the MR state.
In Appendix~\ref{sec:boson-Coulomb-gas}, we specify the normalization
conventions that we use for free bosons. In Appendix~\ref{sec:Mathur}, we review Mathur's procedure~\cite{Mathur1992} for relating products of contour integrals to 2D integrals in the Coulomb gas representation of CFTs. This relation plays a crucial role in our analysis.
In Appendix~\ref{sec:2-order-disorder},
we use the Coulomb gas representation to compute the (multi-valued) correlation function of two order and
two disorder operators in the Ising model.
In Appendix~\ref{sec:two-comp-screening}, we discuss the behavior
of electric and magnetic operators in the plasma phase
of a two-component Coulomb gas. In Appendix~\ref{sec:Debye}, we review and generalize the Debye-H\"{u}ckel theory for application to the plasmas that arise in this paper. In
Appendix~\ref{sec:many-qh-wavefunctions}, we compute
the $2^{\frac{n}{2}-1}$ conformal blocks of $n$ $\sigma$ fields (where $n$ is even)
and an arbitrary number $N$ of $\psi$ fields in the Ising model;
this gives a preferred basis for the $q=2^{\frac{n}{2}-1}$ degenerate
states of $n$ quasiholes in the MR Pfaffian state. The Berry's
matrix is trivial in this basis and braiding properties are
given explicitly by the analytic continuation properties of these wavefunctions.
In Appendix~\ref{sec:direct}, we give an incomplete argument that would allow one to compute the braiding statistics for an arbitrary number of quasiparticles directly from the wavefunctions with arbitrary numbers of quasiparticles.
Although, as we show in Section~\ref{sec:many-quasiparticles}, this
is not necessary, it would nevertheless be a particularly simple and elegant
route to deriving quasiparticle statistics, if it could be completed. In Appendix~\ref{sec:orthog_examples}, we provide two explicit examples demonstrating the orthogonality of wavefunctions that do not have matching quasiparticle types at the same positions.

\section{Berry's Matrix}
\label{sec:Berry}

In this section, we review the derivation of Berry's matrix~\cite{Berry84,Simon83,Wilczek1984} for an adiabatic process when there are energy degeneracies. We consider the Hamiltonian $\hat{H} \left( R_1(t) , \ldots, R_n (t) \right) $, which depends on a set of parameters $R_\mu (t)$ that are varied in time $t$. All states evolve according to the Schr\"{o}dinger equation
\begin{equation}
\label{eq:Schroedinger}
i \hbar \frac{d}{dt} \left| \Psi(t) \right\rangle = \hat{H} \left( t \right) \left| \Psi(t) \right\rangle
.
\end{equation}
One can define orthonormal energy eigenstates $\left| \alpha (R_1 , \ldots , R_n) \right\rangle$ for the Hamiltonian at particular values of the parameters $R_\mu$, such that
\begin{equation}
\hat{H} \left( R_\mu \right) \left| \alpha (R_\mu) \right\rangle = E_\alpha \left( R_\mu \right) \left| \alpha (R_\mu) \right\rangle
\end{equation}
and $\left< \alpha (R_\mu) \right| \left. \beta (R_\mu) \right\rangle = \delta_{\alpha \beta}$. When the parameters $R_\mu$ are varied with $t$, we will leave the $R_\mu$ dependence of quantities implicit, e.g. writing $\hat{H} \left(t \right)$ and $\left| \alpha (t) \right\rangle$. We consider a Hamiltonian such that the Hilbert space splits into subspaces of degenerate energies $\mathcal{H}(t) = \sum_{E(t)} \mathcal{H}_{E(t)}$. We now focus on one of these subspaces $\mathcal{H}_{E_0 (t)}$ (e.g. the subspace of ground-states), and assume that the energy gap between it and the other subspaces does not close during the adiabatic process. The adiabatic theorem tells us that if we start at $t=0$ with a basis state $\left| \psi_\alpha (0) \right\rangle =\left| \alpha (0)) \right\rangle \in \mathcal{H}_{E_0 (0)}$, then the time evolved state $\left| \psi_\alpha (t) \right\rangle$ will be in the $\mathcal{H}_{E_0 (t)}$ subspace, and can thus be written in the form
\begin{equation}
\label{eqn:U_0-def}
\left| \psi_\alpha (t) \right\rangle = e^{-\frac{i}{\hbar} \int_{0}^{t} E_0 (t') dt'} U_0(t) \left| \alpha (t) \right\rangle
\end{equation}
where $U_0$ is the Berry's matrix, which is a generalization of Berry's phase. It is a unitary transformation in the $E_0$ subspace, i.e. $U_0 (t) : \mathcal{H}_{E_0 (t)} \rightarrow \mathcal{H}_{E_0 (t)}$, such that $U_0(0) = \openone$, and the dynamical phase $\exp\left[-\frac{i}{\hbar} \int_{0}^{t} E_0 (t') dt'\right]$ has been separated from the Berry's matrix term. Since it is a matrix, the Berry's matrix can potentially exhibit non-Abelian behavior.
Taking the time-derivative of Eq. \ref{eqn:U_0-def} and taking an inner product with another time-evolved state in $\mathcal{H}_{E_0 (t)}$, we have:
\begin{equation}
\label{eq:Berry_1}
i \hbar  \left\langle \psi_\alpha (t) \right| \frac{d}{dt} \left| \psi_\beta (t) \right\rangle = E_0 (t) \left\langle \alpha (t) \right| \left. \beta (t) \right\rangle + i \hbar \left\langle \alpha (t) \right| U^{-1}_{0}(t) \frac{d U_0 (t)}{dt} \left| \beta (t) \right\rangle + i \hbar \left\langle \alpha (t) \right| \frac{d}{dt} \left| \beta (t) \right\rangle
\end{equation}
Re-writing the left-hand-side by using Eq.~\rf{eq:Schroedinger}, one finds
\begin{equation}
\label{eq:Berry_2}
i \hbar  \left\langle \psi_\alpha (t) \right| \frac{d}{dt} \left| \psi_\beta (t) \right\rangle = \left\langle \psi_\alpha (t) \right| \hat{H} \left( t \right) \left| \psi_\beta (t) \right\rangle = E_0 (t) \left\langle \alpha (t) \right| \left. \beta (t) \right\rangle
.
\end{equation}
Combining Eqs.~\rf{eq:Berry_1} and \rf{eq:Berry_2}, we obtain
\begin{equation}
\left\langle \alpha (t) \right| U^{-1}_{0}(t) \frac{d U_0 (t)}{dt} \left| \beta (t) \right\rangle =-  \left\langle \alpha (t) \right| \frac{d}{dt} \left| \beta (t) \right\rangle
.
\end{equation}
Solving this expression for $U_0$, one finds
\begin{eqnarray}
U_0(t) &=& \mathcal{P} \exp \left[ i \int_{0}^{t} \mathcal{A} (t') dt' \right] \notag \\
&=& \openone + \sum_{n=1}^{\infty} i^n \int_{0}^{t} dt_n \int_{0}^{t_n} dt_{n-1} \ldots \int_{0}^{t_2} dt_{1} \mathcal{A} (t_1) \ldots \mathcal{A} (t_n)
\end{eqnarray}
where $\mathcal{P}$ stands for path-ordering (putting operators to the right of those with smaller $t$ and to the left of those with larger $t$), and we have defined the Berry's connection for the $\mathcal{H}_{E_0 (t)}$ subspace
\begin{eqnarray}
\mathcal{A}_{\alpha,\beta} (t) &\equiv& i \left\langle \alpha (t) \right| \frac{d}{dt} \left| \beta (t) \right\rangle = \sum_{\mu=1}^{n} \mathcal{A}_{\alpha,\beta}^{R_\mu} (t) \frac{d R_\mu(t) }{dt} \\
\mathcal{A}_{\alpha,\beta}^{R_\mu} (t) &\equiv& i\left\langle \alpha \left(R_1 , \ldots , R_n \right) \right| \frac{\partial}{\partial R_\mu} \left| \beta \left(R_1 , \ldots , R_n\right) \right\rangle
.
\end{eqnarray}
Defined this way, $\mathcal{A}$ is Hermitian.

The term $U_0(t)$ only has a gauge-invariant meaning if the Hilbert space is the same as the original one. For this, one must make a closed circuit in configuration space. Let us consider an adiabatic process with $t$ running from $0$ to $t_f$
(where $t_f$ is large enough compared to the inverse of the energy gap that the process is adiabatic),
where $\mathcal{H}_{E_0 ({t_f})} = \mathcal{H}_{E_0 (0)}$ and the path in configuration space is a closed loop, which includes processes that exchange identical (quasi-)particles. Even though
$\mathcal{H}_{E_0 ({t_f})} = \mathcal{H}_{E_0 (0)}$,
it is possible to have $\left| \alpha ({t_f}) \right\rangle \neq \left| \alpha (0) \right\rangle$, e.g. if we have defined $\left| \alpha \left(R_1 , \ldots , R_n \right) \right\rangle$ which is multi-valued as a function of the $R_\mu$.
However, they must be related through a transformation $B: \mathcal{H}_{E_0 (0)} \rightarrow \mathcal{H}_{E_0 (0)}$, defined by
\begin{equation}
B_{\alpha,\beta} \equiv \left\langle \alpha (0)  \right|\left. \beta ({t_f}) \right\rangle
\end{equation}
so that $\left| \alpha ({t_f}) \right\rangle = B \left| \alpha (0) \right\rangle$.
For such an adiabatic process, we can now write the time-evolved state at $t=t_f$ in terms of operators acting on the initial state $\left| \psi_\alpha (0) \right\rangle = \left| \alpha (0) \right\rangle \in \mathcal{H}_{E_0 (0)}$
\begin{equation}
\left| \psi_\alpha ({t_f}) \right\rangle = e^{-\frac{i}{\hbar} \int_{0}^{t_f} E_0 (t') dt'} \mathcal{P} \exp \left[ i \int_{0}^{t_f} \mathcal{A} (t') dt' \right]  B \left| \psi_\alpha (0) \right\rangle
,
\end{equation}
and thus, it may be applied to an arbitrary initial state $\left| \Psi (0) \right\rangle = \sum_{\alpha} c_\alpha \left| \psi_\alpha (0) \right\rangle = \sum_{\alpha} c_\alpha \left| \alpha (0) \right\rangle$ in the subspace $\mathcal{H}_{E_0 (0)}$
\begin{equation}
\left| \Psi (t_f) \right\rangle =
e^{-\frac{i}{\hbar} \int_{0}^{t_f} E_0 (t') dt'}
\mathcal{P} \exp \left[ i \int_{0}^{t_f} \mathcal{A} (t') dt' \right]  B \left| \Psi (0) \right\rangle
.
\end{equation}
If we never consider states outside the $\mathcal{H}_{E_0 (t)}$ subspace, we can obviously ignore the common dynamical phase.
Thus, we see that the evolution of the initial state in the $\mathcal{H}_{E_0 (0)}$ subspace under an adiabatic process is (apart from the common dynamical phase) composed of the Berry's matrix and the wavefunction transformation $B$.

\section{Quasihole wavefunctions and non-Abelian statistics}
\label{sec:two}

In this paper, we will be discussing a set of wavefunctions
and their braiding properties, i.e. the evolution under adiabatic exchange of quasiparticles in $2$D systems.
We will make little reference to
the Hamiltonian of the system, other than to assume that
the Hamiltonian has a gap above its ground state(s).
The wavefunctions which we discuss can be regarded as trial wavefunctions
for the Hamiltonian of electrons in a magnetic field
interacting through the Coulomb interaction. Alternatively,
they can be viewed as exact eigenstates of
electrons in the lowest Landau level at filling fraction
$\nu=1/M$ interacting through
a special model Hamiltonian with three-body interactions,
\begin{equation}
\label{eqn:three-body-generic}
H = {H_3^M}.
\end{equation}
For the case of bosons at $\nu=1$,
the Hamiltonian has the form
\be
\label{eqn:three-body-Ham-bosonic}
{H_3^1} = \lambda \sum_{i<j<k}^{N} \delta^{2}(z_i-z_j) \delta^{2}(z_i-z_k).
\ee
where $\lambda>0$.
For the case of fermions at $\nu=1/2$,
Fermi statistics dictates a more complicated form \cite{Greiter1992,Rezayi00}:
\be
\label{eqn:three-body-Ham}
{H_3^2} = \lambda \sum_{i<j<k}^{N} {\cal S}_{ijk}\{{\partial_i^4}{\partial_j^2}\}
\delta^2 (z_i-z_j) \delta^2 (z_j-z_k).
\ee
where ${\cal S}_{ijk}$ is a symmetrizer.
Our focus in this paper will be wavefunctions with an even number $n$
of quasiholes. For the model Hamiltonians in Eq.~\rf{eqn:three-body-generic}, the $n$
quasihole wavefunctions which we will discuss
are zero-energy eigenstates. (This is typical for
such ultra-local Hamiltonians; quasiparticles
cost finite energy, so there is a finite energy cost
for a quasiparticle-quasihole pair.) As we will
see, when we fix the positions ${\eta_1},\ldots,
\eta_{n}$ of these quasiholes,
we will still have a $2^{\frac{n}{2}-1}$-fold degenerate space of states spanned by
$\Psi_\alpha$, $\alpha=0,1,\ldots,2^{\frac{n}{2}-1}-1$.
For the sake of precision, let us momentarily assume that
the system is on a sphere of fixed area and that
the number of electrons is fixed (and that the magnetic field
is tuned to accommodate $n$ quasiparticles).
Then the only assumption that we will need about the spectrum of the
Hamiltonian of Eq.~\rf{eqn:three-body-generic} is that all other
states with quasiholes at ${\eta_1}$, $\ldots$, ${\eta_n}$
will be separated from $\text{span}(\Psi_\alpha)$
by a finite energy gap.

When we consider states with quasiholes, we will
need to augment this Hamiltonian with a potential
which pins the quasiholes at fixed positions:
\begin{equation}
\label{eqn:3-body+pinning}
H = {H_3^M} + H_{\rm Pinning}.
\end{equation}
This is necessary to guarantee that there is a gap
in the multi-quasihole case; otherwise, it would
cost no energy to move the quasiholes to other positions.
An elegant choice of pinning potentials is constructed
in Ref.~\onlinecite{Prodan09}. However, the Berry's
matrix is computed solely from a set of wavefunctions,
with no explicit reference to the Hamiltonian, apart from
the assumption that it provides a gap. Thus,
the pinning potential, though important as a matter
of principle, is not, as a practical matter, important in its details
for our calculation.

The MR Pfaffian ground state wavefunction for an even number $N$ of particles is given by~\cite{Read1991}:
\be
\label{eq:pf}
\Psi \left(z_1,\dots, z_{N} \right)  =
{\rm Pf}\left(\frac{1}{z_i-z_j}\right)\,\prod_{i<j}^{N}
\left( z_i-z_j \right)^M e^{  - \frac{1}{4} \sum\limits_{i=1}^{N}  \left| z_i \right|^2 }.
\ee
$M$ is a positive integer, taking odd values if the particles are bosons (which may occur, e.g. for neutral bosons in a rapidly rotating trap~\cite{Cooper01,Rezayi05}) and even values if they are fermions (e.g. electrons in the quantum Hall effect).
Throughout most of the paper, we set the magnetic length $\ell_B = \sqrt{\hbar c / e B}$ to $1$, as we have done in Eq.~\rf{eq:pf}, and will only reconstitute it when it provides necessary clarification.
The symbol Pf stands for Pfaffian:
\be
\label{eq:pfaffian}
{\rm Pf}\left(A_{i,j}\right) \equiv \frac{1}{N!!} \sum_{\sigma \in S_{N}} \text{sgn} (\sigma) \prod_{k=1}^{N/2} A_{ \sigma (2k-1), \sigma (2k)}
\ee
where $A$ is an antisymmetric $N \times N$ matrix (where $N$ is even).
The square of the Pfaffian of an antisymmetric matrix is equivalent to the determinant, i.e. $\left[ {\rm Pf}\left(A_{i,j}\right) \right]^2 = \det \left(A_{i,j}\right)$. This wavefunction has the same form as
the BCS wavefunction in real-space~\cite{Read1991,Greiter1992}
multiplied by a Laughlin-Jastrow factor.

The wavefunction in Eq.~\rf{eq:pf} is the unique exact ground state of the
Hamiltonian in Eq.~\rf{eqn:three-body-generic}.
The $M=2$ case is an approximate ground state
for electrons with Coulomb interactions at $\nu=5/2$
(assuming that the lowest Landau level of
both spins is filled and the wavefunction in Eq.~\rf{eq:pf} is transposed
from the lowest Landau level to the second Landau level)
\cite{Morf98,Rezayi00,Feiguin08}.
The $M=1$ case is an approximate ground state for
neutral ultra-cold bosons in a rotating trap~\cite{Cooper01,Rezayi05}.

The wavefunction in Eq.~\rf{eq:pf} can be written as a conformal block in a CFT, as was first proposed in Ref.~\onlinecite{Read1991}. The relevant CFT is a (restricted) product of two theories, one at central charge $c=1/2$ describing the Pfaffian part of the wavefunction, and the other at $c=1$ describing the Jastrow factor $\prod_{i<j} (z_i-z_j)^M$ of the wavefunction, as well as the Gaussian factor. Specifically, one
writes
\begin{eqnarray}
\label{eq:pfcft}
\Psi \left(z_1, \ldots, z_{N} \right) &=&
\left< \psi(z_1)  \dots \psi(z_{N})  \right> \,\times\,
\left< e^{i \sqrt{\frac M 2} \varphi(z_1)}  \dots e^{i \sqrt{\frac M 2} \varphi(z_{N})} e^{-i\frac{1}{2 \pi \sqrt{2M}} \int d^2z \, \varphi(z)} \right>\cr
&=& {\rm Pf}\left(\frac{1}{z_i-z_j}\right) \,\times\, \prod_{i<j}^{N}
\left( z_i-z_j \right)^M e^{  - \frac{1}{4} \sum\limits_{i=1}^{N} \left| z_i \right|^2 }.
\end{eqnarray}
Here $\psi$ represents the holomorphic free Majorana fermion (the operator with conformal dimension $h_{\psi}=1/2$) of the $c=1/2$ Ising CFT, and $\varphi$ is the free boson of a U(1) CFT. Various conventions can be used to describe the free boson. We adopt the one presented in Appendix~\ref{sec:boson-Coulomb-gas}, with Eq.~\rf{eq:phiholcorful} and $g=1/4$.

For future reference, let us note that the $c=1$ correlator is charge neutral, that is, it is invariant under the change $\varphi \rightarrow \varphi+$const. Indeed, under such a change, the exponential factor acquires a term
$ N\sqrt{M/2}  - A/(2 \pi \sqrt{2M})$,
where $A$ is the total area. However, $M = A/2\pi N$ is the inverse filling fraction of the quantum Hall state, since $A/2\pi$ is the total number of available states in a Landau level which we fill with $N$ particles, and so $ N\sqrt{M/2}  - A/(2 \pi \sqrt{2M})=0$.

An excited state wavefunction depends on the positions $z_i$ of
the electrons as well as the positions $\eta_\mu$ of the quasiparticles.
It is important to recognize that the quasiparticles' coordinates are simply parameters of the
electrons' wavefunction (and underlying Hamiltonian), not to be treated on the same footing as the electrons' coordinates.
These wavefunctions were constructed as eigenstates of Eq.~\rf{eqn:three-body-generic}
in Refs.~\onlinecite{Nayak1996,Read96}. Given that the ground state
can be expressed as a conformal block in the $c=\frac{1}{2}+1$
CFT, it is natural to try to construct wavefunctions with $n$ (fundamental) quasiholes
in the same CFT. The natural guess~\cite{Read1991}
is that they are given by:
\begin{multline}
\label{eq:holes}
\Psi_\alpha \left(\eta_1, \dots, \eta_{n}; z_1, \dots, z_{N} \right) =
\left< \sigma(\eta_1) \dots \sigma(\eta_{n}) \, \psi(z_1) \dots \psi(z_{N})  \right>_\alpha
\\
\times \left< e^{i \frac{1}{2 \sqrt{2M}} \varphi(\eta_1)}
\dots e^{i\frac{1}{2 \sqrt{2M}}\varphi(\eta_{n})}
e^{i \sqrt{\frac M 2} \varphi(z_1)} \dots e^{i \sqrt{\frac M 2} \varphi(z_{N})}
e^{- i \frac{1}{2 \pi \sqrt{2M}} \int d^2z \, \varphi(z)} \right>.
\end{multline}
Here $\sigma$ are the holomorphic spin operators of the Ising CFT, with conformal dimension $h_{\sigma}=1/16$. The bosonic part of the correlation function is chosen in such a way that the wavefunction is a polynomial function of the $z_i$.

Notice the index $\alpha$ in Eq.~\rf{eq:holes}.
The holomorphic spin operators of
the Ising CFT have many conformal blocks,
which we label by the index $\alpha$. In fact, it is well-known that the total number of conformal blocks is $2^{\frac{n}{2}-1}$, thus
\be \alpha = 0,1, \dots, 2^{\frac{n}{2}-1}-1.
\ee
The wavefunctions $\Psi_\alpha$ represent the set of degenerate wavefunctions at fixed positions of the quasiholes and form the basis for their non-Abelian statistics.

To find the wavefunctions of Eq.~\rf{eq:holes} explicitly, we need to evaluate
the appropriate conformal blocks of the CFT.
For $n=2$, there is only a single conformal block for Eq.~\rf{eq:holes};
evaluating it for $N$ even, we obtain the two-quasihole wavefunction:
\begin{equation}
\label{eqn:two-quasiholes}
\Psi(\eta_1, \eta_2; z_1, \dots, z_{N}) =
({\eta_1}-{\eta_2})^{\frac{1}{4M}-\frac{1}{8}}\,\,
{\rm Pf}\!\left(
  \frac{\left({\eta_1}-{z_i}\right)\left({\eta_2}-{z_j}\right)+ \left( i \leftrightarrow j \right)}{z_i - z_j}\right)
\prod_{i<j}^{N} (z_i - z_j)^M e^{-\frac{1}{8M}\left( |\eta_1|^2 + |\eta_2|^2 \right) - \frac{1}{4} \sum\limits_{i=1}^{N} |z_i|^2 }~.
\end{equation}
This wavefunction is, indeed, a zero-energy eigenstate
of the Hamiltonian in Eq.~\rf{eqn:three-body-generic}
(see Refs.~\onlinecite{Nayak1996,Read96} for details).
Since there is only a single generator for the two-particle
braid group, a counterclockwise exchange of the two particles,
non-Abelian effects cannot be seen -- they require at least
two different braids which do not commute with each other.
The effect of braiding can, therefore, only be a phase
which is acquired by the wavefunction in Eq.~\rf{eqn:two-quasiholes}.
This wavefunction is single-valued in electron coordinates,
as it must be, but is multi-valued in the quasihole coordinates.
Taking the analytic continuation of this wavefunction at face value,
we would conclude that the effect of a counterclockwise
exchange of two quasiholes in this state is a phase
$\exp \left[i \pi \left( \frac{1}{4M}-\frac{1}{8} \right)\right]$. However, this conclusion is premature,
because we must also take into account the Berry's matrix
(which, in this case, is simply a phase).

Before discussing the Berry's matrix, let us consider
the four-quasihole wavefunctions and, briefly, the general
$n$ quasihole wavefunctions (with $n$ even). In the four-quasihole
case, we are faced with evaluating
Eq.~\rf{eq:holes} for $n=4$. This calculation is more difficult, but
was accomplished in Ref.~\onlinecite{Nayak1996}.
For $N$ even, it results in the following two wavefunctions
\begin{eqnarray}
\label{eq:wave}
\Psi_0 \left(\eta_1, \eta_2, \eta_3, \eta_4; z_1,
\dots, z_{N} \right) &=& \prod_{ \mu < \nu }^{4} \eta_{\mu \nu}^{\frac {1}{4M} - \frac 1 8}
\frac{\left(\eta_{13} \eta_{24} \right)^{\frac 1 4}}{ \sqrt{1+\sqrt{1-x}}} \left(\Psi_{(13)(24)} +
\sqrt{1-x} \, \Psi_{(14)(23)} \right) e^{- \frac{1}{8M} \sum\limits_{\mu=1}^4 \left| \eta_\mu \right|^2}, \cr
\Psi_1 \left(\eta_1, \eta_2, \eta_3, \eta_4; z_1, \dots, z_{N} \right) &=&  \prod_{\mu < \nu }^{4} \eta_{\mu \nu}^{\frac {1}{4M} - \frac 1 8}  \frac{\left(\eta_{13} \eta_{24} \right)^{\frac 1 4}}{ \sqrt{1-\sqrt{1-x}}}  \left(\Psi_{(13)(24)}-
\sqrt{1-x}
\, \Psi_{(14)(23)} \right) e^{- \frac{1}{8M} \sum\limits_{\mu=1}^4 \left| \eta_\mu \right|^2}
\end{eqnarray}
where the so-called anharmonic ratio $x$, well-known in CFT, is given by
\be x = \frac{\eta_{12} \eta_{34}}{\eta_{13} \eta_{24}} \equiv
 \frac{
\left(\eta_1-\eta_2 \right)\left( \eta_3-\eta_4 \right)}{\left(\eta_1-\eta_3\right)\left(\eta_2-\eta_4 \right)}.
\ee
Here, we have introduced the notation $\eta_{\mu \nu }\equiv{\eta_\mu}-{\eta_\nu}$, and the shorthand $\Psi_{(ab)(cd)}$ for
\begin{equation}
\label{eqn:Psi_(13)(24)-def}
{\Psi_{(ab)(cd)}} =
{\rm  Pf}\!\left( \frac{({\eta_a} - z_i) ({\eta_b} - z_i  )({\eta_c}-z_j ) ({\eta_d}-z_j  ) + (i \leftrightarrow j )}
{z_i - z_j}\right)\,
\prod_{i<j}^{N} (z_i - z_j)^M \,
e^{  - \frac{1}{4} \sum\limits_{i=1}^{N} \left| z_i \right|^2}
\end{equation}

The wavefunctions $\Psi_{(13)(24)}$ and
$\Psi_{(14)(23)}$ are zero-energy eigenstates
of Eq.~\rf{eqn:three-body-generic} and they form a basis of the
two-dimensional space of states with four quasiholes at
fixed positions \cite{Nayak1996}. The state
$\Psi_{(12)(34)}$ is not linearly-independent of these
two because of the identity \cite{Nayak1996}:
\begin{equation}
\label{eqn:linear-identity}
{\Psi_{(12)(34)}}-{\Psi_{(13)(24)}}=
(1-x)\left({\Psi_{(12)(34)}}-{\Psi_{(14)(23)}}\right).
\end{equation}
Even though $\Psi_{(13)(24)}$ and
$\Psi_{(14)(23)}$ form a basis of the four quasihole
Hilbert space, they do not provide an orthonormal basis.
In this paper, we demonstrate that
the linear combinations $\Psi_0$ and $\Psi_1$ defined in
Eq.~\rf{eq:wave} are, in fact, an orthogonal basis. Moreover, we show that $\Psi_0$ and $\Psi_1$ have the same norms (though we do not compute the precise value of their overall normalization constant), and thus can provide an orthonormal basis by dividing by a common normalization constant.

It has been argued since Ref.~\onlinecite{Read1991} that using the wavefunctions in Eq.~\rf{eq:wave} allows us to read off the non-Abelian statistics of the quasiparticles in a straightforward manner. Indeed, if the quasihole at $\eta_1$ is exchanged with the quasihole at $\eta_2$ in a counterclockwise fashion (or, equivalently, if the quasiholes at $\eta_3$ and $\eta_4$ undergo counterclockwise exchange),
a straightforward analytic continuation of the wavefunctions
leads to the transformation rules:
\be \label{eq:op1} \eta_1 \leftrightarrows \eta_2 \ {\rm or} \ \eta_3 \leftrightarrows \eta_4 : \Psi_0 \mapsto e^{i \pi \left(\frac{1}{4M} - \frac{1}{8} \right)} \Psi_0, \ \Psi_1 \mapsto e^{i \pi \left(\frac{1}{4M} - \frac{1}{8} \right)} i \Psi_1.
\ee
To see this, we note that $1-x = \eta_{14} \eta_{23} / \eta_{13} \eta_{24} \mapsto \frac{1}{1-x}$ and $\Psi_{(13)(24)} \leftrightarrow \Psi_{(14)(23)}$ under this exchange. We see that the phase $\exp \left[i \pi \left(\frac{1}{4M} - \frac{1}{8} \right)\right]$ acquired by $\Psi_0$ is the same as that acquired from counterclockwise exchange of the two quasiholes in the $n=2$, $N$ even case.

On the other hand, if the quasiparticles at $\eta_2$ and $\eta_3$ undergo counterclockwise exchange (or if the ones at $\eta_1$ and $\eta_4$ are exchanged), then we get
\be
\label{eq:op2}
\eta_2 \leftrightarrows \eta_3  \ {\rm or} \ \eta_1 \leftrightarrows \eta_4: \Psi_0 \mapsto e^{i \pi \left(\frac{1}{4M} + \frac{1}{8} \right)} \frac{\Psi_0 -i \Psi_1}{\sqrt{2}} , \ \Psi_1 \mapsto e^{i \pi \left(\frac{1}{4M} + \frac{1}{8} \right)} \frac{-i \Psi_0 +  \Psi_1}{\sqrt{2}} .
\ee
Finally, if the quasiparticles at $\eta_1$ and $\eta_3$ undergo counterclockwise exchange (or if the ones at $\eta_2$ and $\eta_4$ are exchanged), then we get
\be \label{eq:op3} \eta_1 \leftrightarrows \eta_3 \ {\rm or} \ \eta_2 \leftrightarrows \eta_4: \Psi_0 \mapsto e^{i \pi \left(\frac{1}{4M} + \frac{1}{8} \right)} \frac{ \Psi_0 + \Psi_1}{\sqrt{2}}, \ \Psi_1 \mapsto e^{i \pi \left(\frac{1}{4M} + \frac{1}{8} \right)} \frac{- \Psi_0 + \Psi_1}{\sqrt{2}}.
\ee
These exchange transformations are more difficult to show, but can be checked using algebraic manipulations as in Ref.~\onlinecite{Nayak1996}.
These three exchange operations Eqs.~\rf{eq:op1}, \rf{eq:op2}, and \rf{eq:op3} constitute the building blocks of the non-Abelian statistics of states with four quasiholes.

The explicit form of the conformal block
wavefunctions for $n>4$ was not previously calculated.
In Appendix \ref{sec:many-qh-wavefunctions},
we show that they have the following form:
\begin{eqnarray}
\Psi^{}_{({p^{}_1},{p^{}_2},\ldots,{p^{}_{n/2}})} &=&
\left(\frac{\prod\limits_{i<j}^{n/2} \eta^{}_{2i-1,2j-1}\, \eta^{}_{2i,2j}}
{\prod\limits_{i,j}^{n/2} \eta^{}_{2i-1,2j}}\right)^{\!\frac{1}{8}}
\left\{
\sum_{{r^{}_i}=0,1} (-1)^{r\cdot p}
\prod_{k<l}^{n/2} x^{|{r^{}_k}-{r^{}_l}|/2}_{k,l}
\right\}^{-1/2}
\notag \\
&& \qquad \times \left\{
\sum_{{r^{}_i}=0,1} (-1)^{r\cdot p}
\prod_{k<l}^{n/2} x^{|{r^{}_k}-{r^{}_l}|/2}_{k,l}\,\,
\Psi_{(1+{r^{}_1},3+{r^{}_2},\ldots)(2-{r^{}_1},4-{r^{}_2},\ldots)}
\right\} \prod_{\mu<\nu}^{n} \eta_{\mu \nu}^{\frac{1}{4M} } e^{- \frac{1}{8M} \sum\limits_{\mu=1}^{n} \left| \eta_\mu \right|^2 }
\label{eqn:2n-qp-wvfn-preview}
\end{eqnarray}
The indices take the values $p_i=0,1$, with the constraint that $\sum_{i=1}^{n/2} p_{i}$ is even,
so there are $2^{\frac{n}{2}-1}$ such wavefunctions. (If we were to consider the case where the number of electrons $N$ was odd instead of even, then we would instead require $\sum_{i=1}^{n/2} p_{i}$ to be odd.)
Here, $x_{k,l} \equiv \frac{ \eta^{}_{2k-1,2l} \, \eta^{}_{2l-1,2k}}{\eta^{}_{2k-1,2l-1} \, \eta^{}_{2k,2l}}$ and $\Psi_{(1+{r^{}_1},3+{r^{}_2},\ldots)
(2-{r^{}_1},4-{r^{}_2},\ldots)}$ is
a generalization of the notation in Eq.~\rf{eqn:Psi_(13)(24)-def}
which was introduced in Ref.~\onlinecite{Nayak1996} and is
explained in Appendix~\ref{sec:many-qh-wavefunctions}.
As discussed there, the wavefunctions of Eq.~\rf{eqn:2n-qp-wvfn-preview}
form a basis of the $2^{\frac{n}{2}-1}$-dimensional space of
zero-energy $n$ quasihole eigenstates of the
Hamiltonian in Eq.~\rf{eqn:three-body-generic}.
For the special case $n=4$, Eq.~\rf{eqn:2n-qp-wvfn-preview}
is identical to Eq.~\rf{eq:wave}.
The analytic continuation properties of wavefunctions with an arbitrary number
of quasiholes can be read off from Eq.~\rf{eqn:2n-qp-wvfn-preview}.

However, calculating the explicit analytic continuation of the wavefunctions is not,
in principle, sufficient to establish the statistics
of quasiholes. One also needs to calculate the Berry's connection. It is defined as
\begin{eqnarray}
\label{eq:Berry}
\mathcal{A}_{\alpha, \beta} (t) &=& \sum_{\mu = 1}^{n} \left( \mathcal{A}^{\eta_\mu}_{\alpha, \beta} \frac{d\eta_{\mu}}{dt} + \mathcal{A}^{\bar{\eta}_\mu}_{\alpha, \beta} \frac{d\bar{\eta}_{\mu}}{dt} \right) \\
\label{eq:Berry_eta}
\mathcal{A}^{\eta_\mu}_{\alpha, \beta} &=& i \int \prod\limits_{k=1}^{N} d^2 z_k \, \frac{\bar \Psi_\alpha}{ G_{\alpha, \alpha}^{1/2} } \frac{\partial}{\partial \eta_\mu} \left( \frac{ \Psi_\beta}{ G_{\beta, \beta}^{1/2}} \right) \\
\label{eq:Berry_etabar}
\mathcal{A}^{\bar{\eta}_\mu}_{\alpha, \beta} &=& i \int \prod\limits_{k=1}^{N} d^2 z_k \, \frac{\bar \Psi_\alpha}{ G_{\alpha, \alpha}^{1/2} } \frac{\partial}{\partial \bar{\eta}_\mu} \left( \frac{ \Psi_\beta}{ G_{\beta, \beta}^{1/2}} \right)
,
\end{eqnarray}
where the overlap matrix is defined by:
\be
\label{eq:overlap}
G_{\alpha,\beta}(\bar{\eta}_\mu , \eta_\mu) \equiv \int \prod_{k=1}^{N} d^2 z_k \, \bar \Psi_\alpha(\bar \eta_\mu; \bar z_i)   \Psi_\beta(\eta_\mu;z_i).
\ee
We have allowed for the wavefunctions in Eqs.~\rf{eq:Berry_eta} and \rf{eq:Berry_etabar} to be un-normalized, since we will not determine the overall normalization constant of the wavefunctions we work with in this paper.
When the quasiparticles are adiabatically transported along the coordinate paths $\eta_{\mu}(t)$, forming a closed circuit in parameter space as $t$ goes from $0$ to $t_f$, an arbitrary state $\Psi$ in the ($2^{\frac{n}{2}-1}$-dimensional) degenerate ground-state space is transformed under the following unitary evolution, combining the explicit transformation of the wavefunctions resulting from analytic continuation with the Berry's matrix transformation resulting from the Berry's connection (see Section~\ref{sec:Berry} for more details)
\begin{equation}
\left| \Psi [\eta_{\mu} (t_f) ] \right\rangle  = \mathcal{P} \exp \left[i \int_{0}^{t_f} \mathcal{A}(t) dt \right] \, B \left| \Psi [\eta_{\mu} (0) ] \right\rangle,
\end{equation}
where $\mathcal{P}$ stands for path-ordering, and $B$ is the unitary transformation describing the analytic continuation of orthonormal states
\begin{equation}
\left| \Psi_\alpha [\eta_{\mu} (t_f) ] \right\rangle = \sum_{\beta = 0}^{q-1} B_{\beta , \alpha} \left| \Psi_\beta [\eta_{\mu} (0) ] \right\rangle
.
\end{equation}
(We have dropped the overall dynamical phase, since it is the same for all states in the ground-state space.)
For example, the analytic continuation matrices corresponding to the exchanges in Eqs.~\rf{eq:op1}, \rf{eq:op2}, and \rf{eq:op3} (assuming the wavefunctions have equal norms) are, respectively, given by
\begin{equation}
\label{eq:4qhB}
B^{(1 \leftrightarrows 2)} = e^{i \pi \left( \frac{1}{4M} - \frac{1}{8} \right)} \left[
\begin{array}{cc}
1 & 0 \\
0 & i
\end{array}
\right], \quad
B^{(2 \leftrightarrows 3)} = e^{i \pi \left( \frac{1}{4M} + \frac{1}{8} \right)} \frac{1}{\sqrt{2}} \left[
\begin{array}{cc}
1 & -i \\
-i & 1
\end{array}
\right], \quad
B^{(1 \leftrightarrows 3)} = e^{i \pi \left( \frac{1}{4M} + \frac{1}{8} \right)} \frac{1}{\sqrt{2}} \left[
\begin{array}{cc}
1 & -1 \\
1 & 1
\end{array}
\right]
.
\end{equation}

We will show that the wavefunctions in Eq.~\rf{eq:wave} are
orthogonal for large separations $|{\eta_\mu}-{\eta_\nu}|\rightarrow\infty$, such that
\be
\label{eq:norm}
G_{\alpha, \beta} = C\, \delta_{\alpha \beta}
\,+ \, O\bigl(e^{-|{\eta_\mu}-{\eta_\nu}|/\ell}\bigr),
\ee
where $C$ and $\ell$ are $\eta$-independent constants.
This implies that the Berry's connection is zero, up to terms
that give the Abelian Aharonov-Bohm phase, as may be seen
from the following calculation:
\begin{eqnarray}
\label{eqn:Berry-vanishes-arg}
\mathcal{A}_{\alpha,\beta}^{\eta_\mu} &=& i \int \prod_{k=1}^{N} d^2 z_k \, \frac{\bar \Psi_\alpha}{ G_{\alpha, \alpha}^{1/2} } \frac{\partial}{\partial \eta_\mu} \left( \frac{ \Psi_\beta}{ G_{\beta, \beta}^{1/2}} \right) \cr
&=& i \pp{\eta_\mu} \left( \int  \prod_{k=1}^{N} d^2 z_k\, \frac{\bar \Psi_\alpha}{ G_{\alpha, \alpha}^{1/2} } \frac{ \Psi_\beta}{ G_{\beta, \beta}^{1/2}} \right)
-  \, i \int \prod_{k=1}^{N} d^2 z_k \, \frac{\partial}{\partial \eta_\mu} \left(\frac{\bar \Psi_\alpha}{ G_{\alpha, \alpha}^{1/2} }\right)  \frac{ \Psi_\beta}{ G_{\beta, \beta}^{1/2}} \cr
&=& i \pp{\eta_\mu} \left( G_{\alpha, \alpha}^{-1/2} G_{\beta, \beta}^{-1/2} G_{\alpha, \beta}  \right)
-  \, i G_{\alpha, \alpha}^{-1/2} G_{\beta, \beta}^{-1/2} \int \prod_{k=1}^{N} d^2 z_k \, \left(\frac{ -\bar{\eta}_\mu }{8M}\right) \bar \Psi_\alpha \Psi_\beta + \, O\bigl(e^{-|{\eta_\mu}-{\eta_\nu}|/\ell}\bigr) \cr
&=& i \frac{\bar{\eta}_\mu}{8M} \delta_{\alpha\beta}
 \,+ \, O\bigl(e^{-|{\eta_\mu}-{\eta_\nu}|/\ell}\bigr)
\end{eqnarray}
We have integrated by parts to go from the first line to the second.
Similarly, we have:
\begin{eqnarray}
\label{eqn:barBerry}
\mathcal{A}_{\alpha,\beta}^{\bar{\eta}_\mu} &=& i \int \prod_{k=1}^{N} d^2 z_k \, \frac{\bar \Psi_\alpha}{ G_{\alpha, \alpha}^{1/2} } \frac{\partial}{\partial \bar{\eta}_\mu} \left( \frac{ \Psi_\beta}{ G_{\beta, \beta}^{1/2}} \right) \cr
&=&  i G_{\alpha, \alpha}^{-1/2} G_{\beta, \beta}^{-1/2} \int \prod_{k=1}^{N} d^2 z_k \, \left(\frac{ -\eta_\mu }{8M}\right) \bar \Psi_\alpha \Psi_\beta + \, O\bigl(e^{-|{\eta_\mu}-{\eta_\nu}|/\ell}\bigr) \cr
&=& -i \frac{\eta_\mu}{8M} \delta_{\alpha\beta}
 \,+ \, O\bigl(e^{-|{\eta_\mu}-{\eta_\nu}|/\ell}\bigr)
\end{eqnarray}
In Eqs. \rf{eqn:Berry-vanishes-arg} and \rf{eqn:barBerry},
we have used the fact that the dependence of $\bar \Psi_{\alpha}$ on $\eta_\mu$ and $\Psi_{\alpha}$ on $\bar{\eta}_\mu$ comes only through the Gaussian factor $\exp(- \frac{1}{8M} \left| \eta_{\mu} \right|^2)$ ($\eta_\mu$ and $\bar \eta_\mu$ are considered independent of each other for these purposes). The resulting Berry's connection is
diagonal in the space of wavefunctions, giving rise to the Berry's matrix
\begin{equation}
\mathcal{P} \exp \left[i \int_{0}^{t_f} dt \mathcal{A}(t) \right] = \exp \left[- \frac{1}{8M} \sum_{\mu=1}^{n} \int_{0}^{t_f} dt \left( \bar{\eta}_\mu \frac{ d \eta_\mu}{dt} -  \eta_\mu \frac{ d \bar{\eta}_\mu}{dt} \right)  \right] \, \openone + \, O\bigl(e^{-|{\eta_\mu}-{\eta_\nu}|/\ell}\bigr),
\end{equation}
which is the same for all
wavefunctions in the degenerate subspace.
When the quasiparticle coordinates are taken around a closed loop (or the exchange paths of identical quasiparticles form a closed loop), this term
is equal to the phase $\exp \left(- i \frac{A}{2M} \right)$, which is proportional
to the total enclosed area $A$ encircled by the quasiparticles in the counter-clockwise sense (area encircled in the clockwise sense contributes negatively to $A$). This is unlike particle braiding statistics
which depends only on the enclosed particles and not
on the area. By reconstituting the the magnetic length $\ell_B = \sqrt{\hbar c / e B}$ (which we set equal to $1$) in this expression, we see that this phase is simply the Aharonov-Bohm phase
\begin{equation}
\exp \left[- \frac{1}{8M \ell_B^2} \sum_{\mu=1}^{n} \int_{0}^{t_f} dt \left( \bar{\eta}_\mu \frac{ d \eta_\mu}{dt} -  \eta_\mu \frac{ d \bar{\eta}_\mu}{dt} \right)  \right] = \exp \left(- i \frac{A}{2M \ell_B^2} \right) = \exp\left(-i \frac{e}{2M} \frac{BA}{\hbar c} \right) = \exp\left(i \frac{q \Phi}{\hbar c} \right)
\end{equation}
acquired by a charge $q=e/2M$ particle encircling a total flux $\Phi = -BA$ due to the background magnetic field $\overrightarrow{B} = -B \hat{z}$. (We use the convention where $z_j = x_j +i y_j$, which corresponds to holomorphic wavefunctions for electrons of charge $-e$ in a background magnetic field $\overrightarrow{B} = -B \hat{z}$.) This reconfirms the interpretation of the given wavefunctions as corresponding to charge $e/2M$ quasiholes.
As long as Eq.~\rf{eq:norm} is fulfilled, however, no other contributions arise in the Berry's matrix. In particular, it does not affect the non-Abelian statistics, which comes from the explicit analytic continuation of quasiparticle coordinates in the wavefunctions.

There are other length scales one should be aware of when considering non-Abelian quasiparticles. In general, topologically non-trivial excitations can tunnel between non-Abelian quasiparticles, which has the effect of splitting the degeneracy of their states~\cite{Bonderson09b}. Such tunneling is exponentially suppressed with separation distance, and thus introduces correlation length scales associated with the tunneling of topological excitations and determined by the (non-universal) microscopic physics of the system. As long as the quasiparticles are farther apart than these correlation lengths, the topological degeneracies are preserved (up to exponentially suppressed corrections), but otherwise the notion of the non-Abelian state space and braiding statistics transformations upon it breaks down. For $\sigma$ non-Abelian quasiparticles in Ising-type topologically ordered systems, the relevant correlation length is $\xi_\psi$, which corresponds to tunneling of the $\psi$ excitation, i.e. a Majorana fermion. For $p$-wave superconductors, $\xi_\psi$ is identified as the superconducting coherence length~\cite{Cheng09}. For the MR state, $\xi_\psi$ corresponds to tunneling of the neutral fermion ($\psi_0$ in the notation of Section~\ref{sec:general-consid}). Numerical studies~\cite{Baraban09,Bonderson10c} provide an estimate of $\xi_\psi \approx 0.8 \ell_B - 2.3 \ell_B$ for the $\nu=5/2$ MR state.

The wavefunctions in Eq.~\rf{eq:wave} were derived as correlators of some CFT. There is no reason {\sl a priori} to expect that they will form an orthogonal basis obeying Eq.~\rf{eq:norm} with respect to the inner product of nonrelativistic
electrons in a magnetic field. It is the goal of this paper to show
that this is indeed so.

\section{Laughlin's Plasma Argument}
\label{sec:three}

We proceed by first recalling an argument due to Laughlin~\cite{Laughlin-Lecture87} which he used to deduce the normalization of the
Laughlin wavefunction with $N$ electrons and $n$ quasiholes in the $\nu=1/M$ quantum Hall effect. Such a wavefunction has the form
\be
\label{eq:La}
\Psi_{\frac{1}{M}}(\eta_1, \ldots , \eta_n; z_1, \ldots, z_N) = \prod_{\mu < \nu}^{n} (\eta_\mu-\eta_\nu)^{\frac 1 M} \prod_{\mu=1}^n \prod_{i=1}^N (\eta_\mu-z_i) \prod_{i<j}^{N} (z_i-z_j)^M e^{- \frac{1}{4M} \sum\limits_{\mu=1}^{n} \left| \eta_\mu \right|^2 - \frac 1 4 \sum\limits_{i=1}^N \left| z_i \right|^2 }.
\ee
Note that the prefactor $\prod_{\mu < \nu} (\eta_\mu-\eta_\nu)^{\frac 1 M}$
depends only on the quasihole coordinates $\eta_{\mu}$ and is independent of
the electron coordinates $z_i$s. Therefore, it can be regarded as part of the normalization
of the wavefunction. By including it explicitly in the
definition of the wavefunction, we are anticipating that
it will result in a norm of the wavefunction that is independent of the quasihole positions.
Laughlin proved that
\be
\label{eqn:norm-integral-def}
\left\| \Psi_{\frac{1}{M}} (\eta_\mu; z_i) \right\|^2 \equiv \int \prod_{k=1}^N d^2 z_k \, \bar{\Psi}_{\frac{1}{M}} \Psi_{\frac{1}{M}} = C_1
 \,+ \, O\bigl(e^{-|{\eta_\mu}-{\eta_\nu}|/\ell_1}\bigr),
\ee
where $C_1$ and $\ell_1$ are constants independent of $\eta_\mu$. (We use the subscript $1$ here to indicate quantities that correspond to the one-component plasma and to differentiate them from similar quantities occurring elsewhere in the paper.)
The proof proceeds as follows. One observes that the normalization integral Eq.~\rf{eqn:norm-integral-def} can be rewritten as
\begin{eqnarray}
\label{eq:pl}
\left\| \Psi_{\frac{1}{M}} \right\|^2 &=& \int \prod_{k=1}^N d^2 z_k \, \bar{\Psi}_{\frac{1}{M}} \Psi_{\frac{1}{M}} = \int \prod_{k=1}^N d^2 z_k \, e^{-\Phi_1 /T} = e^{-F_1/T}\\
\Phi_1 &=& - \sum_{\mu < \nu}^n \frac{Q^2}{M^2} \log \left| \eta_\mu -\eta_\nu \right| - \sum_{\mu=1}^n \sum_{i=1}^N
\frac{Q^2}{M} \log \left| \eta_\mu - z_i \right| - \sum_{i<j}^N Q^2 \log \left| z_i - z_j \right| \notag \\
&& \qquad + \frac{Q^2}{4M^2} \sum_{\mu=1}^{n} \left| \eta_\mu \right|^2 + \frac{Q^2}{4M} \sum_{i=1}^N \left| z_i \right|^2
,
\end{eqnarray}
where $T = Q^2/2M$. We note that the $2$D Coulomb interaction between two charges $q_1$ and $q_2$ separated by a distance $R$ is $-q_1 q_2 \log R$. Thus, $\Phi_1$ can be interpreted as the $2$D Coulomb-interaction potential energy for $N$ charge $Q$ particles at $z_i$ and $n$ charge $Q/M$ particles at $\eta_\mu$, together with a uniform neutralizing background of charge density $\rho = -\frac{Q}{2 \pi M \ell_B^2}$ [which is the uniformly negatively charged disk, represented by the Gaussian terms in Eq.~\rf{eq:La}]~\footnote{The self-energy for the neutralizing background charge density are not included in the Coulomb potential energy $\Phi_1$. This can be done safely without altering any of the subsequent arguments, since it simply contribute a constant to this energy.}. Consequently, $F_1$ can be interpreted as the free energy of a classical $2$D one-component plasma at temperature $T$ of $N$ charge $Q$ particles in the presence of $n$ additional test particles of charge $Q/M$ at the fixed positions $\eta_\mu$ and a uniform neutralizing background. Clearly, one can ascribe different charge values $Q$ to the plasma particles, as long as one similarly alters the test charges and temperature in a compensating manner. One convenient choice is to take $T=g$ and $Q = \sqrt{2Mg}$. (Another typical choice is $T=M/2$ and $Q=M$, which gives the test particles unit charge.) In any case, the coupling constant $\Gamma = Q^{2}/T = 2M$ remains invariant under such redefinitions, and it is known from Monte Carlo simulations~\cite{Caillol82} that the freezing point of such a classical 2D plasma is at $\Gamma_{c_1} \approx 140$ (i.e. $T_{c_1} \approx Q^2 / 140$). Hence, the plasma is a screening fluid for $M \lesssim 70$, whereas it freezes into a crystal for $M \gtrsim 70$.

It is important to distinguish this $M_{c_1} \approx 70$ transition point between the fluid and crystal phases of the analogous $2$D one-component plasma from the $M_c \approx 9$ transition point between the quantum Hall fluid and Wigner-crystal phases of the physical electron systems~\cite{Pan02}. The later determines the physical range where quantum Hall states exist, while the former indicates that the plasma analogy argument indeed applies to the Laughlin wavefunctions for all the physically relevant filling fractions.

When the plasma screens, the free energy $F_1$ in Eq.~\rf{eq:pl} cannot depend on the positions
$\eta_{\mu}$ of the $Q/M$ test charges, so long as $\left| \eta_\mu - \eta_\nu \right| \gg \ell_1$, where $\ell_1$ is the Debye length of this plasma, since they are screened by the elementary charges. The Debye length can be estimated using Debye-H\"{u}ckel theory (see Appendix~\ref{sec:Debye}) to be $\ell_1 = \ell_B /\sqrt{2}$, where $\ell_B = \sqrt{ \hbar c / eB}$ is the magnetic length (which we have set to $1$) of the quantum Hall system. Thus, the overlap integral is indeed a constant, as long as the test charges are sufficiently far away from each other.

It follows that the Berry's connection for adiabatically transporting Laughlin quasiholes using the wavefunction as normalized in Eq.~\rf{eq:La} is given by
\begin{eqnarray}
\label{eqn:Berry-Laughlin}
\mathcal{A}^{\eta_{\mu}} &=& i \int \prod_{k=1}^{N} d^2 z_k \,  \frac{\bar\Psi}{ \left\| \Psi \right\|} \frac{\partial}{\partial \eta_{\mu}} \left( \frac{\Psi}{ \left\| \Psi \right\|} \right) \cr
&=& i \pp{\eta_\mu} \left( \left\| \Psi \right\|^{-2} \int  \prod_{k=1}^{N} d^2 z_k\, \bar \Psi \Psi \right) \,
-  \, i \int \prod_{k=1}^{N} d^2 z_k \, \frac{\partial}{\partial \eta_{\mu}} \left( \frac{\bar \Psi}{ \left\| \Psi \right\|} \right) \frac{\Psi}{\left\| \Psi \right\|} \cr
&=& i \frac{{\bar\eta}_\mu}{4M} + O\bigl(e^{-|{\eta_\mu}-{\eta_\nu}|/\ell_1}\bigr),
\end{eqnarray}
and
\begin{eqnarray}
\label{eqn:Berry-Laughlin2}
\mathcal{A}^{\bar{\eta}_{\mu}} &=& i \int \prod_{k=1}^{N} d^2 z_k \,  \frac{\bar\Psi}{ \left\| \Psi \right\|} \frac{\partial}{\partial \bar\eta_{\mu}} \left( \frac{\Psi}{ \left\| \Psi \right\|} \right) \cr
&=& -i \frac{{\eta}_\mu}{4M} + O\bigl(e^{-|{\eta_\mu}-{\eta_\nu}|/\ell_1}\bigr).
\end{eqnarray}
This gives a Berry's phase of
\begin{eqnarray}
\mathcal{P} \exp \left[i \int_{0}^{t_f} dt \mathcal{A}(t) \right] &=& \exp \left[- \frac{1}{4M} \sum_{\mu=1}^{n} \int_{0}^{t_f} dt \left( \bar{\eta}_\mu \frac{ d \eta_\mu}{dt} -  \eta_\mu \frac{ d \bar{\eta}_\mu}{dt} \right)  \right] + \, O\bigl(e^{-|{\eta_\mu}-{\eta_\nu}|/\ell_1}\bigr) \notag \\
&=& \exp \left( -i \frac{e}{M} \frac{BA}{\hbar c} \right) + \, O\bigl(e^{-|{\eta_\mu}-{\eta_\nu}|/\ell_1}\bigr),
\end{eqnarray}
where $A$ is the area encircled by the quasiholes in the counter-clockwise sense. This contributes only the Aharonov-Bohm phase $\exp(i q\Phi / \hbar c)$ acquired by charge $q=e/M$ encircling an area $A$ containing flux $\Phi = -BA$ from the background magnetic field $\overrightarrow{B} = -B \hat{z}$. The remaining contribution to the unitary evolution resulting from adiabatic transport of the quasiparticles comes from explicit analytic continuation of the wavefunction, which is thus the braiding statistics of the quasiparticles.
This proves that the Laughlin quasiholes are anyons that accumulate a
statistical phase $\theta = \pi/M$ as the positions of two of them are exchanged in a counterclockwise fashion, as can be explicitly seen from analytic continuation of the term $(\eta_\mu-\eta_\nu)^{\frac 1 M}$ in the wavefunction of Eq.~\rf{eq:La}.

\section{The Coulomb Gas Construction}
\label{sec:four}

\subsection{Intuitive Approach}

In the previous section, we saw that, although we could not
explicitly evaluate the norm of the Laughlin wavefunction,
we could make a strong statement about its dependence on
quasihole coordinate by appealing to the screening property
of a Coulomb plasma.
We would now like to construct such an argument to
prove Eq.~\rf{eq:norm}, but we must first note that, taken
at face value, the overlap integrals of $\Psi_\alpha$, defined in Eq.~\rf{eq:overlap}, have little to do with the partition function of a plasma. Indeed, the plasma argument seems to be custom tailored for wavefunctions which can be written as products of differences, such as Eq.~\rf{eq:La}. The MR Pfaffian ground state wavefunction Eq.~\rf{eq:pf}, or the wavefunctions with quasihole excitations, such as Eqs.~\rf{eq:holes}, \rf{eqn:two-quasiholes}, and \rf{eq:wave},
are in fact sums of products and cannot be written as exponentials of logarithms.
Nevertheless, there exists an approach,
called the Coulomb gas construction, which allows one
to represent conformal blocks in terms of a plasma.
Let us review this approach, in its particular application
for the $c=1/2$ Ising CFT of interest here.

In the next subsection, we will follow
the logic and notation of Feigin and Fuchs~\cite{Feigin1983},
Dotsenko and Fateev~\cite{Dotsenko1984, Dotsenko1985},
Felder~\cite{Felder1989}, and Mathur~\cite{Mathur1992},
which is essentially an algebraic approach to the
Coulomb gas construction of the minimal models~\cite{Belavin1984}. However,
there is another approach to the Coulomb gas construction
which is more intuitive; we briefly describe it here (see Refs.~\onlinecite{Nienhuis87,Cardy1994,Kondev96}).

The basic question which we answer in this subsection
is: Why does the Ising model, which does not
have a conserved U(1) charge, have anything
at all to do with a gas of electric and magnetic
charges interacting logarithmically?
One way of answering this question
lies in the following steps.

\begin{enumerate}

\item Write the Ising model partition function on the
honeycomb lattice (we choose this lattice for convenience)
in the form
\begin{equation}
\label{eqn:Ising-partition}
Z = \sum_{\{{\sigma_i}=\pm1 \}}e^{-\beta H} = \sum_{\{{\sigma_i}=\pm1 \}} \prod_{i,j} \cosh\beta J\left[
1 + \tanh\left(\beta J \right) {\sigma_i}{\sigma_j}
\right]
\end{equation}
where the Ising Hamiltonian has the form
$H=-J\sum_{<i,j>}{\sigma_i} {\sigma_j}$.

\item In any term in the expansion of the product
in Eq.~\rf{eqn:Ising-partition}, each lattice bond
can either receive a $1$ or an $x{\sigma_i}{\sigma_j}$,
where $x=\tanh\beta J$.
Notice that any term in the expansion vanishes
upon summation over $\{{\sigma_i}=\pm1 \}$
unless every spin $\sigma_i$ appears either zero
times or twice. Consequently, the bonds which
receive an $x{\sigma_i}{\sigma_j}$ form closed loops,
and the partition function takes the form:
\begin{equation}
  \label{eq:Ising-partition}
  Z(x) ={\sum_{\{\alpha\}}} {x}^{b(\alpha)}
  \end{equation}
where $\{\alpha\}$ is
a configuration of loops on the honeycomb lattice
and $b(\alpha)$ is the total length of all of the loops
in the configuration $\{\alpha\}$. The critical point of the Ising model
occurs at $x=1/\sqrt{3}$.

\item Observe that this partition function can be obtained
from the following {\it local rules} for the Boltzmann weights.
When a loop turns left, it acquires a factor $xe^{i\chi}$ and when
it turns right, it acquires a factor $xe^{-i\chi}$.
Since the number of left turns minus the number of
right turns is $\pm 6$ for any closed loop,
every such loop receives a factor
$2{x^b}\cos 6\chi$ after summing over both
orientations of the loop, where $b$ is the number
of bonds in the loop. We obtain the Ising
model partition function in Eq.~\rf{eq:Ising-partition}
provided $2\cos 6\chi=1$.

\item Write the critical partition function (as defined by these local weights) as
a height model on the honeycomb lattice. A height model
is a model of a fluctuating interface which is specified by its
local height $z=\phi(x,y)$. The loops are interpreted as
domain walls between regions with different heights
(the heights live on the plaquettes and the domain walls on the links).
In the continuum limit, the energetic penalty for domain walls
between different heights becomes a gradient
energy $(\nabla \phi)^2$ so that the partition function for
the interface can be viewed as a quantum field theory for
a scalar field $\phi$.

\item Write the height model as a free bosonic field $\phi$
(i.e. as a Coulomb gas) with stiffness $g=1-\frac{6\chi}{\pi}=4/3$
together with a coupling $(1-g)=-1/3$ to the scalar curvature $R$.
(Note that we had to take $-\pi<6\chi<0$ in order to
obtain the critical point of the Ising model; taking $0<6\chi<\pi$
would give us the low-temperature fixed point.)
The stiffness gives the correct energy penalty for a domain wall.
A background charge is necessary to correctly describe the coupling of
the bosonic field to the curvature, because the number of left turns minus the number of
right turns will be different from $\pm 6$
around a point of non-zero curvature.
For the honeycomb lattice on the plane, this
reduces to a background charge at infinity $2(1-g)=-2/3$.

\item In terms of the bosonic field, $\phi$,
the effective action of the height model has a marginal
operator $w\,e^{-2i\phi}$ which enforces the
fact that the heights take values that are integral multiples
of $\pi$ (which would otherwise
be lost in the passage to the continuum limit). In fact,
the term which enforces the integrality of the heights
is more complicated. We have kept only the most relevant
term in its Fourier expansion, which is marginal; the other
terms are irrelevant.
Thus, the effective action
takes the form:
\begin{equation}
S = \frac{g}{4\pi} \int {d^2}x \,{(\nabla \phi)^2}
\,+\,  \frac{i(1-g)}{4\pi} \int {d^2}x \, R\, \phi \, + \,
w\! \int {d^2}x \,e^{-2i\phi} + \ldots
\end{equation}
The $\ldots$ denotes other (irrelevant) terms in the Fourier
expansion of the potential term which enforces integral
heights; we have kept only the marginal term.
Rescaling the field $\phi\rightarrow 2\sqrt{g}\phi$,
the effective action takes the form:
\begin{equation}
S = \frac{1}{16\pi} \int {d^2}x \,{(\nabla \phi)^2}
\,+\,  \frac{i(1-g)}{8\pi\sqrt{g}} \int {d^2}x \, R\, \phi \, + \,
w\! \int {d^2}x \,e^{-i\phi/\sqrt{g}} + \ldots
\end{equation}

\item When we compute an Ising correlation function
with $n$ spin fields in this model,
only the term of order $w^n$ is non-zero,
i.e. this correlation function has $n$ insertions of
the marginal operator $e^{-i\phi\sqrt{3}/2}$
(here, we have substituted $g=4/3$), which
we call a screening operator.


\end{enumerate}

The preceding logic makes it seem natural for
correlation functions in the Ising model (and, in fact,
a large class of models which have a height model
representation) to have a Coulomb gas representation.
It is, thus, helpful for understanding our results
intuitively. However, it is not the most convenient way to
derive the Coulomb gas representation for the conformal
blocks which we need. For that, we use a more technical approach,
described in the next subsection. We note that, although
the two approaches are very similar, there is not really a
one-to-one correspondence between them, although
the results which we find in this paper strengthen
the connection.

\subsection{Algebraic Approach}

The algebraic approach to the Coulomb gas
takes, as its starting point, the action for
a free boson with a background charge $\alpha_0$
at infinity from step 5 of the previous subsection.
This can be re-written as a total
derivative term with imaginary coefficient $i\alpha_0$.
This total derivative term changes the
energy-momentum tensor, thus shifting
the central charge from its free boson value, $c=1$,
to $c=1-24\alpha_0^2$. Since the
added term is imaginary, the theory is not unitary.
However, for certain values of $\alpha_0$, including
the one relevant to the Ising model, the
theory has a unitary subspace.

This approach was introduced by
Feigin and Fuchs~\cite{Feigin1983}, and developed for the minimal models~\cite{Belavin1984} by
Dotsenko and Fateev in Refs.~\onlinecite{Dotsenko1984, Dotsenko1985}.
The method was subsequently refined by
Felder~\cite{Felder1989}, who both elucidated its BRST cohomological structure
and extended the results to the torus.
The advantage of Felder's approach is that it holds at the operator
level, not merely at the level of correlation functions, allowing a more systematic description.
This leads to a simple prescription which
can be applied in a uniform manner.
Thus, we adopt Felder's notation.
The next few paragraphs are a short review of the procedure,
whose full details can be found in Ref.~\onlinecite{Felder1989}.

The approach consists of taking a holomorphic free boson field $\varphi(z)$,
whose two-point correlation function is given by
\be \VEV{\varphi(z_1) \varphi(z_2) } = - 2 \log \left(z_1 - z_2 \right).
\ee
This corresponds to  Eq.~\rf{eq:phihol} with $g=1/4$.
This field can be used to construct the vertex operators of charge $\alpha$,
$e^{i \alpha \varphi(z)}$. If the $\alpha_j$ satisfy charge neutrality,
$\sum_{j=1}^{N}{\alpha_j}=0$, then a collection of vertex operators
has correlation function
\be
\label{eq:corr}
\VEV{ e^{i \alpha_1 \varphi(z_1)} e^{i \alpha_2 \varphi(z_2)} \dots e^{i \alpha_N \varphi(z_N)} } = \prod_{i<j}^{N} \left( z_i - z_j \right)^{2 \alpha_i \alpha_j}.
\ee
If ${\sum_j}{\alpha_j}\neq 0$, then this correlation function vanishes. This matches up precisely with Laughlin's plasma analogy, where the vertex operators in the CFT Coulomb gas formalism correspond to the particles that comprise the plasma, and charge neutrality must be obeyed (though, for the Laughlin states' plasmas, charge neutrality is achieved through a uniform background charge density).
However, one can deviate from this simple Coulomb gas in two ways: (i) If a set of vertex operators violates charge neutrality,
one can place an additional compensating vertex operator of charge $2\alpha_0 = -{\sum_j}{\alpha_j}$ at $\infty$
to obtain a non-vanishing correlation function, and (ii) one can introduce ``screening charges'' which modify the vertex operators. As mentioned before, this will produce a unitary theory only for special values of $\alpha_0$, $\alpha_j$, and screening charges.

The minimal model CFTs with central charge $c = 1 - 6 \frac{(p-p^{\prime})^2}{p p^{\prime}}$ are denoted $\mathcal{M}(p,p^{\prime})$ for positive integers $p$ and $p^{\prime}$, where $p>p^{\prime}$ and the unitary theories are given by $p=p^{\prime}+1$. (The Coulomb gas formulation also applies to the non-unitary minimal models with $p \neq p^{\prime}+1$, so we leave these integers arbitrary in the following expressions.) The $\mathcal{M}(p,p^{\prime})$ CFT has two screening charges given by
\begin{equation}
\alpha_- = - \sqrt{\frac{p^{\prime}}{p}} , \quad \alpha_+ = \sqrt{\frac{p}{p^{\prime}}}
,
\end{equation}
a possible charge at infinity of
\begin{equation}
2\alpha_0 = \alpha_- + \alpha_+ = \frac{p-p^{\prime}}{\sqrt{p p^{\prime}}}
,
\end{equation}
and allowed vertex operator charges
\begin{equation}
\alpha_{nm} = \frac{1}{2} \left(1-n\right) \alpha_- + \frac{1}{2} \left(1-m \right) \alpha_+,
\end{equation}
for $n=1,\ldots,p-1$ and $m=1,\ldots,p^{\prime}-1$. The vertex operator $e^{i \alpha_{nm} \varphi(z)}$ for these charges are used to describe the minimal model's primary field $\phi_{(n,m)}$ with conformal scaling dimension
\begin{equation}
h_{n,m} = \frac{1}{4} \left(n \alpha_- + m \alpha_+ \right)^{2} - \alpha_0^2 = \frac{\left( m p - n p^{\prime}\right)^{2} - \left( p - p^{\prime} \right)^{2}}{4 p p^{\prime} }
,
\end{equation}
which can be arranged into the conventional Kac table. These fields obey the identification $\phi_{(n,m)} \equiv \phi_{(p-n , p^{\prime} - m)}$, so there are $(p-1)(p^{\prime}-1)/2$ distinct fields in the theory, each of which can be represented by two distinct vertex operators. Of course, these vertex operators by themselves cannot correctly represent the minimal model fields, so screening operators
\begin{equation}
\label{eq:screeningoperator}
Q_{\pm} = \oint dz e^{i \alpha_{\pm} \varphi(z)}
\end{equation}
must also be introduced.

\begin{figure}  [t!]
\includegraphics[width=6cm]{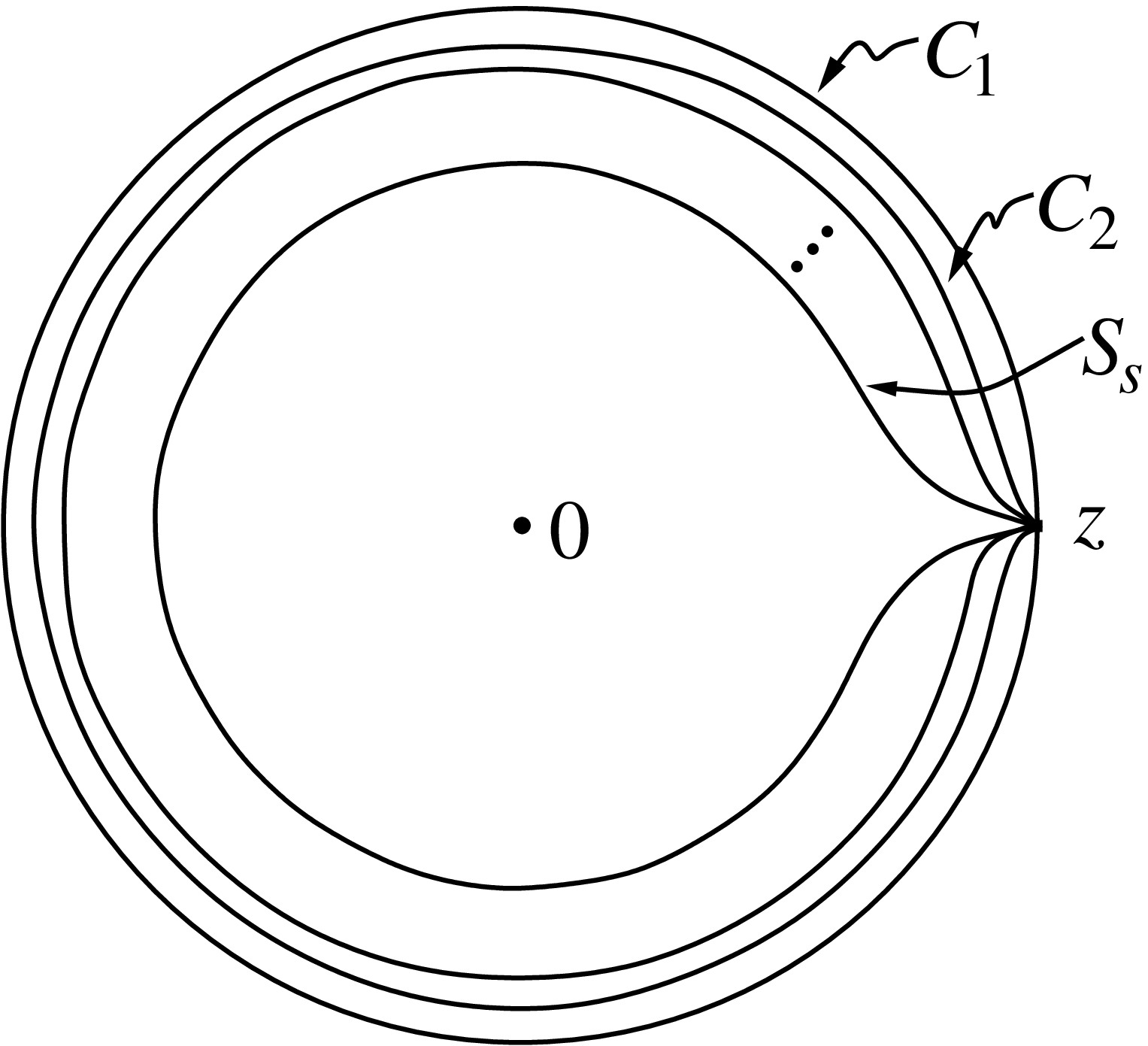}
\caption{\label{space} The integration contours in
Eq.~(\ref{eq:screening}). }
\label{fig:Felder-contours}
\end{figure}

In this way, one can generate the correlation functions of the minimal model CFTs as a Coulomb gas with screening charges. However, the specification of the contour integrals of the screening operators of Eq.~\rf{eq:screeningoperator} is a crucial matter. For this, we follow Felder's prescription of combining the screening operators with the vertex operators to form screened vertex operators~\cite{Felder1989}
\begin{equation}
\label{eq:screening}
V_{nm}^{rs}(z) \equiv\prod_{k=1}^r \oint_{C_k} dw_k \, \prod_{l=1}^s \oint_{S_l} du_l \,  e^{i \alpha_{nm} \varphi(z)} \, e^{i \alpha_- \varphi(w_k)} \, e^{i \alpha_+ \varphi(u_l)}
,
\end{equation}
where the screening charges' integration contours in Eq.~\rf{eq:screening} are taken to be concentric circles of radius $\left| z \right|$ centered at the origin (with the $\alpha_+$ contours $S_l$ inside the $\alpha_-$ contours $C_k$), as shown in Fig.~\ref{space}. These contour integrals have divergences that must be regularized in some manner, i.e. either by an appropriate point-splitting at $z$ or through analytic continuation.

The full conformal block of the CFT operators is represented by the correlation function
\begin{equation}
\label{eq:CGrep}
\mathcal{F}_{\alpha} (z_1, \ldots, z_N) = \VEV{ \phi_{(n_1, m_1) }(z_1) \ldots \phi_{(n_N , m_N)}(z_N)}_{\alpha} = \VEV{ V_{n_1 m_1 }^{r_1 s_1 }(z_1) \ldots V_{n_N m_N}^{r_N s_N }(z_N)}
\end{equation}
for a set of screened vertex operators $V_{n_k m_k }^{r_k s_k}(z_k)$, where the indices $n_k$ and $m_k$ are chosen to represent the minimal model field $\phi_{(n_k , m_k)}$ at $z_k$. The conformal blocks are labeled by the intermediate states in the fusion sequence of fields
\begin{equation}
\alpha = \left\{ \left( \nu_1 , \mu_1 \right), \ldots, \left( \nu_N , \mu_N \right) \right\}
\end{equation}
which must obey the fusion algebra of the CFT, i.e. $\phi_{(\nu_k , \mu_k)} \in \phi_{(\nu_{k-1} , \mu_{k-1})} \times \phi_{(n_{k} , m_{k})}$. This means that
\begin{eqnarray}
&& 1+ \left| n_k - \nu_{k-1} \right| \leq \nu_k \leq \min \left\{ n_k + \nu_{k-1} - 1 , 2p -n_k - \nu_{k-1} - 1 \right\}, \notag \\
&& 1+ \left| m_k - \mu_{k-1} \right| \leq \mu_k \leq \min \left\{ m_k + \mu_{k-1} - 1 , 2p^{\prime} -m_k - \mu_{k-1} - 1 \right\}, \notag \\
&& \left( n_k + \nu_{k-1} +\nu_{k} \right) \text{mod } 2 = 1 , \notag \\
&& \left( m_k + \mu_{k-1} +\mu_{k} \right) \text{mod } 2 = 1 .
\end{eqnarray}
Here, we have $\left( \nu_0 , \mu_0 \right) = \left( \nu_N , \mu_N \right) = (1,1)$, indicating that the entire collection of fields must fuse to vacuum [the identity field $I = \phi_{(1,1)}$]. These different conformal blocks correspond to different choices of the numbers of screening charges $r_k$ and $s_k$ assigned to each screened vertex operator, which must therefore satisfy the conditions~\cite{Felder1989}
\begin{eqnarray}
\mu_{k-1} &=& m_k + \mu_k - 2 s_k - 1 \\
\nu_{k-1} &=& n_k + \nu_k - 2 r_k - 1
.
\end{eqnarray}
When these rules are not satisfied, the correlation functions will evaluate to zero.
It is important to recognize that these rules require that the sum of all Coulomb gas representation charges (vertex operator charges $\alpha_{n m}$ and screening operator charges $\alpha_{\pm}$) is zero.

One can alternatively represent the same conformal block as in Eq.~\rf{eq:CGrep} by using the fact that $\phi_{(n,m)} \equiv \phi_{(p-n , p^{\prime} - m)}$ and, in particular, that the identity field can also be represented by $I = \phi_{(p-1,p^{\prime}-1)}$. With this in mind, it becomes clear that we should obtain the same conformal block if we replace $V_{n_1 m_1 }^{r_1 s_1 }(z_1)$ with $V_{p-n_1, p^{\prime}- m_1}^{0 0}(z_1)$, require $\left( \nu_N , \mu_N \right) = (1,1)$ and $\left( \nu_0 , \mu_0 \right) = (p-1 , p^{\prime}-1)$, and place a charge $2 \alpha_0$ at infinity. In this case, the sum of all Coulomb gas representation charges together with the charge $2 \alpha_0$ at infinity is zero.

We now focus on the Coulomb gas representation of the $c=1/2$ Ising CFT, which is the $\mathcal{M}(4,3)$ minimal model and has three primary fields $I$, $\psi$, and $\sigma$. The two screening charges for this CFT are given by
\begin{equation}
\alpha_- = - \frac{\sqrt{3}}2, \quad \alpha_+ = \frac{2}{\sqrt{3}}.
\end{equation}
The six vertex operator charges are constructed out of $\alpha_-$ and $\alpha_+$, according to
\begin{equation}
\alpha_{nm} = \frac{1}{2} \left(1-n\right) \alpha_- + \frac{1}{2} \left(1-m \right) \alpha_+,
 {\hskip 0.3 cm} n=1,2,3, \ m=1,2.
\end{equation}

It is convenient to put these together into the ``Kac table'':
\renewcommand{\arraystretch}{2}
\begin{equation}
\begin{tabular}{c||c|c|c|}
\hline
2&$-\frac 1 {\sqrt {3}}$ & $-\frac 1 {4 \sqrt{3}}$ & $\frac 1 {2 \sqrt{3}}$ \\
\hline
1 & $0$ & $\frac {\sqrt{3}} {4}$ & $\frac {\sqrt{3}}  2$ \\ \hline\hline
$m$/$n$ & 1& 2& 3\\
\end{tabular} \label{eq:table1}
\end{equation}
\renewcommand{\arraystretch}{1}
The columns of the table are labeled by the index $n=1,2,3$, and the rows by $m=1,2$. The entries of the table are the charges $\alpha_{nm}$ of the vertex operators. They represent the operators of the $c=1/2$ Ising CFT according to the identification
\begin{equation}
\begin{tabular}{c||c|c|c|}
\hline
2&$\psi$ & $\sigma $ & $I$ \\
\hline
1 & $I$ & $\sigma$ & $\psi$ \\ \hline\hline
$m$/$n$ & 1& 2& 3\\
\end{tabular} \label{eq:table2}
\end{equation}
Here $I$ is the identity field, and just as in Eqs.~\rf{eq:pfcft} and \rf{eq:holes}, $\sigma$ is the dimension $1/16$ operator, and $\psi$ is the dimension $1/2$ operator of the $c=1/2$ Ising CFT.

We can now examine in detail a concrete example of how the same conformal block can have several different but equivalent representations using the screened vertex operators. Consider, for example
\be
\VEV{\psi(z) \, \psi(0)} = \frac{1}{z}.
\ee
Since both $V_{12}$ and $V_{31}$ correspond to $\psi$,
this correlation function can be represented in three different ways:
\be \label{eq:cor1} \VEV{ V_{31}^{00}(z) \, V_{12}^{00}(0)} = \frac{1}{z},
\ee
\be \label{eq:cor2}  \VEV{V_{12}^{01}(z) \, V_{12}^{00}(0)} = \oint  du \, \frac{z^{\frac 2 3}}{(z-u)^{\frac 4 3} u^{\frac 4 3}},
\ee
and
\be \label{eq:cor3}
\VEV{V_{31}^{20}(z) \, V_{31}^{00}(0)} = \oint dw_1 \oint dw_2 \, \frac{z^{\frac 3 2 } (w_1-w_2)^{\frac 3 2}}{(z-w_1)^{\frac 3 2} w_1^{\frac 3 2} (z-w_2)^{\frac 3 2} w_2^{\frac 3 2}}.
\ee
It should be clear that all three methods give the same answer, $1/z$, up to an overall unimportant constant (as may be verified
by making the change of variables: $\zeta = u/z $ and $\xi_k = w_k / z$).
The correlation function in Eq.~\rf{eq:cor1} has total charge $-2 \alpha_0 = 1/2 \sqrt{3}$ (which is canceled by a charge at infinity) while the other two correlation functions, Eqs.~\rf{eq:cor2} and \rf{eq:cor3} have the total charge $0$.

Now we can use these techniques to represent the conformal block
corresponding to the Pfaffian as
\be
\label{eq:electronblock}
{\rm Pf}\left(\frac{1}{z_i-z_j}\right) = \VEV{ \psi(z_1) \dots \psi(z_{N})} = \VEV{ V_{31}^{20} (z_1) V_{31}^{00}(z_2) \dots V_{31}^{20}(z_{N-1}) V_{31}^{00}(z_{N})},
\ee
(where $N$ is even). This is not the only way to construct this conformal block, but it is
the most convenient for subsequent generalizations.

Now consider a conformal block with four $\sigma$ operators
(which correspond to four quasiholes). There are two
such conformal blocks [as we saw, for instance,
in Eq.~\rf{eq:wave}], which we denote by:
\be
\label{eq:confblocks}
\mathcal{F}_{\alpha}(\eta_\mu; z_i) =
\left< \sigma(\eta_1) \sigma(\eta_2) \sigma(\eta_{3}) \sigma(\eta_{4}) \psi(z_1) \dots \psi(z_{N})  \right>_\alpha
\ee
where $\alpha=0,1$ corresponds to the block in which
the first two $\sigma$ fields fuse to $I$ or $\psi$, respectively.
We can represent $\mathcal{F}_{0}(\eta_\mu; z_i)$ in the following way:
\be
\label{eq:consfour2}
\mathcal{F}_{0}(\eta_\mu; z_i) =
\VEV{ V_{21}^{10}(\eta_1) V_{21}^{00} (\eta_2) V_{21}^{10}(\eta_3) V_{21}^{00}(\eta_4) \, V_{31}^{20} (z_1) V_{31}^{00}(z_2) \dots V_{31}^{20}(z_{N-1}) V_{31}^{00}(z_{N})}.
\ee
This representation mirrors that of
Eq.~\rf{eq:electronblock} in that it only uses vertex operators
from the $m=1$ row of the Kac table in Eq.~\rf{eq:table1}.
The total charge of all the operators involved in Eq.~\rf{eq:consfour2} is equal to zero.
Furthermore, the total charge of the first two screened vertex
operators is also zero, $V_{21}^{10} V_{21}^{00} \sim I$,
which is the reason for the identification of this Coulomb gas
correlation function with the $c=1/2$ Ising conformal block
$\mathcal{F}_{0}(\eta_\mu; z_i)$. If we wish, instead, to compute $\mathcal{F}_{1}(\eta_\mu; z_i)$,
then we need a Coulomb gas correlation function in which
the first two screened vertex operators have total charge
$\sqrt{3}/2$ corresponding to the $\psi$ field:
\be
\label{eq:consfour2-psi}
\mathcal{F}_{1}(\eta_\mu; z_i) =
\VEV{ V_{21}^{10}(\eta_1) V_{21}^{10} (\eta_2) V_{21}^{00}(\eta_3) V_{21}^{00}(\eta_4) \, V_{31}^{20} (z_1) V_{31}^{00}(z_2) \dots V_{31}^{20}(z_{N-1}) V_{31}^{00}(z_{N})}.
\ee
Since the screening operators are attached to the
first two vertex operators, rather than the first and third,
the construction Eq.~\rf{eq:consfour2-psi} can be interpreted
as simply a different choice of contour for one of the
screening operators in Eq.~\rf{eq:consfour2}.

We note, for later use, that we can also represent $\mathcal{F}_{0}$ and $\mathcal{F}_{1}$
in an alternative way:
\be
\label{eq:consfour1}
\mathcal{F}_{0}(\eta_\mu; z_i) =
\VEV{ V_{22}^{00}(\eta_1) V_{21}^{00} (\eta_2) V_{21}^{10}(\eta_3) V_{21}^{00}(\eta_4) \, V_{31}^{20} (z_1) V_{31}^{00}(z_2) \dots V_{31}^{20}(z_{N-1}) V_{31}^{00}(z_{N})}
,
\ee
\be
\label{eq:consfour1-psi}
\mathcal{F}_{1}(\eta_\mu; z_i) =
\VEV{ V_{22}^{00}(\eta_1) V_{21}^{10} (\eta_2) V_{21}^{00}(\eta_3) V_{21}^{00}(\eta_4) \, V_{31}^{20} (z_1) V_{31}^{00}(z_2) \dots V_{31}^{20}(z_{N-1}) V_{31}^{00}(z_{N})}.
\ee
Unlike in Eqs.~\rf{eq:consfour2} and \rf{eq:consfour2-psi}, the total charge of the
vertex operators involved in Eqs.~\rf{eq:consfour1} and \rf{eq:consfour1-psi} is equal to $1/(2 \sqrt{3})$ (which is another
representation of the identity).

Finally, we can also construct conformal blocks with any
even number $n$ of $\sigma$ fields (corresponding to wavefunctions
with $n$ quasiholes), e.g.
\begin{eqnarray}
\label{eq:consfour4}
&& \mathcal{F}_{(0,0,\ldots,0)}(\eta_\mu; z_i)
= \VEV{ V_{21}^{10}(\eta_1) V_{21}^{00} (\eta_2) \dots V_{21}^{10}(\eta_{n-1}) V_{21}^{00}(\eta_{n})\, V_{31}^{20} (z_1) V_{31}^{00}(z_2) \dots V_{31}^{20}(z_{N-1}) V_{31}^{00}(z_{N})} \\
&&\quad = \VEV{ V_{22}^{00}(\eta_1) V_{21}^{00} (\eta_2) V_{21}^{10}(\eta_3) V_{21}^{00}(\eta_4)  \dots V_{21}^{10}(\eta_{n-1}) V_{21}^{00}(\eta_{n})\, V_{31}^{20} (z_1) V_{31}^{00}(z_2) \dots V_{31}^{20}(z_{N-1}) V_{31}^{00}(z_{N})}
\end{eqnarray}
The subscript $(0,0,\ldots,0)$ denotes that this is the conformal
block in which the first and second $\sigma$ fields fuse to $I$,
the third and fourth $\sigma$ fields fuse to $I$, $\ldots$, the
$(n-1)^{\rm th}$ and $n^{\rm th}$ $\sigma$ fields fuse to $I$.

Similarly, the Coulomb gas construction gives the general conformal block
\begin{eqnarray}
\label{eq:consfourn}
&& \mathcal{F}_{(\pi_1,\pi_2,\ldots,\pi_{n/2})}(\eta_\mu; z_i)
= \VEV{ \prod_{j=1}^{n/2} \left[ V_{21}^{1-\pi_{j-1},0}(\eta_{2j-1}) V_{21}^{\pi_j ,0} (\eta_{2j}) \right] \, V_{31}^{20} (z_1) V_{31}^{00}(z_2) \dots V_{31}^{20}(z_{N-1}) V_{31}^{00}(z_{N})} \\
&&\quad = \VEV{ V_{22}^{00}(\eta_{1}) V_{21}^{\pi_1 ,0} (\eta_{2}) \prod_{j=2}^{n/2} \left[V_{21}^{1-\pi_{j-1},0}(\eta_{2j-1}) V_{21}^{\pi_j ,0} (\eta_{2j}) \right] \, V_{31}^{20} (z_1) V_{31}^{00}(z_2) \dots V_{31}^{20}(z_{N-1}) V_{31}^{00}(z_{N})}
,
\end{eqnarray}
in which the $1^{\rm st}$ through $2j^{\rm th}$ $\sigma$ fields collectively fuse to $I$ if $\pi_{j}=0$ and to $\psi$ if $\pi_j =1$, and where $\pi_0 = \pi_{n/2}=0$, indicating the overall parity constraint that the $\sigma$ fields must collectively fuse to $I$ since there are an even number $N$ of $\psi$ fields. This is presented in the ``standard basis,'' where fusion channels are specified by fusing in the anyons one at a time from left to right~\footnote{More generally, in the standard basis, one should label every fusion channel, i.e. also have a label for the collective topological charge of the $1^{\rm st}$ through $(2j+1)^{\rm th}$ fields. However, for the Ising $\sigma$ fields, these fusion label are always equal to $\sigma$ as a consequence of the simple fusion algebra, so we do not bother to label them.}. For the Ising CFT, we can trivially transform between the standard basis and the qubit basis
\be
\mathcal{F}^{\text{qubit}}_{(p_1,p_2,\ldots,p_{n/2})} = \mathcal{F}^{\text{standard}}_{(\pi_1,\pi_2,\ldots,\pi_{n/2})}
\ee
in which the $(2j-1)^{\rm th}$ and $2j^{\rm th}$ $\sigma$ fields fuse to $I$ if $p_{j}=0$ and to $\psi$ if $p_j =1$, by simply using the conversions
\begin{eqnarray}
\pi_{j} &\equiv& \left( \sum_{k=1}^{j} p_k \right) \text{mod }2 \\
p_{j} &\equiv& \left( \pi_j - \pi_{j-1} \right) \text{mod }2 .
\end{eqnarray}
Since this is a trivial change of basis (i.e. it is just a different way of presenting the subscript label), we can interchange between the two freely. For the purposes of describing conformal blocks using the Coulomb gas formalism, the standard basis is more natural. For describing the explicit evaluation of the conformal blocks using bosonization methods, the qubit basis is more natural. Henceforth, we differentiate the use of these two bases through context.

A similar expression can also be used if the number $N$ of electrons is odd. Specifically, for $N$ odd, one would use
\begin{eqnarray}
\label{eq:consfournNodd}
&& \mathcal{F}_{(\pi_1,\pi_2,\ldots,\pi_{n/2})}(\eta_\mu; z_i)
= \VEV{ \prod_{j=1}^{n/2} \left[ V_{21}^{1-\pi_{j-1},0}(\eta_{2j-1}) V_{21}^{\pi_j ,0} (\eta_{2j}) \right] V_{31}^{00}(z_1) V_{31}^{20} (z_2) V_{31}^{00}(z_2) \dots V_{31}^{20}(z_{N-1}) V_{31}^{00}(z_{N})} \\
&&\quad = \VEV{ V_{22}^{00}(\eta_{1}) V_{21}^{\pi_1 ,0} (\eta_{2}) \prod_{i=2}^{n/2} \left[ V_{21}^{1-\pi_{i-1},0}(\eta_{2i-1}) V_{21}^{\pi_i ,0} (\eta_{2i}) \right] V_{31}^{00}(z_1) V_{31}^{20} (z_2) V_{31}^{00}(z_2) \dots V_{31}^{20}(z_{N-1}) V_{31}^{00}(z_{N})}
\end{eqnarray}
with $\pi_0 =0$ and $\pi_{n/2} = 1$, which indicates that the $n$ $\sigma$s have overall fusion channel $\psi$. We note that the number of screening charges in both Eqs.~\rf{eq:consfourn} and \rf{eq:consfournNodd} is $N+\frac{n}{2}$.

The explicit expressions for correlation functions such as
Eq.~\rf{eq:consfour4} involve products of powers of differences of coordinates, and integrals over some of them, as in the simple examples of Eqs.~\rf{eq:cor2} and \rf{eq:cor3}. This has a reasonably similar structure to the Laughlin states, such as Eq.~\rf{eq:La}, so it brings us closer to the goal of constructing an effective plasma
describing Eq.~\rf{eq:norm}.

\section{Plasma representation for the norm of the ground state wavefunction}

Using the preceding expressions for the conformal blocks
to construct the overlap integrals Eq.~\rf{eq:norm},
we see that they do appear superficially
similar to the plasma construction of Eq.~\rf{eq:pl}.
The difference is that the screening operators need to be integrated over their holomorphic and antiholomorphic coordinates
along some specially chosen contours. As a result,
Eq.~\rf{eq:overlap} no longer takes the form of the
partition function of a classical plasma.
In what follows, we construct the overlap integrals
in a slightly different way which leads to an
expression which does take the form of
a classical plasma's partition function.
For this, we crucially utilize the method invented by Mathur in Ref.~\onlinecite{Mathur1992}
of relating expressions involving products of holomorphic and antiholomorphic screening charge contour integrals to expressions involving $2$D integrals over screening charge positions. (We review this method in Appendix~\ref{sec:Mathur}.)

We will begin by considering the case with no
quasiholes, ie. the ground-state wavefunction.
We will construct a representation of the norm
of the ground-state wavefunction which takes
the form of the partition function for a classical
plasma.
We begin by ignoring the charge part of the wavefunction
and focusing on the Pfaffian:
\be
\left|{\rm Pf}\left(\frac{1}{z_i-z_j}\right) \right|^2
\ee
In order to represent this as a plasma,
we take the conformal block represented by Eq.~\rf{eq:electronblock},
and multiply it by its complex conjugate. Then, instead of integrating the screening operators over the contours in the complex plane of their respective holomorphic and antiholomorphic coordinates,
{\it we integrate the screening operators
over the entire $2$D plane}.

To see why this procedure is valid, we first consider the expression
\begin{eqnarray}
{\rm Pf}\left(\frac{1}{z_i-z_j}\right) &=&
 \VEV{ V_{31}^{20} (z_1) V_{31}^{00}(z_2) \dots V_{31}^{20}(z_{N-1}) V_{31}^{00}(z_{N})}\cr
&=& \left< \oint_{C_{z_1}} dw_1 \, \oint_{C_{z_1}} dw_2 \, e^{ i \alpha_{31} \varphi(z_1)} e^{i \alpha_- \varphi(w_1)} e^{i \alpha_- \varphi(w_2)} e^{ i \alpha_{31} \varphi(z_2)} \, \times \ldots \right.\cr
&& \qquad \left. \ldots \times \oint_{C_{z_{N-1}}} dw_{N-1} \, \oint_{C_{z_{N-1}}} dw_{N} \,e^{ i \alpha_{31} \varphi(z_{N-1})} e^{i \alpha_- \varphi(w_{N-1})} e^{i \alpha_- \varphi(w_{N})} e^{ i \alpha_{31} \varphi(z_{N})} \right>
\end{eqnarray}
where $C_{x}$ is used to indicate a contour of radius $|x|$ centered on the origin (with appropriate regularization, i.e. taking contours at the same radius to be infinitesimally concentric and point-split at the $z_i$ coordinates).
To obtain the norm squared of this wavefunction,
we multiply this expression by its complex conjugate:
\begin{multline}
{\rm Pf}\left(\frac{1}{\bar z_i - \bar z_j}\right) =
\left< \oint_{C_{z_1}} d\bar w_1 \, \oint_{C_{z_1}} d\bar w_2 \, e^{ i \alpha_{31} \varphi(\bar z_1)} e^{i \alpha_- \varphi(\bar w_1)} e^{i \alpha_- \varphi(\bar w_2)} e^{ i \alpha_{31} \varphi(\bar z_2)}
\,\times \ldots  \right. \\
\left. \ldots \times \oint_{C_{z_{N-1}}} d\bar w_{N-1} \, \oint_{C_{z_{N-1}}} d\bar w_{N} \,
e^{ i \alpha_{31} \varphi(\bar z_{N-1})} e^{i \alpha_- \varphi(\bar w_{N-1})} e^{i \alpha_- \varphi(\bar w_{N})} e^{ i \alpha_{31} \varphi(\bar z_{N})} \right>.
\end{multline}
Evaluating the correlation functions of vertex
operators, noting that $\alpha_{31} = -\alpha_{-} = \sqrt{3}/2$, we obtain
\begin{equation}
\label{eq:abs2Pf}
\left| {\rm Pf}\!\left(\frac{1}{z_i-z_j}\right)\right|^2
=   \prod_{k=1}^{N/2} \oint_{C_{z_{2k-1}}} {\hskip -0.5 cm}d{w_{2k-1}} \oint_{C_{z_{2k-1}}} {\hskip -0.5 cm}d\bar{w}_{2k-1} \oint_{C_{z_{2k-1}}}
{\hskip -0.5 cm}d{w_{2k}} \oint_{C_{z_{2k-1}}} {\hskip -0.5 cm}d\bar{w}_{2k}\,
\prod_{i<j}^{N} \left| w_i-w_j \right|^3
\prod_{i,j}^{N} \left| w_i-z_j \right|^{-3}
\prod_{i<j}^{N} \left| z_i-z_j \right|^3.
\end{equation}
It is important to emphasize that $w_i$ and $\bar{w}_i$ are independent variables in this expression, so terms such as $\left| w_i-w_j \right|^3$ should really be understood as shorthand for $\left( w_i-w_j \right)^{3/2} \left( \bar{w}_i-\bar{w}_j \right)^{3/2}$.
Retracing Mathur's steps, as explained in Appendix~\ref{sec:Mathur},
we rewrite the product of $w_i$ and $\bar{w}_i$
contour integrals in Eq.~\rf{eq:abs2Pf} in terms of 2D integrals:
\begin{eqnarray}
\label{eqn:contour-to-2D-gs}
&& \int  \prod_k^{N} d^2 w_k  \prod_{i<j}^{N} \left| w_i-w_j \right|^3
\prod_{i,j}^{N} \left| w_i-z_j \right|^{-3} \prod_{i<j}^{N} \left| z_i-z_j \right|^3 \notag \\
&& \qquad  =\prod_{k=1}^{N/2} \oint_{C_{z_{2k-1}}} {\hskip -0.5 cm}d{w_{2k-1}} \oint_{C_{z_{2k-1}}} {\hskip -0.5 cm}d\bar{w}_{2k-1} \oint_{C_{z_{2k-1}}}
{\hskip -0.5 cm}d{w_{2k}} \oint_{C_{z_{2k-1}}} {\hskip -0.5 cm}d\bar{w}_{2k}\,
\prod_{i<j}^{N} \left| w_i-w_j \right|^3
\prod_{i,j}^{N} \left| w_i-z_j \right|^{-3}
\prod_{i<j}^{N} \left| z_i-z_j \right|^3
\end{eqnarray}

Therefore, we can write the square of the Pfaffian in the form:
\begin{equation}
\label{eqn:ground-plasma}
\left| {\rm Pf}\!\left(\frac{1}{z_i-z_j}\right)\right|^2
= \int  \prod_k^{N} d^2 w_k  \prod_{i<j}^{N}  \left| w_i-w_j \right|^3
\prod_{i,j}^{N} \left| w_i-z_j \right|^{-3} \prod_{i<j}^{N} \left| z_i-z_j \right|^3.
\end{equation}
Note that the right-hand-side of this equation is divergent
as any $w_i$ approaches any $z_j$. It can be made well-defined
by analytic continuation. In other words, we define this
expression by evaluating the integral
\begin{equation}
\label{eqn:ground-plasma-analytic}
\int  \prod_{k}^{N} d^2 w_k  \prod_{i<j}^{N}  \left| w_i-w_j \right|^{3\alpha}
\prod_{i,j}^{N} \left| w_i-z_j \right|^{-3\alpha} \prod_{i<j}^{N} \left| z_i-z_j \right|^{3 \alpha}.
\end{equation}
for $\alpha<2/3$, where the integral is convergent,
and analytically continuing to $\alpha=1$.
This analytic continuation gives the right-hand-side
of Eq.~\rf{eqn:ground-plasma}.
As we will discuss, the associated plasma does not
go through a phase transition as $\alpha$ is varied from
$\alpha<2/3$ to $1$, so the right-hand-side of
Eq.~\rf{eqn:ground-plasma} is a useful representation
of the left-hand-side.

If, instead, we modify the right-hand-side at short distances
by, for instance, introducing a short-ranged repulsion (e.g. a hard-core cutoff),
then the right-hand side will be modified for
$z_i \rightarrow z_j$ but will be unchanged at long-distances.
This will produce a wavefunction in the same universality
class as the Pfaffian. However, rather than introduce a cutoff
and work with a modified wavefunction,
we prefer to define Eq.~\rf{eqn:ground-plasma} by analytic
continuation, as described earlier.

Now we can interpret the norm of the Pfaffian in terms of a two-component plasma. Specifically, we can write
\begin{eqnarray}
\label{eq:pl2}
\left\| {\rm Pf}\!\left(\frac{1}{z_i-z_j}\right)\right\|^2 &\equiv& \int \prod_{k=1}^N d^2 z_k \, \left| {\rm Pf}\!\left(\frac{1}{z_i-z_j}\right)\right|^2 = \int \prod_{k=1}^N d^2 z_k \, d^2 w_k \, e^{-\Phi_2 /T} = e^{-F_2/T}\\
\Phi_2 &=& - \sum_{i < j}^N Q^2 \log \left| w_i -w_j \right| + \sum_{i,j}^N
Q^2 \log \left| w_i - z_j \right| - \sum_{i<j}^N Q^2 \log \left| z_i - z_j \right|
,
\end{eqnarray}
where $T=Q^2/3$. Now $\Phi_2$ is the $2$D Coulomb-interaction potential energy for $N$ charge $Q$ particles at $z_i$ and $N$ charge $-Q$ particles at $w_i$. Thus, $F_2$ is the free energy of a classical $2$D two-component plasma of charges $\pm Q$ at temperature $T$. (We use the subscript $2$ to indicate the two-component plasma.) Again, we can let $T$ take any value as long as $Q$ is adjusted accordingly. One convenient choice is to take $T=g$ and $Q = \sqrt{3g}$.

It is known that such a two-component plasma with coupling constant $\Gamma = Q^2/T$ is a screening fluid for $\Gamma < 4$, i.e. the condensation temperature is $T_{c_2} = Q^2/4$, but that it needs a short-ranged repulsive interaction, such as a hard-core cutoff, in order to be stable against collapse into neutral bound pairs for $\Gamma >2$ (i.e. $T < 2T_{c_2}$)~\cite{May67,Knorr68,Hauge71,Kosterlitz73}. For $\Gamma >4$ all particles are bounded into neutral pairs, while for $\Gamma <2$ all pairs are broken. The Pfaffian wavefunction's corresponding plasma is precisely in the range $2 < \Gamma < 4$ where it is a screening fluid as long as a short-ranged repulsion is introduced. This fits with the preceding discussion regarding the need for a short-distance repulsion or analytic continuation, and is intuitively clear from the fact that ${\rm Pf}\!\left(\frac{1}{z_i-z_j}\right)$ diverges as $z_i \rightarrow z_j$. We discuss the screening properties of this plasma in more detail using field theoretic methods in Appendix~\ref{sec:two-comp-screening}. Its Debye screening length can be estimated (see Appendix~\ref{sec:Debye}) to be $\ell_2 = (12 \pi n_f)^{-1/2}$, where $n_f$ is the electron density.

Adding the charge part of the MR ground-state, we have:
\begin{eqnarray}
\label{eqn:GS-Coulomb-Gas}
{\left| \Psi({z_1},\ldots,{z_N}) \right|}^2
&=& \left| {\rm Pf}\left(\frac{1}{z_i-z_j}\right) \right|^2 \,\prod_{i<j}^{N}
\left| z_i-z_j \right|^{2M} e^{  - \frac{1}{2} \sum\limits_{i=1}^{N}  \left| z_i \right|^2 } \notag \\
&=& \int  \prod_{k}^{N} d^2 w_k  \prod_{i<j}^{N}  \left| w_i-w_j \right|^3  \prod_{i,j}^{N} \left| w_i-z_j \right|^{-3}\,
\prod_{i<j}^{N} \left| z_i-z_j \right|^{2M+3} e^{ - \frac{1}{2} \sum\limits_{i=1}^{N} \left|z_i \right|^2 }.
\end{eqnarray}
This expression is antisymmetric under exchange of
$z_i$ with $z_j$ while holding $\bar z_i$ and $\bar z_j$
fixed. It is also of degree $M(N-1)-1$ in any of the $z_i$s and
degree $M(N-1)-1$ in any of the $\bar z_i$s. Indeed, there
is a unique polynomial satisfying these properties,
so it is clear that once the right-hand-side is
computed by analytic continuation, it will give
the squared modulus of the MR Pfaffian ground-state wavefunction.

Now we can write the norm of the MR ground-state wavefunction in terms of a classical plasma by writing
\begin{eqnarray}
\label{eq:MRpl}
\left\| \Psi({z_1},\ldots,{z_N}) \right\|^2 &\equiv& \int \prod_{k=1}^N d^2 z_k \, \left| \Psi({z_1},\ldots,{z_N}) \right|^2 = \int \prod_{k=1}^N d^2 z_k \, d^2 w_k \, e^{-\left( \Phi_{1} + \Phi_{2} \right)/T} = e^{-F/T}  \\
\Phi_{1} &=& - \sum_{i<j}^{N} Q_1^2 \log \left| z_i - z_j \right| + \frac{Q_1^2}{4M} \sum_{i=1}^N \left| z_i \right|^2 \\
\Phi_{2} &=& - \sum_{i < j}^N Q_2^2 \log \left| w_i -w_j \right| + \sum_{i,j}^N
Q_2^2 \log \left| w_i - z_j \right| - \sum_{i<j}^N Q_2^2 \log \left| z_i - z_j \right|
,
\end{eqnarray}
where $T=g$, $\Phi_{1}$ corresponds to the $2$D Coulomb potential for $N$ charge $Q_1=\sqrt{2Mg}$ particles at $z_i$ in a uniform neutralizing background of charge density $\rho_1 = -\frac{Q_1}{2 \pi M \ell_B^2}$, and $\Phi_{2}$ corresponds to the $2$D Coulomb potential for $N$ charge $Q_2=\sqrt{3g}$ particles at $z_i$ and $N$ charge $-Q_2=-\sqrt{3g}$ particles at $w_i$ (and no neutralizing background charge density). Thus, $F$ is the free energy of a classical $2$D plasma (at temperature $T$) in which its particles can carry charges corresponding to two independent types of Coulomb interactions, differentiated using the subscripts 1 and 2. In particular, the plasma described here consists of $N$ particles at $w_i$ carrying charge $-Q_2$, $N$ particles at $z_i$ carrying charge $Q_1$ and $Q_2$, and a uniform background of charge density $\rho_1 = -\frac{Q_1}{2 \pi M \ell_B^2}$ and $\rho_2 =0$ that neutralizes the charges of type 1.

When plasmas 1 and 2 are independently in the screening liquid phase for the corresponding values of $T$, $Q_1$, and $Q_2$, we expect that the combined plasma should also be in the screening liquid phase (except, perhaps, as these parameters become close to their critical values for either of the two plasmas). A recent numerical study~\cite{Herland-unpublished} of this combined plasma for $M=2$ found that the behavior is very similar to the $M=0$ case (i.e. the two-component plasma) in that it is in the screening phase for $Q_2^2 / T < 4$. In particular, this study verifies the MR state's analogous plasma (which has $Q_2^2 / T = 3$) is in the screening phase for the most important case $M=2$. (For additional results on similar plasmas
in the context of vortices in multi-component superconductors,
which support the idea that the combined plasma is likely to be in
the screening liquid phase, see Refs.~\onlinecite{Smiseth2005,Dahl2008,Herland2010}.)
We estimate the Debye screening length of this plasma (see Appendix~\ref{sec:Debye}) to be
\begin{equation}
\ell = \left( \frac{M}{M+3 - \sqrt{M^2 +9}}\right)^{1/2} \ell_B
.
\end{equation}
For $M=2$, this gives $\ell \approx 1.2 \ell_B$.

\section{Plasma representation for the trace of the overlap matrix}
\label{sec:five}

The situation gets more complicated when we turn to wavefunctions with multiple quasiholes. We would again like to be able to treat quasiparticles as test charges in the analogous plasma. However, this is not as straightforward to do as for the Laughlin case. There are multiple degenerate wavefunctions
(corresponding to multiple conformal blocks) in such a case.
These different conformal blocks are distinguished in the Coulomb
gas formalism by the location of the screening charge operators' contours.
Thus, if we exchange a pair of screening contour integrals,
one holomorphic and one antiholomorphic, for a 2D integral
too naively, we would elide the distinction between the
different conformal blocks, which would clearly be incorrect.

Thus, we must proceed with greater caution.
To do this, it is useful to recall that the $c=1/2$ Ising CFT with its conformal blocks is but a mathematical tool to construct the correlation functions of the Ising model at its critical point~\cite{McCoy1973,Belavin1984}. These correlation functions are real, not complex, and they depend on the $2$D coordinates of the operators of the Ising model, not just on the holomorphic part of these coordinates.
In particular, consider a correlation function of four Ising spins (order operators) $\sigma$, as well as $N$ Ising energy operators
$\epsilon$:
\be
\label{eq:Ising-non-chiral-def}
\VEV{ \sigma(\eta_1,\bar \eta_1)  \sigma(\eta_2,\bar \eta_2)  \sigma(\eta_3,\bar \eta_3)  \sigma(\eta_4,\bar \eta_4) \, \epsilon(z_1,\bar z_1) \dots \epsilon(z_{N},\bar z_{N}) }.
\ee
Note that these are non-chiral operators.
For instance, $\epsilon={\bar \psi} \psi$, where $\psi$
is the chiral Majorana fermion field introduced earlier
and ${\bar \psi}$ is its antiholomorphic counterpart.
This correlation function can be written in terms
of the two conformal blocks, $\mathcal{F}_0$ and $\mathcal{F}_1$.
These conformal blocks are the chiral part(s) of the correlation function
Eq.~\rf{eq:Ising-non-chiral-def}, which we denoted
in the previous section as:
\be
\mathcal{F}_{\alpha}(\eta_\mu; z_i) \equiv
\VEV{ \sigma(\eta_1)  \sigma(\eta_2) \sigma(\eta_3) \sigma(\eta_{4})
\, \psi(z_1) \dots \psi(z_{N})}_{\alpha}
\ee
Note that these are now chiral operators $\sigma(\eta_{\mu})$ and $\psi(z_i)$.
The subscript $\alpha =0, 1$ denotes whether the first two
$\sigma$ fields fuse to $I$ or $\psi$, respectively. The explicit
forms of $\mathcal{F}_0$ and $\mathcal{F}_1$ are:
\begin{equation}
\label{eqn:conformal-blocks-eg}
\mathcal{F}_{0,1}=
\left(\frac{\eta_{13} \eta_{24}}{\eta_{12}\eta_{23} \eta_{34} \eta_{41}} \right)^{\frac 1 8}
\frac{1}{ \sqrt{1 \pm \sqrt{1-x}}} \left({\tilde \Psi}_{(13)(24)} \pm
\sqrt{1-x}
\, {\tilde \Psi}_{(14)(23)} \right),
\end{equation}
where ${\tilde \Psi}_{(13)(24)}$, ${\tilde \Psi}_{(14)(23)}$
are defined in Eq.~\rf{eqn:tilde-Psi-def}.
Note that $\mathcal{F}_0$ and $\mathcal{F}_1$ are clearly multi-valued functions;
they transform under the braiding of coordinates in exactly the same way as the functions $\Psi_0$ and $\Psi_1$ of Eq.~\rf{eq:wave}, up to an overall phase [which is due to the $c=1$ CFT present in Eq.~\rf{eq:wave}].
Indeed, $\Psi_{0,1}$ were constructed by multiplying $\mathcal{F}_{0,1}$ in Eq.~\rf{eqn:conformal-blocks-eg} by
a Laughlin wavefunction-like
factor coming from the $c=1$ CFT.

The antiholomorphic
part of the correlation function is similarly given by
$\bar{\mathcal{F}}_0$ and $\bar{\mathcal{F}}_1$.
However, the non-chiral
correlation function $\VEV{ \sigma(\eta_1,\bar \eta_1)  \dots
\sigma(\eta_4,\bar \eta_4) \, \epsilon(z_1,\bar z_1) \dots \epsilon(z_{N},\bar z_{N}) }$ must combine holomorphic and
antiholomorphic sectors in such a way as to be single-valued. There is a unique way to do this, which is the trace:
\be
\label{eq:Ising}
\VEV{ \sigma(\eta_1,\bar \eta_1)  \sigma(\eta_2,\bar \eta_2)  \sigma(\eta_3,\bar \eta_3)  \sigma(\eta_4,\bar \eta_4) \, \epsilon(z_1,\bar z_1) \dots \epsilon(z_{N},\bar z_{N}) } = \mathcal{F}_0(\eta_\mu;z_i) \bar{\mathcal{F}}_0(\bar \eta_\mu; \bar z_i) + \mathcal{F}_1(\eta_\mu; z_i) \bar{\mathcal{F}}_1(\bar \eta_\mu; \bar z_i).
\ee
Indeed, this is the only combination of the conformal blocks which is single valued as $\eta_1$ and $\bar \eta_1 = \eta_1^{\ast}$ are taken all over the complex plane, and similarly for the other
$\eta_\mu$ and $\bar \eta_\mu$. This may be checked by using the analytic continuation properties of $\mathcal{F}_0$ and $\mathcal{F}_1$, which are exactly the same as Eqs.~\rf{eq:op1}, \rf{eq:op2}, and \rf{eq:op3} for $\Psi_0$ and $\Psi_1$ (up to the overall phase, which obviously cancels between holomorphic and antiholomorphic terms anyway). This expression is also real, as expected for a real correlation function of the Ising model.

Since the sum of the squares of the four-quasihole
conformal blocks is single-valued, we can form a plasma
representation for the sum of overlap integrals:
\be
\label{eq:diag}
\int \prod_{k=1}^{N} d^2 z_k \left[ \mathcal{F}_0(\eta_\mu;z_i) \bar{\mathcal{F}}_0(\bar \eta_\mu; \bar z_i) + \mathcal{F}_1(\eta_\mu; z_i) \bar{\mathcal{F}}_1(\bar \eta_\mu; \bar z_i) \right].
\ee
We cannot do this for each of the individual
terms in this sum. In order to express Eq. \ref{eq:diag}
in terms of a classical plasma, we begin with the Coulomb
gas representation for the conformal blocks of
Eqs.~\rf{eq:consfour2} and \rf{eq:consfour2-psi} [we could have equally well chosen the representations in Eqs.~\rf{eq:consfour1} and \rf{eq:consfour1-psi}, but this choice is more suitable for subsequent generalizations, as we will see later], and multiply them by their complex conjugates.
The conformal block $\mathcal{F}_0(\eta_\mu;z_i)$
is precisely the expression
which we defined in Eq.~\rf{eq:consfour2}: the conformal
block in which the first two $\sigma$s fuse to $I$.
Written explicitly in terms of the vertex operators, this is:
\begin{eqnarray}
\label{eqn:chiral-part-trace1}
\mathcal{F}_0(\eta_\mu;z_i) &=& \VEV{ V_{21}^{10}(\eta_1) V_{21}^{00} (\eta_2) V_{21}^{10}(\eta_3) V_{21}^{00}(\eta_4) \, V_{31}^{20} (z_1) V_{31}^{00}(z_2) \dots V_{31}^{20}(z_{N-1}) V_{31}^{00}(z_{N})} \notag \\
&=& \left< \oint_{C_{\eta_1}} dw_1 \, e^{ i \alpha_{21} \varphi(\eta_1)} e^{i \alpha_- \varphi(w_1)} e^{ i \alpha_{21} \varphi(\eta_2)} \oint_{C_{\eta_3}} dw_2 \, e^{ i \alpha_{21} \varphi(\eta_3)} e^{i \alpha_- \varphi(w_2)} e^{ i \alpha_{21} \varphi(\eta_4)} \right. \notag \\
&& \qquad \qquad \times \oint_{C_{z_1}} dw_3 \, \oint_{C_{z_1}} dw_4 \, e^{ i \alpha_{31} \varphi(z_1)} e^{i \alpha_- \varphi(w_3)} e^{i \alpha_- \varphi(w_4)} e^{ i \alpha_{31} \varphi(z_2)} \, \times \ldots \notag \\
&& \qquad \qquad \left. \ldots \times \oint_{C_{z_{N-1}}} dw_{N+1} \,\oint_{C_{z_{N-1}}} dw_{N+2} \,e^{ i \alpha_{31} \varphi(z_{N-1})} e^{i \alpha_- \varphi(w_{N+1})} e^{i \alpha_- \varphi(w_{N+2})} e^{ i \alpha_{31} \varphi(z_{N})} \right>.
\end{eqnarray}
We multiply this expression by its complex conjugate
\begin{eqnarray}
\label{eqn:chiral-part-trace2}
\bar{\mathcal{F}}_0(\bar \eta_\mu; \bar z_i) &=&
\left< \oint_{C_{\bar{\eta}_1}} d\bar w_1 \, e^{ i \alpha_{21} \varphi(\bar \eta_1)} e^{i \alpha_- \varphi(\bar w_1)} e^{ i \alpha_{21} \varphi(\bar \eta_2)}
\oint_{C_{\bar{\eta}_3}} d\bar w_2 \, e^{ i \alpha_{21} \varphi(\bar \eta_3)} e^{i \alpha_- \varphi(\bar w_2)} e^{ i \alpha_{21} \varphi(\bar \eta_4)} \right. \notag \\
&& \qquad \qquad \times \oint_{C_{\bar{z}_1}} d\bar w_3 \, \oint_{C_{\bar{z}_1}} d\bar w_4 \, e^{ i \alpha_{31} \varphi(\bar z_1)} e^{i \alpha_- \varphi(\bar w_3)} e^{i \alpha_- \varphi(\bar w_4)} e^{ i \alpha_{31} \varphi(\bar z_2)}
\,\times \ldots  \notag \\
&& \qquad \qquad \left. \ldots \times \oint_{C_{\bar{z}_{N-1}}} d\bar w_{N+1} \, \oint_{C_{\bar{z}_{N-1}}} d\bar w_{N+2} \,
e^{ i \alpha_{31} \varphi(\bar z_{N-1})} e^{i \alpha_- \varphi(\bar w_{N+1})} e^{i \alpha_- \varphi(\bar w_{N+2})} e^{ i \alpha_{31} \varphi(\bar z_{N})} \right>.
\end{eqnarray}
In Eqs.~\rf{eqn:chiral-part-trace1} and \rf{eqn:chiral-part-trace2},
there are $N$ coordinates $z_i$ (electrons),
four coordinates $\eta_\mu$ (quasiholes), and $N+2$ coordinates $w_a$
(screening charges). The correlation functions of vertex operators in Eq.~\rf{eqn:chiral-part-trace1} and \rf{eqn:chiral-part-trace2} can be evaluated using Eq.~\rf{eq:corr}:
\begin{eqnarray}
\label{eqn:product-conf-blocks}
\mathcal{F}_0(\eta_\mu;z_i) \bar{\mathcal{F}}_0(\bar \eta_\mu; \bar z_i) &=& \prod_{c=1}^{N+2} \oint dw_c \, \oint d\bar w_c \,
\prod_{a<b}^{N+2}  \left( w_a-w_b \right)^{\frac 3 2} \prod_{a=1}^{N+2} \prod_{\mu=1}^{4} \left( w_a - \eta_\mu \right)^{-\frac 3 4} \prod_{a=1}^{N+2} \prod_{i=1}^{N} \left( w_a-z_i \right)^{-\frac 3 2} \notag \\
&& \qquad \qquad \qquad \qquad \times \prod_{i<j}^{N} \left( z_i-z_j \right)^{\frac 3 2} \prod_{\mu=1}^{4}\prod_{i=1}^{N} \left(\eta_\mu-z_i \right)^{\frac 3 4} \prod_{\mu<\nu}^{4} \left( \eta_\mu-\eta_\nu \right)^{\frac 3 8} \notag \\
&& \qquad \qquad \qquad \qquad \times \prod_{a<b}^{N+2}  \left( \bar w_a- \bar w_b \right)^{\frac 3 2} \prod_{a=1}^{N+2} \prod_{\mu=1}^{4} \left(\bar  w_a - \bar \eta_\mu \right)^{-\frac 3 4} \prod_{a=1}^{N+2} \prod_{i=1}^{N} \left( \bar w_a- \bar z_i \right)^{-\frac 3 2} \notag \\
&& \qquad \qquad \qquad \qquad \times \prod_{i<j}^{N} \left( \bar z_i- \bar z_j \right)^{\frac 3 2} \prod_{\mu=1}^{4}\prod_{i=1}^{N} \left(\bar \eta_\mu- \bar z_i \right)^{\frac 3 4} \prod_{\mu<\nu}^{4} \left(\bar  \eta_\mu- \bar \eta_\nu \right)^{\frac 3 8} .
\end{eqnarray}
In these expressions, the appropriate choice of integration contours (which we left implicit here)
tells us that we are computing $\mathcal{F}_0(\eta_\mu;z_i) \bar{\mathcal{F}}_0(\bar \eta_\mu; \bar z_i)$.
However, by choosing a different contour for one of the screening
charges in Eqs.~\rf{eqn:chiral-part-trace1} and \rf{eqn:chiral-part-trace2},
as per Eq.~\rf{eq:consfour2-psi} (specifically, if the contour $C_2$ corresponding to $w_2$ was a circle of radius $\left|\eta_2\right|$ rather than radius $\left|\eta_3\right|$), we would obtain
$\mathcal{F}_1(\eta_\mu;z_i) \bar{\mathcal{F}}_1(\bar \eta_\mu; \bar z_i)$ instead. To obtain the non-chiral
correlation function, we should add the right-hand-side
of Eq.~\rf{eqn:product-conf-blocks} to the corresponding expression for $\mathcal{F}_1(\eta_\mu;z_i) \bar{\mathcal{F}}_1(\bar \eta_\mu; \bar z_i)$, with these different integration contours.

Instead, following Mathur~\cite{Mathur1992} once again,
we replace the integrations over pairs of contours
by integrations over the plane as described in Appendix
\ref{sec:Mathur}. This replacement gives us neither
$\mathcal{F}_0(\eta_\mu;z_i) \bar{\mathcal{F}}_0(\bar \eta_\mu; \bar z_i)$
nor $\mathcal{F}_1(\eta_\mu;z_i) \bar{\mathcal{F}}_1(\bar \eta_\mu; \bar z_i)$
but, rather, the combination
$\mathcal{F}_0(\eta_\mu;z_i) \bar{\mathcal{F}}_0(\bar \eta_\mu; \bar z_i)+
\mathcal{F}_1(\eta_\mu;z_i) \bar{\mathcal{F}}_1(\bar \eta_\mu; \bar z_i)$. Thus, we obtain
\begin{eqnarray}
\label{eq:pros}
\mathcal{F}_0(\eta_\mu;z_i) \bar{\mathcal{F}}_0(\bar \eta_\mu; \bar z_i)+
\mathcal{F}_1(\eta_\mu;z_i) \bar{\mathcal{F}}_1(\bar \eta_\mu; \bar z_i) &=&
\int \prod_{c=1}^{N+2} d^2 w_c \prod_{a<b}^{N+2}  \left| w_a-w_b \right|^3 \prod_{a=1}^{N+2} \prod_{\mu=1}^{4} \left| w_a - \eta_\mu \right|^{-\frac 3 2} \prod_{a=1}^{N+2} \prod_{i=1}^{N} \left| w_a-z_i \right|^{-3} \notag \\
&& \qquad \qquad \times \prod_{i<j}^{N} \left| z_i-z_j \right|^3 \prod_{\mu=1}^{4} \prod_{i=1}^{N} \left|\eta_\mu-z_i \right|^{\frac 3 2} \prod_{\mu<\nu}^{4} \left| \eta_\mu-\eta_\nu \right|^{\frac 3 4}.
\end{eqnarray}
The reason that the particular combination
$\mathcal{F}_0(\eta_\mu;z_i) \bar{\mathcal{F}}_0(\bar \eta_\mu; \bar z_i)+\mathcal{F}_1(\eta_\mu;z_i) \bar{\mathcal{F}}_1(\bar \eta_\mu; \bar z_i)$
appears on the right-hand-side is, as shown by
Mathur~\cite{Mathur1992}, that when the contour integrals
are replaced by 2D integrals, as described in Appendix
\ref{sec:Mathur}, this has the effect of computing a sum
of holomorphic and antiholomorphic conformal blocks, such that the entire combination is single-valued as a function of all variables.

We now define
\begin{equation}
\label{eq:without}
G^{\mathcal{F}}_{\alpha,\beta}( \bar \eta_\mu , \eta_\mu) \equiv \int \prod_{k=1}^{N} d^2 z_k \bar{\mathcal{F}}_\alpha(\bar \eta_\mu; \bar z_i) \mathcal{F}_\beta ( \eta_\mu;  z_i)  .
\end{equation}
We denote this overlap matrix as $G^{\mathcal{F}}$ to distinguish it from the closely related $G$ defined
in Eq.~\rf{eq:overlap}, which is the overlap matrix of the MR wavefunctions $\Psi_\alpha$.
If we take the integral of Eq.~\rf{eq:pros} over the coordinates $z_i$, we obtain
\begin{eqnarray}
\label{eq:pros2}
\text{Tr} G^{\mathcal{F}} &=& G_{0,0}^{\mathcal{F}} + G_{1,1}^{\mathcal{F}} \notag \\
&=& \int \prod_{k=1}^{N} d^2 z_k \left[ \bar{\mathcal{F}}_0(\bar \eta_\mu; \bar z_i) \mathcal{F}_0(\eta_\mu;z_i) +
\bar{\mathcal{F}}_1(\bar \eta_\mu; \bar z_i) \mathcal{F}_1(\eta_\mu;z_i) \right] \notag \\
&=& \int \prod_{c=1}^{N+2} d^2 w_c \prod_{a<b}^{N+2}  \left| w_a-w_b \right|^3 \prod_{a=1}^{N+2} \prod_{\mu=1}^{4} \left| w_a - \eta_\mu \right|^{-\frac 3 2} \prod_{a=1}^{N+2} \prod_{i=1}^{N} \left| w_a-z_i \right|^{-3} \notag \\
&& \qquad \qquad \qquad \times \prod_{i<j}^{N} \left| z_i-z_j \right|^3 \prod_{\mu=1}^{4} \prod_{i=1}^{N} \left|\eta_\mu-z_i \right|^{\frac 3 2} \prod_{\mu<\nu}^{4} \left| \eta_\mu-\eta_\nu \right|^{\frac 3 4}.
\end{eqnarray}
Comparing with Eq.~\rf{eq:applasma}, we can rewrite this in terms of the partition function of a plasma
\begin{eqnarray}
\label{eq:pl2_4qh}
\text{Tr} G^{\mathcal{F}} &=& \int \prod_{k=1}^{N} d^2 z_k \prod_{c=1}^{N+2} d^2 w_c \, e^{-\Phi_2 /T} = e^{-F_2/T} \\
\label{eq:Phi_2_4qh}
\Phi_2 &=& - \sum_{a < b}^{N+2} Q^2 \log \left| w_a -w_b \right| + \sum_{a=1}^{N+2} \sum_{\mu=1}^{4}
\frac{Q^2}{2} \log \left| w_a - \eta_\mu \right| + \sum_{a=1}^{N+2} \sum_{i=1}^{N}
Q^2 \log \left| w_a - z_i \right|  \notag \\
&& - \sum_{i<j}^{N} Q^2 \log \left| z_i - z_j \right| - \sum_{\mu=1}^{4} \sum_{i=1}^{N} \frac{Q^2}{2} \log \left| \eta_{\mu} - z_i \right| - \sum_{\mu<\nu}^{4} \frac{Q^2}{4} \log \left| \eta_{\mu} - \eta_{\nu} \right|
,
\end{eqnarray}
where, for temperature $T=g$, $\Phi_2$ is the $2$D Coulomb-interaction potential for $N$ particles of charge $Q = \sqrt{3g}$ at positions $z_i$ and $N+2$ particles of charge $-Q$ at positions $w_a$, in the presence of four fixed test particles of charge $Q/2$ at positions $\eta_\mu$. This plasma obeys overall charge neutrality, as can be seen by adding up all the charges. As previously mentioned, it is known that the $2$D two-component classical plasma comprised of particles of opposite charge $\pm Q$ is in the screening fluid phase for $\Gamma = Q^2 /T = 3$, though a short-ranged repulsion (e.g. a hard-core cutoff) is needed. (We discuss the screening properties of this plasma in more detail in Appendix~\ref{sec:two-comp-screening}). Since this plasma screens, the free energy $F_2$ in Eq.~\rf{eq:pl2_4qh} is independent of the
positions $\eta_\mu$, as long as they are farther apart than the screening length $\ell_2$ of the plasma. This proves, for the case of $n=4$ quasiholes, that
\begin{equation}
\label{eq:tracewithout}
{\rm Tr}\, G^{\mathcal{F}} = 2C_2  \, + \, O(e^{-|{\eta_\mu}-{\eta_\nu}|/\ell_2}),
\end{equation}
where $C_2$ is a constant independent of $\eta_\mu$.

We now turn to the overlap matrix $G$ of the MR wavefunctions $\Psi_\alpha$.
The wavefunctions $\Psi_0$ and $\Psi_1$ differ from $\mathcal{F}_0$ and $\mathcal{F}_1$ by an additional $c=1$ correlation function, as is clear from Eqs.~\rf{eq:wave} and \rf{eq:holes}. This additional correlation function is straightforward to calculate [as before, use Eq.~\rf{eq:phiholcor} with $g=1/4$]
\begin{multline}
\label{eqn:4qp-charge-part}
\left< e^{i \frac{1}{2 \sqrt{2M}} \varphi(\eta_1)}e^{i \frac{1}{2 \sqrt{2M}} \varphi(\eta_2)} e^{i \frac{1}{2 \sqrt{2M}} \varphi(\eta_3)} e^{i\frac{1}{2 \sqrt{2M}}\varphi(\eta_{4})} \, e^{i \sqrt{\frac M 2} \varphi(z_1)}  \dots e^{i \sqrt{\frac M 2} \varphi(z_{N})} e^{- i\frac{1}{2 \pi \sqrt{2M}} \int d^2z \, \varphi(z)} \right>  \\
= \prod_{\mu<\nu}^{4} \left( \eta_\mu-\eta_\nu \right)^{\frac{1}{4M}} \prod_{\mu=1}^{4} \prod_{i=1}^{N} \left( \eta_\mu-z_i \right)^{\frac 1 2} \prod_{i<j}^{N} \left( z_i-z_j \right)^{M} e^{-\frac{1}{8M} \sum\limits_{\mu=1}^{4} \left| \eta_\mu \right|^2 -\frac{1}{4} \sum\limits_{i=1}^{N} \left| z_i \right|^2}.
\end{multline}
Consequently, the analog of Eq.~\rf{eq:pros2} for these functions is given by
\begin{eqnarray}
\label{eq:pros1}
\text{Tr} G &=& G_{0,0} + G_{1,1} \notag \\
&=& \int \prod_{k=1}^{N} d^2 z_k \left[ \bar{\Psi}_0(\bar \eta_\mu; \bar z_i) \Psi_0(\eta_\mu;z_i) + \bar{\Psi}_1(\bar \eta_\mu; \bar z_i) \Psi_1(\eta_\mu;z_i) \right] \notag \\
&=& \int \prod_{k=1}^{N} d^2 z_k \int \prod_{c=1}^{N+2} d^2 w_c \prod_{a<b}^{N+2} \left| w_a-w_b \right|^3 \prod_{a=1}^{N+2} \prod_{\mu=1}^{4} \left| w_a - \eta_\mu \right|^{-\frac 3 2} \notag \\
&& \qquad \qquad \qquad \qquad \times \prod_{a=1}^{N+2} \prod_{i=1}^{N} \left| w_a-z_i \right|^{-3} \prod_{i<j}^{N} \left| z_i-z_j \right|^3 \prod_{\mu=1}^{4} \prod_{i=1}^{N} \left|\eta_\mu-z_i \right|^{\frac 3 2} \prod_{\mu<\nu}^{4} \left| \eta_\mu-\eta_\nu \right|^{\frac 3 4}  \notag \\
&& \qquad \qquad \qquad \qquad \qquad \times \prod_{i<j}^{N} \left| z_i-z_j \right|^{2M} \prod_{\mu=1}^{4} \prod_{i=1}^{N} \left| \eta_\mu- z_i \right| \prod_{\mu<\nu}^{4} \left| \eta_\mu-\eta_\nu \right|^{\frac 1 {2M}} e^{-\frac{1}{4M} \sum\limits_{\mu=1}^{4} \left| \eta_\mu \right|^2 - \frac{1}{2} \sum\limits_{i=1}^{N} \left|z_i \right|^2}.
\end{eqnarray}
Similarly, this can be interpreted in terms of a plasma for which there are two independent Coulomb interactions, denoted using subscripts 1 and 2, by rewriting Eq.~\rf{eq:pros1} as
\begin{eqnarray}
\label{eq:pl1_4qh}
\text{Tr} G &=& \int \prod_{k=1}^{N} d^2 z_k \prod_{c=1}^{N+2} d^2 w_c \, e^{-\left(\Phi_1 + \Phi_2 \right) /T} = e^{-F/T} \\
\label{eq:Phi1_4qh}
\Phi_1 &=& - \sum_{\mu < \nu}^{4} \frac{Q_1^2}{4 M^2} \log \left| \eta_\mu -\eta_\nu \right| - \sum_{\mu=1}^{4} \sum_{i=1}^{N}
\frac{Q_1^2}{2M} \log \left| \eta_\mu - z_i \right| - \sum_{i<j}^{N} Q_1^2 \log \left| z_i - z_j \right| \notag \\
&& \qquad + \frac{Q_1^2}{8M^2} \sum_{\mu=1}^{4} \left| \eta_\mu \right|^2 + \frac{Q_1^2}{4M} \sum_{i=1}^{N} \left| z_i \right|^2 \\
\label{eq:Phi2_4qh}
\Phi_2 &=& - \sum_{a < b}^{N+2} Q_2^2 \log \left| w_a -w_b \right| + \sum_{a=1}^{N+2} \sum_{\mu=1}^{4}
\frac{Q_2^2}{2} \log \left| w_a - \eta_\mu \right| + \sum_{a=1}^{N+2} \sum_{i=1}^{N}
Q_2^2 \log \left| w_a - z_i \right|  \notag \\
&& - \sum_{i<j}^{N} Q_2^2 \log \left| z_i - z_j \right| - \sum_{\mu=1}^{4} \sum_{i=1}^{N} \frac{Q_2^2}{2} \log \left| \eta_{\mu} - z_i \right| - \sum_{\mu<\nu}^{4} \frac{Q_2^2}{4} \log \left| \eta_{\mu} - \eta_{\nu} \right|
,
\end{eqnarray}
where, for $T=g$, we have $\Phi_1$ corresponding to a $2$D Coulomb potential for $N$ charge $Q_1 = \sqrt{2Mg}$ particles at $z_i$ and four fixed test particles with charge $Q_1 / 2M = \sqrt{g/2M}$ at $\eta_\mu$, in a uniform neutralizing background of charge density $\rho_1 = -\frac{Q_1}{2 \pi M \ell_B^2}$, and $\Phi_2$ corresponding to a $2$D Coulomb potential for $N$ charge $Q_2 = \sqrt{3g}$ particles at $z_i$, $N+2$ charge $-Q_2$ particles at $w_a$, and four fixed test particles with charge $Q_2 / 2$ at $\eta_\mu$. Hence, this plasma consists of $N$ particles (corresponding to the electrons) at $z_i$ which carry charges $Q_1$ and $Q_2$, $N+2$ particles (screening operators) at $w_a$ which carry charges $-Q_2$, four fixed test charges (quasiholes) at $\eta_\mu$ which carry charges $Q_1 / 2M$ and $Q_2 / 2$, and a uniform neutralizing background of charge density $\rho_1 = -\frac{Q_1}{2 \pi M \ell_B^2}$ (and $\rho_2 =0$). As previously mentioned, we expect such a plasma to be in the screening phase for roughly $T> T_{c_1},T_{c_2}$, where $T_{c_1} = Q_1^2 / 140$ and $T_{c_2} = Q_2^2 / 4$ are the critical temperatures above which plasmas 1 and 2 are individually in their screening fluid phase. Therefore, this plasma at temperature $T=g$ with $Q_1 = \sqrt{2Mg}$ and $Q_2 = \sqrt{3g}$ should be in the screening phase for $M \lesssim 70$. This has been numerically confirmed~\cite{Herland-unpublished} for $M=2$. When the plasma is in the screening phase, the free energy $F$ will not depend on the positions of the test charges $\eta_\mu$, as long as their separations are larger than the screening length $\ell$ of the combined plasma.

Thus, for sufficiently small $M$, we have proved that the trace of the overlap matrix $G$ [defined in Eq.~\rf{eq:overlap}] for $n=4$ quasiholes is an $\eta_\mu$-independent constant for large separations, or
\begin{equation}
\label{eq:norm3}
\text{Tr} G = \sum_{\alpha=0,1} \int \prod_{k=1}^{N} d^2 z_k \bar{\Psi}_\alpha (\bar \eta_\mu; \bar z_i) \Psi_\alpha (\eta_\mu;z_i) = 2C \, + \, O(e^{-|{\eta_\mu}-{\eta_\nu}|/\ell}).
\end{equation}
Hence, we have established that both the trace of $G$, which includes the charge sector as in Eq.~\rf{eq:pros1}, as well as the trace of $G^{\mathcal{F}}$, without the charge sector, as in Eq.~\rf{eq:pros2}, are constants.

The preceding derivation can be generalized
to an arbitrary even number $n$ of quasiholes,
for which the formulas analogous to Eqs.~\rf{eq:pl2_4qh}, \rf{eq:pl1_4qh}, \rf{eq:Phi1_4qh}, \rf{eq:Phi2_4qh} are
\begin{eqnarray}
\label{eq:arb-qhs-plasma2}
\text{Tr} G^{\mathcal{F}} &=& \int \prod_{k=1}^{N} d^2 z_k \sum_{\alpha=0}^{q-1} \bar{ \mathcal{F}}_{\alpha}(\bar \eta_\mu; \bar z_i) \mathcal{F}_{\alpha}(\eta_\mu;z_i) = \int \prod_{k=1}^{N} d^2 z_k \prod_{c=1}^{N+\frac{n}{2}} d^2 w_c \, e^{- \Phi_2  /T} = e^{-F_2/T} \\
\label{eq:arb-qhs-plasma}
\text{Tr} G &=& \int \prod_{k=1}^{N} d^2 z_k \sum_{\alpha=0}^{q-1} \bar \Psi_{\alpha}(\bar \eta_\mu; \bar z_i) \Psi_{\alpha}(\eta_\mu;z_i) = \int \prod_{k=1}^{N} d^2 z_k \prod_{c=1}^{N+\frac{n}{2}} d^2 w_c \, e^{-\left(\Phi_1 + \Phi_2 \right) /T} = e^{-F/T} \\
\label{eq:Phi1_arb_qhs}
\Phi_1 &=& - \sum_{\mu < \nu}^{n} \frac{Q_1^2}{4 M^2} \log \left| \eta_\mu -\eta_\nu \right| - \sum_{\mu=1}^{n} \sum_{i=1}^{N}
\frac{Q_1^2}{2M} \log \left| \eta_\mu - z_i \right| - \sum_{i<j}^{N} Q_1^2 \log \left| z_i - z_j \right| \notag \\
&& \qquad + \frac{Q_1^2}{8M^2} \sum_{\mu=1}^{n} \left| \eta_\mu \right|^2 + \frac{Q_1^2}{4M} \sum_{i=1}^{N} \left| z_i \right|^2
\end{eqnarray}
\begin{eqnarray}
\label{eq:Phi2_arb_qhs}
\Phi_2 &=& - \sum_{a < b}^{N+\frac{n}{2}} Q_2^2 \log \left| w_a -w_b \right| + \sum_{a=1}^{N+\frac{n}{2}} \sum_{\mu=1}^{n}
\frac{Q_2^2}{2} \log \left| w_a - \eta_\mu \right| + \sum_{a=1}^{N+\frac{n}{2}} \sum_{i=1}^{N}
Q_2^2 \log \left| w_a - z_i \right|  \notag \\
&& - \sum_{i<j}^{N} Q_2^2 \log \left| z_i - z_j \right| - \sum_{\mu=1}^{n} \sum_{i=1}^{N} \frac{Q_2^2}{2} \log \left| \eta_{\mu} - z_i \right| - \sum_{\mu<\nu}^{n} \frac{Q_2^2}{4} \log \left| \eta_{\mu} - \eta_{\nu} \right|
,
\end{eqnarray}
where $q=2^{\frac{n}{2}-1}$. The sum over $\alpha$ can be replaced by a sum over $\pi_j = 0,1$ or $p_j = 0,1$, for $j=1,\ldots,n/2$ with the parity constraint $\pi_{n/2} = \left( \sum_{j} p_j \right)\text{mod }2=0$ (for $N$ even).
Summing the diagonal product of holomorphic and antiholomorphic
conformal blocks over all conformal blocks, we obtain the
single-valued expression on the right-hand-side.

The arguments discussed thus far are a carefully worked-out
version of the arguments presented in Ref.~\onlinecite{Gurarie1997}.
Their main drawback is that they do not prove Eq.~\rf{eq:norm}. They only prove the weaker statements
\begin{eqnarray}
\text{Tr} G^{\mathcal{F}} &=& \sum_{\alpha=0}^{q-1} \int \prod_{k=1}^{N} d^2 z_k \bar{\mathcal{F}}_\alpha (\bar \eta_\mu; \bar z_i) \mathcal{F}_\alpha (\eta_\mu;z_i)  = q C_2 \, + \, O(e^{-|{\eta_\mu}-{\eta_\nu}|/\ell_2}) \\
\text{Tr} G &=& \sum_{\alpha=0}^{q-1} \int \prod_{k=1}^{N} d^2 z_k \bar{\Psi}_\alpha (\bar \eta_\mu; \bar z_i) \Psi_\alpha (\eta_\mu;z_i)  = q C \, + \, O(e^{-|{\eta_\mu}-{\eta_\nu}|/\ell})
\end{eqnarray}
which is necessary but not sufficient, with one exception,
for the (nontrivial part of) Berry's connections Eq.~\rf{eq:Berry} to vanish.
In the next section we extend the proof to show that the stronger
statement Eq.~\rf{eq:norm} is true.

The noted exception is the case of two quasiholes.
Since there is only a single conformal block in this case,
equal to either of the Coulomb gas expressions
\begin{eqnarray}
\mathcal{F}_{0}(\eta_\mu; z_i) &=&
\VEV{ V_{21}^{10}(\eta_1) V_{21}^{00} (\eta_2)
 \, V_{31}^{20} (z_1) V_{31}^{00}(z_2) \dots V_{31}^{20}(z_{N-1})
 V_{31}^{00}(z_{N})} \cr
&=&
\VEV{ V_{22}^{00}(\eta_1) V_{21}^{00} (\eta_2)  \, V_{31}^{20} (z_1) V_{31}^{00}(z_2) \dots V_{31}^{20}(z_{N-1}) V_{31}^{00}(z_{N})},
\end{eqnarray}
the overlap matrix is a $1\times 1$ matrix which is equal to
its trace, and so, for the $n=2$ quasiholes case, we have
\begin{eqnarray}
G_{0,0}^{\mathcal{F}} &=& \int \prod_{k=1}^{N} d^2 z_k \bar{\mathcal{F}}_0 (\bar \eta_\mu; \bar z_i) \mathcal{F}_0 (\eta_\mu;z_i)  = C_2 \, + \, O(e^{-|{\eta_\mu}-{\eta_\nu}|/\ell_2}) \\
G_{0,0} &=& \int \prod_{k=1}^{N} d^2 z_k \bar{\Psi}_0 (\bar \eta_\mu; \bar z_i) \Psi_0 (\eta_\mu;z_i)  = C \, + \, O(e^{-|{\eta_\mu}-{\eta_\nu}|/\ell}).
\label{eq:plasma2qhI}
\end{eqnarray}
Thus, as concluded in Ref.~\onlinecite{Gurarie1997} by
the same logic, when there are only two quasiholes,
the effect of a counterclockwise exchange is the accrual
of a statistical phase $\exp \left[ i \pi \left(\frac{1}{4M}-\frac{1}{8}\right) \right]$ and an Aharonov-Bohm phase $\exp\left(-i \frac{e}{2M} \frac{BA}{\hbar c} \right)$. In fact, we should be more precise:
This is the phase that is accrued when there is an even number
of electrons in the system. When there is an odd number of electrons, one can repeat these steps using
\begin{eqnarray}
\mathcal{F}_{1}(\eta_\mu; z_i) &=&
\VEV{ V_{21}^{10}(\eta_1) V_{21}^{10} (\eta_2)
 \, V_{31}^{00} (z_1) V_{31}^{20}(z_2) V_{31}^{00}(z_3) \dots V_{31}^{20}(z_{N-1})
 V_{31}^{00}(z_{N})} \cr
&=&
\VEV{ V_{22}^{00}(\eta_1) V_{21}^{10} (\eta_2)  \, V_{31}^{00} (z_1) V_{31}^{20}(z_2) V_{31}^{00}(z_3) \dots V_{31}^{20}(z_{N-1}) V_{31}^{00}(z_{N})},
\end{eqnarray}
which has the two quasiholes fusing into the $\psi$ channel. The overlap matrix is again a $1\times 1$ matrix which is equal to
its trace, but in this case it gives:
\begin{eqnarray}
G_{1,1}^{\mathcal{F}} &=& \int \prod_{k=1}^{N} d^2 z_k \bar{\mathcal{F}}_1 (\bar \eta_\mu; \bar z_i) \mathcal{F}_1 (\eta_\mu;z_i)  = C_2 \, + \, O(e^{-|{\eta_\mu}-{\eta_\nu}|/\ell_2}) \\
G_{1,1} &=& \int \prod_{k=1}^{N} d^2 z_k \bar{\Psi}_1 (\bar \eta_\mu; \bar z_i) \Psi_1 (\eta_\mu;z_i)  = C \, + \, O(e^{-|{\eta_\mu}-{\eta_\nu}|/\ell}).
\label{eq:plasma2qhpsi}
\end{eqnarray}
(We note that the number of screening charge coordinates $w_a$ is $N+1$ for both $N$ even and odd.) Thus, when there are only two quasiholes and an odd number of electrons, the effect of a counterclockwise exchange is the accrual
of a statistical phase $\exp\left[ i \pi \left( \frac{1}{4M}+\frac{3}{8}\right) \right]$ and an Aharonov-Bohm phase $\exp\left( -i \frac{e}{2M} \frac{BA}{\hbar c} \right)$. This difference in the resulting phase is an indication of non-Abelian braiding statistics, specifically due to the fact that the two quasiholes must be in different fusion channels $I$ and $\psi$ when $N$ is even and odd, respectively. This is is discussed further in Section~\ref{sec:many-quasiparticles}.

\section{Plasma representation of the overlap matrix for four quasiholes}
\label{sec:six}

To represent all the entries of the overlap matrix as a plasma, we need to find Coulomb gas representations for arbitrary products of conformal blocks. While it does not seem possible to find such a representation for an arbitrary product $\bar{\mathcal{F}}_\alpha \mathcal{F}_\beta$ (see Appedix~\ref{sec:direct} for an incomplete approach), it turns out to be possible to do so for
particular linear combinations. These combinations, in turn, are nothing but the correlation functions of the order and disorder operators in the Ising model. The
disorder operator $\mu(\eta, \bar \eta)$ in the Ising model has the same scaling properties as the order operator $\sigma(\eta, \bar \eta)$,
but it changes sign as it is taken around the order operator.
The analog of Eq.~\rf{eq:Ising} for correlation functions of two
disorder operators and two order operators is:
\begin{eqnarray}
\label{eq:threeentries-1}
\VEV{ \mu(\eta_1, \bar \eta_1)  \mu(\eta_2, \bar \eta_2)   \sigma(\eta_3, \bar \eta_3)   \sigma(\eta_4, \bar \eta_4)  \, \epsilon(z_1, \bar z_1) \dots
\epsilon (z_{N}, \bar z_{N}) } &=& \mathcal{F}_0(\eta_\mu; z_i) \bar{\mathcal{F}}_0 (\bar \eta_\mu; \bar z_i) - \mathcal{F}_1 (\eta_\mu; z_i) \bar{\mathcal{F}}_1 (\bar \eta_\mu; \bar z_i), \\
\label{eq:threeentries-2}
\VEV{ \sigma(\eta_1, \bar \eta_1)  \mu(\eta_2, \bar \eta_2)   \sigma(\eta_3, \bar \eta_3)   \mu(\eta_4, \bar \eta_4)  \, \epsilon(z_1, \bar z_1) \dots
\epsilon (z_{N}, \bar z_{N}) } &=& \mathcal{F}_0(\eta_\mu; z_i) \bar{\mathcal{F}}_1 (\bar \eta_\mu; \bar z_i) - \mathcal{F}_1 (\eta_\mu; z_i) \bar{\mathcal{F}}_0 (\bar \eta_\mu; \bar z_i), \\
\label{eq:threeentries-3}
\VEV{ \mu(\eta_1, \bar \eta_1)  \sigma(\eta_2, \bar \eta_2)   \sigma(\eta_3, \bar \eta_3)   \mu(\eta_4, \bar \eta_4)  \, \epsilon(z_1, \bar z_1) \dots
\epsilon (z_{N}, \bar z_{N}) } &=& \mathcal{F}_0(\eta_\mu; z_i) \bar{\mathcal{F}}_1 (\bar \eta_\mu; \bar z_i) + \mathcal{F}_1 (\eta_\mu; z_i) \bar{\mathcal{F}}_0 (\bar \eta_\mu; \bar z_i).
\end{eqnarray}
The expressions for these correlation functions without the energy operators $\epsilon$ were given in Ref.~\onlinecite{Belavin1984}.
Since the transformation laws of the conformal blocks $\mathcal{F}_\alpha$ are the same as those of the wavefunctions $\Psi_\alpha$ (up to an irrelevant phase),
we can use Eqs.~\rf{eq:op1}, \rf{eq:op2}, \rf{eq:op3} to verify that
the expressions in Eqs.~\rf{eq:threeentries-1}, \rf{eq:threeentries-2}, and \rf{eq:threeentries-3} indeed change sign
when an order operator is taken around a disorder operator.
We can similarly verify that the right-hand-sides of
Eqs.~\rf{eq:threeentries-1}, \rf{eq:threeentries-2}, and \rf{eq:threeentries-3} remain invariant if an order operator is taken
around the other order operator or a disorder operator is taken
around the other disorder operator.

If we can prove that the integrals $\int{\prod_k} {d^2}z_k$
of the three expressions in Eqs.~\rf{eq:threeentries-1}, \rf{eq:threeentries-2}, and \rf{eq:threeentries-3} are equal to zero,
then we will have proved that the overlap matrix
$G^{\mathcal{F}}$ defined in Eq.~\rf{eq:without} is proportional to the
identity matrix, since we would know that
\be
\label{eq:overlapsp}
G^{\mathcal{F}}_{0,0}-G^{\mathcal{F}}_{1,1}=0, {\hskip 0.5 cm}
G^{\mathcal{F}}_{0,1}-G^{\mathcal{F}}_{1,0}=0, {\hskip 0.5 cm} G^{\mathcal{F}}_{0,1}+G^{\mathcal{F}}_{1,0}=0.
\ee
Combined with the already proven identity,
$\sum_\alpha G^{\mathcal{F}}_{\alpha, \alpha}=2C_2$, this would prove that
$G^{\mathcal{F}}_{\alpha,\beta} = C_2 \delta_{\alpha,\beta}$.

To do this, let us construct the Coulomb gas representation for
Eqs.~\rf{eq:threeentries-1}, \rf{eq:threeentries-2}, and \rf{eq:threeentries-3}. To the best of our knowledge,
such a representation has not previously
been constructed in the literature. It is straightforward to
construct it, however, using what we have learned so far.
First, let us consider Eq.~\rf{eq:threeentries-1}.
We take Eqs.~\rf{eq:consfour1}, \rf{eq:consfour1-psi} for the holomorphic part of our representation [unlike Eq.~\rf{eq:pros} where we used the alternative Eqs.~\rf{eq:consfour2}, \rf{eq:consfour2-psi}]:
\begin{eqnarray}
\label{eq:consfour-again}
\mathcal{F}_{0}(\eta_\mu; z_i) &=&
\VEV{ V_{22}^{00}(\eta_1) V_{21}^{00} (\eta_2) V_{21}^{10}(\eta_3) V_{21}^{00}(\eta_4) \, V_{31}^{20} (z_1) V_{31}^{00}(z_2) \dots V_{31}^{20}(z_{N-1}) V_{31}^{00}(z_{N})} , \\
\label{eq:consfour-psi-again}
\mathcal{F}_{1}(\eta_\mu; z_i) &=&
\VEV{ V_{22}^{00}(\eta_1) V_{21}^{10} (\eta_2) V_{21}^{00}(\eta_3) V_{21}^{00}(\eta_4) \, V_{31}^{20} (z_1) V_{31}^{00}(z_2) \dots V_{31}^{20}(z_{N-1}) V_{31}^{00}(z_{N})}.
\end{eqnarray}
For the antiholomorphic part, we take Eq.~\rf{eq:consfour-again}
but with the first two operators exchanged, which gives
\begin{equation}
\label{eq:consfour4a}
\bar{\mathcal{F}}_{0}(\bar{\eta}_\mu; \bar{z}_i) = \VEV{ V_{21}^{00}(\bar \eta_1) V_{22}^{00} (\bar \eta_2) V_{21}^{10}(\bar \eta_3) V_{21}^{00}(\bar \eta_4) \, V_{31}^{20} (\bar z_1) V_{31}^{00}(\bar z_2) \dots V_{31}^{20}(\bar z_{N-1}) V_{31}^{00}(\bar z_{N})}.
\end{equation}
For the antiholomorphic sector's other conformal block, we use this same string of vertex operators, except with the screening charge moved from the $\eta_3$ vertex operator to the $\eta_2$ vertex operator. This gives
\begin{equation}
\label{eq:consfour4b}
-\bar{\mathcal{F}}_{1}(\bar{\eta}_\mu; \bar{z}_i) = \VEV{ V_{21}^{00}(\bar \eta_1) V_{22}^{10} (\bar \eta_2) V_{21}^{00}(\bar \eta_3) V_{21}^{00}(\bar \eta_4) \, V_{31}^{20} (\bar z_1) V_{31}^{00}(\bar z_2) \dots V_{31}^{20}(\bar z_{N-1}) V_{31}^{00}(\bar z_{N})},
\end{equation}
so that combining these different representations gives the correlation function in Eq.~\rf{eq:threeentries-1}.

In other words, we use a representation for the Ising disorder operators in which we use the vertex operators, not including possible screening charges, given by
\begin{equation}
\label{eq:opp1}
\mu(\eta_1,\bar \eta_1) = V_{22}^{00}(\eta_1) V_{21}^{00} (\bar \eta_1) =
e^{- i \frac{1}{4\sqrt{3}} \varphi( \eta_1) + i \frac{\sqrt{3}}4
{\bar \varphi}(\bar \eta_1) }
=e^{ i \frac{1}{4\sqrt{3}} \phi( \eta_1 , \bar{\eta}_1) } \mathcal{O}_{-\frac{1}{2\sqrt{3}g }} ( \eta_1 , \bar{\eta}_1)
\end{equation}
for the first disorder operator (at the position $\eta_1,\bar \eta_1$) and
\begin{equation}
\label{eq:opp2}
\mu(\eta_2,\bar \eta_2) = V_{21}^{00}(\eta_2) V_{22}^{00} (\bar \eta_2) = e^{i \frac{\sqrt{3}}4 \varphi( \eta_2) - i \frac{1}{4\sqrt{3}} {\bar \varphi}(\bar  \eta_2)}
=e^{ i \frac{1}{4\sqrt{3}} \phi( \eta_2 , \bar{\eta}_2) } \mathcal{O}_{\frac{1}{2\sqrt{3}g }} ( \eta_2 , \bar{\eta}_2)
\end{equation}
for the second disorder operator (at the position $\eta_2,\bar \eta_2$). The two charges appearing in Eqs.~\rf{eq:opp1} and \rf{eq:opp2} in the holomorphic and antiholomorphic parts of the vertex operators are the two allowed charges which can represent the Ising spin, as can be seen from the table in Eq.~\rf{eq:table1}.
Such objects (whose left and right charges are distinct from each other)
can be viewed as particles in the 2D plasma which carry
not only electric charge, but also magnetic charge, in the sense of Eqs.~\rf{eq:applasma}, \rf{eq:vert_holo}, \rf{eq:vert_antiholo}
in Appendix~\ref{sec:boson-Coulomb-gas}.

Meanwhile, the Ising order operators without their screening charges are represented by purely electric operators
\be \label{eq:opp3}
\sigma(\eta_{3,4},\bar \eta_{3,4}) =
V_{21}^{00}(\eta_{3,4})V_{21}^{00}(\bar \eta_{3,4}) =
\  e^{i \frac{\sqrt{3}}4 \varphi( \eta_{3,4})+ i\frac{\sqrt{3}}4 {\bar \varphi}( \bar \eta_{3,4})}
\ee
as are the Ising energy operators (without screening charges)
\be
\epsilon(z_{i}, \bar z_{i}) =
V_{31}^{00} (z_{i})V_{31}^{00} (\bar z_{i})
= e^{i \frac{\sqrt{3}}2 \varphi(z_i) + i \frac{\sqrt{3}}2 {\bar \varphi}(\bar z_i)}
\ee
and the screening operators
\be \label{eq:opp5}
e^{i{\alpha_-} \varphi(w)} \, e^{i{\alpha_-}{\bar \varphi}(\bar w)}
 = e^{-i \frac{\sqrt{3}}2 \varphi(w) - i \frac{\sqrt{3}}2 \varphi(\bar w)}.
\ee
In this construction we exclusively use the $\alpha_-=-\sqrt{3}/2$ screening operators. The crucial part of the proposed construction is that almost all
of the operators used here are mutually local,
i.e. are single-valued when any one is taken
around any other. The exception is when an order operator is
taken around a disorder operator (or vice versa), which results in a $-1$.
This is easy to check if one uses
\be e^{i \alpha_r \varphi(z) + i \alpha_l \varphi(\bar z)} \, e^{i \beta_r \varphi(w) + i \beta_l \varphi(\bar w)} \sim (z-w)^{2 \alpha_r \beta_r} (\bar z - \bar w)^{2 \alpha_l \beta_l }.
\ee
Thus, when one is taken around the other, a phase $4 \pi (\alpha_r \beta_r - \alpha_l \beta_l)$ is acquired. For example, if a disorder operator represented
either by $\alpha_r=-\frac{1}{4\sqrt{3}}$,
$\alpha_l= \frac{\sqrt{3}}{4}$ or by $\alpha_r= \frac{\sqrt{3}}{4}$,
$\alpha_l=-\frac{1}{4\sqrt{3}}$ is taken around the order operator represented by
$\beta_r=\frac{\sqrt{3}}4$, $\beta_l=\frac{\sqrt{3}}4$,
this phase is $\pi$. (Similarly, when a disorder operator is taken around an energy operator, represented by $\beta_r=\frac{\sqrt{3}}2$, $\beta_l=\frac{\sqrt{3}}2$, it produces a phase of $2\pi$.)

Of course, we would have obtained the same analytic continuation properties
if we had switched the representations of the first and second
disorder operators, i.e. had taken:
\be
\mu(\eta_1,\bar \eta_1) = V_{21}^{00}(\eta_1) V_{22}^{00} (\bar \eta_1) = e^{i \frac{\sqrt{3}}4 \varphi( \eta_1) - i \frac{1}{4\sqrt{3}} {\bar \varphi}(\bar  \eta_1)}
\ee
for the first disorder operator (at the position $\eta_1,\bar \eta_1$) and
\be
\mu(\eta_2,\bar \eta_2) = V_{22}^{00}(\eta_2) V_{21}^{00} (\bar \eta_2) =
e^{- i \frac{1}{4\sqrt{3}} \varphi( \eta_2) + i \frac{\sqrt{3}}4
{\bar \varphi}(\bar \eta_2) }
\ee
In fact, since this correlation function must be the same if we exchange the
two disorder operators (or exchange the two order operators),
we must take an equal linear combination of both
representations for the disorder operators.
Thus, we conclude that we obtain the correct
analytic continuation properties when operators are taken around other operators
or when identical operators are exchanged if we write
\begin{eqnarray}
\label{eqn:multi-val-comb}
&& \mathcal{F}_0(\eta_\mu; z_i) \bar{\mathcal{F}}_0 (\bar \eta_\mu; \bar z_i) - \mathcal{F}_1 (\eta_\mu; z_i) \bar{\mathcal{F}}_1 (\bar \eta_\mu; \bar z_i)
= \VEV{ \mu(\eta_1, \bar \eta_1)  \mu(\eta_2, \bar \eta_2)   \sigma(\eta_3, \bar \eta_3)   \sigma(\eta_4, \bar \eta_4)  \, \epsilon(z_1, \bar z_1) \dots
\epsilon (z_{N}, \bar z_{N}) } \notag \\
&&  \qquad \qquad =\VEV{ V_{22}^{00}(\eta_1) V_{21}^{00} (\eta_2) V_{21}^{10}(\eta_3) V_{21}^{00}(\eta_4) \, V_{31}^{20} (z_1) V_{31}^{00}(z_2) \dots V_{31}^{20}(z_{N-1}) V_{31}^{00}(z_{N})} \notag \\
&&  \qquad \qquad \qquad \qquad \times \VEV{ V_{21}^{00}(\bar \eta_1) V_{22}^{00} (\bar \eta_2) V_{21}^{10}(\bar \eta_3) V_{21}^{00}(\bar \eta_4) V_{31}^{20} (\bar z_1) V_{31}^{00}(\bar z_2) \dots V_{31}^{20}(\bar z_{N-1}) V_{31}^{00}(\bar z_{N})} \notag \\
&& \qquad \qquad \qquad + \VEV{ V_{22}^{00}(\eta_1) V_{21}^{10} (\eta_2) V_{21}^{00}(\eta_3) V_{21}^{00}(\eta_4) \, V_{31}^{20} (z_1) V_{31}^{00}(z_2) \dots V_{31}^{20}(z_{N-1}) V_{31}^{00}(z_{N})} \notag \\
&&  \qquad \qquad \qquad \qquad \times \VEV{ V_{21}^{00}(\bar \eta_1) V_{22}^{10} (\bar \eta_2) V_{21}^{00}(\bar \eta_3) V_{21}^{00}(\bar \eta_4) V_{31}^{20} (\bar z_1) V_{31}^{00}(\bar z_2) \dots V_{31}^{20}(\bar z_{N-1}) V_{31}^{00}(\bar z_{N})} \notag \\
&&  \qquad \qquad \qquad + \VEV{ V_{21}^{00}(\eta_1) V_{22}^{00} (\eta_2) V_{21}^{10}(\eta_3) V_{21}^{00}(\eta_4) \, V_{31}^{20} (z_1) V_{31}^{00}(z_2) \dots V_{31}^{20}(z_{N-1}) V_{31}^{00}(z_{N})} \notag \\
&& \qquad \qquad \qquad \qquad \times \VEV{ V_{22}^{00}(\bar \eta_1) V_{21}^{00} (\bar \eta_2) V_{21}^{10}(\bar \eta_3) V_{21}^{00}(\bar \eta_4)  V_{31}^{20} (\bar z_1) V_{31}^{00}(\bar z_2) \dots V_{31}^{20}(\bar z_{N-1}) V_{31}^{00}(\bar z_{N})} \notag \\
&&  \qquad \qquad \qquad + \VEV{ V_{21}^{00}(\eta_1) V_{22}^{10} (\eta_2) V_{21}^{00}(\eta_3) V_{21}^{00}(\eta_4) \, V_{31}^{20} (z_1) V_{31}^{00}(z_2) \dots V_{31}^{20}(z_{N-1}) V_{31}^{00}(z_{N})} \notag \\
&& \qquad \qquad \qquad \qquad \times \VEV{ V_{22}^{00}(\bar \eta_1) V_{21}^{10} (\bar \eta_2) V_{21}^{00}(\bar \eta_3) V_{21}^{00}(\bar \eta_4)  V_{31}^{20} (\bar z_1) V_{31}^{00}(\bar z_2) \dots V_{31}^{20}(\bar z_{N-1}) V_{31}^{00}(\bar z_{N})}
.
\end{eqnarray}

If we again pursue Mathur's strategy and
replace the integrations over pairs of contours, such
as $\oint dw_c \oint d\bar w_c$,
by integrations over the plane $\int d^2 w_c$,
we obtain the following expression for Eq.~\rf{eqn:multi-val-comb}.
\begin{eqnarray}
\label{eq:disorderco}
&& \mathcal{F}_0(\eta_\mu; z_i) \bar{\mathcal{F}}_0 (\bar \eta_\mu; \bar z_i) - \mathcal{F}_1 (\eta_\mu; z_i) \bar{\mathcal{F}}_1 (\bar \eta_\mu; \bar z_i)
= \VEV{ \mu(\eta_1, \bar \eta_1)  \mu(\eta_2, \bar \eta_2)   \sigma(\eta_3, \bar \eta_3)   \sigma(\eta_4, \bar \eta_4)  \, \epsilon(z_1, \bar z_1) \dots
\epsilon (z_{N}, \bar z_{N}) } \notag \\
&& \qquad \qquad = \int \prod_{c=1}^{N+1} d^2 w_c \, \left|\eta_1-\eta_2\right|^{-\frac 1 4} \, \prod_{\mu=3}^{4} \left[ \left( \eta_1 -\eta_\mu \right)^{-\frac 1 8} \left( \bar \eta_1 -\bar \eta_\mu \right)^{\frac 3 8} \left( \eta_2 -\eta_\mu \right)^{\frac 3 8} \left( \bar \eta_2 -\bar \eta_\mu \right)^{-\frac 1 8} \right] \left| \eta_3-\eta_4 \right|^{\frac 3 4}  \notag \\
&& \qquad \qquad \qquad \qquad \times \prod_{a=1}^{N+1} \left[ (\eta_1-w_a)^{\frac 1 4} \left(\bar \eta_1-\bar w_a \right)^{-\frac 3 4}
(\eta_2-w_a)^{-\frac 3 4} \left( \bar \eta_2-\bar w_a \right)^{\frac 1 4} \left| \eta_3-w_a \right|^{-\frac 3 2} \left| \eta_4-w_a \right|^{-\frac 3 2} \right] \notag \\
&& \qquad \qquad \qquad \qquad \qquad \qquad \times \prod_{i=1}^{N} \left[ (\eta_1-z_i)^{-\frac 1 4}
\left(\bar \eta_1-\bar z_i \right)^{\frac 3 4}
(\eta_2-z_i)^{\frac 3 4} \left( \eta_2-z_i \right)^{-\frac 1 4}
 \left| \eta_3-z_i \right|^{\frac  3 2} \left| \eta_4-z_i \right|^{\frac  3 2} \right] \notag \\
&& \qquad \qquad \qquad \qquad \qquad \qquad \qquad \qquad \times \prod_{a<b}^{N+1} \left| w_a-w_b \right|^{3} \prod_{a=1}^{N+1} \prod_{i=1}^{N} \left|w_a-z_i \right|^{-3} \prod_{i<j}^{N} \left| z_i-z_j \right|^{3} + \ {\rm c. c.}
\end{eqnarray}
We emphasize that Mathur's procedure can be applied to this case because the screening charges have trivial monodromy with all other vertex operators, including, in particular, those of the disorder operators. Eq.~\rf{eq:disorderco} is one of the main results of this work.
It is the correlation function of two order and two disorder operators in the Ising model obtained via the Coulomb gas approach.
In order to understand why this expression is
correct, it is helpful to observe, first of all,
that the holomorphic and the antiholomorphic
parts of Eq.~\rf{eq:disorderco} indeed reduce to the
second through fifth lines of Eq.~\rf{eqn:multi-val-comb}
while the holomorphic and the antiholomorphic
parts of the complex conjugate [the ``c.c.'' at the end of
Eq.~\rf{eq:disorderco}] reduce to the sixth through ninth lines
of Eq.~\rf{eqn:multi-val-comb}. (The complex conjugate is also necessary
to make the correlation function symmetric.)
More importantly, one should note that the expression inside the integral in
Eq.~\rf{eq:disorderco} is single valued if any of $w_a, \bar w_a$ is taken around any other variable, or if any $z_i, \bar z_i$ is taken around any other variable, so that the integrals over these variables are well-defined. However, it changes sign if $\eta_1, \bar \eta_1$ is taken around $\eta_3, \bar \eta_3$, as well as if $\eta_1, \bar \eta_1$ is taken around $\eta_4, \bar \eta_4$, if $\eta_2, \bar \eta_2$ is taken around $\eta_3, \bar \eta_3$, or if $\eta_2, \bar \eta_2$ is taken around $\eta_4, \bar \eta_4$. This is exactly as we would expect for the correlation function of Eq.~\rf{eq:threeentries-1}. Thus, when the $d^2 w_c$
integrals are decomposed into sums of products of conformal blocks,
following Ref.~\onlinecite{Mathur1992} as outlined in
Appendix \ref{sec:Mathur} and the discussion following it,
the analytic continuation properties
automatically select the correct combination of conformal blocks.

Similarly, the two correlation functions in Eqs.~\rf{eq:threeentries-2} and \rf{eq:threeentries-3} can be obtained by a simple permutation of the variables $\eta_\mu, \bar \eta_\mu$. Thus, all three correlation functions from Eqs.~\rf{eq:threeentries-1}, \rf{eq:threeentries-2}, and \rf{eq:threeentries-3} can be constructed in this way.

In fact, the analytic continuation properties noted previously are sufficient to
conclude that if Eq.~\rf{eq:disorderco} is nonzero, it is equal to the
correlation function we need to compute.
Thus, if one were simply handed Eq.~\rf{eq:disorderco},
one could verify it without the arguments of
Ref.~\onlinecite{Mathur1992} (although, of course,
one would probably not discover this equation without
Ref.~\onlinecite{Mathur1992}) by appealing to these analytic continuation
properties and showing that the expression
Eq.~\rf{eq:disorderco} is non-zero.
In Appendix~\ref{sec:2-order-disorder} we explicitly
evaluate Eq.~\rf{eq:disorderco} in the absence of the energy operators.
In this case, there is only one screening operator involved, and only one integral over $w$, which we calculate. The result of the evaluation, given in Eq.~\rf{eq:apcans}, explicitly
produces the correct combination of blocks as given in Eq.~\rf{eq:disorderco}.

It further follows that when the energy operators are included, the representation Eq.~\rf{eq:disorderco} cannot simply vanish. Indeed, we can always take the four Ising order and disorder operators far away from the energy operators, and the correlation function factorizes.
It is clearly non-zero in this limit; by analyticity it will remain nonzero
at finite separation between them.

Thus, we have the following expression for the overlap integral of
the difference of the product of conformal blocks, a generalization of Eq.~\rf{eq:pros2},
\begin{eqnarray}
\label{eq:disorder}
&&  G^{\mathcal{F}}_{0,0}-G^{\mathcal{F}}_{1,1}= \int \prod_{k=1}^{N} d^2 z_k \left[ \bar{\mathcal{F}}_0 (\bar \eta_\mu; \bar z_i) \mathcal{F}_0(\eta_\mu; z_i) - \bar{\mathcal{F}}_1 (\bar \eta_\mu; \bar z_i) \mathcal{F}_1 (\eta_\mu; z_i) \right] \cr
&& \quad =
\int \prod_{k=1}^{N} d^2 z_k \prod_{c=1}^{N+1} d^2 w_c \left|\eta_1-\eta_2\right|^{-\frac 1 4} \, \prod_{\mu=3}^{4} \left[ \left( \eta_1 -\eta_\mu \right)^{-\frac 1 8} \left( \bar \eta_1 -\bar \eta_\mu \right)^{\frac 3 8} \left( \eta_2 -\eta_\mu \right)^{\frac 3 8} \left( \bar \eta_2 -\bar \eta_\mu \right)^{-\frac 1 8} \right] \prod_{3\leq \mu < \nu \leq 4} \left| \eta_\mu-\eta_\nu \right|^{\frac 3 4}  \notag \\
&& \qquad \qquad \qquad \qquad \times \prod_{a=1}^{N+1} \left[ (\eta_1-w_a)^{\frac 1 4} \left(\bar \eta_1-\bar w_a \right)^{-\frac 3 4}
(\eta_2-w_a)^{-\frac 3 4} \left( \bar \eta_2-\bar w_a \right)^{\frac 1 4} \prod_{\mu =3}^{4} \left| \eta_\mu-w_a \right|^{-\frac 3 2} \right] \notag \\
&& \qquad \qquad \qquad \qquad \qquad \times \prod_{i=1}^{N} \left[ (\eta_1-z_i)^{-\frac 1 4}
\left(\bar \eta_1-\bar z_i \right)^{\frac 3 4}
(\eta_2-z_i)^{\frac 3 4} \left( \bar{\eta}_2-\bar{z}_i \right)^{-\frac 1 4}
\prod_{\mu =3}^{4}  \left| \eta_\mu-z_i \right|^{\frac  3 2} \right] \notag \\
&& \qquad \qquad \qquad \qquad \qquad \qquad \qquad \times \prod_{a<b}^{N+1} \left| w_a-w_b \right|^{3} \prod_{a=1}^{N+1} \prod_{i=1}^{N} \left|w_a-z_i \right|^{-3} \prod_{i<j}^{N} \left| z_i-z_j \right|^{3} + \ {\rm c. c.}
\end{eqnarray}
This particular expression gives us the integral over $d^2 z_k$ of the correlation function in Eq.~\rf{eq:threeentries-1}. The correlation functions in Eqs.~\rf{eq:threeentries-2} and \rf{eq:threeentries-3} can be obtained by the simple permutations of the $\eta_\mu$ and $\bar \eta_\mu$ coordinates.

Now we reinterpret Eq.~\rf{eq:disorder} as a partition function of a plasma by matching it against Eq.~\rf{eq:applasma} and rewriting it as
\begin{equation}
\label{eq:pl2_4qh_2dis}
G^{\mathcal{F}}_{0,0} - G^{\mathcal{F}}_{1,1} = \int \prod_{k=1}^{N} d^2 z_k \prod_{c=1}^{N+1} d^2 w_c \left[ e^{-\Phi_2 /T} + e^{-\bar{\Phi}_2 /T} \right] = e^{-F_2/T} + e^{-\bar{F}_2/T}
\end{equation}
\begin{eqnarray}
\label{eq:Phi_2_4qh_2dis}
\Phi_2 &=& - \sum_{a < b}^{N+1} Q^2 \log \left| w_a -w_b \right| + \sum_{a=1}^{N+1} \sum_{\mu=3}^{4}
\frac{Q^2}{2} \log \left| w_a - \eta_\mu \right| + \sum_{a=1}^{N+1} \sum_{i=1}^{N}
Q^2 \log \left| w_a - z_i \right|  \notag \\
&& - \sum_{i<j}^{N} Q^2 \log \left| z_i - z_j \right| - \sum_{\mu=3}^{4} \sum_{i=1}^{N} \frac{Q^2}{2} \log \left| \eta_{\mu} - z_i \right| - \sum_{3\leq \mu<\nu \leq 4} \frac{Q^2}{4} \log \left| \eta_{\mu} - \eta_{\nu} \right| \notag \\
&& + \sum_{a=1}^{N+1} \sum_{\mu=1}^{2} \frac{Q^2}{6} \log \left| w_a - \eta_\mu \right| - \sum_{\mu=1}^{2} \sum_{i=1}^{N} \frac{Q^2}{6} \log \left| \eta_{\mu} - z_i \right| - \sum_{\nu=1}^{2} \sum_{\mu=3}^{4} \frac{Q^2}{12} \log \left| \eta_{\mu} - \eta_{\nu} \right| - \left( \frac{Q^2}{36} - g^2 m^2 \right) \log \left| \eta_{1} - \eta_{2} \right| \notag \\
&& +i g \sum_{a=1}^{N+1} Qm \arg \left( \eta_1 - w_a \right) -i g \sum_{i=1}^{N} Qm \arg \left( \eta_1 - z_i \right) -i g \sum_{\mu=3}^{4} \frac{Q m }{2} \arg \left( \eta_1 - \eta_{\mu} \right) \notag \\
&& -i g \sum_{a=1}^{N+1} Qm \arg \left( \eta_2 - w_a \right) +i g \sum_{i=1}^{N} Qm \arg \left( \eta_2 - z_i \right) +i g \sum_{\mu=3}^{4} \frac{Q m }{2} \arg \left( \eta_2 - \eta_{\mu} \right)
,
\end{eqnarray}
where $T=g$, $Q=\sqrt{3g}$, and $m=1/ \sqrt{3g}$. We see that $\Phi_2$ is the $2$D Coulomb interaction potential for $N$ particles with electric charge $Q$ at positions $z_i$, $N+1$ particles of electric charge $-Q$ at positions $w_a$, two particles of electric charge $Q/2$ at positions $\eta_3$ and $\eta_4$, a particle with electric charge $Q/6$ and magnetic charge $m$ at position $\eta_1$, and a particle with electric charge $Q/6$ and magnetic charge $-m$ at position $\eta_2$. Additionally, there is (implicitly) an electric charge of $-Q/3$ at infinity, corresponding to the background charge needed to maintain charge neutrality. Thus, we can interpret $F_2$ as the free energy of a plasma at temperature $T$, comprised of $N$  particles of electric charge $Q$ and $N+1$ particles of electric charge $-Q$, in the presence of two test particles with electric charge $Q/2$ at positions $\eta_3$ and $\eta_4$, and two ``special'' test particles which carry electric charge $Q/6$ and magnetic charge $\pm m$ at positions $\eta_1$ and $\eta_2$, respectively (as well as a test particle of electric charge $-Q/3$ at infinity).

We observe that this plasma screens, just as the plasma described in Eqs.~\rf{eq:pl2_4qh}, \rf{eq:Phi_2_4qh} screened, due to a large number of $\pm Q$ electric charges at $T>T_{c_2} = Q^2 /4$ present in it. However, when there are magnetic, in addition to electric,
test charges that are at at sufficiently large distances from each other in the plasma, the free energy $F_2$ will diverge due to
the confinement of magnetic charges, as explained in
Appendix~\ref{sec:two-comp-screening} and in Eq.~\rf{eq:mag}.
As a result, we find that $G^{\mathcal{F}}_{0,0} - G^{\mathcal{F}}_{1,1}=0$
(up to corrections that vanish exponentially as separations
between $\eta_\mu$ become larger than the screening length),
just as we expected.

Constructing a similar plasma mapping for $G^{\mathcal{F}}_{0,1} \pm G^{\mathcal{F}}_{1,0}$ also results in a corresponding plasma that has test particles carrying magnetic charges. Thus, we can similarly show that $G^{\mathcal{F}}_{0,1} \pm G^{\mathcal{F}}_{1,0} =0$, and, consequently, we can conclude that the overlap integrals for the case of $n=4$ satisfy
\begin{equation}
\label{eq:GF}
G^{\mathcal{F}}_{\alpha, \beta} = C_2 \delta_{\alpha, \beta}
+ O(e^{-|{\eta_\mu}-{\eta_\nu}|/\ell_2}).
\end{equation}

This result is relevant to $2$D chiral $p$-wave superconductors, whose real space
wavefunction is the Pfaffian without the charge sector (Laughlin type factors)~\cite{Read2000}, up to short-range modifications.
Thus, Eq.~\rf{eq:GF} presents another approach to
computing the non-Abelian statistics of this state, distinct
from that of Refs.~\onlinecite{Read2000,Ivanov2001,Stern04,Stone06,Read2008}.

By a simple extension, we can now prove that the matrix of overlap integrals for the full wavefunctions are also proportional to the identity. Indeed, all we need to do is to multiply the integrand of Eq.~\rf{eq:disorder} by the charge sector terms from Eq.~\rf{eqn:4qp-charge-part} to obtain
\begin{eqnarray}
\label{eq:pl1_4qh_2dis}
G_{0,0} - G_{1,1} &=& \int \prod_{k=1}^{N} d^2 z_k \left[ \bar{\Psi}_0 (\bar \eta_\mu; \bar z_i) \Psi_0(\eta_\mu; z_i) - \bar{\Psi}_1 (\bar \eta_\mu; \bar z_i) \Psi_1 (\eta_\mu; z_i) \right] \notag \\
&=& \int \prod_{k=1}^{N} d^2 z_k \prod_{c=1}^{N+1} d^2 w_c  \left[ e^{-\left(\Phi_1 + \Phi_2 \right) /T} + e^{-\left( \bar{\Phi}_1 + \bar{\Phi}_2 \right)/T} \right] = e^{-F/T} + e^{-\bar{F}/T} \\
\label{eq:Phi1_4qh_2dis}
\Phi_1 &=& - \sum_{\mu < \nu}^{4} \frac{Q_1^2}{4 M^2} \log \left| \eta_\mu -\eta_\nu \right| - \sum_{\mu=1}^{4} \sum_{i=1}^{N}
\frac{Q_1^2}{2M} \log \left| \eta_\mu - z_i \right| - \sum_{i<j}^{N} Q_1^2 \log \left| z_i - z_j \right| \notag \\
&& \qquad + \frac{Q_1^2}{8M^2} \sum_{\mu=1}^{4} \left| \eta_\mu \right|^2 + \frac{Q_1^2}{4M} \sum_{i=1}^{N} \left| z_i \right|^2
\end{eqnarray}
\begin{eqnarray}
\label{eq:Phi2_4qh_2dis}
\Phi_2 &=& - \sum_{a < b}^{N+1} Q_2^2 \log \left| w_a -w_b \right| + \sum_{a=1}^{N+1} \sum_{\mu=3}^{4}
\frac{Q_2^2}{2} \log \left| w_a - \eta_\mu \right| + \sum_{a=1}^{N+1} \sum_{i=1}^{N}
Q_2^2 \log \left| w_a - z_i \right|  \notag \\
&& - \sum_{i<j}^{N} Q_2^2 \log \left| z_i - z_j \right| - \sum_{\mu=3}^{4} \sum_{i=1}^{N} \frac{Q_2^2}{2} \log \left| \eta_{\mu} - z_i \right| - \sum_{3\leq \mu<\nu \leq 4} \frac{Q_2^2}{4} \log \left| \eta_{\mu} - \eta_{\nu} \right| \notag \\
&& + \sum_{a=1}^{N+1} \sum_{\mu=1}^{2} \frac{Q_2^2}{6} \log \left| w_a - \eta_\mu \right| - \sum_{\mu=1}^{2} \sum_{i=1}^{N} \frac{Q_2^2}{6} \log \left| \eta_{\mu} - z_i \right| - \sum_{\nu=1}^{2} \sum_{\mu=3}^{4} \frac{Q_2^2}{12} \log \left| \eta_{\mu} - \eta_{\nu} \right| - \left( \frac{Q_2^2}{36} - g^2 m_2^2 \right) \log \left| \eta_{1} - \eta_{2} \right| \notag \\
&& +i g \sum_{a=1}^{N+1} Q_2 m_2 \arg \left( \eta_1 - w_a \right) -i g \sum_{i=1}^{N} Q_2 m_2 \arg \left( \eta_1 - z_i \right) -i g \sum_{\mu=3}^{4} \frac{Q_2 m_2 }{2} \arg \left( \eta_1 - \eta_{\mu} \right) \notag \\
&& -i g \sum_{a=1}^{N+1} Q_2 m_2 \arg \left( \eta_2 - w_a \right) +i g \sum_{i=1}^{N} Q_2 m_2 \arg \left( \eta_2 - z_i \right) +i g \sum_{\mu=3}^{4} \frac{Q_2 m_2 }{2} \arg \left( \eta_2 - \eta_{\mu} \right)
,
\end{eqnarray}
with $T=g$, $Q_1 = \sqrt{2Mg}$, $Q_2 =\sqrt{3g}$, and $m_2 = 1/\sqrt{3g}$. Again, $F$ can be interpreted as the free energy of a $2$D two-component plasma with two independent types of Coulomb interactions (denoted by subscripts $1$ and $2$) comprised of $N$ particles (corresponding to the electrons) at $z_i$ which carry electric charges $Q_1$ and $Q_2$, $N+1$ particles (screening operators) at $w_a$ which carry electric charge $-Q_2$, two fixed test charges (quasiholes) at $\eta_3$ and $\eta_4$ which carry electric charges $Q_1 / 2M$ and $Q_2 / 2$, two fixed test charge (quasiholes) at $\eta_1$ and $\eta_2$ which carry electric charges $Q_1 / 2M$ and $Q_2 / 6$ and magnetic charge $\pm m_2 $, and a uniform neutralizing background of charge density $\rho_1 = -\frac{Q_1}{2 \pi M \ell_B^2}$ (and $\rho_2 =0$, but an electric charge of $-Q_2/3$ at infinity). As previously mentioned, we expect such a plasma to be in the screening phase for $T > T_{c_1},T_{c_2}$, where $T_{c_1} = Q_1^2 / 140$ and $T_{c_2} = Q_2^2 / 4$ are the critical temperatures above which plasmas 1 and 2 are individually in their screening fluid phase, so this plasma at temperature $T=g$ with $Q_1 = \sqrt{2Mg}$ and $Q_2 = \sqrt{3g}$ should be in the screening phase for $M \lesssim 70$. This has been numerically confirmed~\cite{Herland-unpublished} for $M=2$. When the plasma is in the screening phase, the free energy $F$ will diverge due to confinement of magnetic charges, as long as the separation between the magnetic charges (at $\eta_1$ and $\eta_2$) are larger than the screening length $\ell$ of the plasma.

Thus, for sufficiently small $M$, we have now proved that
$G_{0,0} -G_{1,1} = 0$. By the suitable permutation of $\eta_\mu, \bar \eta_\mu$, we similarly obtain
$G_{0,1} \pm G_{1,0}=0$. Combining this with the already proven relation $G_{0,0}+G_{1,1} = 2C$,
we have finally proven Eq.~\rf{eq:norm}
\begin{equation}
G_{\alpha, \beta} = C \delta_{\alpha, \beta} + O(e^{-|{\eta_\mu}-{\eta_\nu}|/\ell})
\end{equation}
for the four-quasihole wavefunctions.

\section{$n$-Quasiparticle Fusion and Braiding}
\label{sec:many-quasiparticles}

We would now like to go beyond the four quasihole case considered so far. At first glance, it might appear that going beyond four quasiholes is easy. All we need is to consider correlation functions of a larger number of order and disorder operators. Indeed, consider a wavefunction with $N$ electrons and $n$ quasiholes. As we know, there are $2^{\frac{n}{2}-1}$ such wavefunctions. Therefore, there are  ${2^{\frac{n}{2}-1}}\cdot 2^{\frac{n}{2}-1}  = 2^{n-2}$ overlap integrals [the number of entries in the  $G_{\alpha, \beta}$ overlap matrix of Eq.~\rf{eq:overlap}].
On the other hand, we can imagine computing the correlation function of $n$ Ising spins (as well as $N$ Ising energy operators), as well as correlations with $n-2$ order and $2$ disorder operators, $n-4$ order and $4$ disorder operators and so on. If $n/2$ is odd, we should stop at $\frac{n}{2}-1$ disorder and $\frac{n}{2}+1$ order operators. If $n/2$ is even, then we should stop at $n/2$ disorder and $n/2$ order operators. The total number of such correlation functions (which also depend on which operators are chosen to be order, and which are chosen to be disorder) is
\be
\sum_{k=0}^{(\frac{n}{2}-1)/2} \frac{n!}{(n-2k)! (2k)!} = 2^{n-2},
\ee
if $n/2$ is odd and
\be
\sum_{k=0}^{(\frac{n}{2}-2)/2} \frac{n!}{(n-2k)! (2k)!} + \oh \cdot \frac{n!}{\frac{n}{2}! \frac{n}{2}!} = 2^{n-2}
\ee
if $n/2$ is even. Either way, the total number of combinations of conformal blocks one can get in this way (a generalization of Eq.~\rf{eq:threeentries-1}-\rf{eq:threeentries-3} to $n$ $\sigma$-operators) is exactly equal to the number of entries in the $G_{\alpha, \beta}$
matrix, so computing the integrals over $d^2z_k$ via the plasma construction would allow us to deduce what every entry of $G_{\alpha, \beta}$ is, for
the general case of $n$ quasiholes.

Suppose we pursue the strategy of the previous sections
to compute $G_{\alpha, \beta}$ for an arbitrary number of
quasiholes. Eq.~\rf{eq:pros} is easy to generalize to an arbitrary number $n$ of quasiholes. Replacing the four quasihole operators in that expression with $n$ of these operators resulted in Eqs.~\rf{eq:arb-qhs-plasma2}, \rf{eq:arb-qhs-plasma}, \rf{eq:Phi1_arb_qhs}, \rf{eq:Phi2_arb_qhs}, which led us to conclude that
\begin{eqnarray}
\text{Tr} G^{\mathcal{F}} &=& q C_2 \, + \, O(e^{-|{\eta_\mu}-{\eta_\nu}|/\ell_2}) \\
\text{Tr} G &=& q C \, + \, O(e^{-|{\eta_\mu}-{\eta_\nu}|/\ell})
,
\end{eqnarray}
as discussed in Ref.~\onlinecite{Gurarie1997}.
Likewise, Eqs.~\rf{eq:disorder}, \rf{eq:pl2_4qh_2dis}, \rf{eq:Phi_2_4qh_2dis}, \rf{eq:pl1_4qh_2dis}, \rf{eq:Phi1_4qh_2dis}, \rf{eq:Phi2_4qh_2dis}
can also be generalized to $n$ quasiholes by extending
the product or sum over $3\leq \mu < \nu \leq 4$ to $3 \le \mu < \nu \le n$
and the products or sums over $k=3,4$ to $k=3,\ldots, n$.
However, this only corresponds to the case of
$n-2$ order and $2$ disorder operators. There
are $n \choose 2$ such correlation functions in addition to the correlation function with $n$ order operators.
However, for $n=2$ quasiholes, the correlation function
with $2$ order operators is equal to the correlation function
with $2$ disorder operators by Kramers-Wannier duality.
Similarly, for $n=4$ quasiholes, Kramers-Wannier duality reduces the number
of distinct correlation functions with $2$ order and $2$ disorder
operators from ${4 \choose 2} = 6$ to $3$,
as we have seen in the previous section.
For higher $n$, Kramers-Wannier duality relates distinct
correlation functions and therefore does not lead
to any such reduction.

For $n=2$, we have a plasma representation
for the only non-trivial correlation function
and $G_{\alpha, \beta}$ is a $1\times 1$ matrix,
so we can compute it. For $n=4$, we have a plasma representation
for the $4$ non-trivial correlation functions
and $G_{\alpha, \beta}$ is a $2\times 2$ matrix,
so we can compute all of its entries.
For $n=6$, we have a plasma representation
for the $1+ {6 \choose 2} = 16$ non-trivial correlation functions
and $G_{\alpha, \beta}$ is a $4\times 4$ matrix,
so we can compute all of its entries.
However, for $n=8$, we have a plasma representation
for the $1+ {8 \choose 2} = 29$ non-trivial correlation functions
but $G_{\alpha, \beta}$ is an $8\times 8$ matrix,
so we can compute fewer than half of its entries.
The situation gets worse with increasing $n$ since
$2^{n-2}$ grows much faster than $1+ {n \choose 2}$.

In order to compute all of the entries in $G_{\alpha, \beta}$,
we clearly need to be able to compute correlation
functions in which there are more than two order and more than two
disorder operators.
Unfortunately, we do not know how to generalize Eq.~\rf{eq:disorder} to more than two Ising disorder operators, so we cannot
compute $G_{\alpha, \beta}$ by this strategy.
The problem is that we use operators with opposite magnetic charges, Eqs.~\rf{eq:opp1} and \rf{eq:opp2}, to represent two disorder operators. If we have more than two disorder operators, then we need more than one operator of each type, Eq.~\rf{eq:opp1} as well as \rf{eq:opp2}. However, Eq.~\rf{eq:opp1} is not local with respect to itself, and neither is Eq.~\rf{eq:opp2}. This prevents us from using Eq.~\rf{eq:opp1} or \rf{eq:opp2} more than once in any correlation function.

Another approach is to separate the screening operators into ones that are applied to $\psi$ vertex operators and ones that are applied to $\sigma$ vertex operators, and then attempt to apply a procedure like Mathur's to change holomorphic-antiholomorphic pairs of $\psi$ screening operator contour integrals into $2$D integrals over the complex plane. If this could be done, then the result before performing the $\sigma$ screening operator contour integrations would be a plasma with test particles. Among these test particles are ones corresponding to the holomorphic and antiholomorphic $\sigma$ vertex operators' screening operators, which, in addition to carrying electric charge $-Q_2/2 = - \sqrt{3g/4}$, respectively carry magnetic charge $\mp 3 m_2/2 = \mp \sqrt{3/4g}$. Thus, if the holomorphic and antiholomorphic $\sigma$ screening operators are not paired up properly, they will give a vanishing result because of confinement of magnetic charge. This would give the sought after orthogonality (exponential in the separation of quasiparticles) for arbitrary $n$-quasiparticle wavefunctions, since the different conformal blocks (degenerate wavefunctions) correspond to assigning the screening operators to different $\sigma$ vertex operators. Unfortunately, there is a barrier in this approach to applying a Mathur-style argument, which is that there is a branch cut which prevents the $J$-terms from canceling in a simple way, and we have not managed to push the argument past this barrier. We provide more details on this incomplete approach in Appendix~\ref{sec:direct}.

Fortunately, neither of these approaches is really needed. From the basic underlying
structure of a topological phase, we know that,
given a few basic assumptions (which rest on the
assumption of an energy gap), we can deduce
the braiding statistics of arbitrary numbers of quasiparticles,
given much less information about quasiparticle statistics.
This will be made precise in the following subsections.

\subsection{Braiding in the ``Qubit Basis''}
\label{sec:qubit-basis}

In this section, we will use some features special to
the MR Pfaffian state to deduce the braiding
properties of an arbitrary number of quasiholes,
given the Berry's matrices which can be computed from
the $2$-, $4$-, and $6$-quasihole wavefunctions by
the methods described earlier. We will assume that
the system is governed by a $3$-body Hamiltonian
with pinning potentials as in Eq.~\rf{eqn:3-body+pinning}.
The only assumption we make is that there is a
gap between the degenerate set of ground states with
$n$ quasiparticles at fixed positions (determined by
the pinning potentials) and all higher excited states.
So long as the gap remains open, the braiding properties
that we discuss cannot change if the Hamiltonian
is modified from Eq.~\rf{eqn:3-body+pinning} to something
more realistic.

From Ref.~\onlinecite{Nayak1996}, we know that when there
are $n$ fundamental (charge $e/2M$)
quasiholes at fixed positions, there is
a degenerate set of states, rather than a unique state, and
the following is a basis of these states:
\begin{multline}
\Psi_{(1+{r^{}_1},3+{r^{}_2},\ldots,n-1+r^{}_{\frac{n}{2}})(2-{r^{}_1},4-{r^{}_2},\ldots,n-r^{}_{\frac{n}{2}})}
\equiv
\\
\mbox{${\rm  Pf}\!\left( \frac{(\eta^{}_{1+{r^{}_1}} - z_i ) (\eta^{}_{3+{r^{}_2}} -z_i)
\ldots ( \eta^{}_{n-1+r^{}_{\frac{n}{2}}} - z_i) \,( \eta^{}_{2-{r^{}_1}} - z_j) ( \eta^{}_{4-{r^{}_2}} -z_j )\ldots
( \eta^{}_{n-r_{\frac{n}{2}}} -z_j) + (i \leftrightarrow j )}
{z_i - z_j}\right)$}
\prod_{i<j}^{N} (z_i - z_j)^M \,
e^{  - \frac{1}{4} \sum\limits_{i=1}^{N} \left| z_i \right|^2}
\end{multline}
where ${r^{}_j}=0,1$. This double-counts the number of states, since these wavefunctions are invariant under the interchange of indices: $(1+{r^{}_1},3+{r^{}_2},\ldots,n-1+r^{}_{\frac{n}{2}}) \leftrightarrow (2-{r^{}_1},4-{r^{}_2},\ldots,n-r^{}_{\frac{n}{2}})$. Thus, there are $2^{\frac{n}{2}-1}$
such states. One can think of these states as each pair $(1,2)$,
$(3,4)$, $\ldots$, $(n-1,n)$ of quasiholes being a two-state system, i.e. a qubit,
with an overall parity constraint on the $n/2$ qubits.
For the sake of concreteness, we will fix this parity constraint
by taking $r^{}_{\frac{n}{2}}=0$.
Note that this is a special feature of this particular topological
phase; in a generic non-Abelian topological phase,
the $n$-particle Hilbert space will not decompose into
such a tensor product of two-state systems.
Each of these two-state systems can be measured
in the following way when the corresponding pair of quasiparticles
is far from all of the others. Suppose you
want to know if $r_j$ is $0$ or $1$.
Take $\eta_{2j-1}$ to $\eta_{n-1}$. If
$\Psi_{(1+{r^{}_1},3+{r^{}_2},\ldots,n-1)(2-{r^{}_1},4-{r^{}_2},\ldots,n)}$ now
vanishes when any $z_i$ approaches $\eta_{2j-1}=\eta_{n-1}$,
then $r_j = 1$. If, instead, we take $\eta_{2j}$ to $\eta_{n-1}$
and $\Psi_{(1+{r^{}_1},3+{r^{}_2},\ldots,n-1)(2-{r^{}_1},4-{r^{}_2},\ldots,n)}$
now vanishes when any $z_i$ approaches $\eta_{2j}=\eta_{n-1}$,
then $r_j = 0$.

Consider, in this basis, the effect of a braid
group generator $\tau_{2i-1}$,
which executes a counterclockwise
exchange of quasiholes $2i-1$
and $2i$.
This can be done with all of the other particles far away. Since the
state in each of those other two-level systems can be measured
while keeping those pairs far away (as described earlier), exchanging
quasiparticle $2i-1$ and $2i$ must, by locality (which follows
from the assumption of a gap), act as the identity within
each of those two-dimensional vector spaces.
Therefore, it must be of the form:
\begin{equation}
\label{eqn:t_2i-1}
\tau_{2i-1} = \openone_2 \otimes \ldots \otimes \openone_2  \otimes
{B_2}  \otimes  \openone_2 \otimes \ldots \otimes \openone_2
\end{equation}
so that it only acts non-trivially on the $i^{\rm th}$ pair.
Thus, the computation of $\tau_{2i-1}$ reduces
to the computation of $B_2$. As we discuss in the
next subsection, the eigenvalues of $B_2$
are the numbers $R^{\sigma_1 \sigma_1}_{I_2}$ and $R^{\sigma_1 \sigma_1}_{\psi_2}$.

The braid group generator $\tau_{2i}$ affects pairs $(2i-1,2i)$ and $(2i+1,2i+2)$
and must, therefore, take the form
\begin{equation}
\label{eqn:t_2i}
\tau_{2i} = \openone_2 \otimes \ldots \otimes \openone_2  \otimes
{B_4}  \otimes  \openone_2 \otimes \ldots \otimes \openone_2
\end{equation}
so that it only acts non-trivially on the $i^{\rm th}$ and $(i+1)^{th}$ pair. Once again, by locality, we can ignore all of the other
quasiparticle pairs.

Indeed, locality further guarantees that $B_2$ and
$B_4$ cannot depend on the number of other
quasiparticle pairs in the system (since all of the other
pairs can be taken far away), so long as there
are enough so that there is, indeed, a two-state
system on which $B_2$ can act and a four-state system
on which $B_4$ can act. If there is only a single
pair of quasiholes, then there is a unique state,
so at least four quasiparticles are needed in order to
compute $B_2$. The cognoscenti may object at this point
by noting that there is a second state of two quasiholes,
namely the state with an odd number of electrons.
However, these states do not lie within the same
Hilbert space, since they require their wavefunctions to have different electron number. In order to show that $B_2$ can be computed
from the combination of a $N$ even electron number computation
and a $N$ odd electron number computation, we need to use
more powerful consistency arguments, which
are discussed within the next section. Put slightly differently,
unless we know in advance that the MR Pfaffian state is
a state of Ising-type anyons (i.e. unless we assume the answer),
there is no reason to assert that the two eigenvalues
of $B_2$ are given by the Berry's matrices of two
quasiholes with an even or odd number of electrons.
Fortunately, we do not need to make any such
assumptions and can compute $B_2$ from
the four-quasihole case (or the six-quasihole case, though this is overkill).
Similarly, $B_4$ can be computed in a system with six
quasiholes, for which there are the needed four
degenerate states and which is the largest number of
quasiholes for which our order-disorder operator strategy
allows us to compute the full Berry's matrix.

Thus, the representation matrices for all of the braid group generators
can be obtained from the wavefunctions with four and six quasiholes
and, therefore, the representation of the entire braid group can be obtained. (As mentioned earlier, if we use
consistency conditions more effectively, we can reduce this
to two- and four-quasihole wavefunctions, as discussed
in the next section.)

In order to actually compute the desired matrices
$B_2$ and $B_4$, we need to go into the
qubit basis defined by the appropriate
conformal blocks in the $c=1/2+1$ theory,
which are computed in Appendix
\ref{sec:many-qh-wavefunctions}:
\begin{eqnarray}
\Psi^{}_{({p^{}_1},{p^{}_2},\ldots,{p^{}_{n/2}})} &=&
\left(\frac{\prod\limits_{i<j}^{n/2} \eta^{}_{2i-1,2j-1}\, \eta^{}_{2i,2j}}
{\prod\limits_{i,j}^{n/2} \eta^{}_{2i-1,2j}}\right)^{\frac{1}{8}} \, \left\{
\sum_{{r^{}_i}=0,1} (-1)^{r\cdot p}
\prod_{k<l}^{n/2} x^{|{r^{}_k}-{r^{}_l}|/2}_{k,l}
\right\}^{-1/2} \notag \\
&& \qquad \times \left\{
\sum_{{r^{}_i}=0,1} (-1)^{r\cdot p}
\prod_{k<l}^{n/2} x^{|{r^{}_k}-{r^{}_l}|/2}_{k,l} \,\,
\Psi_{(1+{r^{}_1},3+{r^{}_2},\ldots)(2-{r^{}_1},4-{r^{}_2},\ldots)}
\right\} \prod_{\mu<\nu}^{n} \eta_{\mu \nu}^{\frac{1}{4M}} e^{- \frac{1}{8M} \sum\limits_{\mu=1}^{n} \left| \eta_\mu \right|^2 }
\end{eqnarray}
On the left-hand-side of this expression,
the indices take the values $p_i=0,1$, respectively, and obey the overall parity constraint that $\sum_{i} p_i$ be even.
The preceding arguments hold for this basis as well,
for which, given the plasma screening arguments for the four-
and six-quasiparticle wavefunctions, the braiding matrices can be computed from analytic continuation to be:
\begin{equation}
{B_2} = e^{i \pi \left( \frac{1}{4M} -\frac{1}{8} \right)} \left[\begin{matrix}
1 &0 \cr
0 & i
\end{matrix}\right]
\qquad \text{ and } \qquad
{B_4} = \frac{e^{i \pi \left( \frac{1}{4M} + \frac{1}{8} \right)}}{\sqrt{2}}\left[\begin{matrix}
 1 & 0 & 0 &  -i \\
  0 & 1 &  -i & 0 \\
  0 & -i & 1 &  0   \\
   -i & 0 & 0 & 1
\end{matrix}\right]
.
\end{equation}

Alternatively, if one used the standard basis, one would similarly have (for $N$ even)
\begin{eqnarray}
\tau_{1} &=& {B_2} \otimes \openone_2 \otimes \ldots \otimes \openone_2 \\
\tau_{2i} &=& \openone_2 \otimes \ldots \otimes \openone_2  \otimes
{B_2^{\prime}} \otimes  \openone_2 \otimes \ldots \otimes \openone_2 \\
\tau_{2i-1} &=& \openone_2 \otimes \ldots \otimes \openone_2  \otimes
{B_4^{\prime}} \otimes  \openone_2 \otimes \ldots \otimes \openone_2 \qquad \text{for } i\neq 1,n/2 \\
\tau_{n-1} &=& \openone_2 \otimes \ldots \otimes \openone_2  \otimes
{B_2}
\end{eqnarray}
with
\begin{equation}
{B_2} = e^{i \pi \left( \frac{1}{4M} -\frac{1}{8} \right)} \left[\begin{matrix}
1 &0 \cr
0 & i
\end{matrix}\right],
\qquad
{B_2^{\prime}} = \frac{e^{i \pi \left( \frac{1}{4M} + \frac{1}{8} \right)}}{\sqrt{2}} \left[\begin{matrix}
1 &-i \cr
-i & 1
\end{matrix}\right],
\quad \text{ and } \quad
{B_4^{\prime}} = e^{i \pi \left( \frac{1}{4M} -\frac{1}{8} \right)} \left[\begin{matrix}
 1 & 0 & 0 & 0 \\
  0 & i &  0 & 0 \\
  0 & 0 & i &  0   \\
  0 & 0 & 0 & 1
\end{matrix}\right]
,
\end{equation}
where again we can use the four-quasiparticle wavefunctions to compute ${B_2}$ and ${B_2^{\prime}}$ (which are, respectively, $B^{(1 \leftrightarrows 2)}$ and $B^{(2 \leftrightarrows 3)}$ of Eq.~\rf{eq:4qhB}), but must use the six-quasiparticle wavefunctions to compute ${B_4^{\prime}}$.

Comparing the braiding generators in the qubit and standard bases, we see that $\tau_{2i-1}$ acts on a two-dimensional subspace in the qubit basis, while $\tau_{2i}$ (as well as $\tau_{1}$ and $\tau_{n-1}$) acts on a two-dimensional subspace in the standard basis. Thus, we can imagine that if we had a way of ensuring {\it a priori} that braiding operations are consistent through changes of basis (which we can, in fact, show, but only after the six-quasiparticle
computation of this subsection), then we would only need to compute braiding transformations on two-dimensional subspaces, and hence could avoid the need to compute anything using wavefunctions with more than four quasiparticles.

\subsection{General Considerations}
\label{sec:general-consid}

In this section, we use more powerful and general
arguments to show that the braiding properties
of anyons can be deduced from the two- and four-
quasiparticle wavefunctions, provided we can
compute the corresponding Berry's matrices.
Although these arguments are
more general in the sense that they do not use any
special properties of the MR Pfaffian state, the assumed locality properties, which follow from the existence of an energy gap, are equivalent to those used in the previous argument.

The long-distance, low-energy properties of quasiparticles (e.g. their braiding statistics) in a topological phase are assumed to
be completely specified by an ``anyon model,'' a.k.a. a unitary braided tensor category~\cite{Turaev94,Bakalov01,Preskill04,Kitaev06a,Bonderson07b}.
An anyon model is characterized by:

\begin{enumerate}

\item A finite set $\mathcal{C}$ of ``topological charges'' $a,b,c,\ldots$, which are conserved quantum numbers specifying the different types of quasiparticle excitations.

\item Fusion rules specifying how these topological charges may combine and split, as described by a commutative and associative fusion algebra
\begin{equation}
a \times b = \sum_{c\in \mathcal{C}} N_{ab}^{c } c
,
\end{equation}
where the integer $N_{ab}^{c}$ indicates the distinct number of ways the charges $a$ and $b$ can combine to produce charge $c$. For simplicity, we will restrict our attention to the case where there are no fusion multiplicities, i.e. $N_{ab}^{c}=0,1$, since this is all that is needed for this paper, but there are more general cases~\footnote{For example, tensor product (fusion) of two adjoint representations of SU(3) give
$8\otimes 8 = 27\oplus10\oplus\overline{10}\oplus8\oplus8\oplus1$, which therefore has $N_{88}^{8}=2$.}. There is a vector space $\mathcal{V}_{c}^{ab}$ with $\text{dim}\mathcal{V}_{c}^{ab} = N_{ab}^{c}$ assigned to each fusion product, and one can represent the basis states of these spaces diagrammatically as
\begin{equation}
\pspicture[shift=-0.65](-0.1,-0.2)(1.5,1.2)
  \small
  \psset{linewidth=0.9pt,linecolor=black,arrowscale=1.5,arrowinset=0.15}
  \psline(0.7,0)(0.7,0.55)
  \psline(0.7,0.55) (0.25,1)
  \psline(0.7,0.55) (1.15,1)	
  \rput[bl]{0}(0.4,0){$c$}
  \rput[br]{0}(1.4,0.8){$b$}
  \rput[bl]{0}(0,0.8){$a$}
  \endpspicture
=\left| a,b;c \right\rangle \in
\mathcal{V}_{c}^{ab}.
\label{eq:ket}
\end{equation}
The anyonic states describing a collection of quasiparticles can be represented by fusion diagrams such as
\begin{equation}
 \pspicture[shift=-0.65](-0.2,-0.4)(2,2)
  \small
  \psset{linewidth=0.9pt,linecolor=black,arrowscale=1.5,arrowinset=0.15}
  \psline(0.0,1.75)(1,0.5)
  \psline(2.0,1.75)(1,0.5)
  \psline(0.4,1.25)(0.8,1.75)
   \rput[bl]{0}(-0.15,1.85){$a_1$}
   \rput[bl]{0}(0.75,1.85){$a_2$}
   \rput[bl]{0}(1.95,1.85){$a_n$}
\rput[bl](1.25,1.85){$\cdots$}
\rput{-45}(0.9,1.05){$\cdots$}
   \rput[bl]{0}(0.25,0.65){$c_2$}
  \psline(1,-0.2)(1,0.5)
\   
   \rput[bl]{0}(1.15,0.0){$c_n$}
  \endpspicture
\end{equation}
where the topological charge $a_j$ of the $j$th quasiparticle is assigned to the $j$th endpoint at the top of the diagram.

\item Associativity of fusion within the fusion state spaces, which is specified by unitary isomorphisms written diagrammatically as
\begin{equation}
  \pspicture[shift=-1.0](0,-0.45)(1.8,1.8)
  \small
  \psset{linewidth=0.9pt,linecolor=black,arrowscale=1.5,arrowinset=0.15}
  \psline(0.2,1.5)(1,0.5)
  \psline(1,0.5)(1,0)
  \psline(1.8,1.5) (1,0.5)
  \psline(0.6,1) (1,1.5)
   \rput[bl]{0}(0.05,1.6){$a$}
   \rput[bl]{0}(0.95,1.6){$b$}
   \rput[bl]{0}(1.75,1.6){${c}$}
   \rput[bl]{0}(0.5,0.5){$e$}
   \rput[bl]{0}(0.9,-0.3){$d$}
  \endpspicture
= \sum_{f \in \mathcal{C} } \left[F_d^{abc}\right]_{ef}
 \pspicture[shift=-1.0](0,-0.45)(1.8,1.8)
  \small
  \psset{linewidth=0.9pt,linecolor=black,arrowscale=1.5,arrowinset=0.15}
  \psline(0.2,1.5)(1,0.5)
  \psline(1,0.5)(1,0)
  \psline(1.8,1.5) (1,0.5)
  \psline(1.4,1) (1,1.5)
   \rput[bl]{0}(0.05,1.6){$a$}
   \rput[bl]{0}(0.95,1.6){$b$}
   \rput[bl]{0}(1.75,1.6){${c}$}
   \rput[bl]{0}(1.25,0.45){$f$}
   \rput[bl]{0}(0.9,-0.3){$d$}
  \endpspicture
.
\end{equation}
The $F$-symbols are analogous to $6j$-symbols, providing a change of basis between the basis in which topological charges $a$ and $b$ are first fused and then their result is fused with $c$ to the basis in which topological charges $b$ and $c$ are first fused and then their result is fused with $a$.

\item Braiding of topological charges enacting unitary transformations on the state space, which are written diagrammatically as
\begin{eqnarray}
\pspicture[shift=-0.65](-0.1,-0.2)(1.5,1.2)
  \small
  \psset{linewidth=0.9pt,linecolor=black,arrowscale=1.5,arrowinset=0.15}
  \psline(0.7,0)(0.7,0.5)
  \psarc(0.8,0.6732051){0.2}{120}{240}
  \psarc(0.6,0.6732051){0.2}{-60}{35}
  \psline (0.6134,0.896410)(0.267,1.09641)
  \psline(0.7,0.846410) (1.1330,1.096410)	
  \rput[bl]{0}(0.4,0){$c$}
  \rput[br]{0}(1.35,0.85){$a$}
  \rput[bl]{0}(0.05,0.85){$b$}
  \endpspicture
= R_{c}^{ab}
\pspicture[shift=-0.65](-0.1,-0.2)(1.5,1.2)
  \small
  \psset{linewidth=0.9pt,linecolor=black,arrowscale=1.5,arrowinset=0.15}
  \psline(0.7,0)(0.7,0.55)
  \psline(0.7,0.55) (0.25,1)
  \psline(0.7,0.55) (1.15,1)	
  \rput[bl]{0}(0.4,0){$c$}
  \rput[br]{0}(1.4,0.8){$a$}
  \rput[bl]{0}(0,0.8){$b$}
  \endpspicture
\end{eqnarray}

\end{enumerate}

It is also worth mentioning that there is a unique ``vacuum'' charge, denoted $0$ or $I$, for which fusion and braiding is trivial. Furthermore, each topological charge $a$ has a unique conjugate $\bar{a}$ with which it is allowed to fuse to vacuum (in a unique way), i.e. $N_{ab}^{0} = \delta_{\bar{a} b}$.

In summary, $\mathcal{C}$, $N_{ab}^{c}$, $\left[F_d^{abc}\right]_{ef}$, and $R_{c}^{ab}$ comprise the basic data which completely specifies an anyon model. Given this basic data, one can describe the operation representing an arbitrary fusion and braiding process by using a series of applications of associativity and braiding ($F$ and $R$). Hence, once this basic data of a system's anyon model is obtained, they can be used to compute the complete fusion and braiding statistics of (arbitrary configurations and exchanges of) its quasiparticles.

In order to provide a coherent description of the fusion and braiding, this basic data must satisfy certain consistency relations known as the ``polynomial equations'' (i.e. the pentagon and hexagon equations)~\cite{Moore88}, which ensure that any two series of applications of $F$ and $R$ starting in the same state space and ending in the same state space are equivalent~\cite{MacLane63}. These consistency relations put strong constraints on the basic data (up to gauge transformations of the trivalent basis states). Consequently, it is often possible to start with an incomplete subset of the basic data and derive the rest using nothing more than these consistency relations.

There are several important gauge invariant quantities worth describing here. The first is the quantum dimension of charge $a$, given by
\begin{equation}
d_{a} = d_{\bar{a}} = \left| \left[F_a^{a \bar{a} a}\right]_{00} \right|^{-1}
.
\end{equation}
Quantum dimensions can be shown to satisfy the relation
\begin{equation}
\label{eq:fusion_d}
d_a d_b = \sum_{c\in \mathcal{C}} N_{ab}^{c } d_c
.
\end{equation}
The second is the topological spin of charge $a$, given by
\begin{equation}
\theta_{a} = \theta_{\bar{a}} =\sum_{c\in \mathcal{C}} \frac{d_c}{d_a} R^{aa}_{c} = d_a \left[F_a^{a \bar{a} a}\right]_{00} \left(R^{\bar{a} a}_{0}\right)^{\ast}
.
\end{equation}
These two (gauge invariant) quantities are particularly important, because it is often the case that the fusion algebra, together with the quantum dimensions and topological spins (or even a subset of them) can uniquely identify an anyon model.

Once it is known that the braiding statistics of quasiparticles is given by analytic continuation of the wavefunctions given by CFT conformal blocks, we do not need to explicitly perform analytic continuation directly on the wavefunctions. Rather, one can instead obtain all the basic data directly from known properties of the CFT. For example, the topological spin is simply given by $\theta_{a} = e^{i 2\pi h_a}$, where $h_a$ is the (holomorphic) conformal scaling dimension of the primary field corresponding to topological charge $a$, while the $F$- and $R$-symbols can be obtained from the operator product expansions (the $R$-symbols can simply be read off). For a complementary discussion of how the structure of topological phases can be decoded from wavefunctions -- in particular, from their pattern of zeros -- see Refs.~\onlinecite{Wen08,Bernevig08}.

\subsubsection{Ising}

As an example of an application of this consistency, we consider starting with nothing more than the Ising fusion algebra
\begin{equation}
\begin{array}{lll}
I \times I = I, \quad & I \times \psi = \psi, \quad & I \times \sigma = \sigma, \\
\psi \times \psi = I, \quad & \psi \times \sigma = \sigma, \quad & \sigma \times \sigma = I+\psi .
\end{array}
\label{eq:Isingfusion}
\end{equation}
This fusion algebra is believed to describe the $\nu=1$ bosonic MR state (i.e. $M=1$, for which one has the identifications $\psi_2 = I_0 = I$, $\sigma_1 = \sigma$, and $I_2 = \psi_0 = \psi$) and also the closely related SU$(2)_2$ Chern-Simons theory (equating $I=0$, $\sigma = \frac{1}{2}$, and $\psi=1$). In the next subsection,
we will show that this identification is correct.
Given this fusion algebra, one can solve the consistency conditions to find exactly eight different anyon models (up to gauge transformations)~\cite{Kitaev06a,Bonderson07b}. These eight different anyon models are completely distinguished by their values of $\theta_{\sigma} = e^{i 2 \pi (2j+1)/16}$, where $j= 0,\ldots,7$ for the eight different theories [e.g. Ising anyons have $j=0$ and SU$(2)_2$ has $j=1$]. Hence, knowing only the fusion algebra one can use consistency to determine the theory up to this eight-fold degeneracy of theories.

Since we only need to supplement the Ising fusion algebra with the value of $\theta_{\sigma}$ in order to completely identify the anyon model describing such a system, we should examine it more closely to determine what is left to compute. However, we first note that one can easily determine the quantum dimensions to be $d_I = d_\psi = 1$ and $d_\sigma = \sqrt{2}$ from Eq.~\rf{eq:fusion_d}. Now we can write out the definition
\begin{equation}
\theta_{\sigma} = \frac{1}{\sqrt{2}} \left( R^{\sigma \sigma }_{I} + R^{\sigma \sigma }_{\psi} \right) = \sqrt{2} \left[F_\sigma^{\sigma \sigma \sigma}\right]_{II} \left(R^{\sigma \sigma}_{I}\right)^{\ast}
.
\end{equation}
From this we see that all we need is $R^{\sigma \sigma }_{I}$ and either $R^{\sigma \sigma }_{\psi}$ or $\left[F_\sigma^{\sigma \sigma \sigma}\right]_{II}$.

\subsubsection{Putative Anyon Model for
the Moore-Read Pfaffian State}

One can similarly analyze the anyon model
which is generally assumed to correspond to the MR Pfaffian
state. In the next subsection, we will compute
the $F$ and $R$ matrices to show that this identification
is correct.
We define the topological charges through the
corresponding CFT operators of the quasiparticles, which are the product of an Ising sector operator with a U(1) bosonic charge sector vertex operator
\begin{equation}
I_{j} = e^{i \frac{j }{\sqrt{2M}}\varphi}, \qquad \psi_{j} = \psi e^{i \frac{j }{\sqrt{2M}}\varphi}, \qquad \sigma_{j} = \sigma e^{i \frac{j }{\sqrt{2M}}\varphi}
,
\end{equation}
where $j$ must be even for $I_j$ and $\psi_j$, and odd for $\sigma_j$.
From the construction of wavefunctions from CFT correlation functions, we can identify
the charge $I_0$ as corresponding to vacuum (an insertion of this operator leaves the wavefunctions unchanged), while $\psi_{2M}$ corresponds to the underlying particle of the system, e.g. the electron or atom. Consequently, when $M$ is odd, we identify the bosonic underlying particle with vacuum, i.e. $\psi_{2M} = I_0$. When $M$ is even, the underlying particle (the electron) is a fermion, so we cannot identify it with vacuum (though one can introduce a $\mathbb{Z}_2$ grading and put it in a doublet with vacuum). Instead, for $M$ even we identify a pair of electrons with vacuum, i.e. $\psi_{2M} \times \psi_{2M} = I_{4M}=I_0$.

Furthermore, we can identify $\sigma_1$ as the label corresponding to the charge $e/2M$ fundamental quasihole. One can think of $I_2$ as the $e/M$ Laughlin quasihole obtained by inserting one flux. Alternatively, on can think of $I_2$ and $\psi_2$ as the quasiparticles obtained at $\eta_1$ when one takes $\eta_2 \rightarrow \eta_1$ in the four quasihole wavefunctions $\Psi_0$ and $\Psi_1$, respectively. The labels $I_{2k+2}$, $\psi_{2k+2}$, and $\sigma_{2k+1}$ can then be thought of as describing quasiparticles obtained by similarly combining $k$ charge $e/M$ $I_2$ Laughlin quasiholes with quasiparticles of type $I_{2}$, $\psi_{2}$, and $\sigma_{1}$, respectively.
Thus, we have the fusion rules
\begin{equation}
\begin{array}{lll}
I_{j} \times I_{k} = I_{[j+k]}, \quad & I_j \times \psi_k = \psi_{[j+k]}, \quad & I_{j} \times \sigma_{k} = \sigma_{[j+k]}, \\
\psi_{j} \times \psi_{k} = I_{[j+k]}, \quad & \psi_{j} \times \sigma_{k} = \sigma_{[j+k]}, \quad & \sigma_{j} \times \sigma_{k} = I_{[j+k]}+\psi_{[j+k]} .
\end{array}
\label{eq:MRfusion}
\end{equation}
where we have defined the short-hand $[j] \equiv j \text{mod} 4M$. For $M$ odd, one has the additional identifications $I_{j} = \psi_{[j+2M]}$, $\psi_{j} = I_{[j+2M]}$, and $\sigma_j = \sigma_{[j+2M]}$. In this way, there are $6M$ distinct topological charge types for $M$ even and $3M$ topological charge types for $M$ odd. Given these fusion rules, the quantum dimensions can be found to be $d_{I_j} = d_{\psi_j} = 1$ (for $j$ even) and $d_{\sigma_j} = \sqrt{2}$ (for $j$ odd) from Eq.~\rf{eq:fusion_d}. The topological spins are
\begin{eqnarray}
\theta_{I_j} &=& R^{I_j I_j }_{I_{2j}} = \left[F_{I_j}^{I_j I_{-j} I_j}\right]_{I_0 I_0} \left(R^{I_{-j} I_j}_{I_0}\right)^{\ast} \\
\theta_{\psi_j} &=& R^{\psi_j \psi_j }_{I_{2j}} = \left[F_{\psi_j}^{\psi_j \psi_{-j} \psi_j}\right]_{I_0 I_0} \left(R^{\psi_{-j} \psi_j}_{I_0}\right)^{\ast}\\
\theta_{\sigma_j} &=& \frac{1}{\sqrt{2}} \left( R^{\sigma_j \sigma_j }_{I_{2j}} + R^{\sigma_j \sigma_j }_{\psi_{2j}} \right) = \sqrt{2} \left[F_{\sigma_j}^{\sigma_j \sigma_{-j} \sigma_j}\right]_{I_0 I_0} \left(R^{\sigma_{-j} \sigma_j}_{I_0}\right)^{\ast}
.
\end{eqnarray}

As mentioned earlier, the eight consistent anyon models permitted for the $M=1$ MR fusion algebra can be uniquely distinguished by their value of $\theta_{\sigma_1}$. For $M=2$ (corresponding to the $\nu=1/2$ fermionic MR state), it was found in Refs.~\onlinecite{Bonderson07b,BondersonWIP} that given the corresponding fusion algebra, there are $32$ different possible anyon models that satisfy the consistency conditions and that these $32$ different anyon models can be uniquely identified by their topological spins. In fact, closer inspection reveals that they can be uniquely identified merely by their values of $\theta_{\sigma_1}$ and $\theta_{I_2}$. Moreover, it can be shown~\cite{BondersonWIP} that this is true for general $M$, i.e. that $\theta_{\sigma_1}$ and $\theta_{I_2}$ uniquely identifies the anyon models corresponding to the MR fusion algebra.

\subsection{Identifying the anyon model}

Once the anyon model corresponding to a state is known, one can use it to compute the braiding statistics of an arbitrary $n$-quasiparticle wavefunction. Thus, we now turn to the wavefunctions of the MR
Pfaffian state, to extract the quantities necessary to identify the corresponding anyon model.

We now give the charge $e/2M$ quasiholes in
this system the label $\sigma_1$. At the moment,
this labeling is completely innocent, but we note that
$\sigma_1$ has the same quantum dimension
as the similarly-named quasiparticle of the previous subsection
since $d_{\sigma_1}=\sqrt{2}$ and, as discussed
in Section \ref{sec:qubit-basis}, there are
$(\sqrt{2})^{n}$ states for $n$ even quasiparticles.
Now consider the wavefunctions $\Psi_0$ and $\Psi_1$
defined in Eq. \ref{eq:wave}.
These wavefunctions are linear combinations of
the wavefunctions in Eq. \ref{eqn:Psi_(13)(24)-def}.
Suppose that, in $\Psi_0$, we bring
the $\sigma_1$ quasiholes at $\eta_1$ and $\eta_2$
close together and take the quasiholes at
$\eta_3$ and $\eta_4$ far away. Then there is
a localized excitation at $\eta_1\approx \eta_2$.
Let us call this localized excitation $I_2$.
Suppose we do the same thing with $\Psi_1$;
we call the resulting excitation $\psi_2$. \emph{A priori},
we do not know if $I_2$ and $\psi_2$ are distinct
excitations or are topologically-equivalent
(or, perhaps, are different superpositions of
excitations). However, from the four-quasiparticle braiding
matrices computed in Section \ref{sec:six} through
plasma analogy arguments,
we know that if the quasihole at $\eta_3$ is taken
around the excitation at $\eta_1\approx \eta_2$, the resulting
phase in $\Psi_1$ differs from that in $\Psi_0$ by $-1$.
Thus, it is correct to assign two different labels to the
corresponding excitations; the labels $I_2$ and $\psi_2$
are as good a choice as any.

We now compute the $F$-symbols and the fusion algebra.
This can be done with four-quasihole wavefunctions.
The $F$-symbol is a unitary transformation between
two different bases of the two-dimensional Hilbert space
of four quasiparticle states. One basis is given
by $\Psi_0$ and $\Psi_1$, in which the $\sigma_1$ quasiparticles at
$\eta_1$ and $\eta_2$ fuse to $I_2$ and $\psi_2$, respectively.
From the computation of matrix elements in
Section~\ref{sec:six}, we see that these wavefunctions
provide an orthonormal basis:
\begin{equation}
\left\langle \Psi_0 | \Psi_0\right\rangle =
\left\langle \Psi_1 | \Psi_1\right\rangle = 1\,,
\hskip 0.5 cm \left\langle \Psi_0 | \Psi_1\right\rangle = \left\langle \Psi_1 | \Psi_0\right\rangle = 0
.
\end{equation}
(Strictly speaking, we have shown that $\Psi_0$ and $\Psi_1$
are orthogonal and have the same norm, which is simply
an $\eta_\mu$-independent constant in the limit that
the $\eta_\mu$ are all far apart. Thus, to normalize them,
we need to divide both wavefunctions by their common norm.)
The second orthonormal basis is given by $\left| {\Psi}_{0,1}^{\prime} \right\rangle$, in which the $\sigma_1$ quasiparticles
at $\eta_1$ and $\eta_4$ fuse to $I_2$, $\psi_2$.
Since $1$, $2$, $3$, and $4$ are an arbitrary labeling
of the quasiparticles, the states in which $1$ and $4$ fuse
to $I_2$, $\psi_2$ are given by changing the labels to obtain:
\begin{equation}
\label{eq:wave-other-basis}
{\Psi}_{0,1}^{\prime} \left(\eta_1, \eta_2, \eta_3, \eta_4; z_1,
\dots, z_{N} \right) =
\prod_{ \mu < \nu }^{4} \eta_{\mu \nu }^{\frac{1}{4M} - \frac{1}{8}}
\frac{\left(\eta_{13}\eta_{24} \right)^{\frac 1 4}}
{ \sqrt{1\pm\sqrt{x}}} \left[\Psi_{(13)(24)} \pm
\sqrt{x}
\, \Psi_{(12)(34)} \right]e^{- \frac{1}{8M} \sum\limits_{\mu=1}^{4} \left| \eta_\mu \right|^2}.
\end{equation}
Then, using the identity Eq.~\rf{eqn:linear-identity}, we have:
\begin{equation}
\label{eq:4qhwf_prime}
{\Psi}_{0,1}^{\prime} = \prod_{\mu <\nu}^{4}
\eta_{\mu \nu}^{\frac{1}{4M} - \frac{1}{8}}\,
\frac{\left(\eta_{13}\eta_{24} \right)^{\frac 1 4}}
{ \sqrt{1\pm\sqrt{x}}} \left[ \left(1\pm\frac{1}{\sqrt{x}}\right)
\Psi_{(13)(24)} \mp \left(\frac{1-x}{\sqrt{x}}\right)
\, \Psi_{(14)(23)} \right]
e^{- \frac{1}{8M} \sum\limits_{\mu=1}^{4} \left| \eta_\mu \right|^2}
\end{equation}
Therefore,
\begin{eqnarray}
\label{eqn:F-linear-rel}
\Psi_{0}^{\prime} \pm \Psi_{1}^{\prime}
&=& \prod_{\mu < \nu}^{4} \eta_{\mu \nu}^{\frac{1}{4M} - \frac{1}{8}}\,
\left(\eta_{13}\eta_{24} \right)^{\frac 1 4}
\left[\left(\frac{1+\frac{1}{\sqrt{x}}}{\sqrt{1+\sqrt{x}}}
\pm \frac{1-\frac{1}{\sqrt{x}}}{\sqrt{1-\sqrt{x}}}\right) \Psi_{(13)(24)} \right.\cr
& &\left. \hskip 5 cm - \frac{(1-x)}{\sqrt{x}}\left(\frac{1}{\sqrt{1+\sqrt{x}}}
\mp \frac{1}{\sqrt{1-\sqrt{x}}}\right)
\Psi_{(14)(23)}
\right] e^{- \frac{1}{8M} \sum\limits_{\mu=1}^{4} \left| \eta_\mu \right|^2}\cr
&=& \prod_{\mu < \nu}^{4} \eta_{\mu \nu}^{\frac{1}{4M} - \frac{1}{8}}\,
\left(\eta_{13}\eta_{24} \right)^{\frac 1 4}
\left[\left(\frac{\sqrt{1+\sqrt{x}}}{\sqrt{x}} \mp
\frac{\sqrt{1-\sqrt{x}}}{\sqrt{x}}\right)\Psi_{(13)(24)} \right.\cr
& &\left. \hskip 5 cm
- \frac{(1-x)}{\sqrt{x}}\left(
\frac{\sqrt{1-\sqrt{x}}  \mp \sqrt{1+\sqrt{x}}}{\sqrt{1-x}}\right)
\Psi_{(14)(23)}
\right] e^{- \frac{1}{8M} \sum\limits_{\mu=1}^{4} \left| \eta_\mu \right|^2}\cr
&=& \prod_{\mu < \nu}^{4} \eta_{\mu \nu}^{\frac{1}{4M} - \frac{1}{8}}\,
\left(\eta_{13}\eta_{24} \right)^{\frac 1 4}
\left(\frac{\sqrt{1+\sqrt{x}}  \mp \sqrt{1-\sqrt{x}}}{\sqrt{x}}\right)
\left[\Psi_{(13)(24)} \pm \sqrt{1-x}\,\Psi_{(14)(23)}
\right] e^{- \frac{1}{8M} \sum\limits_{\mu=1}^{4} \left| \eta_\mu \right|^2}\cr
&=& \sqrt{2} \prod_{\mu< \nu}^{4} \eta_{\mu \nu}^{\frac{1}{4M} - \frac{1}{8}}\,
\frac{\left(\eta_{13}\eta_{24} \right)^{\frac 1 4}}{\sqrt{1\pm\sqrt{1-x}}}
\left[\Psi_{(13)(24)} \pm \sqrt{1-x}\,\Psi_{(14)(23)}
\right] e^{- \frac{1}{8M} \sum\limits_{\mu=1}^{4} \left| \eta_\mu \right|^2}\cr
&=& \sqrt{2} \, \, \Psi_{0,1}
\end{eqnarray}
Thus, from Eq.~\rf{eqn:F-linear-rel}, we conclude that:
\begin{equation}
\left[F^{\sigma_1 \sigma_ 1 \sigma_1 }_{\sigma_3}\right]_{ab}=
\frac{1}{\sqrt{2}}
\left[\begin{matrix}
1 &1 \cr
1 & -1
\end{matrix}\right]_{ab}
\end{equation}
where $a,b = I_2, \psi_2$. One can similarly compute all the $F$-symbols directly from wavefunctions with no more than four
quasiparticles.

We note that we computed the $F$-symbols directly from the wavefunctions, with no appeal to orthogonality or the plasma analogy, so it might at first appear that orthogonality played no role here. However, it is important to establish that the associativity encoded in the $F$-symbols is unitary, since we are describing quantum mechanical systems. For this, the orthogonality result (obtained from the plasma analogy) is crucial, since it establishes that the wavefunctions in question provide orthonormal bases that are related by this $F$-symbol, which can thus be interpreted as a unitary change of basis transformation.

We now compute the $R$-symbols.
With the definitions of $I_2$ and $\psi_2$
given earlier, we can read these off from the
four-quasihole wavefunctions $\Psi_0$ and $\Psi_1$,
since our plasma analogy has shown that the braiding statistics is simply given by analytic continuation of the wavefunctions.
Thus, as found in Eq.~\rf{eq:op1} for the counterclockwise
exchange $\eta_1 \leftrightarrows \eta_2$, we have
$R^{\sigma_1 \sigma_1 }_{I_{2}} =
e^{i \pi \left(\frac{1}{4M} - \frac{1}{8} \right) }$,
$R^{\sigma_1 \sigma_1 }_{\psi_{2}} =
e^{i \pi \left(\frac{1}{4M} + \frac{3}{8} \right) }$.
We note that these $R$-symbols can also be obtained from the analytic continuation of the wavefunctions $\Psi_0^{\prime}$ and $\Psi_1^{\prime}$ in Eq.~\rf{eq:4qhwf_prime} corresponding the counterclockwise
exchange $\eta_2 \leftrightarrows \eta_3$.
However, once we know that the braiding statistics is given by explicit analytic continuation of the wavefunctions, we can save ourselves the trouble of explicitly computing the $R$-symbols from wavefunctions, since we know the analytic continuation of the wavefunctions is determined by the CFT through its operator product expansion.

Generalizing this to quasiparticles of type $\sigma_{j}$ by appropriately modifying the U$(1)$ charge sector vertex operators, we similarly obtain
\begin{eqnarray}
R^{\sigma_j \sigma_k }_{I_{j+k}} &=& e^{i \pi \left(\frac{jk}{4M} - \frac{1}{8} \right) } \\
R^{\sigma_j \sigma_k }_{\psi_{j+k}} &=& e^{i \pi \left(\frac{jk}{4M} + \frac{3}{8} \right) }
\end{eqnarray}
Similarly, one can compute the remaining $R$-symbols from analytic continuation by additionally introducing quasiparticles carrying topological charge $I_j$ and/or $\psi_k$ into the wavefunction, since the screening properties can easily be shown to still apply (as long as $M$ is not too large). The results are
\begin{eqnarray}
R^{I_j I_k }_{I_{j+k}} &=& e^{i \pi \frac{jk}{4M} } \\
R^{\psi_j \psi_k }_{I_{j+k}} &=& -e^{i \pi \frac{jk}{4M}} \\
R^{I_j \psi_k }_{\psi_{j+k}} &=& R^{\psi_k I_j }_{\psi_{j+k}} = e^{i \pi \frac{jk}{4M} } \\
R^{\sigma_j I_k }_{\sigma_{j+k}} &=& R^{I_k \sigma_j }_{\sigma_{j+k}} =  e^{i \pi \frac{jk}{4M} } \\
R^{\sigma_j \psi_k }_{\sigma_{j+k}} &=& R^{\psi_k \sigma_j }_{\sigma_{j+k}} =  e^{i \pi \left(\frac{jk}{4M} - \frac{1}{2} \right) }
.
\end{eqnarray}
As previously mentioned, there is some ambiguity in these quantities, since they are not all gauge invariant. However, we can use them to obtain
\begin{eqnarray}
\theta_{I_j} &=& e^{i 2 \pi \frac{j^2}{8M} } \\
\theta_{\psi_j} &=& e^{i 2 \pi \left( \frac{j^2}{8M} + \frac{1}{2} \right) } \\
\theta_{\sigma_j} &=& e^{i 2 \pi \left( \frac{j^2}{8M} + \frac{1}{16} \right)}
.
\end{eqnarray}

This can be used to specify the corresponding anyon model. In fact, as previously mentioned, only a subset of this information is absolutely necessary to uniquely identify the anyon model. Hence, the fusion and braiding of the MR state can be completely determined through consistency using only the fusion algebra and the plasma argument applied to the wavefunctions with at most four $\sigma$ quasiparticles.

In fact, wavefunctions with at most two $\sigma$ quasiparticles are sufficient to
compute the $R$-symbols, provided that they are allowed to also include quasiparticles that can carry either of the Ising charges $I$ or $\psi$ (which will not change any of the plasma arguments). In particular, consider the wavefunctions (for $N$ even electrons) with two $\sigma_1$ quasiholes at $\eta_1$ and $\eta_2$ and either quasiparticle of type $I_2$ or $\psi_2$ at $\eta_3$. These two wavefunctions can, respectively, be obtained from the four $\sigma_1$ quasihole wavefunctions $\Psi_0$ and $\Psi_1$ by taking $\eta_4 \rightarrow \eta_3$. Clearly, since $\eta_3$ and $\eta_4$ can be taken to be far away from $\eta_1$ and $\eta_2$, locality dictates that the braiding statistics factors $R^{\sigma_1 \sigma_1 }_{I_{2}}$ and $R^{\sigma_1 \sigma_1 }_{\psi_{2}}$ obtained by exchanging $\eta_1 \leftrightarrows \eta_2$ in the wavefunctions with two $\sigma_1$ quasiholes will be the same as those obtained from the wavefunctions with four $\sigma_1$ quasiholes.

We saw an effect similar to this at the end of Section~\ref{sec:five}, where we considered wavefunctions with two $\sigma_1$ quasiholes (and no other quasiparticles) when the number of electrons $N$ was either even or odd. There we found the two different braiding factors $R^{\sigma_1 \sigma_1 }_{I_{2}}$ and $R^{\sigma_1 \sigma_1 }_{\psi_{2}}$ for $N$ even and odd, respectively. This is, of course, related to the same properties discussed here. In particular, each electron has topological charge $\psi_{2M}$, so changing between even and odd $N$ for wavefunctions with two $\sigma_1$ quasiholes has a similar effect as changing between the type $I_2$ and $\psi_2$ (at fixed $N$) of an additional quasiparticle, as described in the previous paragraph. However, the cases of $N$ even and odd necessarily belong to different Hilbert spaces, since the number of electrons in the system is different. Hence, one could not have simply taken those results as proof of the non-Abelian statistics, since doing so would have required making additional assumptions about the nature of the state, essentially equivalent to assuming the answer.

\section{Orthogonality For Unmatched Quasiparticles}
\label{sec:qporthog}

It is often assumed that the overlap between two wavefunctions that do not have the same types of quasiparticles at (nearly) the same locations should vanish. For example, such an orthogonality postulate is used~\cite{Read90} (sometimes implicitly) in the determination of the braiding statistics and other properties of hierarchical states, such as the Abelian Haldane-Halperin (HH) hierarchy states~\cite{Haldane83,Halperin84}, which are built on the $\nu=1/m$ Laughlin states, and the Bonderson-Slingerland (BS) hierarchy states~\cite{Bonderson08}, which can be built on arbitrary states, notably providing a non-Abelian hierarchy over the $\nu=1/2$ MR and anti-Pfaffian states. In fact, this orthogonality is of paramount importance, since it is necessary to establish the interpretation of the wavefunctions as describing anyonic quasiparticles. Without it, we would be missing the notion of a specific, distinguishable localized conserved quantum number (topological charge) associated with the quasiparticle coordinate, i.e. property $1$ of anyon models in Section~\ref{sec:general-consid}, and thus the rest of the properties (fusion, braiding, etc.) that we can derive would lack a proper interpretation beyond some algebraic relations between certain special wavefunctions. In this section, we will prove this orthogonality using the plasma analogy.

First we consider the wavefunctions in terms of their Coulomb gas CFT formalism. Next we recognize that, as described in Appendices~\ref{sec:boson-Coulomb-gas} and \ref{sec:two-comp-screening}, the chiral vertex operators can be expressed in terms of electric and magnetic Coulomb charges. Specifically,
\begin{equation}
e^{i \alpha \varphi \left( z \right)} = e^{i \frac{\alpha}{2} \phi \left( z,\bar{z} \right)} \mathcal{O}_{\frac{\alpha}{2g}}\left( z,\bar{z} \right)
\end{equation}
corresponds to a particle carrying electric charge $q = \alpha/2$ and magnetic charge $m = \alpha/2g$. Similarly,
\begin{equation}
e^{i \alpha \bar{\varphi} \left( \bar{z} \right)} = e^{i \frac{\alpha}{2} \phi \left( z,\bar{z} \right)} \mathcal{O}_{-\frac{\alpha}{2g}}\left( z,\bar{z} \right)
\end{equation}
corresponds to a particle carrying electric charge $q = \alpha/2$ and magnetic charge $m=-\alpha/2g$. If the corresponding holomorphic and antiholomorphic vertex operators have coinciding positions, they produce an operator $e^{i \alpha \varphi \left( z \right)} e^{i \alpha \bar{\varphi} \left( \bar{z} \right)} = e^{i \alpha \phi \left( z,\bar{z} \right)}$ corresponding to a particle carrying electric charge $\alpha$ and no magnetic charge. If quasiparticle coordinates are not matched up appropriately between two wavefunctions, then, in the plasma analogy, they will leave stray (uncanceled) magnetic charges in the plasma. Since magnetic charge is confined in screening plasmas, this makes it clear that the overlap between wavefunctions described by a plasma analogy will vanish unless they have matching quasiparticles at nearly coinciding positions. More specifically, the overlaps will be zero up to $O\left( e^{- r / \ell} \right)$ corrections, where $r$ is the largest separation between unmatched quasiparticles (i.e. between magnetic charges in the plasma).

With this orthogonality established, one might wonder whether a stronger result can be established. Indeed, one can attain stronger results for holes in the $\nu=1$ filled Landau level and for single Laughlin-type quasihole wavefunctions. We explain this in more detail in Appendix~\ref{sec:orthog_examples}, but note the results here that
\begin{equation}
\label{eq:LLdelta}
G \left( \bar{\eta}_{\mu} , \eta_{\mu}^{\prime}  \right) = C_1 \sum_{\pi\in S_{n}} {(-1)^\pi} {\prod_{\mu=1}^{n}} \,
e^{-\frac{1}{4} \left( {|{\eta_{\pi(\mu)}}|^2 + |{\eta'_{\mu}}|^2 - 2 {\bar{\eta}_{\pi(\mu)}}{\eta^{\prime}_{\mu}} } \right) }
= C_1 (2\pi)^n \sum_{\pi\in S_{n}} {(-1)^\pi} {\prod_{\mu=1}^{n}} \, \delta_{\text{LLL}}^{2} \left( \eta_{\pi(\mu)} - \eta'_{\mu} \right)
\end{equation}
for $n$ holes in the $\nu=1$ filled Landau level, and
\begin{equation}
\label{eq:Laughlin_delta}
G \left( \bar{\eta} , \eta^{\prime}  \right) = C_1 e^{-\frac{1}{4M} \left( {|{\eta}|^2 + |{\eta'}|^2 - 2 {\bar{\eta}}{\eta^{\prime}} } \right) }
= C_1 \left( 2 \pi M \right) \delta_{\text{LLL}_M}^{2} \left( \eta - \eta' \right)
\end{equation}
for a Laughlin-type quasihole in any $\nu = 1/M$ quantum Hall state with plasma analogy (for example, this holds for the $I_2$ excitation of the MR state).

Based on these examples, we conjecture that the overlap of wavefunctions that are not properly matched up in the U$(1)$ boson charge sector will actually vanish with Gaussian, rather than exponential, falloff. This is expected to result from the neutralizing background that gives rise to the Gaussian terms in the wavefunctions. We also expect that the overlaps will behave effectively as delta-functions, projected into the appropriate subspace, while keeping track of the braiding statistics due to analytic continuation of the wavefunctions. This leads us to a general orthogonality postulate for quantum Hall states of the form
\begin{equation}
\label{eq:orthog_postulate}
G_{\alpha, \beta} \left( \bar{\eta}_{\mu_a} , \eta_{\mu_a}^{\prime}  \right) \sim
B_{\alpha, \beta} \left( \bar{\eta}_{\mu_a} , \eta_{\mu_a}^{\prime} \right) \sum_{a \in \mathcal{C} } \sum_{\pi\in S_{n_a}} {\prod_{\mu=1}^{n_a}} \, \delta^{2} \left( \eta_{\pi(\mu_a)} - \eta'_{\mu_a} \right)
,
\end{equation}
where there are $n_a$ quasiparticles of type (topological charge) $a$ at coordinates $\eta_{\mu_a}$ in one wavefunction and at $\eta_{\mu_a}^{\prime}$ in the other wavefunction (the delta-functions are only between quasiparticles of matching topological charge), and $B_{\alpha, \beta} \left( \bar{\eta}_{\mu_a} , \eta_{\mu_a}^{\prime} \right)$ is a term that only keeps track of the braiding statistics due to exchange of the quasiparticles.
For example, Laughlin-type quasiholes would have
\begin{equation}
B \left( \bar{\eta}_{\mu} , \eta_{\mu}^{\prime} \right) = \prod_{\mu < \nu}^{n} \left[ \frac{ \left( \bar{\eta}_{\mu} - \bar{\eta}_{\nu} \right) \left( \eta_{\mu}^{\prime} - \eta_{\nu}^{\prime} \right)}{ \left| {\eta}_{\mu} - {\eta}_{\nu} \right| \left| \eta_{\mu}^{\prime} - \eta_{\nu}^{\prime} \right| } \right]^{\frac{1}{M} }
.
\end{equation}
Eq.~\rf{eq:orthog_postulate} clearly cannot be exact, since the right-hand side does not obey the necessary analytic properties mandated by the wavefunctions one is taking overlaps between, but we expect the answer with the correct analytic properties to have this effective form with respect to quasiparticle wavefunctions, up to exponentially suppressed corrections.

\section{Braiding in the anti-Pfaffian and Bonderson-Slingerland Hierarchy States}
\label{sec:anti-Pfaffian}

In Refs.~\onlinecite{Lee07,Levin07}, it was pointed out
that the particle-hole conjugate of the MR Pfaffian
state is a distinct state and that, in the absence of
Landau level mixing, it must be equal in energy
to the MR state. Upon inclusion of Landau level mixing, the anti-Pfaffian state
appears to be lower in energy~\cite{Bishara09b,Rezayi09}.
A candidate wavefunction for this state can be obtained
by particle-hole conjugation~\cite{Girvin84} of the MR wavefunction.

In Ref.~\onlinecite{Bonderson08}, it was shown that one can construct hierarchical states over the MR and anti-Pfaffian $\nu=1/2$ states, and that these Bonderson-Slingerland (BS) hierarchy states provide candidates for all the (remaining) observed second Landau level plateaus. Moreover, it was numerically demonstrated~\cite{Bonderson09a} that the BS candidate for $\nu=12/5$ is a competitive state, along with the RR and HH candidates.

Here we will show that the wavefunction orthogonality results obtained in this paper for the MR wavefunctions imply the same orthogonality for the anti-Pfaffian and BS hierarchy wavefunctions obtained from them. Hence, the braiding statistics of these states are similarly obtained through analytic continuation of the wavefunctions.

We begin by demonstrating that orthogonality of any two wavefunctions implies the orthogonality of their particle-hole conjugate wavefunction. For a general wavefunction $\Psi (\eta_\mu; z_i)$ with quasiholes, one generates its particle-hole conjugate wavefunction~\cite{Girvin84} by filling one Landau level, introducing $N_1$ holes at coordinates $\xi_a$, and projecting the holes onto the wavefunctions $\Psi$ by multiplying by $\bar\Psi (\bar{\eta}_\mu;\bar{\xi}_a)$ and integrating over the holes' coordinates:
\begin{equation}
\Psi^{(\text{p-h})} (\bar \eta_\mu; z_i) = \int\prod_{c=1}^{N_1} {d^2}{\xi_c}\, \bar{\Psi}(\bar{\eta}_{\mu};{\bar \xi}_a) \times
\prod_{a<b}^{N_1} \left({\xi_a} - {\xi_b}\right)\,  \prod_{a=1}^{N_1} \prod_{i=1}^{N} \left({\xi_a} - {z_i}\right)\,
\prod_{i<j}^{N} \left({z_i} - {z_j}\right)\,
e^{- \frac{1}{4} \sum\limits_{a=1}^{N_1} \left| {\xi}_a \right|^2 }\, e^{- \frac{1}{4} \sum\limits_{i=1}^{N} {\left|{z_i}\right|^2}} \,
\end{equation}
In this expression, one must obey the constraint
\begin{equation}
N = \left( \nu^{-1} -1 \right) N_1 - S+1 + N_{\phi}^{(qh)}
\end{equation}
where $\nu$ and $S$ are respectively the filling fraction and shift of $\Psi$, while $N_{\phi}^{(qh)}$ is the number of fluxes associated with the quasiholes of $\Psi$.

Consider the overlap for two arbitrary particle-hole conjugate wavefunctions with quasiparticles (at matching positions):
\begin{eqnarray}
&& G_{\alpha,\beta}^{(\text{p-h})}(\eta_\mu,\bar\eta_\mu) = \int {\prod_{k=1}^{N} }{d^2}{z_k}\,
\bar{\Psi}^{(\text{p-h})}_{\alpha} ( \eta_\mu; \bar z_i) \Psi^{(\text{p-h})}_{\beta} (\bar \eta_\mu; z_i) \\
&& = \int {\prod_{k=1}^{N} }{d^2}{z_k}\,{\prod_{c=1}^{N_1}}{d^2}{\xi_c}\,{d^2}{\xi'_c}\,
{\Psi_\alpha}(\eta_\mu;\xi_{a}){\bar{\Psi}_\beta}( \bar{\eta}_\mu ; \bar \xi_{a}^{\prime})
\, \prod_{a<b}^{N_1} \left({\bar \xi}_{a} - {\bar \xi}_{b}\right) \left({\xi_a^{\prime}} - {\xi_b^{\prime}}\right) \notag \\
&& \qquad \qquad \times \prod_{a=1}^{N_1} \prod_{i=1}^{N} \left({\bar \xi}_{a} - {\bar z}_{i}\right) \left({\xi_a^{\prime}}-{z_i}\right)
\prod_{i<j}^{N} \left| {z_i} - {z_j}\right|^2
e^{- \frac{1}{4} \sum\limits_{a=1}^{N_1} \left| {\xi}_a \right|^2}\,
e^{- \frac{1}{4} \sum\limits_{a=1}^{N_1} \left| \xi^{\prime}_{a} \right|^2 }\,
e^{- \frac{1}{2} \sum\limits_{i=1}^{N} {\left|{z_{i} }\right|^2}}\,
.
\end{eqnarray}
Now we use Eq.~\rf{eq:LLdelta} to re-write this as
\begin{eqnarray}
G_{\alpha,\beta}^{(\text{p-h})}(\eta_\mu, \bar\eta_\mu) &=& \int {\prod_{c=1}^{N_1}} {d^2}{\xi_c}\,{d^2}{\xi'_c}\,
{\Psi_\alpha}(\eta_\mu;\xi_{a}){\bar{\Psi}_\beta}( \bar{\eta}_\mu ; \bar \xi_{a}^{\prime})
\, \left[ C_1 \left( 2 \pi \right)^{N_1} \sum_{\pi \in S_{N_1} } (-1)^{\pi} \prod_{a=1}^{N_1} \delta^2_{\text{LLL}} \left( {\xi_{\pi(a)}} - {\xi_a^{\prime}} \right) \right] \notag \\
\label{eq:G_anti}
&=& C_1 \left( 2 \pi \right)^{N_1} \int {\prod_{c=1}^{N_1}}{d^2}{\xi_c}
{\Psi_\alpha}(\eta_\mu;\xi_{a}){\bar{\Psi}_\beta}( \bar{\eta}_\mu ; \bar \xi_{a}) = C_1 \left( 2 \pi \right)^{N_1} G_{\beta,\alpha}(\bar\eta_\mu , \eta_\mu)
,
\end{eqnarray}
where $G_{\beta, \alpha}(\bar\eta_\mu , \eta_\mu)$ is the overlap of the original two wavefunctions $\Psi_\beta$ and $\Psi_\alpha$ (with $N_1$ electrons).

It is now trivial to apply Eq.~\rf{eq:G_anti} to the anti-Pfaffian state, using our previously obtained results for $G_{\alpha,\beta}$ of the MR state. (For the MR state at $\nu=1/2$, one has $S=3$ and $N_{\phi}^{(qh)} = n/2$ for $n$ fundamental charge $e/4$ non-Abelian quasiholes.)

A similar, but slightly more complicated, argument applies to hierarchical wavefunctions, such as those of the Haldane-Halperin (HH) states~\cite{Haldane83,Halperin84} and Bonderson-Slingerland (BS) states~\cite{Bonderson08}. In particular, the wavefunctions for these states can be constructed by projecting Abelian quasiparticles of the Laughlin, MR, or anti-Pfaffian state into a Laughlin state, in a manner similar to how holes of the filled Landau level were projected onto a wavefunction to generate its particle-hole conjugate. One can then similarly use the results from Section~\ref{sec:qporthog} on orthogonality of wavefunctions with quasiparticles in different positions and the orthogonality postulate of Eq.~\rf{eq:orthog_postulate}.

For concreteness, let us first examine the $\nu=2/7$ HH hierarchy state. Wavefunctions of this state with $n$ charge $-e/7$ quasiparticles at positions $\eta_\mu$ are given by
\begin{eqnarray}
\label{eq:HH_2_7}
\Psi^{\text{HH}_{2/7}} \left( \bar{\eta}_\mu ; z_i \right) &=& \int \prod \limits_{c = 1}^{N_{1}} d^{2}u_{c} \prod\limits_{\mu<\nu}^{n} \left( \bar{\eta}_{\mu}-\bar{\eta}_{\nu} \right)^{3/7} \prod\limits_{\mu=1}^{n} \prod\limits_{a=1}^{N_1} \left( \bar{\eta}_{\mu} - \bar{u}_{a} \right) \prod\limits_{ a <b }^{N_1} \left( \bar{u}_{a}-\bar{u}_{b} \right)^{7/3} e^{ -\frac{1}{28} \sum\limits_{\mu=1}^{n} \left| \eta_{\mu} \right|^{2} } e^{ -\frac{1}{12} \sum\limits_{a=1}^{N_1} \left| u_{a} \right|^{2} } \notag \\
&& \qquad \qquad \times \prod\limits_{ a <b }^{N_1} \left(  u_{a}-u_{b} \right)^{1/3} \prod\limits_{a=1}^{N_1} \prod\limits_{i=1}^{N} \left( u_{a}-z_{i} \right) \prod\limits_{i<j}^{N} \left( z_{i}-z_{j} \right)^{3} e^{ -\frac{1}{12} \sum\limits_{a=1}^{N_1} \left| u_{a} \right|^{2} -\frac{1}{4} \sum\limits_{i=1}^{N} \left| z_{i} \right|^{2} } \\
&=& \int \prod \limits_{c = 1}^{N_{1}} d^{2}u_{c} \bar{\Phi} \left( \bar{\eta}_\mu ; \bar{u}_a \right) \Psi_{\frac{1}{3}} \left( u_a ; z_i \right)
\end{eqnarray}
where $\Psi_{\frac{1}{3}} \left( u_a ; z_i \right)$ is the $\nu=1/3$ Laughlin wavefunction with $N_1 = \frac{1}{2} (N-n) +1$ charge $e/3$ quasiholes at coordinates $u_a$, which are projected into a Laughlin-type wavefunction
\begin{equation}
\Phi \left( \eta_\mu ; u_a \right) =  \prod\limits_{\mu<\nu}^{n} \left( {\eta}_{\mu}-{\eta}_{\nu} \right)^{3/7} \prod\limits_{\mu=1}^{n} \prod\limits_{a=1}^{N_1} \left( {\eta}_{\mu} - {u}_{a} \right) \prod\limits_{ a <b }^{N_1} \left( u_{a}-u_{b} \right)^{7/3} e^{ -\frac{1}{28} \sum\limits_{\mu=1}^{n} \left| \eta_{\mu} \right|^{2} } e^{ -\frac{1}{12} \sum\limits_{a=1}^{N_1} \left| u_{a} \right|^{2} }
\end{equation}
with $n$ quasiholes at $\eta_\mu$. Taking the inner product and using Eq.~\rf{eq:orthog_postulate}, we obtain
\begin{eqnarray}
\label{eqn:HH-to-Laughlin}
G^{(\text{HH})}( \eta_\mu,\bar\eta_\mu)
&=&\int {\prod_{k=1}^{N}}{d^2}{z_k} \bar{\Psi}^{\text{HH}_{2/7}} \left( \eta_\mu ; \bar{z}_i \right) \Psi^{\text{HH}_{2/7}} \left( \bar{\eta}_\mu ; z_i \right) \notag \\
&=& \int {\prod_{k=1}^{N}}{d^2}{z_k} \prod \limits_{c = 1}^{N_{1}} d^{2}u_{c}^{\prime} d^{2}u_{c} \, \Phi \left( \eta_\mu ; u_a^{\prime} \right) \bar{\Psi}_{\frac{1}{3}} \left( \bar{u}_a^{\prime} ; \bar{z}_i \right) \bar{\Phi} \left( \bar{\eta}_\mu ; \bar{u}_a \right) \Psi_{\frac{1}{3}} \left( u_a ; z_i \right) \notag\\
&=& \int \prod\limits_{c = 1}^{N_{1}} d^{2}u_{c}^{\prime} d^{2}u_{c} \, \Phi \left( \eta_\mu ; u_a^{\prime} \right) \bar{\Phi} \left( \bar{\eta}_\mu ; \bar{u}_a \right) G_{\frac{1}{3}} \left( \bar{u}_a , u_a \right) \notag\\
&\sim& \int \prod\limits_{c = 1}^{N_{1}} {d^2}{u_c^{\prime}}{d^2}{u_c}\, \Phi \left( \eta_\mu ; u^{\prime}_a \right) \bar{\Phi} \left( \bar{\eta}_\mu ; \bar{u}_a \right)
\left[ B \left( \bar{u}_a^{\prime} , {u}_a  \right) \sum_{ \pi \in S_{N_1}} \prod_{a=1}^{N_1} \, \delta^{2} \left( {u_{\pi (a)}^{\prime}}-{u_{a}} \right) \right] \notag \\
&=& \int \prod\limits_{c = 1}^{N_{1}} {d^2}{u_c}\, \Phi \left( \eta_\mu ; u_a \right) \bar{\Phi} \left( \bar{\eta}_\mu ; \bar{u}_a \right)  \notag \\
&=& G^{\Phi} (\bar\eta_\mu , \eta_\mu)
,
\end{eqnarray}
where $G^{\Phi} (\bar\eta_\mu , \eta_\mu)$ is the overlap for the Laughlin-type state's wavefunctions $\Phi$, which, by the plasma analogy argument, is a constant, up to corrections that are exponentially suppressed in the separations $| \eta_\mu - \eta_\nu|$ between quasiholes. Hence, the inner product is equal to a constant, up to exponentially suppressed corrections. From this, it follows that the Berry's connection for transporting the quasiparticles at $\eta_\mu$ is trivial (up to Aharonov-Bohm terms) and the braiding statistics of the HH hierarchy states are similarly obtained from direct analytic continuation of the wavefunction. The braiding statistics phase of $e^{-i 3 \pi /7}$ for a counterclockwise exchange of a pair of the charge $-e/7$ quasiparticles in the $\nu=2/7$ HH state can easily be read off the wavefunction in Eq.~\rf{eq:HH_2_7}.

Now, let us examine the slightly more complicated case of the $\nu=2/5$ BS hierarchy state~\cite{Bonderson08} formed by condensing Laughlin-type quasiparticles (i.e. $I_2$ excitations) of the $\nu=1/2$ MR state. Wavefunctions of this state with $n$ charge $e/5$ non-Abelian quasiholes at positions $\eta_\mu$ are given by
\begin{equation}
\Psi_{\alpha}^{\text{BS}_{2/5}} \left( \eta_\mu ; z_i \right) = \int \prod \limits_{c = 1}^{N_{1}} d^{2}u_{c} \, \bar{\Phi} \left( \bar{\eta}_\mu ; \bar{u}_a \right) \Psi^{\text{MR}}_{\alpha} \left( \eta_\mu ; u_{a} ; z_i \right)
\end{equation}
where
\begin{eqnarray}
\Psi^{\text{MR}}_{\alpha} \left( \eta_\mu ; u_{a} ; z_i \right) &=& \prod\limits_{ a <b }^{N_1} \left(  u_{a}-u_{b} \right)^{1/2} \prod\limits_{\mu=1}^{n} \prod\limits_{a=1}^{N_1} \left( \eta_{\mu} - u_{a} \right)^{1/4} \prod\limits_{a=1}^{N_1} \prod\limits_{i=1}^{N} \left( u_{a}-z_{i} \right) e^{ -\frac{1}{8} \sum\limits_{a=1}^{N_1} \left| u_{a} \right|^{2} } \times \Psi^{\text{MR}}_{\alpha} \left( \eta_\mu ; z_i \right)
\end{eqnarray}
are the $\nu=1/2$ MR wavefunctions with $n$ charge $e/4$ $\sigma_1$ quasiholes at $\eta_\mu$ and $N_1 = \frac{1}{2}N+1$ charge $e/2$ $I_2$ Laughlin quasiholes at $u_a$ [$\Psi^{\text{MR}}_{\alpha} \left( \eta_\mu ; z_i \right)$ are the MR wavefunctions with $n$ charge $e/4$ quasiparticles defined previously in this paper], and the $N_1$ Laughlin quasiholes are projected into the Laughlin-type wavefunction
\begin{equation}
\Phi \left( \eta_\mu ; u_a \right) =  \prod\limits_{\mu<\nu}^{n} \left( {\eta}_{\mu}-{\eta}_{\nu} \right)^{1/40} \prod\limits_{\mu=1}^{n} \prod\limits_{a=1}^{N_1} \left( {\eta}_{\mu} - {u}_{a} \right)^{1/4} \prod\limits_{ a <b }^{N_1} \left( u_{a}-u_{b} \right)^{5/2} e^{ -\frac{1}{80} \sum\limits_{\mu=1}^{n} \left| \eta_{\mu} \right|^{2} } e^{ -\frac{1}{8} \sum\limits_{a=1}^{N_1} \left| u_{a} \right|^{2} }
\end{equation}
with $n$ quasiholes at $\eta_\mu$. We take the inner product and use Eq.~\rf{eq:orthog_postulate} to obtain
\begin{eqnarray}
\label{eqn:BS-to-pfaff}
G_{\alpha,\beta}^{(\text{BS})}(\bar\eta_\mu , \eta_\mu)
&=& \int {\prod_{k=1}^{N}}{d^2}{z_k} \bar{\Psi}_{\alpha}^{\text{BS}_{2/5}} \left( \bar{\eta}_\mu ; \bar{z}_i \right) \Psi_{\beta}^{\text{BS}_{2/5}} \left( \eta_\mu ; z_i \right) \notag \\
&=& \int {\prod_{k=1}^{N}}{d^2}{z_k} {\prod_{c=1}^{N_1}}{d^2}{u'_c}{d^2}{u_c}\, \Phi \left( \eta_\mu ; u^{\prime}_a \right) \bar{\Psi}^{\text{MR}}_{\alpha} \left( \bar{\eta}_\mu ; \bar{u}_{a}^{\prime} ; \bar{z}_i \right) \bar{\Phi} \left( \bar{\eta}_\mu ; \bar{u}_a \right) \Psi^{\text{MR}}_{\beta} \left( \eta_\mu ; u_{a} ; z_i \right) \notag \\
&\sim& \int {\prod_{c=1}^{N_1}}{d^2}{u'_c}{d^2}{u_c}\, \Phi \left( \eta_\mu ; u^{\prime}_a \right) \bar{\Phi} \left( \bar{\eta}_\mu ; \bar{u}_a \right)
\left[ B \left( \bar{u}_a^{\prime} , {u}_a  \right) \sum_{ \pi \in S_{N_1}} \prod_{a=1}^{N_1} \, \delta^{2} \left( {u_{\pi (a)}^{\prime}}-{u_{a}} \right) \right] G^{\text{MR}}_{\alpha,\beta} (\bar\eta_\mu , \eta_\mu) \notag \\
&=& G^{\text{MR}}_{\alpha,\beta} (\bar\eta_\mu , \eta_\mu) \int {\prod_{c=1}^{N_1}}{d^2}{u_c}\, \Phi \left( \eta_\mu ; u_a \right) \bar{\Phi} \left( \bar{\eta}_\mu ; \bar{u}_a \right)  \notag \\
&=& G^{\text{MR}}_{\alpha,\beta} (\bar\eta_\mu , \eta_\mu) G^{\Phi} (\bar\eta_\mu , \eta_\mu)
,
\end{eqnarray}
where $G^{\Phi} (\bar\eta_\mu , \eta_\mu)$ is the overlap for the Laughlin-type state's wavefunctions $\Phi$. Hence, the inner product is equal, up to an overall constant and exponentially suppressed corrections, to that of the MR state times that of the Laughlin-type state (which is simply constant). From this, it follows that the Berry's connection is trivial (up to Aharonov-Bohm terms) and the braiding statistics of the BS hierarchy states are similarly obtained from direct analytic continuation of the wavefunctions.

\section{States based on other CFTs}
\label{sec:Others}

The Coulomb gas formalism~\cite{Dotsenko1984,Felder1989} that we used to describe the Ising CFT in this paper can similarly be used to describe any of the CFT minimal models $\mathcal{M}(p,p^{\prime})$, so we can use the methods developed in our paper to analyze the braiding statistics of candidate quantum Hall states constructed from such CFTs. Unfortunately, this becomes more complicated in general, since the structure of general minimal models is more complex than that of the Ising CFT (which is $\mathcal{M}(4,3)$) and requires both types of screening operators to describe quasiparticles, which can create obstacles to applying Mathur's procedure~\cite{Mathur1992} or result in rather complicated plasmas about which little is known. However, we can make a few preliminary statements regarding certain cases.

\subsection{The $\mathcal{M}(5,4)$ state}

One may attempt to construct a quantum Hall state from the $\mathcal{M}(5,4)$ CFT, using the $\varepsilon^{\prime \prime} = \phi_{(4,1)} = \phi_{(1,3)}$ operator with scaling dimension $h_{\varepsilon^{\prime \prime}} =3/2$ for the particles (electrons or bosons) in the same way the Ising $\psi$ operator was used for the MR state~\cite{Read2008}. Specifically, the $\varepsilon^{\prime \prime}(z_i)$ operator from the $\mathcal{M}(5,4)$ sector is paired with an $e^{i \sqrt{M/2} \varphi (z_i)}$ operator from a U$(1)$ charge sector, and then the correlation function of an even number $N$ of such operators produces a ``ground-state'' wavefunction at $\nu=1/M$. (We note that $M \geq 3$ is necessary for the resulting wavefunction not to diverge as $z_i \rightarrow z_j$.)

In the Coulomb gas formulation of the $\mathcal{M}(5,4)$ CFT, the $\varepsilon^{\prime \prime}$ operator corresponds to vertex operators with $\alpha_{4,1} = -\frac{3}{2} \alpha_{-} = \sqrt{\frac{9}{5}}$ or $\alpha_{1,3} = - \alpha_{+} = - \sqrt{\frac{5}{4}}$ (and appropriate screening operators). Using the $\alpha_{1,3}$ representation, we can apply the plasma analogy for this ground-state exactly as we did for the MR ground-state, resulting in a $2$D two-component plasma of $N$ charge $Q_2 = - \sqrt{5g}$ particles at $z_i$ (corresponding to the electrons/bosons) and $N$ charge $-Q_2$ particles at $w_a$ (corresponding to $\alpha_{+}$ screening operators) at temperature $T=g$. This plasma has coupling constant $\Gamma_2 = Q_2^2 / T = 5$, which is greater than the critical value $\Gamma_{c_2}=4$ above which such plasmas are not in the screening phase (as previously discussed).

Including the U$(1)$ charge sector, we expect the resulting combined plasma to be screening in the U$(1)$ charge sector for sufficiently small $M$, but not screening in the $\mathcal{M}(5,4)$ sector. This can be used to demonstrate that the pair correlation functions of such states exhibit long-ranged correlations, and hence do not describe gapped states. The braiding statistics of quasiparticle excitations for such states is ill-defined, so whether or not we can construct an analogous plasma for wavefunctions with quasiparticles is irrelevant.

\subsection{The $\mathcal{M}(p,p-1)$ states}

One may attempt to construct a quantum Hall state from the unitary $\mathcal{M}(p,p-1)$ CFT for $p>5$, using the Abelian $\phi_{(p-1,1)} = \phi_{(1,p-2)}$ operator with scaling dimension $h_{(p-1,1)} = (p-2)(p-3)/4$ for the particles (electrons or bosons) in the same way the Ising $\psi$ operator was used for the MR state. In this case, this operator from the $\mathcal{M}(p,p-1)$ sector is paired with an $e^{i \sqrt{M/2} \varphi (z_i)}$ operator from a U$(1)$ charge sector, where $M$ is an even or odd integer depending on whether $h_{(p-1,1)}$ is integer or half-integer. The correlation function of an even number $N$ of such operators produces a ``ground-state'' wavefunction at $\nu=1/M$.

In the Coulomb gas formulation of the $\mathcal{M}(p,p-1)$ CFT, the $\phi_{(p-1,1)}$ operator corresponds to vertex operators with $\alpha_{p-1,1} = -\frac{p-2}{2} \alpha_{-} = \frac{p-2}{2} \sqrt{\frac{p-1}{p}}$ or $\alpha_{1,p-2} = - \frac{p-3}{2} \alpha_{+} = - \frac{p-3}{2} \sqrt{\frac{p}{p-1}}$ (and appropriate screening operators). We can apply the plasma analogy for this ground-state similar to how we did for the MR ground-state, except in this case, the plasma analogy construction for the ground-state necessarily results in a two-component plasma where the different particle types carry charges of unequal magnitudes. Specifically, using the $\alpha_{1,p-2}$ representation results in a $2$D two-component plasma of $N$ charge $Q_2 = - \sqrt{\frac{ p (p-3)^2 g}{p-1}}$ particles at $z_i$ (corresponding to the electrons/bosons) and $\frac{p-3}{2}N$ charge $-\frac{2}{p-3} Q_2$ particles at $w_a$ (corresponding to $\alpha_{+}$ screening operators) at temperature $T=g$.

Little appears to be known about the screening properties of plasmas with particles carrying charges of unequal magnitudes. Examining these plasmas coupling constants, the larger charges have $Q_2^2 / T = p(p-3)^2 / (p-1)$ and the smaller charges have $\left( \frac{2}{p-3} \right)^2 Q_2^2 / T = 4p / (p-1)$. Since both of these are greater than $\Gamma_{c_2} =4$ (the critical value of the coupling constant for the two-component plasma with charges of equal magnitude) for $p>4$, we expect (by comparison to the case with charges of equal magnitude) that these plasmas are not in their screening phase at temperature $T=g$. Including the U$(1)$ charge sector, we again expect the resulting combined plasma to be screening in the U$(1)$ charge sector for sufficiently small $M$, but not screening in the $\mathcal{M}(5,4)$ sector, and consequently that these wavefunction do not describe gapped states.

\subsection{The Gaffnian state}

One may also attempt to construct a quantum Hall state from the non-unitary $\mathcal{M}(5,3)$ CFT, using the $\psi = \phi_{(4,1)} = \phi_{(1,2)}$ operator with scaling dimension $h_{\psi} =3/4$ for the particles (electrons or bosons) in a similar way, producing the so-called Gaffnian wavefunctions~\cite{Simon07}. In this case, the $\psi(z_i)$ operator from the $\mathcal{M}(5,3)$ sector is paired with an $e^{i \sqrt{(2M+1)/4} \varphi (z_i)}$ operator from a U$(1)$ charge sector, and then the correlation function of an even number $N$ of such operators produces the ground-state wavefunction at $\nu=2/(2M+1)$.

In the Coulomb gas formulation of the $\mathcal{M}(5,3)$ CFT, the $\psi$ operator corresponds to vertex operators with $\alpha_{4,1} = -\frac{3}{2} \alpha_{-} = \sqrt{\frac{27}{20}}$ or $\alpha_{1,2} = - \frac{1}{2} \alpha_{+} = - \sqrt{\frac{5}{12}}$ (and appropriate screening operators). In this case, a similar plasma analogy construction for the ground-state results in a two-component plasma where the different particle types carry charges of unequal magnitudes. Specifically, using the $\alpha_{4,1}$ representation for the $\psi$ operator results in a $2$D two-component plasma of $N$ charge $Q_2 = \sqrt{27g/5}$ particles at $z_i$ (corresponding to the electrons/bosons) and $3N/2$ charge $- 2 Q_2 /3$ particles at $w_a$ (screening operators) at temperature $T=g$. Alternatively, using the $\alpha_{1,2}$ representation for the $\psi$ operator results in a $2$D two-component plasma of $N$ charge $\tilde{Q}_2 = -\sqrt{5g/3}$ particles at $z_i$ (corresponding to the electrons/bosons) and $N/2$ charge $-2 \tilde{Q}_2$ particles at $w_a$ (screening operators) at temperature $T=g$. We can also construct the analogous plasma for Gaffnian wavefunctions with up to two fundamental quasiholes. These quasiholes carry non-Abelian charge $\sigma$, which corresponds to vertex operators with $\alpha_{2,1} = -\frac{1}{2} \alpha_{-} = \sqrt{\frac{3}{20}}$ in the Coulomb gas formulation. These map to test particles with charge $Q_2 / 2$ in the analogous plasma resulting from the $\alpha_{4,1}$ representation of $\psi$.

Examining these two plasmas coupling constants for their larger charges we have $Q_2^2 / T = 27/5$ and $(2 \tilde{Q}_2)^2 / T = 20/3$, which are greater than $\Gamma_{c_2} =4$ (the critical value of the coupling constant for the two-component plasma with charges of equal magnitude), whereas the smaller charges of these plasmas give $(2 Q_2 / 3)^2 / T = 12/5$ and $\tilde{Q}_2^2 / T = 5/3$, which are less than $\Gamma_{c_2} =4$. Hence, it is difficult to make a prediction by comparing to the two-component plasma with charges of equal magnitude. Whether or not these plasmas are in their screening phase is an interesting question that will be addressed in future research~\cite{Herland-unpublished}.

If these plasmas do not screen, then we expect that including the U$(1)$ charge sector will result in a combined plasma that screens in the U$(1)$ charge sector for sufficiently small $M$, but not in the $\mathcal{M}(5,3)$ sector. This could then be used to demonstrate that pair correlation functions of such states have long-ranged correlations, and hence do not describe gapped states. However, if these plasmas are in the screening phase for parameters corresponding to the Gaffnian wavefunctions, then something must go awry elsewhere, as the braiding statistics resulting from analytic continuation of the wavefunctions with quasiparticles would result in a non-unitary anyon model, which cannot describe a topological phase.

\subsection{The Read-Rezayi States}

The next step in extending our work would be an
adaptation of our plasma analogy construction to the Read-Rezayi $\mathbb{Z}_k$-parafermion states~\cite{Read1999}.
To accomplish this, we need a Coulomb gas formulation of the $\mathbb{Z}_k$-parafermion CFTs.
Such a construction, based on the related Coulomb gas
construction for the SU$(N)_k$ Wess-Zumino-Witten models,
is known rather well~\cite{Wakimoto1986,Nemeschansky1989,Griffin1989,Bilal1989,Gerasimov1990,Dotsenko1990,Dotsenko1991}, but it has
not yet been developed to a point where an explicit representation of the overlaps of the desired wavefunctions (with excitations) can be written as a partition function of a $2$D plasma. We suggest this as a subject for future work.

\section{Discussion}
\label{sec:Discussion}

In this paper, we have constructed a new representation for
the matrix elements between different MR Pfaffian
four-quasihole and six-quasihole states. This representation allowed
us to conclude that the Berry's matrix is trivial
in the basis given by conformal
blocks in a $c=1/2+1$ CFT. This result implies that
this CFT encapsulates the topological properties of
this quantum Hall state -- in other words, that this CFT can be used
to compute braiding, fusion, etc. for quasiparticles in
the MR Pfaffian state~\cite{Read1991,Greiter1992,Nayak1996}
and, by a straightforward extension,
the anti-Pfaffian state \cite{Lee07,Levin07}
and BS hierarchy states built on these~\cite{Bonderson08}.
This was, of course, the hope right from the beginning~\cite{Read1991},
but there was no proof, although there is strong evidence
coming from a variety of arguments and numerical calculations,
as we review below. Our paper provides a proof.

We now consider the relation of our proof to previous results.
In Ref.~\onlinecite{Nayak1996},
a method was proposed to calculate the desired
matrix elements, in terms of correlation
functions in a perturbed CFT.
The basic idea is that the integrals in Eq.~\rf{eqn:intro-Gs}
can be written as the correlation functions of the
operators corresponding to quasiholes if the operators
which correspond to electrons are put into the action
of the CFT and treated as a perturbation.
For instance, ${\rm Tr} G$ for $n=4$ quasiholes, defined in Eq.~\rf{eq:norm3}, is given by the product of the
Ising model correlation function in Eq.~\rf{eq:Ising-non-chiral-def}
and the charge sector correlation function in Eq.~\rf{eqn:4qp-charge-part}:
\begin{multline}
{\rm Tr} G = \int {\prod_{k=1}^{N}}{d^2}z_k\,
\left\langle \sigma(\eta_1,\bar \eta_1)
e^{i \frac{1}{2 \sqrt{2M}} \phi(\eta_1,\bar \eta_1)}
 \sigma(\eta_2,\bar \eta_2) e^{i \frac{1}{2 \sqrt{2M}} \phi(\eta_2,\bar \eta_2)}
   \sigma(\eta_3,\bar \eta_3)  e^{i \frac{1}{2 \sqrt{2M}} \phi(\eta_3,\bar \eta_3)}
   \sigma(\eta_4,\bar \eta_4)e^{i\frac{1}{2 \sqrt{2M}}\phi(\eta_{4},\bar \eta_4)}
 \,\right. \\
 \left. \times \, \epsilon(z_1,\bar z_1) e^{i \sqrt{\frac M 2} \phi(z_1,\bar z_1)}
    \dots \epsilon(z_{N},\bar z_{N}) e^{i \sqrt{\frac M 2} \phi(z_{N},\bar z_N)}
     e^{- i\frac{1}{2 \pi \sqrt{2M}} \int d^2z \, \phi(z, \bar z)}
     \right\rangle
\end{multline}
Here, we have made the abbreviation $\phi(z,\bar z)
\equiv \varphi(z) + \bar \varphi(\bar z)$.
This can be re-written in the form
\begin{equation}
{\rm Tr} G = \int {\prod_{k=1}^{N}}{d^2}z_k \,\left\langle
{{O}_q}(\eta_1,\bar \eta_1) {{O}_q}(\eta_2,\bar \eta_2)
{{O}_q}(\eta_3,\bar \eta_3) {{O}_q}(\eta_4,\bar \eta_4)\,\,
{{O}_e} (z_1,\bar z_1) \ldots {{O}_e}(z_{N},\bar z_{N})
e^{- i\frac{1}{2 \pi \sqrt{2M}} \int d^2z \, \phi(z, \bar z)}
\right\rangle
\end{equation}
where
\begin{equation}
{{O}_q}(\eta,\bar \eta) =
\sigma(\eta,\bar \eta)
e^{i \frac{1}{2 \sqrt{2M}} \phi(\eta,\bar \eta)}
{\hskip 0.5 cm} \mbox{and} {\hskip 0.5 cm}
{{O}_e} (z,\bar z) = \epsilon(z,\bar z)
e^{i \sqrt{\frac M 2} \phi(z,\bar z)}
\end{equation}
This can now be rewritten in the following form
[Eq.~(8.3) of Ref.~\onlinecite{Nayak1996}]:
\begin{eqnarray}
\label{eqn:massive-pert}
{\rm Tr} G &=& \lim_{t\rightarrow 0}
\frac{d^N}{dt^N} \left\langle
{{O}_q}(\eta_1,\bar \eta_1) {{O}_q}(\eta_2,\bar \eta_2)
{{O}_q}(\eta_3,\bar \eta_3) {{O}_q}(\eta_4,\bar \eta_4)\,\,
e^{t \!\int d^2z \, {{O}_e} (z,\bar z)}
e^{- i\frac{1}{2 \pi \sqrt{2M}} \int d^2z \, \phi(z, \bar z)}
\right\rangle\cr
&=& \lim_{t\rightarrow 0}
\frac{d^N}{dt^N} \left\langle
{{O}_q}(\eta_1,\bar \eta_1) {{O}_q}(\eta_2,\bar \eta_2)
{{O}_q}(\eta_3,\bar \eta_3) {{O}_q}(\eta_4,\bar \eta_4)\,\,
e^{- i\frac{1}{2 \pi \sqrt{2M}} \int d^2z \, \phi(z, \bar z)}
\right\rangle_{{\cal L}\rightarrow {\cal L} + t {{O}_e}}
\end{eqnarray}
This equation implies that we should view
$t \!\int d^2z \, {{O}_e} (z,\bar z)$ as a perturbation
of the Lagrangian of the Ising model + a non-chiral boson
and compute the correlation function of four ${{O}_q}$ operators
in this perturbed theory.
By taking $N$ derivatives of the correlation function with respect to the
coupling constant $t$ for this perturbation, a correlation
function with $N$ electron operators is obtained and,
therefore, matrix elements for $N$-electron wavefunctions.
The integrals in Eq.~\rf{eqn:intro-Gs} are exponentially-decaying
at long distances if the renormalization group (RG)
flow of the perturbed CFT is to a massive fixed point.
Although this approach is highly suggestive, it is unclear
how to show that the perturbed theory indeed flows to a massive fixed point.
In Ref.~\onlinecite{Nayak1996}, it was suggested that the
perturbation has negative scaling dimension since correlation
functions of ${{\cal O}_e}$ increase with distance as a result
of the background charge. However, it was pointed out
in Ref.~\onlinecite{Read2008} that ${{\cal O}_e}$ should
be viewed as an operator of positive scaling dimension (which is,
in fact, an irrelevant perturbation for $M\geq 2$),
with the background charge merely shifting the charge neutrality
condition. Thus, it is not clear how to show that the perturbation
in Eq.~\rf{eqn:massive-pert} is relevant. This is even a
problem if this method is applied to the Laughlin states.
Since all quantum Hall wavefunctions have a (Laughlin-like) bosonic charge sector, they all suffer the same complication. Furthermore,
a leading-order RG calculation can show that the initial flow goes away from
the conformal theory, but it is much harder to predict the fixed point
at which the flow ends; it is necessary for the infrared fixed point to
be massive in order for this approach to succeed. For these reasons,
a Coulomb gas approach was initiated in Ref.~\onlinecite{Gurarie1997};
this approach has been brought to fruition here.

The method of Ref.~\onlinecite{Nayak1996}
was recently discussed again by Read in an important
paper~\cite{Read2008}. It was shown there that
this method can be applied more straightforwardly
to the case of a chiral $p$-wave superconductor,
which is described by the Ising CFT.
This state has no charge sector and the perturbed Ising CFT is simply
a free massive Majorana fermion. This result is important because
it allows one to directly compute $R^{\sigma\sigma}_{I}$
and $R^{\sigma\sigma}_{\psi}$,
the two possible phases that result when
two $hc/2e$ vortices with fusion
channel $I$ and $\psi$, respectively, are exchanged in
a counter-clockwise fashion. This is a significant
advance compared to approaches relying on the BCS
wavefunction for this state~\cite{Read2000,Ivanov2001,Stern04,Stone06},
which have only derived the
ratio $R^{\sigma\sigma}_{I}/R^{\sigma\sigma}_{\psi}$.
Ref.~\onlinecite{Read2008} also presents
an alternative calculation, based on a bosonization procedure
related to the one which we use in Appendix~\ref{sec:many-qh-wavefunctions}. This calculation gives
all of the needed matrix elements for two, four, and six quasiholes
for a chiral $p$-wave superconductor (i.e. pure Ising CFT
with no charge sector).

A recent Monte Carlo evaluation~\cite{Baraban09}
of the matrix elements between MR Pfaffian
two-quasihole states with even- and odd-numbers of
electrons is consistent with
$R^{\sigma_1 \sigma_1}_{I_2}=1$ (which was already obtained
in an earlier numerical calculation~\cite{Tserkovnyak03})
and $R^{\sigma_1 \sigma_1}_{\psi_2}=i$ for
relatively large system sizes $N\sim 150$.
Numerical diagonalization~\cite{Prodan09} of the
3-body Hamiltonian in the presence of pinning potentials for the
quasiparticles is also consistent, for small systems
($N\sim 16$), with these $R$-matrices and also with the
fusion rules of the $c=1/2+1$ CFT. (The calculations of
Refs.~\onlinecite{Baraban09} and \onlinecite{Prodan09}
are very similar, in principle. The difference -- aside from
the difference in system sizes --
is that wavefunction overlaps are computed by Monte Carlo
evaluation of overlap integrals in Ref.~\onlinecite{Baraban09}.
In Ref.~\onlinecite{Prodan09}, different wavefunctions
are computed in an orbital basis by exact diagonalization
of the Hamiltonian with pinning potentials;
the overlaps are then obtained from the inner products
of the corresponding vectors in this basis.)

It should be straightforward to extend the
numerical calculations of
Refs.~\onlinecite{Tserkovnyak03,Baraban09,Prodan09}
to compute the $F$-matrices as well and, thereby,
fully determine the braiding properties of quasiholes
using the logic of Section~\ref{sec:general-consid}.
It should also be possible to extend those numerical calculations
to the six-quasihole case and, thereby, fully determine
braiding without making any assumptions beyond the
gap.

Finally, quasiparticle braiding
has recently been computed~\cite{Seidel08} using
coherent states in bases obtained from
the ``thin-torus'' quasi-one-dimensional limit of the
$\nu=1$ bosonic version of the MR Pfaffian state.
The key step in this derivation is changing between the basis obtained from the limit in which the torus is thin in one direction to
the dual basis obtained from the limit in which it is thin in the other direction. The relation between these two bases (which one might recognize as the modular $S$-duality) is constrained by their properties under magnetic translation. The change of basis is not computed directed, but rather is determined by consistency. However, this only determines the change of basis and subsequent braiding relations up to an $8$-fold degeneracy. This is precisely the same $8$-fold degeneracy of anyon models that are consistent with the Ising fusion algebra (i.e. are solutions to the pentagon and hexagon equations)~\footnote{In fact, these are the only modular non-Abelian anyon models with $3$ topological charge types.}, as described in Section~\ref{sec:general-consid}. Thus, the results obtained by this method are equivalent to assuming no more than the fusion algebra (or, equivalently, the modular $S$-matrix~\footnote{We note that when the topological $S$-matrix, $S_{ab} = \left( \sum_{c} d_c^2 \right)^{-1/2} \sum_{c} N_{ab}^c d_c \frac{ \theta_c}{\theta_a \theta_b} $, is unitary, the theory is called ``modular'' and obeys the relation $N_{ab}^c = \sum_{x} S_{ax} S_{bx} S_{cx}^{\ast} / S_{0x} $.}) and locality.

It should be noted that an alternative approach
to quasiparticle braiding in this state relies on
a mapping between the MR Pfaffian state and
a chiral $p$-wave superconductor. Wavefunction
analytic continuation has been computed in the latter state
directly from the BCS wavefunction~\cite{Read2000,Ivanov2001,Stern04,Stone06}, and the Berry's
matrix has been computed~\cite{Stone06} up to an overall phase
(these studies find the ratio
$R^{\sigma\sigma}_{\psi}/R^{\sigma\sigma}_{I}=i$
but cannot determine $R^{\sigma\sigma}_{I}$ itself;
in this respect, our calculation gives more information,
as does the calculation of Ref.~\onlinecite{Read2008}).
However, the mapping between the MR Pfaffian state and
a chiral $p$-wave superconductor~\cite{Greiter1992}
is not exact. While the approximate mapping
between the two is highly suggestive,
the relationship between the two states
is, strictly speaking, established by computing quasiparticle
braiding in both and comparing the result (i.e. comparing
universal quantities in both states). This is done by
comparing the results of this paper with the combined results
of Refs.~\onlinecite{Read2000,Ivanov2001,Stern04,Stone06}.
Finally, the statistics of vortices in the non-Abelian phase
of Kitaev's honeycomb lattice model~\cite{Kitaev06a}, which exhibits Ising topological order,
has recently been computed through an explicit numerical
computation of the Berry's matrix~\cite{Lahtinen09}.

\acknowledgements

We thank E. Babaev, N. Bonesteel, P. Fendley, E. Herland,
M. Hermele, N. Read, A. Seidel, and S. Simon for enlightening discussions. We thank
E. Ardonne and G. Sierra for discussing their unpublished
work with us. PB and CN thank the Aspen Center
for Physics, where part of this work was done.
VG is grateful for support provided by the KITP Santa Barbara,
where part of this work was done.
VG was supported in part by the NSF grants DMR-0449521, PHY-0904017, and PHY-0551164.
CN was supported in part by the DARPA-QuEST program.

\appendix

\section{The free boson and the Coulomb gas}
\label{sec:boson-Coulomb-gas}

This section sets the conventions and the normalizations for
free bosons used throughout this paper.

Consider a free boson field (compact such that $\phi \equiv \phi+2\pi$) with the action
\be  \label{eq:gauss} S = \frac{g}{4\pi} \int d^2 z \left( \nabla \phi \right)^2.
\ee
The value of $g$ is often not particularly important. In the CFT Coulomb gas formalism convention that we follow, we set $g=1/4$.
The correlation function of $\phi$ is given by, up to an additive constant
\be \VEV{ \phi(z, \bar z) \, \phi(w, \bar w) } = - \frac{ \ln \left| z- w \right|}{g}.
\ee
The electric operators are given by $e^{i q_k \phi(z_k, \bar z_k)}$. Their correlation function is
\be
\label{eq:corpla}
\VEV{ \prod_k e^{i q_k \phi(z_k, \bar z_k)} } = \exp \left( {\frac{1}{g} \sum_{k<l} q_k q_l \ln \left| z_k - z_l \right|} \right),
\ee
where one must have charge neutrality $\sum_k q_k =0$.

We can also consider magnetic operators. They can be thought of as vortices in the field $\phi$, such that $\phi$ changes by $2 \pi m$ when going
around a magnetic charge $m$. Let us denote magnetic operators
by ${\cal O}_m$.
For electric operators located at positions $z_k$
and magnetic operators at positions $w_a$,
the correlation function is~\cite{Nienhuis1984}
\begin{eqnarray}
&& \VEV{ \prod_k e^{i q_k \phi(z_k, \bar z_k)}  \prod_a {\cal O}_{m_a}(w_a, \bar w_a)} \notag \\
&& \qquad = \exp \left(  \frac{1}{g} \sum_{k<l} q_k q_l \ln \left| z_k - z_l \right|+ g \sum_{a<b} m_a m_b \ln \left| w_a-w_b \right| + i \sum_{k,a} q_k m_a \, {\rm arg}\,\left(z_k-w_l \right) \right),
\label{eq:applasma}
\end{eqnarray}
where $\sum_k q_k = \sum_a m_a =0$.

Throughout the paper, we also use the holomorphic field $\varphi(z)$, such that
\be \label{eq:phihol} \left< \varphi(z) \varphi(w) \right>  = -\frac{\ln (z-w) }{2g}.
\ee
We can write $\phi(z,\bar z) \equiv \varphi(z)+\bar{\varphi}(\bar z)$ and treat $\phi(z,\bar z)$ as the sum of two independent fields $\varphi(z)$ and $\bar{\varphi}(\bar z)$. As a result,
\be
\label{eq:phiholcor}
\VEV{\prod_k e^{i \alpha_k \varphi(z_k)} } = \prod_{k<l}(z_k-z_l)^{\frac{\alpha_k \alpha_l}{2g}}.
\ee
We notice that, just as $\exp\left[i q \phi(z,\bar z)\right] = \exp\left[i q \left(\varphi(z)+\bar{\varphi}(\bar z)\right)\right]$, we also have $\mathcal{O}_{m} (z,\bar z) = \exp\left[i m g \left(\varphi(z) - \bar{\varphi}(\bar z)\right)\right]$. Hence, we can write
\begin{eqnarray}
\label{eq:vert_holo}
e^{i \alpha \varphi \left( z \right)} &=& e^{i \frac{\alpha}{2} \phi \left( z,\bar{z} \right)} \mathcal{O}_{\frac{\alpha}{2g}}\left( z,\bar{z} \right)
\\
\label{eq:vert_antiholo}
e^{i \alpha \bar{\varphi} \left( \bar{z} \right)} &=& e^{i \frac{\alpha}{2} \phi \left( z,\bar{z} \right)} \mathcal{O}_{-\frac{\alpha}{2g}}\left( z,\bar{z} \right)
.
\end{eqnarray}
In other words, the holomorphic/antiholomorphic vertex operator with coefficient $\alpha$ corresponds to an operator carrying electric charge $q=\alpha/2$ and magnetic charge $m=\pm \alpha/2g$.

For further reference, we also show how to calculate the important correlation function
\begin{equation}
\VEV{\prod_k e^{i \alpha_k \varphi(z_k)}  e^{i \rho \int d^2z \, \varphi(z)}} =  \prod_{k<l}(z_k-z_l)^{\frac{\alpha_k \alpha_l}{2g}} e^{\frac 1 {2 g} \sum\limits_k \alpha_k \rho \int d^2 z
\log \left( z_k-z \right)},
\end{equation}
where now we can have $\sum_{k} \alpha_k \neq 0$, since there is a background charge density $\rho$ that can be used to maintain charge neutrality through $\rho \int d^2 z = - \sum_{k} \alpha_k $. The integral over the logarithm can be calculated under the following assumptions. First of all, we take the domain over which $z$ is integrated to be a disk. This means the imaginary part of the logarithm integrates to zero for symmetry reasons. As for its real part, we observe that
\be
\nabla^2 \log \left| z\right| = 2 \pi \delta^2 (z, \bar z),
\ee
where $\nabla^2$ is the Laplacian and $\delta^2 (z, \bar z)$ is the $2$D delta function. This allows us to calculate this integral by solving the corresponding Laplace equation
\be
\nabla_{w}^2 \int d^2 z \log \left| w-z \right| = 2 \pi ,
\ee
which gives
\be
\int d^2 z \log \left| w-z\right| = \frac{\pi}{2} \left| w \right|^2.
\ee
Taken together, this gives
\be
\label{eq:phiholcorful}
\VEV{\prod_k e^{i \alpha_k \varphi(z_k)}  e^{i \rho \int d^2z \, \varphi(z)}} =  \prod_{k<l}(z_k-z_l)^{\frac{\alpha_k \alpha_l}{2g}} e^{\frac {\pi \rho} {4 g} \sum\limits_k \alpha_k \left| z_k \right|^2}.
\ee
For the charge sector of quantum Hall states, we had electrons with $\alpha = Q_1 = \sqrt{2Mg}$ and a neutralizing background charge density $\rho_1 = - \frac{Q_1}{2\pi M}$ that gives rise to the Gaussian factors.

\section{Mathur's Procedure for Relating Products of Contour Integrals
to 2D Integrals}
\label{sec:Mathur}

The purpose of this appendix is to review Mathur's trick
which expresses sums of products of conformal blocks
in terms of a 2D integral (i.e. as the classical Boltzmann
weight for a plasma). Consider a 2D integral of the form
\begin{equation}
\label{eqn:2D-integral-example}
{\int_D} {d^2}w\, \sum_{\alpha,\beta} {\bar f_\alpha}(\bar w) Q_{\alpha\beta} {f_\beta}(w)
\end{equation}
We will assume that ${\bar f_\alpha}(\bar w) Q_{\alpha\beta} {f_\beta}(w)$
is single-valued in $D$ so that this integral is well-defined.
Let us suppose that $D$ is a simply-connected region with
no singularities. Then, we can re-write the integral
over the interior of $D$ in the form
\begin{equation}
\label{eqn:2D-contour1}
\int_{D} {d^2}w\, {\bar f}_{\alpha}(\bar w) Q_{\alpha\beta} f_{\beta}(w)
= \int_{D} {d^2}w\,
{\bar f}_{\alpha}(\bar w) Q_{\alpha\beta}
\frac{\partial}{\partial w}\left({\int_{P}^w}dw' f_{\beta}(w')\right)
\end{equation}
where $P$ is any point in the interior of $D$. Since there are no
singularities in $D$, the integral from $P$ to $w$ is independent of
the path. Then, since ${\bar f}_{\alpha}(\bar w)$ depends only on
$\bar w$ and not $w$,
\begin{eqnarray}
\label{eqn:2d-to-contour}
\int_{D} {d^2}w\, {\bar f}_{\alpha}(\bar w) Q_{\alpha\beta} f_{\beta}(w)
&=& \int_{D} {d^2}w\, \frac{\partial}{\partial w} \left({\int_P^w}dw'
{\bar f}_{\alpha}(\bar w) Q_{\alpha\beta}
{f_\beta}(w')\,
\right)\cr
&=& \frac{i}{2}\oint_{\partial D} d\bar w {\int_P^w}dw'
{\bar f}_{\alpha}(\bar w) Q_{\alpha\beta}
{f_\beta}(w')\cr
&=& \frac{i}{2}\oint_{\partial D} d\bar w\,
{\bar f}_{\alpha}(\bar w) Q_{\alpha\beta}
 \left({\int_P^w}dw'{f_\beta}(w')\right)
\end{eqnarray}
The penultimate step involves an integration by parts. Since $x=(w+\bar w)/2$,
$y=(w-\bar w)/2i$, the complex derivative is defined by
$\partial g \equiv \partial g / \partial w = (\partial_x g -i \partial_y g)/2$. Thus,
\begin{equation}
\int_{D} {d^2}w\, \partial g = \int_{D} dx\,dy\,(\partial_x g -i \partial_y g)/2 = \frac{i}{2} \oint_{\partial D} d\bar w \, g
\end{equation}

\begin{figure}[t!]
\includegraphics[width=3in]{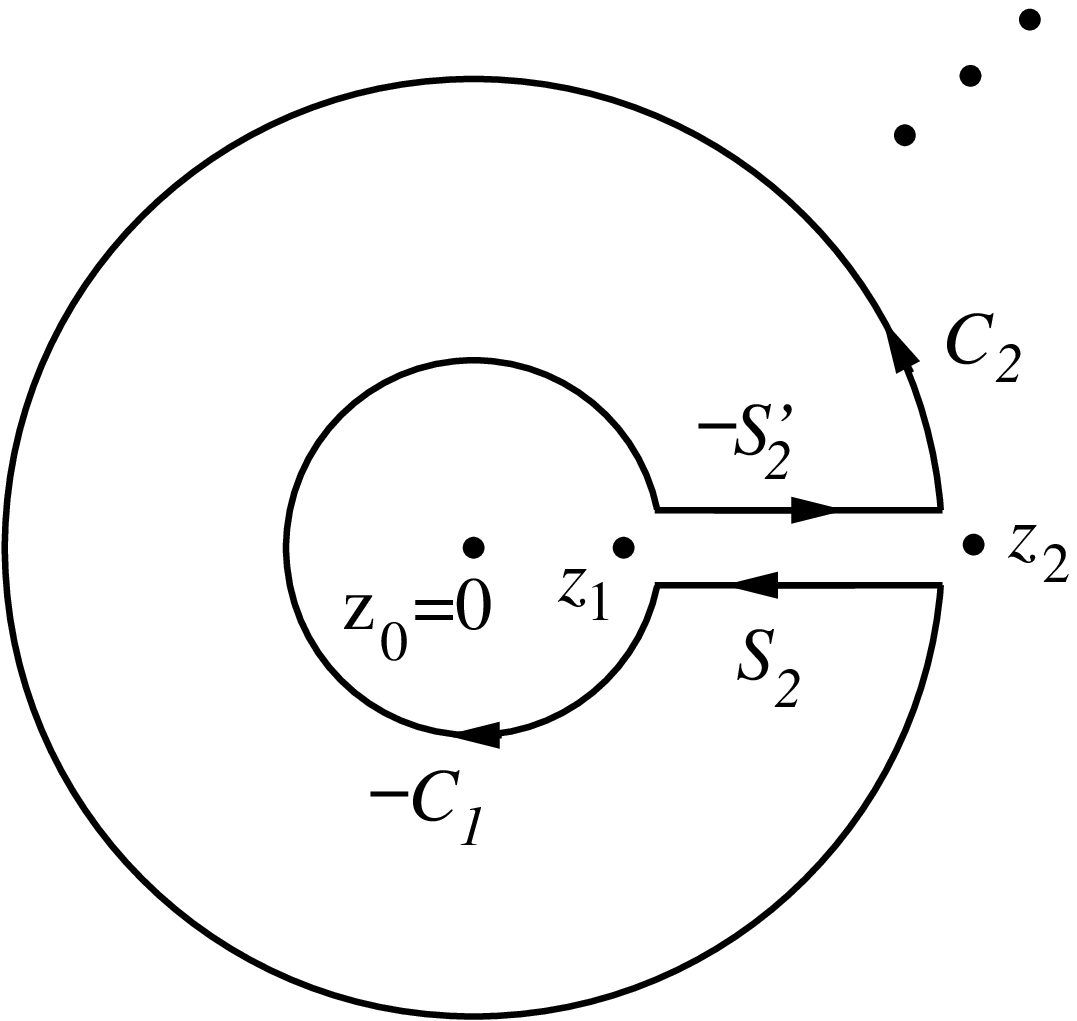}
\caption{The plane is divided into annuli, such as the annulus $A_2$
shown here, bounded by $C_2$, $S_2$, $-C_{2}$, $-S'_2$,
as described in the text. }
\label{fig:Mathur-contours}
\end{figure}

We are interested in integrals of the form Eq.~\rf{eqn:2D-integral-example}
in which ${f_\alpha}(w)$, ${\bar f_\alpha}(\bar w)$ are conformal
blocks. Thus, we expect them to depend on the coordinates
${z_1}, {z_2}, \ldots, {z_m}$ of all of the other fields besides
the one at $w$, and it will be singular when $w$ approaches any $z_k$.
In the expressions that concern us, $w$ is the coordinate of a screening charge and the $z_k$ are the coordinates of the other
screening charges, the electrons, and the quasiholes. In order
to avoid these singularities, we split the complex plane into
annuli $A_k$ with inner and outer radii $|z_{k-1}|$ and $|z_{k}|$.
We additionally define the points $z_0 = 0$ and $z_{m+1}=\infty$, so that the
annuli cover the entire complex plane.
Each of these annuli contains no singularities. However, they are not
simply-connected. Therefore, we cut the annulus $A_k$ open
along a line from $z_{k-1}$ and $z_{k}$, as shown in
Fig.~\ref{fig:Mathur-contours}. We then have a simply-connected region
bounded by the union of the curves $C_k$, $S_k$, $-C_{k-1}$, $-S'_k$.
The circular contour $-C_{k-1}$ runs from $P_k^3$ to $P_k^4$,
while $C_k$ runs from $P_k^1$ to $P_k^2$.
Then according to Eq.~\rf{eqn:2d-to-contour},
\begin{eqnarray}
\label{eqn:2d-to-contour2}
\int_{A_k} {d^2}w\, {\bar f}_{\alpha}(\bar w) Q_{\alpha\beta} f_{\beta}(w)
&=& \frac{i}{2}\oint_{\partial A_k} d\bar w \,
{\bar f}_{\alpha}(\bar w) Q_{\alpha\beta} {\int_{P_k^1}^w}dw' {f_\beta}(w') \cr
&=& \frac{i}{2}\int_{C_k} d\bar w \,
{\bar f}_{\alpha}(\bar w) Q_{\alpha\beta} {\int_{P_k^1}^w}dw' {f_\beta}(w') +
\frac{i}{2}\int_{S_k} d\bar w\,
{\bar f}_{\alpha}(\bar w) Q_{\alpha\beta} {\int_{P_k^1}^w}dw' {f_\beta}(w') \cr
&& + \,
\frac{i}{2}\int_{-C_{k-1}} d\bar w \,
{\bar f}_{\alpha}(\bar w) Q_{\alpha\beta} {\int_{P_k^1}^w}dw' {f_\beta}(w') +
\frac{i}{2}\int_{-S'_k} d\bar w \,
{\bar f}_{\alpha}(\bar w) Q_{\alpha\beta} {\int_{P_k^1}^w}dw' {f_\beta}(w')
\end{eqnarray}
Let us now define
\begin{eqnarray}
J_C &\equiv& \sum_{\alpha,\beta} \int_P^{P'}
d\bar w \, {\bar f}_{\alpha}(\bar w) Q_{\alpha\beta}
{\int_P^w}dw'{f_\beta}(w') \\
({I_C})_\alpha &\equiv& \int_P^{P'}
dw \, f_{\alpha}(w)
\end{eqnarray}
where $C$ is a contour from $P$ to $P'$
Then, we can re-write the four terms on the right-hand side
of the second equality in Eq.~\rf{eqn:2d-to-contour2}
in the form:
\begin{eqnarray}
\sum_{\alpha,\beta} \int_{C_k} d\bar w \, {\bar f}_{\alpha}(\bar w) Q_{\alpha\beta} {\int_{P_k^1}^w}dw' {f_\beta}(w') &=& J_{C_k} \\
\sum_{\alpha,\beta} \int_{S_k} d\bar w\, {\bar f}_{\alpha}(\bar w) Q_{\alpha\beta} {\int_{P_k^1}^w}dw' {f_\beta}(w') &=& \sum_{\alpha,\beta} \int_{S_k} d\bar w\, {\bar f}_{\alpha}(\bar w) Q_{\alpha\beta} \left({\int_{P_k^1}^{P_k^2}}dw' {f_\beta}(w') + {\int_{P_k^2}^w}dw' {f_\beta}(w')\right) \notag \\
&=& \sum_{\alpha,\beta} (\bar I_{S_k})_\alpha Q_{\alpha\beta} (I_{C_k})_\beta + J_{S_k} \\
\sum_{\alpha,\beta} \int_{-C_{k-1}} d\bar w \, {\bar f}_{\alpha}(\bar w) Q_{\alpha\beta} {\int_{P_k^1}^w}dw' {f_\beta}(w') &=& - \sum_{\alpha,\beta} \int_{C_{k-1}} d\bar w \, {\bar f}_{\alpha}(\bar w) Q_{\alpha\beta} \left( {\int_{P_k^1}^{P_k^4}}dw' {f_\beta}(w') + {\int_{P_k^4}^w}dw' {f_\beta}(w')\right) \notag \\
&=& - \sum_{\alpha,\beta} (\bar I_{C_{k-1}})_\alpha Q_{\alpha\beta} (I_{S'_k})_\beta - J_{C_{k-1}} \\
\sum_{\alpha,\beta} \int_{-S'_k} d\bar w\, {\bar f}_{\alpha}(\bar w) Q_{\alpha\beta} {\int_{P_k^1}^w}dw' {f_\beta}(w') &=& - \sum_{\alpha,\beta} \int_{S'_k} d\bar w\, {\bar f}_{\alpha}(\bar w) Q_{\alpha\beta} {\int_{P_k^1}^w}dw' {f_\beta}(w') = - J_{S'_k}
\end{eqnarray}
Thus, we have:
\begin{equation}
\label{eqn:2d-to-contour3}
\sum_{\alpha,\beta} \int_{A_k} {d^2}w\, {\bar f}_{\alpha}(\bar w) Q_{\alpha\beta} f_{\beta}(w)
= \frac{i}{2}\left[ J_{C_k} - J_{C_{k-1}}\right]
+ \frac{i}{2}\left[J_{S_k}- J_{S'_k}\right] +
\frac{i}{2} \sum_{\alpha,\beta} \left[(\bar I_{S_k})_\alpha Q_{\alpha\beta} (I_{C_k})_\beta
- (\bar I_{C_{k-1}})_\alpha Q_{\alpha\beta} (I_{S'_k})_\beta
\right]
\end{equation}

When we sum over the different annuli $A_k$, the $J_{C_k} - J_{C_{k-1}}$
terms will cancel. Now consider the terms in the second set of
square brackets on the right-hand-side of Eq.~\rf{eqn:2d-to-contour3}.
If $f_\alpha(w)$ is taken in a counterclockwise direction
from a point $w$ on $S'_k$ to the corresponding point on $S_k$,
then it is transformed by the monodromy matrix $M$:
\begin{equation}
\left[f_\alpha(w)\right]_{S_k}
= \sum_{\beta} M_{\alpha\beta} \left[f_\beta(w)\right]_{S'_k}
\end{equation}
This monodromy is unitary (at least for the unitary CFTs), so $M^{-1} = M^{\dagger}$. Since ${\bar f_\alpha}(\bar w) Q_{\alpha\beta} {f_\beta}(w)$ is
single-valued, $M$ satisfies
\begin{equation}
M^\dagger Q M = Q.
\end{equation}
Here, we have used matrix notation and suppressed the indices.
Consequently, $J_{S_k} = J_{S'_k}$.
This leaves only the terms in the third set of
square brackets on the right-hand-side of Eq.~\rf{eqn:2d-to-contour3}, to which we now turn.

Since there are no singularities in $A_k$,
\begin{equation}
(I_{C_k})_\alpha+(I_{S_k})_\alpha-(I_{C_{k-1}})_\alpha-(I_{S'_k})_\alpha = 0
\end{equation}
Meanwhile,
\begin{equation}
(I_{S_k})_\alpha = \sum_{\beta}  M_{\alpha\beta} (I_{S'_k})_\beta
\end{equation}
Combining these two equations, we have
\begin{eqnarray}
I_{S'_k} &=& (1-M)^{-1}\,
(I_{C_k}-I_{C_{k-1}})\cr
I_{S_k} &=& (M^\dagger-1)^{-1}\,
(I_{C_k}-I_{C_{k-1}})
\end{eqnarray}
Substituting these expressions into the third set of
square brackets on the right-hand side of Eq.~\rf{eqn:2d-to-contour3},
we find that the cross terms cancel so that we are left with
\begin{eqnarray}
\int {d^2}w\, \sum_{\alpha,\beta} {\bar f}_{\alpha}(\bar w) Q_{\alpha\beta} f_{\beta}(w) &=&
\sum_{k, \alpha, \beta} \int_{A_k} {d^2}w\, {\bar f}_{\alpha}(\bar w) Q_{\alpha\beta} f_{\beta}(w) \notag \\
\label{eqn:2d-to-contour4}
&=& \frac{i}{2}{\sum_k} \left[ -\bar I_{C_k} Q(1-M_k)^{-1} I_{C_k} + \bar I_{C_{k-1}} Q (1-M_k)^{-1} I_{C_{k-1}} \right] \\
\label{eqn:2d-to-contour5}
&=& \frac{i}{2}{\sum_k} \left[ -\bar I_{C_k} Q(1-M_k)^{-1} I_{C_k} + \bar I_{C_{k}} Q (1-M_{k+1})^{-1} I_{C_{k}} \right] .
\end{eqnarray}
We have added a subscript $k$ to $M$ in this equation
to emphasize that this is the monodromy matrix which
results from deforming $S'_k$ to $S_k$, and regrouped terms to obtain the form in the last line.
This expression is Eqs.~(2.4) and (2.19) of Ref.~\onlinecite{Mathur1992}, where there is a slight typo in Mathur's
Eq.~(2.19), in that the index $i-1$ there should actually be $i+1$.
Note that, by construction, both the left- and right-hand sides of this expression are single-valued.
By applying Eqs.~\rf{eqn:2d-to-contour4}, \rf{eqn:2d-to-contour5} repeatedly
to each screening charge integral, we obtain the desired sum of products of contour integrals.

We now specialize to the case of the Ising CFT in order to demonstrate some of the potential technical subtleties of applying Mathur's procedure.
Consider the squared norm of the
correlation function of two $\psi$ fields (or, equivalently,
the correlation function of two energy operators).
The generalization to $N$ $\psi$ fields (where $N$ is even) gives the
square of the Pfaffian, which is a factor in the square
of the ground state wavefunction. The basic structure
is already apparent for just two $\psi$ fields, however,
so we will begin with this simple case. Let us put the
two $\psi$ operators at $0$, $z$. We will call the screening
operator coordinates $w_1$, $w_2$.
There is a single conformal block so, with appropriate
choice of normalization, we can simply take $Q=1$.
When a screening operator $e^{i\alpha^{}_- \varphi}$
is taken around a $\psi$'s vertex operator
$e^{i\alpha_{31} \varphi} = e^{-i\alpha^{}_- \varphi}$
or around another screening operator, it changes by $-1$.
Thus, the monodromy $M_k$ is $\pm 1$, depending
on whether an even or odd number of operators is
contained within $C_k$. Consider the left-hand side
of Eq.~\rf{eqn:contour-to-2D-gs} for two $\psi$ fields
\begin{equation}
\int  d^2 w_1 \, d^2 w_2
\left|\left\langle e^{-i\alpha^{}_- \varphi(z)}
e^{i\alpha^{}_- \varphi({w_1})}\,
e^{i\alpha^{}_- \varphi({w_2})} \, e^{-i\alpha^{}_- \varphi(0)}
\right\rangle\right|^2 =
\int  d^2 w_1 \, d^2 w_2  \left| z \right|^3 \left| w_1-w_2 \right|^3
\prod_{i=1}^2 \left( \left| z - w_i \right|^{-3} \left| w_i \right|^{-3} \right)
.
\end{equation}
We can show that it is equal to the right-hand side
of Eq.~\rf{eqn:contour-to-2D-gs} by
applying Eq.~\rf{eqn:2d-to-contour5}.
First, we use Eq.~\rf{eqn:2d-to-contour5} to reduce
the $w_2$ integral for fixed $w_1$. There are two cases,
$0<|w_1|<|z|$ and $|w_1|>|z|$. For $|w_1|<|z|$,
Eq.~\rf{eqn:2d-to-contour5} tells us that:
\begin{eqnarray}
\int d^2 w_2 \left|\left\langle  e^{-i\alpha^{}_- \varphi(z)} \,
e^{i\alpha^{}_- \varphi({w_1})}\,
e^{i\alpha^{}_- \varphi({w_2})} \, e^{-i\alpha^{}_- \varphi(0)}
\right\rangle\right|^2_{|w_1|<|z|} &=&
\frac{i}{2}\left[ (1-M_z)^{-1}-(1-M_1)^{-1}\right]
\bar I_{C_{w_1}}I_{C_{w_1}} \notag \\
&& + \frac{i}{2}\left[ (1-M_\infty)^{-1} - (1-M_z)^{-1}\right]
\bar I_{C_z}I_{C_z}
\end{eqnarray}
Here, $C_{w_1}$ is the circle $|w|=|w_1|-\epsilon$ and $C_z$ is the circle
$|w|=|z|-\epsilon$.
$M_1$ is the monodromy of the holomorphic conformal block
\begin{equation}
\left\langle  e^{-i\alpha^{}_- \varphi(z)} \,
e^{i\alpha^{}_- \varphi({w_1})}\,
e^{i\alpha^{}_- \varphi({w_2})} \, e^{-i\alpha^{}_- \varphi(0)}
\right\rangle
\end{equation}
when $w_2$ encircles the origin with $0<|w_2|<|w_1|$.
Since such a circle encloses the $\psi$ field at the
origin's vertex operator, $e^{-i\alpha^{}_- \varphi(0)}$,
we have ${M_1} = -1$. $M_z$ is the monodromy when
$w_2$ encircles the origin with $|w_1|<|w_2|<|z|$.
Since such a circle encloses both the $\psi$ field at the
origin's vertex operator, $e^{-i\alpha^{}_- \varphi(0)}$, and the screening operator
$e^{i\alpha^{}_- \varphi({w_1})}$, this monodromy is $M_z=1$.
Finally, $M_\infty$ is the monodromy when $w_2$ encircles the origin
with $|w_2|>|z|$; since such a circle encloses
the the $\psi$ fields' vertex operators, $e^{-i\alpha^{}_- \varphi(0)}$
and $e^{-i\alpha^{}_- \varphi(z)}$, and the screening operator
$e^{i\alpha^{}_- \varphi({w_1})}$, this monodromy is $M_\infty=-1$.

Clearly, there will be divergent terms due to $M_z=1$. However, these divergence terms cancel, and so can be regulated (e.g. by adding small values to the vertex operators' charges that are taken to zero at the end). We will do this by letting $q = e^{-i 4 \pi \alpha_{-}^2} =-1$ and $q^0 = 1 -\delta$ where $\delta$ will be taken to zero at the end. Hence, for $|w_1|<|z|$, we have
\begin{eqnarray}
&&\int d^2 w_2 \left|\left\langle e^{-i\alpha^{}_- \varphi(z)} \,
e^{i\alpha^{}_- \varphi({w_1})}\,
e^{i\alpha^{}_- \varphi({w_2})} \, e^{-i\alpha^{}_- \varphi(0)}
\right\rangle\right|^2_{|w_1|<|z|} \notag \\
&& \qquad = \frac{i}{2}\left[ \delta^{-1} - (1-q)^{-1} \right]
\left|\left\langle \oint_{C_{w_1}}{\hskip -0.2 cm} dw_2 \, e^{-i\alpha^{}_- \varphi(z)}
\, e^{i\alpha^{}_- \varphi({w_1})} \, e^{i\alpha^{}_- \varphi({w_2})} \, e^{-i\alpha^{}_- \varphi(0)}
\right\rangle\right|^2 \notag \\
&& \quad \qquad + \frac{i}{2}\left[ (1-q)^{-1} - \delta ^{-1}\right]
\left|\left\langle
\oint_{C_z} {\hskip -0.2 cm} dw_2 \, e^{-i\alpha^{}_- \varphi(z)}\, e^{i\alpha^{}_- \varphi({w_2})}\,
e^{i\alpha^{}_- \varphi({w_1})}\, e^{-i\alpha^{}_- \varphi(0)}
\right\rangle\right|^2
.
\end{eqnarray}

For $|w_1|>|z|$, we obtain a similar expression,
but with $M_z=M_\infty=-1$ and $M_1=1-\delta$:
\begin{eqnarray}
&& \int d^2 w_2 \left|\left\langle e^{-i\alpha^{}_- \varphi(z)} \, e^{i\alpha^{}_- \varphi({w_1})}\,
e^{i\alpha^{}_- \varphi({w_2})} \, e^{-i\alpha^{}_- \varphi(0)}
\right\rangle\right|^2_{|w_1|>|z|} \notag \\
&& \qquad = \frac{i}{2}\left[ (1-q^2)^{-1} - (1-q)^{-1} \right]
\left| \left\langle \oint_{C_z} {\hskip -0.2 cm} dw_2 \, e^{i\alpha^{}_- \varphi({w_1})} \, e^{-i\alpha^{}_- \varphi(z)} \, e^{i\alpha^{}_- \varphi({w_2})} \, e^{-i\alpha^{}_- \varphi(0)}
\right\rangle\right|^2 \notag \\
&& \quad \qquad + \frac{i}{2} \left[ (1-q)^{-1} - (1-q^2)^{-1} \right]
\left|\left\langle \oint_{C_{w_1}}{\hskip -0.2 cm} dw_2 \, e^{i\alpha^{}_- \varphi({w_1})} \, e^{i\alpha^{}_- \varphi({w_2})}\,e^{-i\alpha^{}_- \varphi(z)}\, e^{-i\alpha^{}_- \varphi(0)}
\right\rangle\right|^2
\end{eqnarray}

We can now perform the $w_1$ integral. For instance,
\begin{eqnarray}
&& \int_{|w_1|<|z|}  d^2 w_1\,
\left|\left\langle \oint_{C_{w_1}}{\hskip -0.2 cm} dw_2\, e^{-i\alpha^{}_- \varphi(z)} \, e^{i\alpha^{}_- \varphi({w_1})}
e^{i\alpha^{}_- \varphi({w_2})}\, e^{-i\alpha^{}_- \varphi(0)}
\right\rangle\right|^2 \notag \\
&& \qquad = \frac{i}{2}\left[ -(1-q)^{-1} \right]
\left|\left\langle \oint_{C_z}{\hskip -0.2 cm} dw_1\, \oint_{C_z}{\hskip -0.2 cm} dw_2 \, e^{-i\alpha^{}_- \varphi(z)} \, e^{i\alpha^{}_- \varphi({w_1})} \, e^{i\alpha^{}_- \varphi({w_2})} \, e^{-i\alpha^{}_- \varphi(0)}
\right\rangle\right|^2 + J \text{-terms}
.
\end{eqnarray}
Since $w_2$ is always enclosed when $w_1$ encircles the
origin (by definition, $C_1$ is the circle at radius $|w_1|-\epsilon$),
the winding of $w_1$ around $w_2$ contributes to the monodromy, in addition to the winding of $w_1$ around $0$ and $w_2$ around $0$, giving the total combined monodromy of $M = q$. When $w_1$ lies on the contour $C_z$, the contour $C_{w_1}$
becomes the same contour (but point split, so that it is at infinitesimally
smaller radius.) The $J$-terms cancel off, and so can be neglected.

Similarly, we get
\begin{eqnarray}
&& \int_{|w_1|>|z|}  d^2 w_1 \,
\left|\left\langle \oint_{C_{w_1}}{\hskip -0.2 cm} dw_2\,  e^{i\alpha^{}_- \varphi({w_1})}
e^{i\alpha^{}_- \varphi({w_2})}\, e^{-i\alpha^{}_- \varphi(z)} \, e^{-i\alpha^{}_- \varphi(0)}
\right\rangle\right|^2 \notag \\
&& \qquad = \frac{i}{2}\left[ (1-q^3 )^{-1} \right]
\left|\left\langle \oint_{C_z}{\hskip -0.2 cm} dw_1\, \oint_{C_z}{\hskip -0.2 cm} dw_2 \, e^{-i\alpha^{}_- \varphi(z)} \, e^{i\alpha^{}_- \varphi({w_1})} \, e^{i\alpha^{}_- \varphi({w_2})} \, e^{-i\alpha^{}_- \varphi(0)}
\right\rangle\right|^2 + J \text{-terms} \\
&& \int_{|w_1|<|z|}  d^2 w_1\,
\left|\left\langle \oint_{C_{z}}{\hskip -0.2 cm} dw_2\, e^{-i\alpha^{}_- \varphi(z)} \, e^{i\alpha^{}_- \varphi({w_2})}\, e^{i\alpha^{}_- \varphi({w_1})} e^{-i\alpha^{}_- \varphi(0)}
\right\rangle\right|^2 \notag \\
&& \qquad = \frac{i}{2}\left[ -(1-q)^{-1} \right]
\left|\left\langle \oint_{C_z}{\hskip -0.2 cm} dw_1\, \oint_{C_z}{\hskip -0.2 cm} dw_2 \, e^{-i\alpha^{}_- \varphi(z)} \, e^{i\alpha^{}_- \varphi({w_1})} \, e^{i\alpha^{}_- \varphi({w_2})} \, e^{-i\alpha^{}_- \varphi(0)}
\right\rangle\right|^2 + J \text{-terms} \\
&& \int_{|w_1|>|z|}  d^2 w_1\,
\left|\left\langle \oint_{C_{z}}{\hskip -0.2 cm} dw_2\, e^{i\alpha^{}_- \varphi({w_1})} e^{-i\alpha^{}_- \varphi(z)} \, e^{i\alpha^{}_- \varphi({w_2})}\, e^{-i\alpha^{}_- \varphi(0)}
\right\rangle\right|^2 \notag \\
&& \qquad = \frac{i}{2}\left[ (1-q)^{-1} \right]
\left|\left\langle \oint_{C_z}{\hskip -0.2 cm} dw_1\, \oint_{C_z}{\hskip -0.2 cm} dw_2 \, e^{-i\alpha^{}_- \varphi(z)} \, e^{i\alpha^{}_- \varphi({w_1})} \, e^{i\alpha^{}_- \varphi({w_2})} \, e^{-i\alpha^{}_- \varphi(0)}
\right\rangle\right|^2 + J \text{-terms}
.
\end{eqnarray}

Putting this all together, we find
\begin{eqnarray}
&& \int  d^2 w_1 \, d^2 w_2
\left|\left\langle e^{-i\alpha^{}_- \varphi(z)}
e^{i\alpha^{}_- \varphi({w_1})}\,
e^{i\alpha^{}_- \varphi({w_2})} \, e^{-i\alpha^{}_- \varphi(0)}
\right\rangle \right|^2 \notag \\
\label{eqn:two-contours}
&& \qquad = \frac{1}{4} \left( q^{1/2} - q^{-1/2} \right)^{-1} \left( q^{3/2} - q^{-3/2} \right)^{-1} \left|\left\langle \oint_{C_z}{\hskip -0.2 cm} dw_1\, \oint_{C_z}{\hskip -0.2 cm} dw_2 \, e^{-i\alpha^{}_- \varphi(z)} \, e^{i\alpha^{}_- \varphi({w_1})} \, e^{i\alpha^{}_- \varphi({w_2})} \, e^{-i\alpha^{}_- \varphi(0)} \right\rangle\right|^2 \\
&& \qquad = \frac{1}{16} \left|\left\langle V_{31}^{20} (z) V_{31}^{00} (0) \right\rangle\right|^2
\end{eqnarray}
where we used $q=-1$ and the definition of screened vertex operators to obtain the last line (notice that the divergent terms canceled prior to taking $\delta \rightarrow 0$). This is precisely the two-$\psi$ version of Eq.~\rf{eqn:contour-to-2D-gs}. We note that we could have derived the same result with the $\psi$ fields at two arbitrary points $z_1$ and $z_2$. In that case, one would produce screening charge contours at both radii $|z_1|$ and $|z_2|$ using this procedure. In the end, the terms with screening charge contours at the smaller radius will vanish due to Felder's rules, so one will again be left with both screening charges attached to the $\psi$ field's vertex operator that is at the larger radius. The derivation here allowed us to avoid some extra steps by situating one of the $\psi$ fields at the origin.

From the preceding derivation, we can now see how the general $N$-$\psi$ case works.
Since the monodromy for taking a screening operator around
a $\psi$ operator is $-1$, there will only be a non-trivial contribution
from the contour associated with every second $\psi$ operator.
Consequently, as in Eq.~\rf{eqn:two-contours},
there will be two screening operator contour integrals
attached to every second $\psi$ operator, precisely as on the
right-hand side of Eq.~\rf{eqn:contour-to-2D-gs}.

Turning to conformal blocks with $\sigma$ fields,
we consider first the combination that gives us the
trace of the overlap matrix, namely the correlation
function of $\sigma(\eta,\bar \eta)$ operators and $N$
energy operators. For this combination,
$Q_{\alpha\beta}=\delta_{\alpha\beta}$.
The monodromy matrices are again diagonal:
the monodromy for a screening charge to go around
a $\psi$ operator is $-1$; to go around a $\sigma$ operator, it is $\pm i$.
In a similar manner to the steps that led to
Eq.~\rf{eqn:two-contours}, the diagonality
of $Q$ and $M$ simplifies matters and leads to
Eq.~\rf{eq:pros}. Correlation functions with
both order and disorder operators, which give us the off-diagonal
elements and the difference between the diagonal elements
of the overlap matrix, are a little more complicated
because $Q$ is no longer diagonal. They
are considered in detail in the next appendix.

\section{Correlation function of two order and two disorder
operators in the Ising model}
\label{sec:2-order-disorder}

As explained in Section~\ref{sec:six} we can take advantage of
Eqs.~\rf{eq:consfour-again}, \rf{eq:consfour-psi-again}, \rf{eq:consfour4a}, \rf{eq:consfour4b}, \rf{eq:opp1}, \rf{eq:opp2}, and \rf{eq:opp3} to represent the correlation function of two order and two disorder operators in the Ising model in terms of the following
Coulomb gas correlator
\begin{eqnarray}
&& \VEV{ \mu(\eta_1, \bar \eta_1) \mu(\eta_2, \bar \eta_2) \sigma(\eta_3, \bar \eta_3) \sigma (\eta_4, \bar \eta_4)} = \\
&& \int d^2 w \VEV{ e^{- \frac{i}{4\sqrt{3}} \varphi(\eta_1) +i \frac{\sqrt{3}}{4} \bar \varphi(\bar \eta_1)} \, e^{ i \frac{\sqrt{3}}4 \varphi(\eta_2) - \frac{i}{4\sqrt{3}} \bar \varphi(\bar \eta_2)}
\, e^{i \frac{\sqrt{3}}{4} \varphi(\eta_3) + i \frac{\sqrt{3}}{4} \bar \varphi(\bar \eta_3)} \, e^{-i \frac{\sqrt{3}}{2} \varphi(w) - i \frac{\sqrt{3}}{2} \bar \varphi(\bar w)} \, e^{i \frac{\sqrt{3}}{4} \varphi(\eta_4) + i \frac{\sqrt{3}}{4} \bar \varphi(\bar \eta_4)} } + \ {\rm c. c.} \nonumber
\end{eqnarray}

Evaluating the correlation function results in the following expression, which is the particular case of Eq.~\rf{eq:disorderco} with $N=0$:
\begin{multline}
\VEV{ \mu(\eta_1, \bar \eta_1) \mu(\eta_2, \bar \eta_2) \sigma(\eta_3, \bar \eta_3) \sigma (\eta_4, \bar \eta_4)} =
\left( \eta_{13} \eta_{14} \right)^{- \frac 1 8} \left( \bar \eta_{13} \bar \eta_{14} \right)^{\frac 3 8} \left( \eta_{23} \eta_{24} \right)^{\frac  3 8}
\left( \bar \eta_{23} \bar \eta_{24} \right)^{-\frac  1 8} \left| \eta_1 - \eta_2 \right|^{- \frac 1 4} \left| \eta_3 - \eta_4 \right|^{\frac 3 4} \\
\times \int d^2 w \left\{ \left[ \left(w - \eta_1 \right) \left( \bar w - \bar \eta_2 \right) \right]^{\frac 1 4} \left[ \left( w- \eta_2 \right) \left( \bar w - \bar \eta_1 \right)\right]^{-\frac 3 4} \left| w-\eta_3  \right|^{-\frac 3 2} \left| w-\eta_4  \right|^{-\frac  3 2} \right\} + \ {\rm c. c.}
\label{eq:www}
\end{multline}
Our goal is to calculate the integral over $w$ and show that the result of integration indeed coincides with this correlation function, as given, for example, in Ref.~\onlinecite{Belavin1984}.

The first step of the calculation, as standard in practical calculations of four point correlation functions, is to use global conformal invariance to assign specific values to three of the variables:
\begin{equation}
\eta_1=\infty , \quad \eta_2=1, \quad \eta_3=x , \quad \eta_4=0.
\end{equation}
$x$ remains arbitrary and is effectively the only free parameter upon which the correlation function depends. Without loss of generality,
we take $x$ to be a real variable (which can be analytically continued to the complex plane later if needed) satisfying
\begin{equation}
0< x< 1.
\end{equation}
When taking the limit $\eta_1 \rightarrow \infty$, we multiply the correlation function Eq.~\rf{eq:www} by
$(\eta_1 \bar{\eta}_1)^{\frac 1 8}$ to ensure a finite result, since the conformal dimension of the order and disorder
operators is $1/16$. This results in the following expression for the correlation function
\begin{eqnarray}
\VEV{ \mu(\infty, \infty) \mu(1, 1) \sigma(x, \bar x) \sigma(0, 0) } &=&
(1-x)^{\frac{3}{8} } (1-\bar x )^{- \frac{1}{8} } \left| x\right|^{\frac{3}{4}}
\int d^2 w \left( 1-w \right)^{-\frac{3}{4}} \left( 1-\bar{w} \right)^{\frac{1}{4}}
\left| x - w \right|^{-\frac 3 2} \left| w \right|^{-\frac 3 2} \notag \\
&& + (1-\bar{x})^{\frac{3}{8} } (1- x )^{- \frac{1}{8} } \left| x\right|^{\frac{3}{4}}
\int d^2 w \left( 1- \bar{w} \right)^{-\frac{3}{4}} \left( 1-{w} \right)^{\frac{1}{4}}
\left| x - w \right|^{-\frac 3 2} \left| w \right|^{-\frac 3 2}
\label{eq:ww1}
\end{eqnarray}
We now introduce the following convenient notation
\begin{eqnarray}
f_1 &=& (1-x)^{\frac{3}{8}}  x^{\frac{3}{8}} (1-w)^{-\frac{3}{4}} (x-w)^{-\frac{3}{4}}  w^{-\frac{3}{4}}  \\
f_2 &=& (1-x)^{-\frac{1}{8}} x^{\frac{3}{8}} (1-w)^{\frac{1}{4}}  (x-w)^{-\frac{3}{4}}  w^{-\frac{3}{4}},
\end{eqnarray}
as well as the following matrix
\begin{equation}
\label{eqq}
Q=\left(
\begin{matrix} 0 & 1 \cr 1 & 0 \end{matrix}
\right).
\end{equation}
Then Eq.~\rf{eq:ww1} can be rewritten in the following compact way using matrix notation
\begin{equation}
\label{eq:matint}
\VEV{ \mu(\infty, \infty) \mu(1, 1) \sigma(x, \bar x) \sigma(0, 0) } = \int d^2 w\,  \bar f Q f.
\end{equation}
We are now in a position to use the techniques developed by Mathur in Ref.~\onlinecite{Mathur1992} specifically to compute integrals of this sort. In that paper, it was shown that an expression in the form of Eq.~\rf{eq:matint} can be rewritten in the form of Eq.~\rf{eqn:2d-to-contour5}. Hence, Eq.~\rf{eq:matint} becomes
\begin{eqnarray}
\label{eq:mathurans}
&& \VEV{ \mu(\infty, \infty) \mu(1, 1) \sigma(x, \bar x) \sigma(0, 0) } = \notag \\
&& \quad \frac{i}{2} \left[ -  \bar I^{(1)} Q (1-M_1)^{-1} I^{(1)} + \bar I^{(1)} Q (1-M_2)^{-1} I^{(1)} - \bar I^{(2)} Q \left( 1-M_2 \right)^{-1} I^{(2)} + \bar I^{(2)} Q \left( 1-M_3 \right)^{-1} I^{(2)} \right]
.
\end{eqnarray}
\begin{figure}[t!]
\includegraphics[width=7cm,angle=-90]{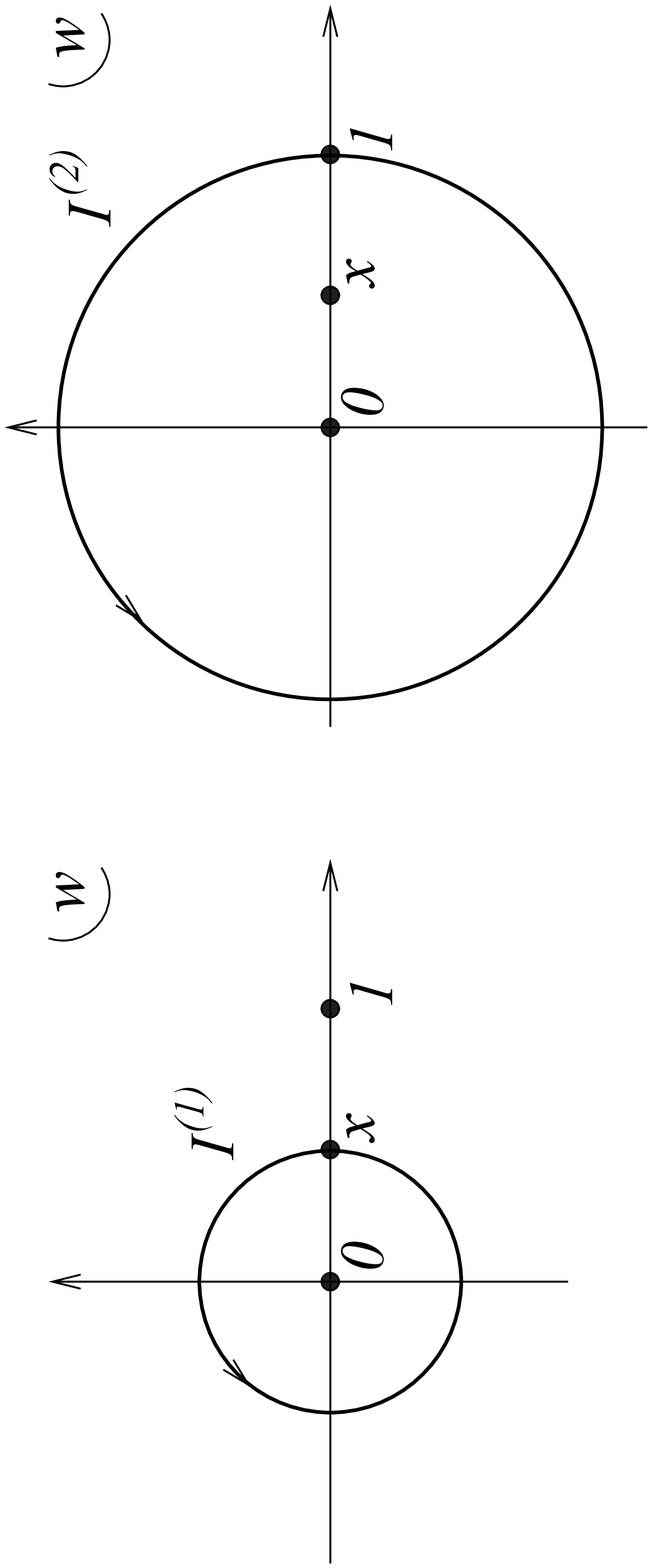}
\caption{The integration contours in Eq.~\rf{eq:maturinteg}.}
\label{mathur}
\end{figure}
Here $I^{(1)}$ and $I^{(2)}$ are contour integrals defined according to
\begin{equation}
\label{eq:maturinteg}
I^{(1)}_\alpha = \oint_{\left| w \right| = x} dw \, f_\alpha \, ,
\qquad
I^{(2)}_\alpha = \oint_{\left| w \right| = 1} dw \, f_\alpha.
\end{equation}
The contours of integration are shown in Fig.~\ref{mathur}.
$M_1$, $M_2$ and $M_3$ are the analytic continuation matrices of the functions $f$, defined in the following way: Take the function $f$ just above the real axis and analytically continue it over $w$ along a big circle centered at zero to the values just below the real axis. The new value $f_C$ is then given by
\begin{equation}
f_C = M_1 f, \quad 0<w<x; \qquad f_C = M_2 f,  \quad x<w<1; \qquad f_C = M_3 f, \quad x>1.
\end{equation}
In this case, these matrices are diagonal and can be calculated in a straightforward fashion, to give
\begin{equation}
\label{eq:Monodromies}
M_1 = \left( \begin{matrix}  i & 0 \cr 0 & i \end{matrix} \right), \quad
M_2 = \left( \begin{matrix} -1 & 0 \cr 0 &  -1 \end{matrix} \right), \quad
M_3 = \left( \begin{matrix}  -i& 0 \cr 0 & -i \end{matrix} \right).
\end{equation}
All that remains  is to compute the integrals of Eq.~\rf{eq:maturinteg} and substitute them into Eq.~\rf{eq:mathurans} to find the answer. The appropriate integrals are computed by deforming the contours from $\left|w \right| = x$ to $0<w<x$, and from $\left| w \right| = 1$ to $1<w<\infty$. The resulting integrals are then standard and can be expressed in terms of hypergeometric functions~\cite{AbramowitzBook,Dotsenko1984}. We find:
\begin{eqnarray}
I_1^{(1)} &=& (1-x)^{\frac{3}{8}}  x^{\frac{3}{8}} \oint_{\left|w \right| = x} dw \, (1-w)^{-\frac{3}{4}} (x-w)^{-\frac{3}{4}}  w^{-\frac{3}{4}} \notag \\
&=& (1-x)^{\frac{3}{8}} x^{\frac{3}{8}} \left(-1 + e^{-i \frac{3\pi}{2} } \right) \int_0^x dw  \, (1-w)^{-\frac{3}{4}} (x-w)^{-\frac{3}{4}}  w^{-\frac{3}{4}} \notag \\
&=& \sqrt{2} e^{i \frac{3 \pi}{4} } (1-x)^{\frac{3}{8}} x^{ - \frac{1}{8}} \int_0^1 d\zeta  \, \zeta^{-\frac{3}{4}}  (1-\zeta)^{-\frac{3}{4}} (1-x \zeta)^{-\frac{3}{4}}  \notag \\
&=& \sqrt{2} e^{i \frac{3 \pi}{4}} (1-x)^{\frac{3}{8}} x^{ - \frac{1}{8}} \frac{\Gamma(\frac{1}{4}) \Gamma(\frac{1}{4})}{\Gamma(\frac{1}{2})} F\left(\frac{3}{4}, \frac{1}{4}; \frac{1}{2}; x \right) \notag \\
&=& \sqrt{2} e^{i \frac{3 \pi}{4} } (1-x)^{-\frac{1}{8}} x^{ - \frac{1}{8}} \frac{\Gamma(\frac{1}{4}) \Gamma(\frac{1}{4})}{\Gamma(\frac{1}{2})} F\left(-\frac{1}{4}, \frac{1}{4}; \frac{1}{2}; x \right) \notag \\
&=& e^{i \frac{3 \pi}{4}} \frac{\Gamma(\frac{1}{4}) \Gamma(\frac{1}{4})}{\Gamma(\frac{1}{2})} (1-x)^{-\frac{1}{8}} x^{ - \frac{1}{8}} \sqrt{1+\sqrt{1-x}},
\end{eqnarray}
\begin{eqnarray}
I_2^{(1)}& =& (1-x)^{-\frac{1}{8}} x^{\frac{3}{8}} \oint_{\left|w \right| = x} dw \, (1-w)^{\frac{1}{4}}  (x-w)^{-\frac{3}{4}}  w^{-\frac{3}{4}} \notag \\
&=& (1-x)^{-\frac{1}{8}} x^{\frac{3}{8}} \left(-1 + e^{-i \frac{3\pi }{2} } \right) \int_0^x dw  \, (1-w)^{\frac{1}{4}} (x-w)^{-\frac{3}{4}}  w^{-\frac{3}{4}} \notag \\
&=& \sqrt{2} e^{i \frac{3 \pi}{4} } (1-x)^{-\frac{1}{8}} x^{-\frac{1}{8}} \int_0^1 d\zeta  \, \zeta^{-\frac{3}{4}}  (1-\zeta)^{-\frac{3}{4}} (1-x \zeta)^{\frac{1}{4}}  \notag \\
&=& \sqrt{2} e^{i \frac{3 \pi}{4} } (1-x)^{-\frac{1}{8}} x^{-\frac{1}{8}} \frac{\Gamma(\frac{1}{4}) \Gamma(\frac{1}{4})}{\Gamma(\frac{1}{2})} F\left(-\frac{1}{4}, \frac{1}{4}; \frac{1}{2}; x \right) \notag \\
&=& e^{i \frac{3 \pi}{4} } \frac{\Gamma(\frac{1}{4}) \Gamma(\frac{1}{4})}{\Gamma(\frac{1}{2})} (1-x)^{-\frac{1}{8}} x^{ - \frac{1}{8}} \sqrt{1+\sqrt{1-x}}
\end{eqnarray}
\begin{eqnarray}
I_1^{(2)} &=& (1-x)^{\frac{3}{8}}  x^{\frac{3}{8}} \oint_{\left|w \right| = 1} dw \, (1-w)^{-\frac{3}{4}} (x-w)^{-\frac{3}{4}}  w^{-\frac{3}{4}} \notag \\
&=& (1-x)^{\frac{3}{8}} x^{\frac{3}{8}} \left(1 - e^{-i \frac{3\pi}{2} } \right) \int_1^\infty dw  \, (1-w)^{-\frac{3}{4}} (x-w)^{-\frac{3}{4}}  w^{-\frac{3}{4}} \notag \\
&=& \sqrt{2} e^{-i \frac{\pi}{4} } (1-x)^{\frac{3}{8}} x^{ \frac{3}{8}} e^{-i \frac{3\pi}{2} } \int_0^1 d\zeta  \, \zeta^{\frac{1}{4}} (1-\zeta)^{-\frac{3}{4}} (1-x \zeta)^{-\frac{3}{4}}  \notag \\
&=& \sqrt{2} e^{i \frac{\pi}{4} } (1-x)^{\frac{3}{8}} x^{\frac{3}{8}} \frac{\Gamma(\frac{5}{4}) \Gamma(\frac{1}{4})}{\Gamma(\frac{3}{2})} F\left(\frac{3}{4}, \frac{5}{4}; \frac{3}{2}; x \right) \notag \\
&=& e^{i \frac{\pi}{4} } \frac{\Gamma(\frac{1}{4}) \Gamma(\frac{1}{4})}{ \Gamma(\frac{1}{2})} (1-x)^{-\frac{1}{8}} x^{-\frac{1}{8}} \sqrt{1 - \sqrt{1-x}},
\end{eqnarray}
\begin{eqnarray}
I_2^{(2)} &=& (1-x)^{-\frac{1}{8}}  x^{\frac{3}{8}} \oint_{\left|w \right| = 1} dw \, (1-w)^{\frac{1}{4}} (x-w)^{-\frac{3}{4}}  w^{-\frac{3}{4}} \notag \\
&=& (1-x)^{-\frac{1}{8}} x^{\frac{3}{8}} \left(1 - e^{-i \frac{3\pi}{2} } \right) \int_1^\infty dw  \, (1-w)^{\frac{1}{4}} (x-w)^{-\frac{3}{4}}  w^{-\frac{3}{4}} \notag \\
&=& \sqrt{2} e^{-i \frac{\pi}{4} } (1-x)^{-\frac{1}{8}} x^{ \frac{3}{8}} e^{-i \frac{\pi}{2} } \int_0^1 d\zeta  \, \zeta^{-\frac{3}{4}} (1-\zeta)^{\frac{1}{4}} (1-x \zeta)^{-\frac{3}{4}}  \notag \\
&=& - \sqrt{2} e^{i \frac{\pi}{4} } (1-x)^{-\frac{1}{8}} x^{\frac{3}{8}} \frac{\Gamma(\frac{1}{4}) \Gamma(\frac{5}{4})}{\Gamma(\frac{3}{2})} F\left(\frac{3}{4}, \frac{1}{4}; \frac{3}{2}; x \right) \notag \\
&=& - e^{i \frac{\pi}{4} } \frac{\Gamma(\frac{1}{4}) \Gamma(\frac{1}{4})}{ \Gamma(\frac{1}{2})} (1-x)^{-\frac{1}{8}} x^{-\frac{1}{8}} \sqrt{1 - \sqrt{1-x}}.
\end{eqnarray}
We define the conformal blocks
\begin{eqnarray}
\mathcal{F}_0 &\equiv& e^{i \frac{3 \pi}{4} } \frac{\Gamma(\frac{1}{4}) \Gamma(\frac{1}{4})}{\Gamma(\frac{1}{2})} \sqrt{2} (1-x)^{-\frac{1}{8}} x^{ - \frac{1}{8}} \sqrt{1+\sqrt{1-x}} = \sqrt{2} I_1^{(1)} = \sqrt{2} I_2^{(1)} , \\
\mathcal{F}_1 &\equiv& e^{i \frac{\pi}{4} } \frac{\Gamma(\frac{1}{4}) \Gamma(\frac{1}{4})}{ \Gamma(\frac{1}{2})} \sqrt{2} (1-x)^{-\frac{1}{8}} x^{-\frac{1}{8}} \sqrt{1 - \sqrt{1-x}} = \sqrt{2} I_1^{(2)} = - \sqrt{2} I_2^{(2)}
.
\end{eqnarray}
This is the conventional definition of the Ising CFT's conformal blocks, up to the overall phases, which are unimportant here since they cancel with their conjugates, and the overall constant $\frac{\Gamma(\frac{1}{4}) \Gamma(\frac{1}{4})}{\Gamma(\frac{1}{2})}\sqrt{2}$, which causes no problems since it enters as a common factor to both blocks. Employing vector notation, we can now write
\begin{equation}
\label{eq:I-F}
I^{(1)} = \sqrt{2} \left( \begin{matrix} \mathcal{F}_0 \cr \mathcal{F}_0 \end {matrix} \right), \quad I^{(2)} = \sqrt{2} \left( \begin{matrix} \mathcal{F}_1 \cr -\mathcal{F}_1 \end{matrix} \right).
\end{equation}
Substituting Eqs.~\rf{eqq}, \rf{eq:Monodromies}, \rf{eq:I-F} into Eq.~\rf{eq:mathurans}, we finally find
\begin{equation}
\label{eq:apcans}
\VEV{ \mu(\eta_1, \bar \eta_1) \mu(\eta_2, \bar \eta_2) \sigma(\eta_3, \bar \eta_3) \sigma (\eta_4, \bar \eta_4)}  =
\bar{\mathcal{F}}_0 \mathcal{F}_0 - \bar{\mathcal{F}}_1 \mathcal{F}_1 .
\end{equation}
This is indeed the right answer for this correlation function.
Not only does it match the expected form in Eqs.~\rf{eq:threeentries-1}, \rf{eq:threeentries-2}, and \rf{eq:threeentries-3},
it also coincides with the explicit expression for this
correlation function worked out in Ref.~\onlinecite{Belavin1984}.

\section{Screening in a two-dimensional two-component plasma}
\label{sec:two-comp-screening}

In this section, we examine the screening properties of a $2$D two-component plasma using field theoretical analysis (see, e.g. Ref.~\onlinecite{Zinn89}).
The interaction energy of a number of electric charges $q_k$ located at positions $z_k$ in 2D is given by
\be
\Phi = - \sum_{i<j} q_i q_j \ln \left| z_i - z_j \right|.
\ee
Suppose we have a 2D plasma of a large but equal number of electric charges of magnitude $q$ and $-q$ at temperature $T$.
The partition function of this plasma can be written as
\be
Z = \int \prod_k d^2 z_k \, e^{- \Phi / T} = \int \prod_k d^2 z_k \, e^{\frac{1}{T} \sum\limits_{i<j} q_i q_j \ln \left| z_i - z_j \right|},
\ee
where $q_i$ are either $q$ or $-q$.

With the help of Eq.~\rf{eq:corpla}, we observe that this same partition function can be rewritten as
\be
\label{eq:partitioncoulomb}
Z = \int  \prod_k d^2 z_k  \VEV{\prod_l e^{i q_l \phi(z_l, \bar z_l)}}
\ee
while making the identification
\be
\label{eq:tg}
T = g.
\ee
This identification of $T$ with $g$ is not particularly necessary. We could have multiplied all the electric charges by an arbitrary
factor, and simultaneously multiplied the temperature by the square of this factor. This would keep the partition
function in Eq.~\rf{eq:partitioncoulomb} exactly the same. However, $T=g$ is a convenient choice that we use in this paper.

The partition function of such a plasma is most easily computed in the grand canonical ensemble, using methods developed originally in the context of the Kosterlitz-Thouless transition~\cite{Kosterlitz73}. For completeness, we present the derivation
here. The partition function in the grand canonical ensemble looks like
\be
Z = \sum_{n=1}^{\infty} \sum_{q_l=\pm q} \frac{1}{n!} \int \prod_{k=1}^n d^2 z_k \, \VEV{ \exp \left[i \sum\limits_{l=1}^n q_l \phi(z_l, \bar z_l) \right]} \lambda^n.
\ee
Here $\lambda$ is the fugacity, the parameter whose logarithm gives the chemical potential of the charges.
We can now sum over $q_k=\pm q$ and over $n$ to find~\cite{Polyakov1987}
\be
\label{eq:KT}
Z = \frac{ \int {\cal D} \phi \, e^{-\frac{g}{4\pi} \int d^2 z \left( \nabla \phi \right)^2 +2 \lambda \int d^2 z \cos \left( q \phi \right)} }{ \int {\cal D} \phi \, e^{-\frac{g}{4\pi} \int d^2 z \left( \nabla \phi \right)^2}}.
\ee

The behavior of this plasma depends crucially on whether $\lambda$ is a relevant or irrelevant perturbation. Since the correlation function
\be
\VEV{e^{i q \phi(z,\bar z)} \, e^{-i q \phi(w, \bar w)}} =  \left| z- w\right|^{-q^2 / g },
\ee
the dimension of this perturbation is
\be
\Delta_\lambda = 2-\frac{q^2}{2g}.
\ee
The perturbation is relevant if $\Delta_\lambda>0$ or, equivalently,
$q<\sqrt{4 g}$. Otherwise, it is irrelevant.

If the perturbation is irrelevant, then the plasma does not screen. If it is relevant, the plasma screens.
In the latter case, the correlation function of two electric operators is a constant if
they are farther away from each other than the correlation (screening) length $\ell_2$ of the plasma
\be
\int {\cal D} \phi \, e^{i q_1 \phi(z_1,\bar z_1)} e^{i q_2 \phi(z_2, \bar z_2)} \, e^{-\frac{g}{4\pi} \int d^2 z \left( \nabla \phi \right)^2 + 2 \lambda \int d^2 z \cos \left( q \phi \right)} \, \sim \, C_2 \,+ O\bigl(e^{-\left|z_1-z_2 \right|/\ell_2}\bigr),
\ee
where $C_2$ is a constant independent of $z_1$ and $z_2$. To see this, we simply observe that if $\lambda$ is a relevant perturbation, typical values of $\phi$ are restricted to the minima of $\cos \left( q \phi \right)$, and so $\phi$ fluctuates very little about this value.

Indeed, upon identifying $g$ with the temperature of the plasma, we see that it is at high temperature that the plasma screens.
At low temperature, charges of opposite magnitude bind strongly pairwise to form neutral dipoles and hence the plasma does not screen. The transition temperature between these two phases is seen to be $T_{c_2} = q^2 /4$.

Additionally, one can consider the behavior of the magnetic operators in the screening phase. Their correlator goes to zero at large distances
\be
\label{eq:mag}
\int {\cal D} \phi \, {\cal O}_{m}(z_1,\bar z_1) {\cal O}_{-m}(z_2,\bar z_2) \, e^{-\frac{g}{4\pi} \int d^2 z \left( \nabla \phi \right)^2 + 2 \lambda \int d^2 z \cos \left( q \phi \right)} \sim  O\!\left( e^{-\left| z_1-z_2 \right|/\ell_2} \right).
\ee
This corresponds to the well known fact that magnetic charges are confined in the electrically-screening phase.
To see why this is so, recall that the operator ${\cal O}_{m}(z,\bar z)$
creates a vortex in the field $\phi$, so that $\phi$ winds by $2\pi m$ in
going around $(z,\bar z)$. As was pointed out earlier, $\phi$ is restricted to the minima of $\cos \left( q \phi \right)$ in the screening phase. Therefore, the two magnetic charges $m$ and $-m$ are necessarily connected by a region across which $\phi$ changes by $2\pi m$. Since $\phi$ must leave the minima of $\cos \left( q \phi \right)$ to do this, it is energetically favorable for this region to be as small as possible. Consequently, this region will take the form of a string connecting the magnetic charges, with $\phi$ rapidly changing by $2\pi m$ as it crosses the string perpendicularly. The energy cost of such a configuration is proportional to the length of the string and the magnetic charges are thus confined.
In order that the integrand of the partition function for the plasma (of electric charges $\pm q$) be single-valued when there are test particles with magnetic charges $\pm m$ included, the charges must satisfy a Dirac quantization condition: $qm \in \mathbb{Z}$. (This condition is satisfied for the plasmas with magnetic charges that arise in our plasma analogy mapping for the MR state.) This makes it clear that $\phi$ can pass from one minima of $\cos \left( q \phi \right)$ to another as $\phi$ changes by $2\pi m$. If one does not satisfy the Dirac quantization condition, one must introduce a branch cut or deal with the multi-valued integrand in some other manner. It is not clear whether this can provide something physically meaningful, but, if so, the confinement argument still applies to these (fractionally quantized) magnetic charges in the plasma.

\section{Debye Screening Length of General Plasmas}
\label{sec:Debye}

The considerations of the previous Appendix tell us whether or
not a two-component plasma screens. When
a plasma does screen (either one or two component,
or even more general plasmas with multiple types of
Coulomb interactions), we would like to know what its screening length
is. Deep within the screening phase, we can use
the Debye-H\"{u}ckel theory~\cite{Debye1923}.
We review this theory here, and generalize it to plasmas with multiple types of Coulomb interactions. We consider the case where there are $m$ different types of Coulomb interactions and $S$ different particle species. The $j$th species particles have density $n_{j}\left( \overrightarrow{r} \right)$ and carry the $k$th type of Coulomb charge $q_{j}^{(k)}$. We start with the Poisson equation
\begin{equation}
\label{eq:poisson}
\nabla^2 \phi^{(k)} \left( \overrightarrow{r} \right) = - \frac{1}{\epsilon_0} \rho^{(k)} \left( \overrightarrow{r} \right)
\end{equation}
for the $k$th electric potential $\phi^{(k)} \left( \overrightarrow{r} \right)$ and charge density $\rho^{(k)} \left( \overrightarrow{r} \right)$. We take the convention in which the Coulomb energy in $D$ spatial dimensions between two point charges $q_1$ and $q_2$ at $\overrightarrow{r}_1$ and $\overrightarrow{r}_2$, respectively, takes the form
\begin{equation}
\Phi = \left\{
\begin{array}{ll}
-q_1 q_2 \log \left| \overrightarrow{r}_1 - \overrightarrow{r}_2 \right| & \qquad \text{ for } D=2 \\
&\\
\frac{q_1 q_2}{ \left| \overrightarrow{r}_1 - \overrightarrow{r}_2 \right|^{D-2}} & \qquad \text{ for } D\ge 3
\end{array}
\right.
,
\end{equation}
which corresponds to
\begin{equation}
\epsilon_0 = \left\{
\begin{array}{ll}
\frac{1}{2 \pi}  & \qquad \text{ for } D=2 \\
&\\
\frac{ \Gamma \left( \frac{D-2}{2} \right)}{ 4 \pi^{D/2}} & \qquad \text{ for } D\ge 3
\end{array}
\right.
.
\end{equation}

Next, we assume that the system is in thermal equilibrium, so the particle densities are given by a Boltzmann distribution with respect to the Coulomb energies
\begin{equation}
\label{eq:boltzmann}
n_{j}\left( \overrightarrow{r} \right) = n_{j}^{(0)} \exp\left[ - \frac{1}{T} \sum_{k=1}^{m} q_{j}^{(k)} \phi^{(k)} \left( \overrightarrow{r} \right)  \right]
,
\end{equation}
where $n_{j}^{(0)}$ is the homogeneous density of the $j$th species of particles (far away from test particles).

Charge neutrality for the $k$th type of Coulomb charge can be obtained either by balancing charge among the different species of particles to sum to zero (i.e. $\sum_{j=1}^{S} q_{j}^{(k)} n_{j}^{(0)}=0$) or through a uniform neutralizing background charge density [i.e. $\rho^{(k)}_{\text{neutralizing}} \left( \overrightarrow{r} \right) = -\sum_{j=1}^{S} q_{j}^{(k)} n_{j}^{(0)}$]. In either case, charge neutrality allows us to write
\begin{equation}
\label{eq:rho}
\rho^{(k)} \left( \overrightarrow{r} \right) = \sum_{j=1}^{S} q_{j}^{(k)} \left[ n_{j}\left( \overrightarrow{r} \right) - n_{j}^{(0)} \right]
,
\end{equation}
giving zero charge density where the potential vanishes.

Combining Eqs.~\rf{eq:poisson}, \rf{eq:boltzmann}, and \rf{eq:rho}, we obtain the generalized Poisson-Boltzmann equation
\begin{equation}
\label{eq:Dm}
\nabla^2 \phi^{(k)} \left( \overrightarrow{r} \right) = \frac{1}{\epsilon_0} \sum_{j=1}^{S} q_{j}^{(k)} n_{j}^{(0)} \left\{ 1-  \exp\left[ - \frac{1}{T} \sum_{l=1}^{m} q_{j}^{(l)} \phi^{(l)} \left( \overrightarrow{r} \right)  \right] \right\}
.
\end{equation}
This differential equation for $\phi^{(k)}$ is obviously non-linear, but is approximately linear when and where the plasma is weakly coupled (i.e. the Coulomb energies are small compared to the temperature) so that $\sum_{l=1}^{m} q_{j}^{(l)} \phi^{(l)} \left( \overrightarrow{r} \right) \ll T$. In this regime, we can expand the exponential
\begin{equation}
\exp\left[ - \frac{1}{T} \sum_{l=1}^{m} q_{j}^{(l)} \phi^{(l)} \left( \overrightarrow{r} \right)  \right] \simeq 1 - \frac{1}{T} \sum_{l=1}^{m} q_{j}^{(l)} \phi^{(l)} \left( \overrightarrow{r} \right)
\end{equation}
to obtain the linear approximation of Eq.~\rf{eq:Dm}
\begin{eqnarray}
\label{eq:DHm}
\nabla^2 \phi^{(k)} \left( \overrightarrow{r} \right) &\simeq&  \sum_{l=1}^{m} \Lambda_{kl} \phi^{(l)} \left( \overrightarrow{r} \right) \\
\Lambda_{kl} &=& \frac{1}{ \epsilon_0 T} \sum_{j=1}^{S} n_{j}^{(0)} q_j^{(k)} q_j^{(l)}
,
\end{eqnarray}
which generalizes the Debye-H\"{u}ckel equation. $\Lambda$ is a symmetric, real, positive-definite matrix, so all of its eigenvalues $\lambda_a$ (where $a=1,\ldots,m$) are positive. It is now straightforward to solve this differential equation by changing to a basis in which $\Lambda$ is diagonal, giving $m$ independent diffusion equations
\begin{equation}
\nabla^2 \tilde{\phi}^{(a)} \left( \overrightarrow{r} \right) \simeq \lambda_a  \tilde{\phi}^{(a)} \left( \overrightarrow{r} \right)
,
\end{equation}
where $\tilde{\phi}^{(a)} = \sum_{k} S_{ak} \phi^{(k)}$, for $S$ the similarity transformation that diagonalizes $\Lambda$, i.e. $S \Lambda S^{-1} = \text{diag} [\lambda_1 , \ldots , \lambda_m]$. Requiring the potentials $\phi^{(k)}$ to go to zero at infinity, we know the solutions must generally have an exponentially decaying behavior, with decay lengths $\ell_{a} = \lambda_{a}^{-1/2}$.
We define the longest decay length to be the Debye screening length of the plasma
\begin{equation}
\label{eq:screeninglength}
\ell_D \equiv \max_{a} \left\{ \lambda_a^{-1/2} \right\}
.
\end{equation}

We now consider several examples relevant to this paper:

\subsection{The One-Component Plasma}

For a one-component plasma, there is one species of particles with charge $Q$ and a neutralizing background, which gives
\begin{equation}
\Lambda = \frac{n^{(0)} Q^2 }{\epsilon_0 T}
.
\end{equation}

For the $\nu=1/M$ Laughlin states, the corresponding plasma is a $2$D one-component plasma with $T=g$ and $Q = \sqrt{2Mg}$. This gives
\begin{equation}
\Lambda = \frac{2}{ \ell_B^2}
\end{equation}
where $n^{(0)} = \nu / 2 \pi \ell_B^2$ is the electron density of the quantum Hall fluid. Thus, the Debye screening length for a Laughlin state is $\ell_1 = \ell_B /\sqrt{2}$.

\subsection{The Two-Component Plasma}

For a two-component plasma, there are two species of particles with charge $Q$ and $-Q$, respectively, which gives
\begin{equation}
\Lambda = \frac{2 n^{(0)} Q^2 }{\epsilon_0 T}
.
\end{equation}

For the $p$-wave superconductor, the corresponding plasma is a $2$D two-component plasma with $T=g$ and $Q = \sqrt{3g}$. This gives
\begin{equation}
\Lambda = 12 \pi n^{(0)}
\end{equation}
where $n^{(0)}$ is the fermion density. Thus, the Debye screening length is $\ell_2 = \left[ 12 \pi n^{(0)} \right]^{-1/2}$.

\subsection{The Moore-Read Pfaffian States' Plasma}

For the $\nu=1/M$ MR Pfaffian states, the corresponding $2$D plasma has temperature $T=g$, two types of Coulomb interactions, and two particle species: the first with charge $Q^{(1)} = \sqrt{2Mg}$ and $Q^{(2)} = \sqrt{3g}$, the second with charge $-Q^{(2)} = -\sqrt{3g}$. There is also a neutralizing background for Coulomb charge of type 1. This gives
\begin{equation}
\Lambda = \frac{2 \pi n^{(0)}}{T}  \left[
\begin{array}{cc}
\left(Q^{(1)}\right)^2 & Q^{(1)} Q^{(2)} \\
Q^{(1)} Q^{(2)} & 2 \left(Q^{(2)}\right)^2
\end{array}
\right]
= \frac{1}{ \ell_B^2}
\left[
\begin{array}{cc}
2 & \sqrt{6/M} \\
\sqrt{6/M} & 6/M
\end{array}
\right]
\end{equation}
where $n^{(0)} = \nu / 2 \pi \ell_B^2$ is the electron density of the quantum Hall fluid, which must also be the density of the screening charges. The two eigenvalues of this $\Lambda$ are $\lambda_{\pm} =\left(M+3 \pm \sqrt{M^2 +9} \right) / M \ell_B^{2}$. Thus, the Debye screening length is
\begin{equation}
\label{eq:MRplasma_screening_length}
\ell_D = \left( \frac{M}{ M+3 - \sqrt{M^2 +9} } \right)^{1/2} \ell_B
.
\end{equation}
For $M=2$ this is $\ell_D \approx 1.2 \, \ell_B $.

\section{Conformal Blocks of $n$ $\sigma$ fields and
$N$ $\psi$ fields: a basis of $n$ quasihole wavefunctions
in which braiding properties are manifest}
\label{sec:many-qh-wavefunctions}

In this section, we compute the conformal blocks which correspond
to $n$-quasihole (and $N$-electron) wavefunctions~\footnote{As we were completing this paper, we
learned that the conformal blocks and associated electron wavefunctions computed in this Appendix were obtained recently
by a slightly different method by E.~Ardonne and G.~Sierra~\cite{Ardonne10}.}. The four-quasihole case was computed in Ref.~\onlinecite{Nayak1996}; here we extend this result to arbitrary even $n$ . From the previous discussion, it is clear that we only need these conformal blocks in the
two-, four-, and six-quasihole cases. However, for completeness,
we compute them for arbitrary numbers of quasiholes. These wavefunctions
have the nice property of furnishing, through their explicit analytic continuation, representations of the $n$-quasihole braid groups.

The basic strategy is to use Refs.~\onlinecite{Fendley06,Fendley07a} to compute
\begin{equation}
\label{eqn:CB1}
\left\langle  {\sigma_1}{\sigma_2} \,{\sigma_1}{\sigma_2}\,\ldots
{\sigma_1}{\sigma_2} \, {\psi_1}{\psi_1}\ldots {\psi_1}\right\rangle
\end{equation}
Here, we have two chiral Majorana fermions $\psi_1$, $\psi_2$
with their two spin fields ${\sigma_1}$, ${\sigma_2}$. Since ${\sigma_1}$
and ${\sigma_2}$ are completely independent, this is equal to the product
of
\begin{equation}
\label{eqn:CB-factor1}
\left\langle  {\sigma_1} \,{\sigma_1}\,\ldots
{\sigma_1} \, {\psi_1}{\psi_1}\ldots {\psi_1}\right\rangle
\end{equation}
and
\begin{equation}
\label{eqn:CB-factor2}
\left\langle {\sigma_2}\ldots{\sigma_2}\right\rangle
\end{equation}
The bosonization formulas derived in Refs.~\onlinecite{Fendley06,Fendley07a}
allow us to compute
all of the conformal blocks of Eqs.~\rf{eqn:CB1} and \rf{eqn:CB-factor2},
thereby giving us the desired conformal blocks of
Eq.~\rf{eqn:CB-factor1}.
According to Refs.~\onlinecite{Fendley06,Fendley07a}, if we have two chiral Majorana fermions,
$\psi_1$, $\psi_2$, we can combine them into a single Dirac
fermion which can be bosonized:
\begin{equation}
e^{i\varphi} = {\psi_1} + i {\psi_2}
\end{equation}
so that ${\psi_1}=\cos\varphi$. Bosonizing chiral spin fields is trickier.
The individual spin fields do not have a simple expression, but the
product of two factors of ${\sigma_1} {\sigma_2}$ can be written in
the form:
\begin{equation}
\label{eqn:sigma-bos-formula}
{\sigma_1}({\eta_1})\, {\sigma_2}({\eta_1}) \,\, {\sigma_1}({\eta_2}) {\sigma_2}({\eta_2})
= e^{i\varphi({\eta_1})/2}\,e^{-i\varphi({\eta_2})/2} \,\pm \, e^{-i\varphi({\eta_1})/2}\,e^{i\varphi({\eta_2})/2}
\end{equation}
In a conformal block in which a given set of fields ${\sigma_1} {\sigma_2}\cdot
{\sigma_1} {\sigma_2}$ fuse to $1\cdot 1$ we take the $+$ sign;
if they fuse to ${\psi_1}\cdot{\psi_2}$, we take the $-$ sign.

Thus, we can compute the square of a conformal block of
$n$ $\sigma$ fields by computing:
\begin{equation}
\label{eqn:2n-sigma-CB}
\left\langle \sigma \sigma \ldots \sigma \right\rangle^2_{({p^{}_1},{p^{}_2},\ldots,{p^{}_{n/2}})} =
\sum_{{r^{}_i}=0,1} (-1)^{r\cdot p}
 \Bigl\langle
e^{i(-1)^{r^{}_1}({\varphi_1}-{\varphi_2})/2}\,e^{i(-1)^{r^{}_2}({\varphi_3}-{\varphi_4})/2}\,\ldots\,
e^{i(-1)^{r^{}_{n/2}}({\varphi_{n-1}}-{\varphi_{n}})/2}
\Bigr\rangle
\end{equation}
We have employed the shorthand ${\varphi_\mu}\equiv \varphi({\eta_\mu})$
and $(-1)^{r\cdot p}\equiv(-1)^{{\sum_j}{r^{}_j}{p^{}_j}}$.
The subscript $({p^{}_1},{p^{}_2},\ldots,{p^{}_{n/2}})$
on the left-hand-side is used to specify the conformal
block of this correlation function which we are computing: ${p^{}_i}=0,1$ denotes that
the $(2i-1)^{\rm th}$ and $2i^{\rm th}$ $\sigma$ fields fuse to $I$ or $\psi$, respectively. There is an overall parity constraint
${\sum_i}{p_i}\equiv 0 (\text{mod}\,2)$ for Eq.~\rf{eqn:2n-sigma-CB}, since there are no additional $\psi$ field insertions in this correlation function.
The two values ${r^{}_i}=0,1$ correspond to whether we have used the first or second term
on the right-hand-side of Eq.~\rf{eqn:sigma-bos-formula} in the bosonic correlation
function on the right-hand-side of Eq.~\rf{eqn:2n-sigma-CB}
(which is equivalent to the usage of ${r^{}_i}=0,1$ in
Section~\ref{sec:many-quasiparticles}).
Thus, we have
\begin{equation}
\label{eqn:2n-sigma-CB-eval}
\left\langle \sigma \sigma \ldots \sigma \right\rangle^2_{({p^{}_1},{p^{}_2},\ldots,{p^{}_{n/2}})} =
\left(\frac{\prod\limits_{i<j}^{n/2}\eta^{}_{2i-1,2j-1}\, \eta^{}_{2i,2j}}
{\prod\limits_{i,j}^{n/2}\eta^{}_{2i-1,2j}}\right)^{\!\frac{1}{4}}\,\times\,
\left\{
\sum_{{r^{}_i}=0,1} (-1)^{r\cdot p}
\prod_{k<l}^{n/2} x^{|{r^{}_k}-{r^{}_l}|/2}_{k,l}
\right\}
\end{equation}
where
\begin{equation}
x_{k,l} \equiv \frac{\eta^{}_{2k-1,2l}\, \eta^{}_{2l-1,2k}}{\eta^{}_{2k-1,2l-1}\,
\eta^{}_{2k,2l}}
\end{equation}
This generalizes the formulas for the four-$\sigma$ conformal blocks
in Ref.~\onlinecite{Belavin1984}, which has only two terms in curly brackets.

Now consider the correlation function
\begin{equation}
\left\langle  {\sigma_1}{\sigma_2} \,{\sigma_1}{\sigma_2}\,\ldots
{\sigma_1}{\sigma_2} \, {\psi_1}{\psi_1}\ldots {\psi_1}\right\rangle
\end{equation}
with $N$ Majorana fermion fields ${\psi_1}=\cos\varphi$.
We will initially consider the even $N$ electron number case
and briefly mention the odd $N$ case at the end of this section.
This can be computed in the same way as before by choosing half of
the fermions to be $e^{i\varphi}$, the other half to be $e^{-i\varphi}$,
and then summing over all permutations, which gives the Pfaffian. Hence,
we obtain:
\begin{multline}
\label{eqn:2n-sigma-psi-CB}
\left\langle {\sigma_1}{\sigma_2} \,{\sigma_1}{\sigma_2}\,\ldots
{\sigma_1}{\sigma_2} \, {\psi_1}{\psi_1}\ldots {\psi_1}
\right\rangle^{}_{({p^{}_1},{p^{}_2},\ldots,{p^{}_{n/2}})} =
\left(\frac{\prod\limits_{i<j}^{n/2} \eta^{}_{2i-1,2j-1}\, \eta^{}_{2i,2j}}
{\prod\limits_{i,j}^{n/2} \eta^{}_{2i-1,2j}}\right)^{\!\frac{1}{4}}\,\times\\
\left\{
\sum_{{r^{}_i}=0,1} (-1)^{r\cdot p}
\prod_{k<l}^{n/2} x^{|{r^{}_k}-{r^{}_l}|/2}_{k,l}\,\,
{\tilde \Psi}_{(1+{r^{}_1},3+{r^{}_2},\ldots)(2-{r^{}_1},4-{r^{}_2},\ldots)}
\right\}
\end{multline}
where
\begin{equation}
\label{eqn:tilde-Psi-def}
{\tilde \Psi}_{(1+{r^{}_1},3+{r^{}_2},\ldots)(2-{r^{}_1},4-{r^{}_2},\ldots)}
\equiv
 \text{Pf} \left\{\frac{1}{{z_i}-{z_j}}\,
\left(\frac{{\eta_1}-{z_i}}{{\eta_2}-{z_i}}\,
\frac{{\eta_2}-{z_j}}{{\eta_1}-{z_j}}\right)^{\!\frac{1}{2}-{r^{}_1}}
\left(\frac{{\eta_3}-{z_i}}{{\eta_4}-{z_i}}\,
\frac{{\eta_4}-{z_j}}{{\eta_3}-{z_j}}\right)^{\!\frac{1}{2}-{r^{}_2}} \,\ldots\,
+ \left( i \leftrightarrow j \right)
 \right\}
\end{equation}
Including the charge sector of the $\nu=1/M$ MR Pfaffian wavefunctions and dividing
by $\left\langle \sigma \sigma \ldots \sigma
\right\rangle_{({p^{}_1},{p^{}_2},\ldots,{p^{}_{n/2}})}$,
we finally obtain:
\begin{multline}
\label{eqn:2n-qp-wvfn}
\Psi^{}_{({p^{}_1},{p^{}_2},\ldots,{p^{}_{n/2}})} =
\left(\frac{\prod\limits_{i<j}^{n/2} \eta^{}_{2i-1,2j-1}\, \eta^{}_{2i,2j}}
{\prod\limits_{i,j}^{n/2} \eta^{}_{2i-1,2j}}\right)^{\!\frac{1}{8}}\, \left\{
\sum_{{r^{}_i}=0,1} (-1)^{r\cdot p}
\prod\limits_{k<l}^{n/2} x^{|{r^{}_k}-{r^{}_l}|/2}_{k,l}
\right\}^{-1/2} \\
\times \left\{
\sum_{{r^{}_i}=0,1} (-1)^{r\cdot p}
\prod_{k<l}^{n/2} x^{|{r^{}_k}-{r^{}_l}|/2}_{k,l}\,\,
{\Psi}_{(1+{r^{}_1},3+{r^{}_2},\ldots)(2-{r^{}_1},4-{r^{}_2},\ldots)}
\right\} \,\, \prod_{\mu<\nu}^{n} \eta_{\mu \nu }^{\frac{1}{4M} } \, e^{- \frac{1}{8M} \sum\limits_{\mu=1}^{n} \left| \eta_\mu \right|^2 }
\end{multline}
where
\begin{eqnarray}
\label{eq:qubit_unnorm}
&& \Psi_{(1+{r^{}_1},3+{r^{}_2},\ldots)(2-{r^{}_1},4-{r^{}_2},\ldots)} \equiv {\tilde \Psi}_{(1+{r^{}_1},3+{r^{}_2},\ldots)(2-{r^{}_1},4-{r^{}_2},\ldots)} \times \prod_{\mu = 1}^{n} \prod_{i = 1}^{N} \left(\eta_\mu - z_i \right)^{1/2} \prod_{i<j}^{N} \left(z_i -z_j \right)^M e^{- \frac{1}{4} \sum\limits_{i=1}^{N} \left| z_i \right|^2 } \notag \\
&&  = \mbox{${\rm  Pf}\!\left( \frac{(\eta^{}_{1+{r^{}_1}} - z_i ) (\eta^{}_{3+{r^{}_2}} -z_i)
\ldots ( \eta^{}_{n-1+r^{}_{\frac{n}{2}}} - z_i) \,( \eta^{}_{2-{r^{}_1}} - z_j) ( \eta^{}_{4-{r^{}_2}} -z_j )\ldots
( \eta^{}_{n-r_{\frac{n}{2}}} -z_j) + (i \leftrightarrow j )}
{z_i - z_j}\right)$}
\prod_{i<j}^{N} (z_i - z_j)^M \,
e^{ - \frac{1}{4} \sum\limits_{i=1}^{N} \left| z_i \right|^2}
\end{eqnarray}
are electron wavefunctions with normalizations that do not explicitly contain the quasiholes' braiding statistics. This equation expresses the $2^{\frac{n}{2} -1}$ basis vectors
$\Psi^{}_{({p^{}_1},{p^{}_2},\ldots,{p^{}_{n/2}})}$ with $N$ even in terms of the
$2^{\frac{n}{2}-1}$ basis vectors ${\Psi}_{(1+{r^{}_1},3+{r^{}_2},\ldots)(2-{r^{}_1},4-{r^{}_2},\ldots)}$.
The basis vectors ${\Psi}_{(1+{r^{}_1},3+{r^{}_2},\ldots)(2-{r^{}_1},4-{r^{}_2},\ldots)}$
are intuitive and easy to write down (and were, therefore, written down
in Ref.~\onlinecite{Nayak1996}).
However, they are not orthonormal and their braiding properties
are complicated. Meanwhile, according to the result shown in
this paper that Berry's matrices are trivial in the conformal block basis,
the basis vectors $\Psi^{}_{({p^{}_1},{p^{}_2},\ldots,{p^{}_{n/2}})}$
have simple braiding properties given by the branch cuts in their
definition Eq.~\rf{eqn:2n-qp-wvfn}.

In order to compute wavefunctions for an odd number $N$ of
electrons, we can compute the wavefunction for $N-1$ electrons and $n+2$ quasiholes as before, and then obtain the desired wavefunction by taking
${\eta_{n+1}}\rightarrow {\eta_{n+2}} \equiv z_{1}$, dividing by the appropriate
power of ${\eta_{n+1}}-{\eta_{n+2}}$, and correcting the charge sector terms so that $z_1$ corresponds to an electron coordinate:
\begin{multline}
\Psi^{}_{({p^{}_2},\ldots,{p^{}_{n/2}})}
({\eta_1},\ldots,{\eta_n};{z_1},\ldots,{z_{N}}) \equiv \prod_{\mu = 1}^{n} \left(\eta_\mu - z_1 \right)^{\frac{M-1}{2M}} \prod_{i= 2}^{N} \left( z_1 - z_i \right)^{M -1} e^{-\frac{M-1}{4M} |z_1|^2 } \\
\times \lim_{\epsilon\rightarrow 0}
\epsilon^{-\left(\frac{1}{4M}+\frac{3}{8}\right)} \, \Psi^{}_{({p^{}_1},\ldots,{p^{}_{n/2}}, 1)}
({\eta_1}, \ldots , \eta_n , \eta_{n+1}=z_{1}+\epsilon,{\eta_{n+2}}=z_{1};{z_2},\ldots,{z_{N}})
\end{multline}
The $p_i$s which index the odd $N$ electron
number wavefunction on the left-hand-side satisfy the parity
constraint $\sum_{i=1}^{n/2} {p_i}~\equiv~1 (\text{mod}\,2)$.

\section{Incomplete Direct Approach to Plasma Analogy for $n$ Quasiparticles}
\label{sec:direct}

In this section, we present a more direct approach to constructing the plasma analogy for $n$ quasiparticle wavefunctions. Unfortunately, the argument is incomplete, but presenting the argument serves to clarify the obstacle in proceeding in this manner, and why we needed to use the methods involving disorder operators.

We consider the conformal blocks with $n$ $\sigma$ operators
(which correspond to $n$ fundamental quasiholes). There are $2^{\frac{n}{2}-1}$
such conformal blocks, which we denote by:
\be
\label{eq:confblocks2}
\mathcal{F}_{\alpha}(\eta_{\mu}; z_i) =
\left< \sigma(\eta_1) \sigma(\eta_2) \ldots \sigma(\eta_{n-1}) \sigma(\eta_{n}) \psi(z_1) \dots \psi(z_{N})  \right>_\alpha
\ee
where $\alpha= \left( \pi_1 , \ldots , \pi_{n/2} \right)$, for $\pi_j = 0,1$ with the overall constraint $\pi_{n/2} = 0$ for $N$ even and $\pi_{n/2} = 1$ for $N$ odd. We can represent these in the following way (defining $\pi_0 = 0$):
\begin{eqnarray}
\label{eq:consfour2a}
&& \mathcal{F}_{(\pi_1,\ldots,\pi_{n/2})}(\eta_\mu; z_i)
= \VEV{ \prod_{j=1}^{n/2} V_{21}^{1-\pi_{j-1},0}(\eta_{2j-1}) V_{21}^{\pi_j ,0} (\eta_{2j}) \, V_{31}^{20} (z_1) V_{31}^{00}(z_2) \dots V_{31}^{20}(z_{N-1}) V_{31}^{00}(z_{N})} \\
&& = \prod_{j=1}^{n/2} \oint_{C_{r_j}} dw_j  \oint_{C_{z_1}} du_1 \oint_{C_{z_1}} du_2 \oint_{C_{z_3}} du_3 \oint_{C_{z_3}} du_4 \ldots \oint_{C_{z_{N-1}}} du_{N-1}  \oint_{C_{z_{N-1}}} du_{N} \notag \\
&& \qquad \left\langle \prod_{j=1}^{n/2} V_{21}^{00}(\eta_{2j-1}) e^{i \left(1-\pi_{j-1}\right) \alpha_- \phi(w_{j})}   V_{21}^{00} (\eta_{2j}) e^{i \pi_{j} \alpha_- \phi(w_{j+1})} \right. \notag \\
&& \qquad \qquad \times \left. V_{31}^{00} (z_1) e^{i \alpha_- \phi(u_1)} e^{i \alpha_- \phi(u_2)} V_{31}^{00}(z_2) \dots V_{31}^{00}(z_{N-1}) e^{i \alpha_- \phi(u_{N-1})} e^{i \alpha_- \phi(u_N)}  V_{31}^{00}(z_{N}) \right\rangle \\
&=& \prod_{j=1}^{n/2} \oint_{C_{r_j}} dw_j \oint_{C_{z_1}} du_1 \oint_{C_{z_1}} du_2 \oint_{C_{z_3}} du_3 \oint_{C_{z_3}} du_4 \ldots \oint_{C_{z_{N-1}}} du_{N-1}  \oint_{C_{z_{N-1}}} du_{N} f_{(\pi_1,\ldots,\pi_{n/2})} \left( w_a ; u_i; \eta_\mu ; z_i  \right)
\end{eqnarray}
where $C_{x}$ is the contour at radius $\left| x \right|$ centered on the origin, $r_j = \eta_{2j-1-\pi_{j-1}}$, and we have defined
\begin{eqnarray}
f_{(\pi_1,\ldots,\pi_{n/2})} &=& \left\langle \prod_{j=1}^{n/2} V_{21}^{00}(\eta_{2j-1}) e^{i \left(1-\pi_{j-1}\right) \alpha_- \phi(w_{j})}   V_{21}^{00} (\eta_{2j}) e^{i \pi_{j} \alpha_- \phi(w_{j+1})} \right. \notag \notag \\
&& \qquad \qquad \times \left. V_{31}^{00} (z_1) e^{i \alpha_- \phi(u_1)} e^{i \alpha_- \phi(u_2)} V_{31}^{00}(z_2) \dots V_{31}^{00}(z_{N-1}) e^{i \alpha_- \phi(u_{N-1})} e^{i \alpha_- \phi(u_N)}  V_{31}^{00}(z_{N}) \right\rangle
.
\end{eqnarray}
There are $N$ screening operators with coordinate $u_i$ for the $\psi$ fields and $n/2$ screening operators with coordinates $w_a$ for the $\sigma$ fields, and the conformal blocks are determined by the placement of the $w_a$ $\sigma$ screening charge contours. Specifically, $\pi_j =0,1$ indicates that the contour for $w_j$ is at radius $\left| \eta_{2j -1 - \pi_{j-1} } \right|$. Strictly speaking, $f_\alpha$ does not encode the fusion channel without the knowledge of these contours, but we will nonetheless use the subscript label to remind us of the contour placements.

We want to compute the overlap
\begin{eqnarray}
&& G_{\alpha,\beta} \left(\bar{\eta}_\mu, \eta_\mu^{\prime}\right) = \int \prod_{k=1}^{N} d^2 z_k \bar{\mathcal{F}}_{\alpha}(\bar{\eta}_\mu; \bar{z}_i) \mathcal{F}_{\beta}(\eta_\mu; z_i) \notag \\
\!\!\!\!&&= \int \prod_{k=1}^{N} d^2 z_k \prod_{j=1}^{n/2}  \oint_{C_{r_j}} d\bar{w}_j \oint_{C_{r_j^{\prime}}} dw_j^{\prime}
\oint_{C_{z_1}} du_1^{\prime} \oint_{C_{z_1}} du_2^{\prime} \oint_{C_{z_3}} du_3^{\prime} \oint_{C_{z_3}} du_4^{\prime} \ldots \oint_{C_{z_{N-1}}} du_{N-1}^{\prime}  \oint_{C_{z_{N-1}}} du_{N}^{\prime} \notag \\
\!\!\!\!&& \quad \times \oint_{C_{z_1}} d\bar{u}_1 \oint_{C_{z_1}} d\bar{u}_2 \oint_{C_{z_3}} d\bar{u}_3 \oint_{C_{z_3}} d\bar{u}_4 \ldots \oint_{C_{z_{N-1}}} d\bar{u}_{N-1}  \oint_{C_{z_{N-1}}} d\bar{u}_{N}
\bar{f}_{\alpha} \left( \bar{w}_a ; \bar{u}_i; \bar{\eta}_\mu ; \bar{z}_i  \right) f_{\beta} \left( w_a^{\prime} ; u_i^{\prime}; \eta_\mu ; z_i  \right) \\
\!\!\!\! &&= \prod_{j=1}^{n/2} \oint_{C_{r_j}} d\bar{w}_j \oint_{C_{r_j^{\prime}}} dw_j^{\prime}
\Gamma_{\alpha,\beta}\left( \bar{w}_a , w_a^{\prime} ; \bar{\eta}_\mu, \eta_\mu \right)
\label{eq:Gamma_alphabeta}
\end{eqnarray}
where $r_j^{\prime} = \eta_{2j-1-\pi_{j-1}^{\prime}}$ corresponds to $\beta = (\pi_1^{\prime},\ldots, \pi_{n/2}^{\prime}) $, and
\begin{eqnarray}
\!\!\!\! && \Gamma_{\alpha,\beta}\left( \bar{w}_a , w_a^{\prime} ; \bar{\eta}_\mu, \eta_\mu \right)
= \int \prod_{k=1}^{N} d^2 z_k \oint_{C_{z_1}} du_1^{\prime} \oint_{C_{z_1}} du_2^{\prime} \oint_{C_{z_3}} du_3^{\prime} \oint_{C_{z_3}} du_4^{\prime} \ldots \oint_{C_{z_{N-1}}} du_{N-1}^{\prime}  \oint_{C_{z_{N-1}}} du_{N}^{\prime} \notag \\
\!\!\!\! && \quad \times \oint_{C_{z_1}} d\bar{u}_1 \oint_{C_{z_1}} d\bar{u}_2 \oint_{C_{z_3}} d\bar{u}_3 \oint_{C_{z_3}} d\bar{u}_4 \ldots \oint_{C_{z_{N-1}}} d\bar{u}_{N-1}  \oint_{C_{z_{N-1}}} d\bar{u}_{N}
\bar{f}_{\alpha} \left( \bar{w}_a ; \bar{u}_i; \bar{\eta}_\mu ; \bar{z}_i  \right) f_{\beta} \left( w_a^{\prime} ; u_i^{\prime}; \eta_\mu ; z_i  \right)
.
\end{eqnarray}

Now let us see if we can convert the pairs of $u_i^{\prime}$, $\bar{u}_i$ contour integrals into $d^{2}u_i$ integrals. If we can do so, then we can
deduce the properties of $\Gamma_{\alpha,\beta}$ and $G_{\alpha,\beta}$ via plasma analogy. We follow Mathur's steps, starting from the integral
\begin{equation}
\int d^{2}u_1 \ldots \int d^{2}u_{N} \bar{f}_{\alpha} \left( \bar{w}_j ; \bar{u}_i; \bar{\eta}_\mu ; \bar{z}_i  \right) f_{\beta} \left( w_j^{\prime} ; u_i; \eta_\mu ; z_i  \right)
.
\end{equation}
In this case, we are not considering screening operators of non-Abelian fields, and thus monodromies will not take one to a different conformal block, so we do not need the bilinear form notation. Also, we are not restricting our attention to diagonal components, so the plasma potential will not be monodromy invariant (i.e. not single-valued), but this is not necessarily a problem as long as we keep track of monodromies. We will show that this expression decomposes into
\begin{eqnarray}
&& \int d^{2}u_1 \ldots \int d^{2}u_{N} \bar{f}_{\alpha} \left( \bar{w}_a ; \bar{u}_i; \bar{\eta}_\mu ; \bar{z}_i  \right) f_{\beta} \left( w_a^{\prime} ; u_i; \eta_\mu ; z_i  \right) = (\text{terms with contours all at radii} \left| z_i \right|) \notag \\
&& \qquad \qquad \qquad \qquad + (\text{terms with at least one contour at radius} \left| \eta_\mu \right|) + J\text{-terms at branch cuts}
.
\end{eqnarray}
The terms with the contours all at the radii $|z_i|$ are equal to $\Gamma_{\alpha,\beta}\left( \bar{w}_a , w_a^{\prime} ; \bar{\eta}_\mu , \eta_\mu \right)$ (up to some constant). This is because the contributions to this term from integrals with contours on the wrong $z_i$ will vanish, leaving only $\Gamma_{\alpha,\beta}$. This is known from Ref.~\onlinecite{Felder1989}, since contours in the unallowed configurations give rise to overall multiplicative terms with canceling phases. The terms with at least one contour at $\eta_\mu$ will vanish after taking the $\prod_{j=1}^{n/2} \oint_{C_{r_j}} d\bar{w}_j \oint_{C_{r_j^{\prime}}} dw_j^{\prime}$ contour integrations, because this results in unallowed configurations of screening contours (i.e. too many screening charges on the $V_{21}$ operators). Most of the $J$-terms will cancel each other, just as in Mathur's construction, except for the ones on either side of a branch cut. The branch cuts occur when $w_a \neq w_a^{\prime}$. (We note that $w_1$ and $w_1^{\prime}$ can actually be treated the same as the $u_i$ screening charges, since its contour placement is uniquely specified). We want to show that, in the end, the $J$-terms vanish or cancel, at least after performing the $d w_a^{\prime}$ and $d \bar{w}_a$ integrations.

If we could make the $J$-terms vanish, then Eq.~\rf{eq:Gamma_alphabeta} would become
\begin{equation}
G_{\alpha,\beta} \left(\bar{\eta}_\mu, \eta_\mu^{\prime}\right)  = \prod_{j=1}^{n/2} \oint_{C_{r_j}} d\bar{w}_j \oint_{C_{r_j^{\prime}}} dw_j^{\prime}  \tilde{\Gamma}_{\alpha,\beta}\left( \bar{w}_a , w_a^{\prime} ; \bar{\eta}_\mu, \eta_\mu \right)
\end{equation}
where one can now apply the plasma analogy to
\begin{equation}
\label{eq:direct_plasma}
\tilde{\Gamma}_{\alpha,\beta}\left( \bar{w}_a , w_a^{\prime} ; \bar{\eta}_\mu, \eta_\mu \right) = \int \prod_{k=1}^{N} d^2 z_k d^2 u_k \bar{f}_{\alpha} \left( \bar{w}_a ; \bar{u}_i; \bar{\eta}_\mu ; \bar{z}_i  \right) f_{\beta} \left( w_a^{\prime} ; u_i; \eta_\mu ; z_i  \right) = \int \prod_{k=1}^{N} d^2 z_k d^2 u_k e^{- \tilde{\Phi}/T} = e^{-F}
.
\end{equation}
Here, $T=g$ and $\tilde{\Phi}$ describes the $2$D Coulomb interaction between $N$ charge $Q = \sqrt{3g}$ particles at $z_i$, $N$ charge $-Q$ particles at $u_i$, $n$ charge $Q/2$ particles at $\eta_\mu$, $\frac{n}{2}$ particles with electric charge $-Q/2$ and magnetic charge $m = \sqrt{3/4g}$ at $w_a$, and $\frac{n}{2}$ particles with electric charge $-Q/2$ and magnetic charge $-m $ at $w_a^{\prime}$. Hence, $F$ is the free energy of a classical $2$D two-component plasma at temperature $T$ of $N$ charge $Q$ particles and $N$ charge $-Q$ particles, with $n$ charge $Q/2$ test particles at $\eta_\mu$, $\frac{n}{2}$ test particles with electric charge $-Q/2$ and magnetic charge $m$ at $w_a$, and $\frac{n}{2}$ test particles with electric charge $-Q/2$ and magnetic charge $-m$ at $w_a^{\prime}$. By confinement of magnetic charge in a screening plasma, we know that Eq.~\rf{eq:direct_plasma} will vanish unless $w_a=w_a^{\prime}$, which shows that the result is proportional to $\delta_{\alpha \beta}$, as desired, since $\alpha \neq \beta$ requires $w_a \neq w_a^{\prime}$ for at least one $a$.

We now proceed by seeing what happens for a single $u$ screening charge. We partition the plane into a number of annular regions $D_l$ such that the positions of the other coordinates where there are potentially singularities or branch cuts are left outside of $D_l$. Leaving the other coordinates implicit, we have for a region $D$
\begin{equation}
\int_{D} d^{2}u \bar{f}_{\alpha} \left( \bar{u}  \right) f_{\beta} \left( u \right) = \frac{i}{2} \int_{\partial D} d\bar{u} \bar{f}_{\alpha} \left( \bar{u}  \right) \hat{f}_{\beta} \left( u \right)
\end{equation}
where
\begin{equation}
\hat{f}_{\beta} \left( u \right) = \int_{P}^{u} du^{\prime} f_{\beta} \left( u^{\prime} \right)
.
\end{equation}
We define
\begin{eqnarray}
J_{\alpha \beta}^{C} &=& \int_{P}^{P^{\prime}} d\bar{u} \bar{f}_{\alpha} \left( \bar{u}  \right) \hat{f}_{\beta} \left( u \right) \\
I_{\beta}^{C} &=& \int_{P}^{P^{\prime}} du f_{\beta} \left( u \right)
\end{eqnarray}
for $C$ a contour running from $P$ to $P^{\prime}$. Now taking the same steps as Mathur, we get
\begin{eqnarray}
&& \int_{D} d^{2}u \bar{f}_{\alpha} \left( \bar{u}  \right) f_{\beta} \left( u \right) = \frac{i}{2} \left[ J_{\alpha \beta}^{C_1}
+ \bar{I}_{\alpha}^{S_1} I_{\beta}^{C_1} + J_{\alpha \beta}^{S_1} - \bar{I}_{\alpha}^{C_2} I_{\beta}^{S_2} - J_{\alpha \beta}^{C_2} - J_{\alpha \beta}^{S_2} \right] \\
&&= \frac{i}{2} \left[ J_{\alpha \beta}^{C_1} - J_{\alpha \beta}^{C_2} + J_{\alpha \beta}^{S_1} - J_{\alpha \beta}^{S_2}
- \left( 1-M_{\alpha} \right)^{-1} \bar{I}_{\alpha}^{C_1} I_{\beta}^{C_1}  - \left( 1-M_{\beta} \right)^{-1} \bar{I}_{\alpha}^{C_2} I_{\beta}^{C_2} +\frac{M_{\alpha}-M_{\beta}}{\left( 1-M_{\alpha} \right) \left( 1-M_{\beta} \right) } \bar{I}_{\alpha}^{C_2} I_{\beta}^{C_1}  \right]
\end{eqnarray}
where we used $I_{\beta}^{C_1} + I_{\beta}^{S_1} -I_{\beta}^{C_2}-I_{\beta}^{S_2} =0$ and $I_{\beta}^{S_1} = M_{\beta} I_{\beta}^{S_2}$, and now $M_\alpha $ and $M_\beta$ are not equal for the regions $D_{l}$ between $w_{j+1}$ to $w_{j+1}^{\prime}$ (i.e. from radius $|\eta_{2j}|-\epsilon$ to $|\eta_{2j+1}|-\epsilon$) when $\pi_{j} \neq \pi_{j}^{\prime}$, but are otherwise equal. This is not a problem, because (as previously mentioned) this only gives us an extra term with a contour at $\eta_{2j}$ and one at $\eta_{2j+1}$, but these give a vanishing result when one evaluates the $d\bar{w}_{j+1}$ and $dw_{j+1}^{\prime}$ contour integrals.

The $J$-terms are however a more difficult problem. We know that these will cancel as long as $w_{j}=w_{j}^{\prime}$, since then $J_{\alpha \beta}^{S_1} = J_{\alpha \beta}^{S_2}$ and $J_{\alpha \beta}^{C_1}$ from region $D_{l}$ is equal to $J_{\alpha \beta}^{C_2}$ from region $D_{l+1}$. However, when $w_{j}\neq w_{j}^{\prime}$, there is a branch cut running between $w_{j+1}$ and $w_{j+1}^{\prime}$ and the $J$-term on the two sides of the cut do not cancel each other. Thus, we are stuck with a $J$-term integrated around these branch cuts, and no obvious way to cancel them out.

\section{Explicit Examples of Orthogonality for Unmatched Quasiparticles}
\label{sec:orthog_examples}

In this appendix, we provide the derivation of the overlap Eq.~\rf{eq:LLdelta} for wavefunctions describing the $\nu=1$ filled Landau level with $n$ holes and of the overlap Eq.~\rf{eq:Laughlin_delta} for wavefunctions describing an arbitrary quantum Hall state (that has a plasma analogy) with one Laughlin-type quasihole.

\subsection{$\nu=1$ Integer Quantum Hall State}

We first consider the $\nu=1$ filled Landau level state, which can be solved exactly. The wavefunction for one filled Landau level of $N$ electrons with $n$ holes is
\begin{equation}
\Psi_1 \left( \eta_{\mu} ; z_i  \right) =
\prod_{\mu<\nu}^{n} \left({\eta_\mu} - {\eta_\nu}\right)
\prod_{\mu = 1}^{n} \prod_{i = 1}^{N} \left({\eta_\mu} -{z_i} \right)\,
\prod_{i<j}^{N} \left({z_i} - {z_j}\right)\,
 e^{- \frac{1}{4} \sum\limits_{\mu=1}^{n} \left| \eta_\mu \right|^2  - \frac{1}{4} \sum\limits_{i=1}^{N} {\left|{z_i}\right|^2}}
.
\end{equation}
Taking the inner product of two such wavefunctions with holes not necessarily at the same positions, one finds
\begin{eqnarray}
\label{eq:LL_orthog}
&& G \left( \bar{\eta}_{\mu} , \eta_{\mu}^{\prime}  \right) =
\int {\prod_{k=1}^{N}}{d^2}{z_k} \bar{\Psi}_1 \left( \bar{\eta}_\mu ; \bar{z}_i  \right) \Psi_1 \left( \eta_{\mu}^{\prime} ; z_i  \right) \notag \\
&& \qquad = \int {\prod_{k=1}^{N}}{d^2}{z_k} \prod_{\mu<\nu}^{n} \left[ \left(\bar{\eta}_\mu - \bar{\eta}_\nu\right) \left(\eta_\mu^{\prime} - \eta_\nu^{\prime} \right) \right]
\prod_{\mu = 1}^{n} \prod_{i = 1}^{N} \left[ \left({\bar \eta_\mu} -{\bar{z}_i} \right) \left({\eta_\mu^{\prime}} -{z_i} \right) \right]
\prod_{i<j}^{N} \left|{z_i} - {z_j}\right|^2\,
e^{- \frac{1}{4} \sum\limits_{\mu=1}^{n} \left( \left| \eta_\mu \right|^2 + \left| \eta_\mu^{\prime} \right|^2 \right) - \frac{1}{2} \sum\limits_{i=1}^{N} {\left|{z_i}\right|^2}} \notag \\
&& \qquad = C_1 \sum_{\pi\in S_{n}} {(-1)^\pi} {\prod_{\mu=1}^{n}} \,
e^{-\frac{1}{4} \left( {|{\eta_{\pi(\mu)}}|^2 + |{\eta'_{\mu}}|^2 - 2 {\bar{\eta}_{\pi(\mu)}}{\eta^{\prime}_{\mu}} } \right) }
\notag \\
&& \qquad = C_1 (2\pi)^n \sum_{\pi\in S_{n}} {(-1)^\pi} {\prod_{\mu=1}^{n}} \, \delta_{\text{LLL}}^{2} \left( \eta_{\pi(\mu)} - \eta'_{\mu} \right)
,
\end{eqnarray}
where $C_1$ is the (unspecified) normalization constant. For this, we note that one can think of wavefunction with $N$ electrons and $n$ holes as one filled Landau level of $N+n$ particles, which is a
Slater determinant state. Thus, the integral
over the $N$ electron coordinates $z_i$ gives the $2n$-point
particle correlation function, which factorizes into a
product of $2$-point functions summed over permutations,
as per Wick's theorem. Finally, we used the fact that the (normalized) two-point function~\cite{Girvin84b}
\begin{equation}
\frac{1}{2 \pi} e^{ -\frac{1}{4} \left( |z|^2 + |z'|^2 - 2 \bar{z} z^{\prime} \right) } = \delta_{\text{LLL}}^{2} \left( z - z' \right)
\end{equation}
is the lowest Landau level projection of the delta-function, in the sense that
\begin{equation}
\int d^2 z \, \, \delta_{\text{LLL}}^{2} \left( z - z' \right) f \left( z \right) e^{- \frac{1}{4} |z|^2} = f \left( z' \right) e^{- \frac{1}{4} |z'|^2}
\end{equation}
for any holomorphic function $f(z)$. We also note that
\begin{equation}
e^{ -\frac{1}{4} \left( |z|^2 + |z'|^2 - 2 \bar{z} z^{\prime} \right) } = e^{ -\frac{1}{4} |z-z'|^2} e^{\frac{1}{4} \left( \bar{z} z^{\prime} - z \bar{z}^{\prime} \right) } = e^{ -\frac{1}{4} |z-z'|^2} e^{i \frac{1}{2} \text{\bf Im} \left[ \bar{z} z^{\prime} \right] }
,
\end{equation}
so the two-point function has Gaussian decay with the distance between $z$ and $z'$. Hence, we find that
\begin{equation}
\left\| \Psi_{1} \left( \eta_\mu ; z_i  \right) \right\|^2 = C_1 + O\left(e^{- |\eta_\mu - \eta_\nu|^2 / 4 \ell_B^2}\right)
.
\end{equation}
This is a somewhat stronger result than given by the plasma analogy, which nominally involves $O(e^{- |\eta_\mu - \eta_\nu|/ \ell_1})$ corrections.

One could also arrive at the result of Eq.~\rf{eq:LL_orthog} by noting that
\begin{equation}
\Gamma \left( \bar{\eta}_\mu , \eta_\mu^{\prime} \right) =
\int {\prod_{k=1}^{N}}{d^2}{z_k} \prod_{\mu = 1}^{n} \prod_{i = 1}^{N} \left[ \left({\bar \eta_\mu} -{\bar{z}_i} \right) \left({\eta_\mu^{\prime}} -{z_i} \right) \right]
\prod_{i<j}^{N} \left|{z_i} - {z_j}\right|^{2} \,
e^{- \frac{1}{2} \sum\limits_{i=1}^{N} {\left|{z_i}\right|^2}}
\end{equation}
is holomorphic in $\eta_\mu^{\prime}$, antiholomorphic in $\eta_\mu$, and
\begin{equation}
\label{eq:propGamma}
\Gamma\left( \bar{\eta}_\mu , \eta_\mu^{\prime} \right) = \Gamma\left( \bar{\eta}_\mu , \eta_{\pi\left(\mu\right)}^{\prime} \right) = \Gamma\left( \bar{\eta}_{\pi\left(\mu\right)} , \eta_\mu^{\prime} \right) = \overline{\Gamma\left( \bar{\eta}^{\prime}_\mu , \eta_\mu \right)}
\end{equation}
for any $\pi \in S_n$. By the plasma analogy, we know that $G \left( \bar{\eta}_{\mu} , \eta_{\mu}  \right) = C_1 + O\left( e^{- \left| \eta_\mu - \eta_\nu \right| / \ell_1 } \right)$ and thus can uniquely obtain the result of Eq.~\rf{eq:LL_orthog}, i.e. that
\begin{eqnarray}
G \left( \bar{\eta}_{\mu} , \eta_{\mu}^{\prime}  \right) &=& \prod_{\mu<\nu}^{n} \left[ \left(\bar{\eta}_\mu - \bar{\eta}_\nu\right) \left(\eta_\mu^{\prime} - \eta_\nu^{\prime} \right) \right] e^{- \frac{1}{4} \sum\limits_{\mu=1}^{n} \left( \left| \eta_\mu \right|^2 + \left| \eta_\mu^{\prime} \right|^2 \right) } \Gamma \left( \bar{\eta}_\mu , \eta_\mu^{\prime} \right) \notag \\
\Gamma \left( \bar{\eta}_\mu , \eta_\mu^{\prime} \right) &=& C_1 \prod_{\mu<\nu}^{n} \left[ \left(\bar{\eta}_\mu - \bar{\eta}_\nu\right) \left(\eta_\mu^{\prime} - \eta_\nu^{\prime} \right)  \right]^{-1} \sum_{\pi \in S_n} (-1)^\pi \prod_{\mu=1}^{n} e^{\frac{1}{2} \bar{\eta}_{\pi(\mu)} \eta_{\mu}^{\prime} }
,
\end{eqnarray}
where $C_1$ can now be identified as the undetermined constant from the plasma analogy, by using only the plasma analogy and the analytic properties of $\Gamma$.

\subsection{Laughlin-type Quasihole}

For the $\nu=1/M$ Laughlin states, Laughlin demonstrated such an orthogonality for the single quasihole wavefunction using the plasma analogy and analyticity~\cite{Laughlin-Lecture87}. Specifically, he showed that
\begin{eqnarray}
\label{eq:Laughlinorthog}
&& G \left( \bar{\eta} , \eta^{\prime}  \right) =
\int {\prod_{k=1}^{N}}{d^2}{z_k} \bar{\Psi}_{\frac{1}{M}} \left( \bar{\eta} ; \bar{z}_i  \right) \Psi_{\frac{1}{M}} \left( \eta^{\prime} ; z_i  \right) \notag \\
&& \qquad = \int {\prod_{k=1}^{N}}{d^2}{z_k} \prod_{i=1}^{N} \left[ \left({\bar \eta} -{\bar{z}_i} \right) \left({\eta^{\prime}} -{z_i} \right) \right]
\prod_{i<j}^{N} \left|{z_i} - {z_j}\right|^{2M}\,
e^{- \frac{1}{4M} \left( \left| \eta \right|^2 + \left| \eta^{\prime} \right|^2 \right) - \frac{1}{4} \sum\limits_{i=1}^{N} {\left|{z_i}\right|^2}} \notag \\
&& \qquad = C_1 e^{-\frac{1}{4M} \left( {|{\eta}|^2 + |{\eta'}|^2 - 2 {\bar{\eta}}{\eta^{\prime}} } \right) }
,
\end{eqnarray}
where $C_1 = \left\| \Psi_{\frac{1}{M}} \left( \eta ; z_i  \right) \right\|^2 $. For this, he noted that, except for the Gaussian factors $\exp \left( - \frac{1}{4M} \left| \eta \right|^2 - \frac{1}{4M} \left| \eta^{\prime} \right|^2 \right)$, the inner product is holomorphic in $\eta^{\prime}$ and antiholomorphic in $\eta$, and the plasma analogy indicates that $G(\bar{\eta},\eta) = C_1$. These properties uniquely determine the result of the inner product. One similarly has that
\begin{equation}
\frac{1}{2 \pi M} e^{ -\frac{1}{4M} \left( |\eta|^2 + |\eta'|^2 - 2 \bar{\eta} \eta^{\prime} \right) } = \delta_{\text{LLL}_M}^{2} \left( \eta - \eta' \right)
\end{equation}
is a projection of the delta-function into a lowest Landau level with a re-scaled magnetic length of $\ell_B^{(M)} = \sqrt{M} \ell_B$, i.e.
\begin{equation}
\int d^2 \eta \, \, \delta_{\text{LLL}_M}^{2} \left( \eta - \eta' \right) f \left( \eta \right) e^{- \frac{1}{4M} |\eta|^2} = f \left( \eta' \right) e^{- \frac{1}{4M} |\eta'|^2}
\end{equation}
for any holomorphic function $f(\eta)$.

The same argument applies to wavefunctions with one Laughlin-type quasihole for any state with a plasma analogy (e.g. the $I_2$ excitation in the MR state). Specifically, one has
\begin{eqnarray}
\label{eq:Laughlintypeorthog}
&& G \left( \bar{\eta} , \eta^{\prime}  \right) =
\int {\prod_{k=1}^{N}}{d^2}{z_k} \bar{\Psi}_{\frac{1}{M}} \left( \bar{\eta} ; \bar{z}_i  \right) \Psi_{\frac{1}{M}} \left( \eta^{\prime} ; z_i  \right) \notag \\
&& \qquad = \int {\prod_{k=1}^{N}}{d^2}{z_k} \prod_{i=1}^{N} \left[ \left({\bar \eta} -{\bar{z}_i} \right) \left({\eta^{\prime}} -{z_i} \right) \right]
e^{- \frac{1}{4M} \left( \left| \eta \right|^2 + \left| \eta^{\prime} \right|^2 \right) }\left| \Psi_{\frac{1}{M}}\left( z_i\right) \right|^{2}\, \notag \\
&& \qquad = C_1 e^{-\frac{1}{4M} \left( {|{\eta}|^2 + |{\eta'}|^2 - 2 {\bar{\eta}}{\eta^{\prime}} } \right) }
,
\end{eqnarray}
where $\Psi_{\frac{1}{M}}\left( z_i\right)$ is the ground-state wavefunction and $\Psi_{\frac{1}{M}} \left( \eta ; z_i  \right)$ the wavefunction with one Laughlin-type quasihole at $\eta$.

It is difficult to generalize these methods of obtaining explicit overlap results that go beyond the qualitative behavior obtained in Section~\ref{sec:qporthog} for cases that involve multiple Laughlin-type quasiparticles or different types of quasiparticles.


\end{document}